\newif\ifdraft
\draftfalse

\ifdefined\DRAFT\drafttrue\fi

\ifdraft
   \documentclass[twocolumn,tighten,times,linenumbers,trackchanges,twocolappendix]{aastex631}
\else
   \documentclass[twocolumn,tighten,times,twocolappendix]{aastex631}    
\fi   
   
\usepackage{amsmath}
\let\tablenum\relax

\usepackage{afterpage}

\usepackage{siunitx}
\DeclareSIUnit{\parsec}{pc}
\DeclareSIUnit{\Mpc}{\mega\parsec}
\DeclareSIUnit{\year}{yr}

\usepackage{textgreek}

\usepackage[utf8]{inputenc}
\usepackage{xspace}
\usepackage{acronym}
\usepackage{booktabs}

\usepackage{color, colortbl}

\usepackage{graphicx}
\usepackage[caption=false]{subfig}

\usepackage{tabularx}
\usepackage{multirow}

\usepackage{makecell}

\newcolumntype{Y}{>{\centering\arraybackslash}X}
\newcolumntype{Z}{>{\raggedright\arraybackslash}X}
\newcolumntype{K}{>{\raggedleft\arraybackslash}X}
\newcolumntype{U}{>{\hsize=1.01\hsize}Y}
\newcolumntype{V}{>{\hsize=1.2\hsize}Y}
\newcolumntype{W}{>{\hsize=0.71\hsize}Y}

\usepackage{scrextend}
\usepackage{textgreek}

\usepackage{microtype}

\AtBeginDocument{}

\newcommand{\checkbar}{\checkmark\kern-1.1ex\raisebox{.7ex}{\rotatebox[origin=c]{135}{--}}}
\newcommand{\checkminus}{\sout{\checkmark}}
\newcommand{\checkprevext}{+}
\newcommand{\checkprevlvk}{++}
\newcommand{\checknodata}{\!\!\nodata\!\!}

\newcommand{\tablefootnote}[1]{}

\usepackage{makerobust}

\renewcommand{\today}{\number\day\space\ifcase\month\or
  January\or February\or March\or April\or May\or June\or
  July\or August\or September\or October\or November\or December\fi
  \space\number\year}

\usepackage{float}

\definecolor{NOTECOLOR}{rgb}{0.4, 0.2, 0.1}
\definecolor{notecolor}{rgb}{0.4, 0.2, 0.1}
\definecolor{danger-red}{rgb}{0.8, 0.4, 0.0}
\definecolor{ok-green}{rgb}{0.0, 0.6, 0.5}

\newcommand{\reviewed}[1]{{#1}}

\newcommand{\macro}[1]{#1}

\definecolor{mgreen}{rgb}{0.1,0.7,0.1}

\newcommand{\subpapersection}[1]{\section{#1}}
\newcommand{\subpapersubsection}[1]{\subsection{#1}}

\newcommand{\paperscommonname}{GWTC-4.0: Tests of General Relativity. }

\newtoggle{fullauthorlist}
\toggletrue{fullauthorlist}

\newtoggle{endauthorlist}
\toggletrue{endauthorlist}

\newcommand{\LVKCollabAuthors}{The LIGO Scientific Collaboration, the Virgo Collaboration, and the KAGRA Collaboration}

\newcommand{\LVKCorrespondence}{The full author list is given at the end.\\
 For correspondence: LSC P\&P Committee, for LVK Publications, via}

\newcommand\gwtc[1][?]{\mbox{GWTC\if#1?\else-#1\fi}}
\newcommand\thisgwtcversionmajor{4}
\newcommand\thisgwtcversionminor{0}
\newcommand\thisgwtcversionfull{\thisgwtcversionmajor.\thisgwtcversionminor}
\newcommand\thisgwtcversion\thisgwtcversionfull

\newcommand{\OoneStartDate}{{{2015~September~12}}}

\newcommand{\OoneEndDate}{{{2016~January~19}}}

\newcommand{\OtwoStartDate}{{{2016~November~30}}}

\newcommand{\OtwoEndDate}{{{2017~August~25}}}

\newcommand{\OthreeAStartDate}{{{2019~April~1}}}

\newcommand{\OthreeAEndDate}{{{2019~October~1}}}

\newcommand{\OthreeBStartDate}{{{2019~November~1}}}

\newcommand{\OthreeBEndDate}{{{2020~March~27}}}

\newcommand{\OfourAStartDate}{{{2023~May~24}}}

\newcommand{\OfourAEndDate}{{{2024~January~16}}}

\newcommand{\GWTCfourENDDate}{{{2024~January~31}}}

\newcommand{\soft}[1]{\textsc{#1}}

\newcommand{\GSTLAL}{\soft{GstLAL}\xspace}

\newcommand{\CWB}{\soft{cWB}\xspace}

\newcommand{\PYCBC}{\soft{PyCBC}\xspace}
\newcommand{\MBTA}{\soft{MBTA}\xspace}

\newcommand{\BAYESWAVE}{\soft{BayesWave}\xspace}
\newcommand{\BILBY}{\soft{Bilby}\xspace}

\newcommand{\LALSUITE}{\soft{LALSuite}\xspace}

\newcommand{\PBILBY}{\soft{ParallelBilby}\xspace}
\newcommand{\ASIMOV}{\soft{Asimov}\xspace}
\newcommand{\PESUMMARY}{\soft{PESummary}\xspace}

\newcommand{\NUMPY}{\soft{NumPy}\xspace}
\newcommand{\SCIPY}{\soft{SciPy}\xspace}

\newcommand{\SEABORN}{\soft{seaborn}\xspace}
\newcommand{\GWPY}{\soft{GWpy}\xspace}
\newcommand{\DYNESTY}{\soft{Dynesty}\xspace}

\newcommand{\IMRPhenomXP}{\soft{IMRPhenomXP}\xspace}
\newcommand{\IMRPhenomXPHM}{\soft{IMRPhenomXPHM}\xspace}
\newcommand{\IMRPhenomXPHMST}{\soft{IMRPhenomXPHM\_SpinTaylor}\xspace}

\newcommand{\SEOBNRFIVEPHM}{\soft{SEOBNRv5PHM}\xspace}
\newcommand{\SEOBNRFIVEHMROM}{\soft{SEOBNRv5HM\_ROM}\xspace}

\newcommand{\IMRPhenomXPNRTidalTWO}{\soft{IMRPhenomXP\_NRTidalv2}\xspace}

\newcommand{\IMRPhenomNSBH}{\soft{IMRPhenomNSBH}\xspace}

\newcommand{\SEOBNRFOURNRtidalTWONSBH}{\soft{SEOBNRv4\_ROM\_NRTidalv2\_NSBH}\xspace}

\newcommand{\SURSEVENDQFOUR}{\soft{NRSur7dq4}\xspace}

\DeclareSIUnit\parsec{pc}
\DeclareSIUnit\Mpc{\mega\parsec}
\DeclareSIUnit\yr{yr}
\DeclareSIUnit\GpcCubedYear{\giga\parsec\cubed\yr}

\newcommand{\OfouraDuration}{{{\qty{237.04166666666666}{\day}}}}

\newcommand{\OfouraDurationHL}{{{\qty{126.47008101851851}{\day}}}}

\newcommand{\Msun}{\ensuremath{\mathit{M_\odot}}}

\newcommand\PEpdfp{\ensuremath{p}}

\newcommand{\PEparameter}{\ensuremath{\boldsymbol{\theta}}}%

\newcommand\PEpdf[2][?]{\ensuremath{\PEpdfp({#2}\ifx#1?\else | {#1}\fi)}}

\newcommand\PEpriorpdfpi{\ensuremath{\pi}}
\newcommand\PEpdfprior[1]{\ensuremath{\PEpriorpdfpi({#1})}}
\newcommand\PEprior[1][\PEparameter]{\PEpdfprior{#1}}
\newcommand\PEpriorpe[1][\PEparameter]{{\let\keepPEpriorpdfpi\PEpriorpdfpi\def\PEpriorpdfpi{\keepPEpriorpdfpi_{\text{PE}}}\PEprior[#1]\let\PEpriorpdfpi\keepPEpriorpdfpi}}

\newcommand{\PN}[0]{\ac{PN}\xspace}
\newcommand{\BBH}[0]{\ac{BBH}\xspace}

\newcommand{\GR}[0]{\ac{GR}\xspace}

\usepackage{amstext}

\makeatletter
\@ifpackageloaded{cleveref}{
  \AtBeginDocument{
    \def\ltx@label#1{\cref@label{#1}}%
    \def\label@in@display@noarg#1{\cref@old@label@in@display{#1}}%
    \def\label@in@mmeasure@noarg#1{%
      \begingroup%
        \measuring@false%
        \cref@old@label@in@display{#1}%
      \endgroup}%
  }
}{}
\makeatother

\protected\def\protectedacused{\acused}

\acrodef{LIGO}[LIGO]{Laser Interferometer Gravitational-Wave Observatory}
\acrodef{LHO}[LHO]{\ac{LIGO} Hanford Observatory}
\acrodef{LLO}[LLO]{\ac{LIGO} Livingston Observatory}
\acrodef{KAGRA}[KAGRA]{KAGRA}\acused{KAGRA}
\acrodef{iKAGRA}[iKAGRA]{initial-phase \ac{KAGRA}}
\acrodef{bKAGRA}[bKAGRA]{baseline-design \ac{KAGRA}}
\acrodef{GEO}[GEO]{GEO\,600 \ac{GW} detector}
\acrodef{aLIGO}{Advanced \ac{LIGO}}
\acrodef{A+}{Advanced+ \ac{LIGO}}
\acrodef{Asharp}[\ensuremath{\text{A}^\sharp}]{\ac{LIGO} \acs{Asharp}}
\acrodef{AdV}{Advanced \acl{Virgo}}
\acrodef{AdV+}{Advanced \acl{Virgo}+}
\acrodef{Virgo}{Virgo}\acused{Virgo}
\acrodef{VirgoNEXT}[Virgo\_nEXT]{Virgo\_nEXT}\acused{VirgoNEXT}

\acrodef{LSC}[LSC]{\acs{LIGO} Scientific Collaboration}
\acrodef{LV}[LV]{\acs{LIGO}--\acs{Virgo} Collaboration\protect\protectedacused{LVC}}
\acrodef{LVC}[LV]{\acs{LIGO}--\acs{Virgo} Collaboration\protect\protectedacused{LV}}
\acrodef{LVK}[LVK]{\acs{LIGO}--\ac{Virgo}--\ac{KAGRA} Collaboration}
\acrodef{IGWN}[IGWN]{International \ac{GWH} Observatory Network}

\acrodef{O1}[O1]{first observing run}
\acrodef{O2}[O2]{second observing run}
\acrodef{O3}[O3]{third observing run}
\acrodef{O3a}[O3a]{first half of the third observing run}
\acrodef{O3b}[O3b]{second half of the third observing run}
\acrodef{O3GK}[O3GK]{observing run}
\acrodef{O4}[O4]{fourth observing run}
\acrodef{O4a}[O4a]{first part of the fourth observing run}
\acrodef{O4b}[O4b]{second part of the fourth observing run}
\acrodef{O4c}[O4c]{third part of the fourth observing run}
\acrodef{O5}[O5]{fifth observing run}

\acrodef{BH}[BH]{black hole}
\acrodef{BBH}[BBH]{binary \ac{BH}}
\acrodef{BNS}[BNS]{binary \ac{NS}}
\acrodef{IMBH}[IMBH]{intermediate-mass \ac{BH}}
\acrodef{NS}[NS]{neutron star}
\acrodef{BHNS}[BHNS]{\ac{BH}--\ac{NS} binary}
\acrodefplural{BHNS}[BHNSs]{\ac{BH}--\ac{NS} binaries}
\acrodef{NSBH}[NSBH]{\ac{NS}--\ac{BH} binary}
\acrodefplural{NSBH}[NSBHs]{\ac{NS}--\ac{BH} binaries}
\acrodef{PBH}[PBH]{primordial \ac{BH}}
\acrodef{CBC}[CBC]{compact binary coalescence}

\acrodef{GW}[GW]{gravitational wave\protect\protectedacused{GWH}}
\acrodef{GWH}[GW]{gravitational-wave\protect\protectedacused{GW}}
\acrodef{IFO}[IFO]{interferometer}
\acrodef{SNR}[SNR]{signal-to-noise ratio}
\acrodef{FAR}[FAR]{false-alarm rate}
\acrodef{IFAR}[IFAR]{inverse false-alarm rate}
\acrodef{FAP}[FAP]{false alarm probability}
\acrodef{PSD}[PSD]{power spectral density}
\acrodef{GR}[GR]{general relativity}
\acrodef{NR}[NR]{numerical relativity}
\acrodef{PN}[PN]{post-Newtonian}
\acrodef{EOB}[EOB]{effective-one-body}
\acrodef{ROM}[ROM]{reduced-order model}
\acrodef{IMR}[IMR]{inspiral--merger--ringdown}
\acrodef{PDF}[pdf]{probability density function}
\acrodef{PE}[PE]{parameter estimation}
\acrodef{CI}[CI]{credible interval}
\acrodef{CL}[CL]{credible level}
\acrodef{EOS}[EoS]{equation of state}
\acrodef{KLD}[KLD]{Kullback--Leibler divergence}
\acrodef{JSD}[JSD]{Jensen--Shannon divergence}
\acrodef{GCN}[GCN]{General Coordinates Network}
\acrodef{GWTC}[GWTC]{Gravitational-Wave Transient Catalog}
\acrodef{GWOSC}[GWOSC]{Gravitational Wave Open Science Center}
\acrodef{WDM}[WDM]{Wilson--Debauchies--Meyer}

\acrodef{CWB}[cWB]{coherent WaveBurst}
\acrodef{LAL}[LAL]{\ac{LIGO} algorithm library}

\acrodef{CHRoCC}{central heating radius of curvature correction}
\acrodef{NonSENS}{non-stationary estimation and noise subtraction}

\acrodef{PTA}{Pulsar Timing Array}

\input{common__event_macros}

\newcommand{\BILBYTGR}{\soft{BilbyTGR}\xspace}

\newcommand{\IMRPhenomXPHMMSA}{\soft{IMRPhenomXPHM\_MSA}\xspace}
\newcommand{\IMRPhenomXPMSA}{\soft{IMRPhenomXP\_MSA}\xspace}

\newcommand{\TGRFARTHRESH}{\reviewed{$\le \qty{e-3}{\yr^{-1}}$}\xspace}
\newcommand{\TGRNUMEVENTS}{\reviewed{42}\xspace}
\newcommand{\TGRNUMTESTS}{\reviewed{19}\xspace}

\newcommand{\TGRNUMEVENTSPREVPLUSOFOURA}{\reviewed{91}\xspace}  

\newcommand{\TGRFigureWidth}{3.375in}

\DeclareRobustCommand{\COMMONNAME}[1]{\IfEqCase{#1}{{GW230529}{\MINIMALNAME{GW230529_181500}{}}{GW230814other}{\MINIMALNAME{GW230814_061920}{}}{GW230814single}{\MINIMALNAME{GW230814_230901}{}}{GW231123}{\MINIMALNAME{GW231123_135430}{}}}[\textcolor{red}{???}]}

\newcommand{\ImrMfHierImprovement}{\reviewed{\ensuremath{2.0}}}
\newcommand{\ImrChifHierImprovement}{\reviewed{\ensuremath{2.5}}}

\DeclareRobustCommand{\TGRImrctGWTCFOURResults}[1]{\IfEqCase{#1}{{DMFGWTC4PHENOM}{\reviewed{\ensuremath{-0.00^{+0.05}_{-0.05}}}}{DCHIFGWTC4PHENOM}{\reviewed{\ensuremath{-0.04^{+0.07}_{-0.06}}}}{GRQUANTGWTC4}{\reviewed{\ensuremath{89.1}}}{GRQUANTGWTC3}{\ensuremath{79.6}}}}

\DeclareRobustCommand{\TGRImrctGWTCFOURResultsWOGWOneNineZeroEightOneFour}[1]{\IfEqCase{#1}{{DMFGWTC4PHENOM}{\reviewed{\ensuremath{0.01^{+0.05}_{-0.05}}}}{DCHIFGWTC4PHENOM}{\reviewed{\ensuremath{-0.03^{+0.06}_{-0.06}}}}{GRQUANTGWTC4}{\reviewed{\ensuremath{86.0}}}}}

\DeclareRobustCommand{\TGRImrctHierPopGWTCFOURResults}[1]{\IfEqCase{#1}{{DMFGWTC4HIERPOP}{\reviewed{\ensuremath{0.00^{+0.07}_{-0.06}}}}{DCHIFGWTC4HIERPOP}{\reviewed{\ensuremath{-0.05^{+0.11}_{-0.11}}}}{GRQUANTDevParamsTwoDGWTC4HIERPOP}{\reviewed{\ensuremath{51.7}}}{GRQUANTMeansTwoDGWTC4HIERPOP}{\reviewed{\ensuremath{94.2}}}{GRQUANTHyperParamsFourDGWTC4HIERPOP}{\reviewed{\ensuremath{73.1}}}}}

\DeclareRobustCommand{\TGRImrctHierPopGWTCFOURResultsWOGWOneNineZeroEightOneFour}[1]{\IfEqCase{#1}{{DMFGWTC4HIERPOP}{\reviewed{\ensuremath{0.02^{+0.06}_{-0.06}}}}{DCHIFGWTC4HIERPOP}{\reviewed{\ensuremath{-0.02^{+0.07}_{-0.07}}}}{GRQUANTDevParamsTwoDGWTC4HIERPOP}{\reviewed{\ensuremath{44.8}}}{GRQUANTMeansTwoDGWTC4HIERPOP}{\reviewed{\ensuremath{78.7}}}{GRQUANTHyperParamsFourDGWTC4HIERPOP}{\reviewed{\ensuremath{38.6}}}}}

\DeclareRobustCommand{\TGRPOLlogBImprovement}[1]{\IfEqCase{#1}{
{S}{\reviewed{10.48}}
{V}{\reviewed{3.63}}
{TS}{\reviewed{0.40}}
{TV}{\reviewed{0.21}}
{VS}{\reviewed{3.48}}
{TVS}{\reviewed{0.23}}
{min}{\reviewed{0.21}}
{max}{\reviewed{10.48}}
}}

\DeclareRobustCommand{\TGRPOLlogBResults}[1]{\IfEqCase{#1}{
{S}{\reviewed{-14.72}}
{Serr}{\reviewed{0.59}}
{V}{\reviewed{-5.33}}
{Verr}{\reviewed{0.58}}
{TS}{\reviewed{-0.20}}
{TSerr}{\reviewed{0.57}}
{TV}{\reviewed{0.10}}
{TVerr}{\reviewed{0.57}}
{VS}{\reviewed{-5.21}}
{VSerr}{\reviewed{0.58}}
{TVS}{\reviewed{-0.31}}
{TVSerr}{\reviewed{0.57}}
{min}{\reviewed{-14.72}}
{minerr}{\reviewed{0.59}}
{max}{\reviewed{0.10}}
{maxerr}{\reviewed{0.57}}
}}

\DeclareRobustCommand{\TGRTIGERBound}[1]{\IfEqCase{#1}{
{dchiminus2}{\reviewed{\num{5.3e-04}}}
{dchi0}{\reviewed{\num{8.6e-02}}}
{dchi1}{\reviewed{0.18}}
{dchi2}{\reviewed{0.10}}
{dchi3}{\reviewed{\num{6.5e-02}}}
{dchi4}{\reviewed{0.57}}
{dchi5l}{\reviewed{0.23}}
{dchi6}{\reviewed{0.38}}
{dchi6l}{\reviewed{1.18}}
{dchi7}{\reviewed{0.93}}
{min}{\reviewed{\num{5.3e-04}}}
{max}{\reviewed{1.2}}
}}

\DeclareRobustCommand{\TGRTIGERBoundPI}[1]{\IfEqCase{#1}{
{db1}{\reviewed{\num{2.8e-02}}}
{db2}{\reviewed{\num{7.7e-03}}}
{db3}{\reviewed{\num{7.9e-03}}}
{db4}{\reviewed{\num{1.7e-02}}}
{dc1}{\reviewed{0.13}}
{dc2}{\reviewed{\num{4.3e-02}}}
{dc4}{\reviewed{0.13}}
{dcl}{\reviewed{0.28}}
{min}{\reviewed{\num{7.7e-03}}}
{max}{\reviewed{0.28}}
}}

\DeclareRobustCommand{\TGRTIGERBoundImprovement}[1]{\IfEqCase{#1}{
{dchiminus2}{\reviewed{3.9}}
{dchi0}{\reviewed{1.6}}
{dchi1}{\reviewed{1.7}}
{dchi2}{\reviewed{2.0}}
{dchi3}{\reviewed{1.6}}
{dchi4}{\reviewed{1.3}}
{dchi5l}{\reviewed{1.4}}
{dchi6}{\reviewed{1.7}}
{dchi6l}{\reviewed{1.4}}
{dchi7}{\reviewed{1.7}}
{min}{\reviewed{1.3}}
{max}{\reviewed{3.9}}
}}

\DeclareRobustCommand{\TGRFTIBound}[1]{\IfEqCase{#1}{
{dchiMinus2}{\reviewed{\num{1.6e-03}}}
{dchi0}{\reviewed{\num{9.5e-02}}}
{dchi1}{\reviewed{0.26}}
{dchi2}{\reviewed{0.14}}
{dchi3NS}{\reviewed{\num{8.0e-02}}}
{dchi4NS}{\reviewed{0.46}}
{dchi5lNS}{\reviewed{0.14}}
{dchi6NS}{\reviewed{0.19}}
{dchi6l}{\reviewed{1.52}}
{dchi7NS}{\reviewed{0.34}}
{min}{\reviewed{\num{1.6e-03}}}
{max}{\reviewed{1.5}}
}}

\DeclareRobustCommand{\TGRFTIBoundImprovement}[1]{\IfEqCase{#1}{
{dchiMinus2}{\reviewed{1.2}}
{dchi0}{\reviewed{1.4}}
{dchi1}{\reviewed{1.2}}
{dchi2}{\reviewed{1.5}}
{dchi3NS}{\reviewed{1.8}}
{dchi4NS}{\reviewed{1.5}}
{dchi5lNS}{\reviewed{3.5}}
{dchi6NS}{\reviewed{2.8}}
{dchi6l}{\reviewed{1.4}}
{dchi7NS}{\reviewed{5.5}}
{min}{\reviewed{1.2}}
{max}{\reviewed{5.5}}
}}

\DeclareRobustCommand{\FirstPCAFTIjointbound}{$-0.01^{+0.04}_{-0.03}$}
\DeclareRobustCommand{\FirstPCATIGERjointbound}{$0.01^{+0.06}_{-0.06}$}

\DeclareRobustCommand{\TGRSIMPhenomImprovement}{\reviewed{$1.4$}}
\DeclareRobustCommand{\TGRSIMPhenomCombinedCI}[1]{\IfEqCase{#1}{{HIER_POP}{\reviewed{\ensuremath{-19^{+28}_{-34}}}}{RESTRICTED_POP}{\reviewed{\ensuremath{-14^{+12}_{-14}}}}{HIER_POP_NEG}{\reviewed{46}}{HIER_POP_POS}{\reviewed{26}}{RESTRICTED_POP_NEG}{\reviewed{24}}{RESTRICTED_POP_POS}{\reviewed{5.7}}{HIER_MU}{\reviewed{\ensuremath{-19^{+20}_{-22}}}}{HIER_SIGMA}{\reviewed{25}}}}
\DeclareRobustCommand{\TGRSIMEOBCombinedCI}[1]{\IfEqCase{#1}{{HIER_POP}{\reviewed{\ensuremath{-49^{+95}_{-176}}}}{RESTRICTED_POP}{\reviewed{\ensuremath{-32^{+29}_{-53}}}}{HIER_POP_NEG}{\reviewed{179}}{HIER_POP_POS}{\reviewed{127}}{RESTRICTED_POP_NEG}{\reviewed{68}}{RESTRICTED_POP_POS}{\reviewed{13}}{HIER_MU}{\reviewed{\ensuremath{-51^{+53}_{-128}}}}{HIER_SIGMA}{\reviewed{119}}}}

\DeclareRobustCommand{\TGRLOSAResults}[1]{\IfEqCase{#1}{
{GW170817}{\reviewed{0.42_{-1.87}^{+2.00}} \times 10^{-6}}
{GW190425}{\reviewed{{-0.01}_{-7.40}^{+6.29}} \times 10^{-6}}
}}

\DeclareRobustCommand{\TGRMDRGravitonBound}[1]{\IfEqCase{#1}{
{gwtc3}{\reviewed{\num{2.23e-23} $\mathrm{eV}/c^2$}}
{gwtc4}{\reviewed{\num{1.92e-23} $\mathrm{eV}/c^2$}}
}}
\DeclareRobustCommand{\TGRMDRGravitonBoundNoUnits}[1]{\IfEqCase{#1}{
{gwtc3}{\reviewed{\num{2.23e-23}}}
{gwtc4}{\reviewed{\num{1.92e-23}}}
}}
\DeclareRobustCommand{\TGRMDRBoundImprovement}[1]{\IfEqCase{#1}{
{mg}{\reviewed{1.16}}
{alpha_0p0}{\reviewed{1.48}}
{alpha_0p5}{\reviewed{1.64}}
{alpha_1p5}{\reviewed{1.68}}
{alpha_2p5}{\reviewed{1.75}}
{alpha_3p0}{\reviewed{2.10}}
{alpha_3p5}{\reviewed{2.60}}
{alpha_4p0}{\reviewed{2.88}}
{alpha_m1p0}{\reviewed{1.69}}
{alpha_m2p0}{\reviewed{1.81}}
{alpha_m3p0}{\reviewed{2.00}}
{amplitude_mean}{\reviewed{1.96}}
{amplitude_min}{\reviewed{1.48}}
{amplitude_max}{\reviewed{2.88}}
}}
\DeclareRobustCommand{\TGRMDRAmplitudeBoundPeV}[1]{\IfEqCase{#1}{
{amplitude_min}{\reviewed{0.01}}
{amplitude_max}{\reviewed{351}}
}}

\DeclareRobustCommand{\TGRSSBKFiveVZeroZero}[1]{\IfEqCase{#1}{
{gwtc4}{\reviewed{\num{1.52e-14} m}}
}}
\DeclareRobustCommand{\TGRSSBKFiveVZeroZeroNoUnits}[1]{\IfEqCase{#1}{
{gwtc4}{\reviewed{\num{1.52e-14}}}
}}

\DeclareRobustCommand{\pyRingHierarchicalDeltaFMedian}{\reviewed{0.10}}
\DeclareRobustCommand{\pyRingHierarchicalDeltaFPlus}{\reviewed{0.23}}
\DeclareRobustCommand{\pyRingHierarchicalDeltaFMinus}{\reviewed{0.18}}
\DeclareRobustCommand{\pyRingHierarchicalDeltaTauMedian}{\reviewed{0.18}}
\DeclareRobustCommand{\pyRingHierarchicalDeltaTauPlus}{\reviewed{0.27}}
\DeclareRobustCommand{\pyRingHierarchicalDeltaTauMinus}{\reviewed{0.26}}

\DeclareRobustCommand{\pyRingHierarchicalQuantileFourDValue}{\reviewed{94.7}}

\DeclareRobustCommand{\pyRingHierarchicalQuantileOneDMuDeltaTauValuelnBcut}{\reviewed{80}}

\DeclareRobustCommand{\pseobImprovementDeltaF}{\reviewed{1.09}}
\DeclareRobustCommand{\pseobImprovementDeltaTau}{\reviewed{1.52}}

\DeclareRobustCommand{\pseobHierarchicalDeltaFMedian}{\reviewed{0.00}}
\DeclareRobustCommand{\pseobHierarchicalDeltaFPlus}{\reviewed{0.06}}
\DeclareRobustCommand{\pseobHierarchicalDeltaFMinus}{\reviewed{0.06}}
\DeclareRobustCommand{\pseobHierarchicalDeltaTauMedian}{\reviewed{0.16}}
\DeclareRobustCommand{\pseobHierarchicalDeltaTauPlus}{\reviewed{0.18}}
\DeclareRobustCommand{\pseobHierarchicalDeltaTauMinus}{\reviewed{0.16}}

\DeclareRobustCommand{\QuantileJointGWTCFour}{\reviewed{98.6}\%}
\DeclareRobustCommand{\QuantileHierGWTCFourTauOneD}{\reviewed{99.3}\%}

\DeclareRobustCommand{\QuantileJointGWTCFourPlusTwoFiveZeroOneOneFour}{\reviewed{92.2}\%}
\DeclareRobustCommand{\QuantileHierGWTCFourTauOneDPlusTwoFiveZeroOneOneFour}{\reviewed{96.2}\%}

\DeclareRobustCommand{\highestDTwoTwoOneoverThreshold}{\reviewed{3.1}}

\DeclareRobustCommand{\highestlogTenBFIMREvsIMR}{\reviewed{1.1}}

\DeclareRobustCommand{\BHPhighestlogTenBFIMREvsIMR}{\reviewed{0.2}}

\DeclareRobustCommand{\BWEchoeshighestlogTenBFSvsN}{\reviewed{-1.8}}

\DeclareRobustCommand{\INSTRUMENTSOONEOTWO}[1]{\IfEqCase{#1}{{GW150914}{HL}{GW151012}{HL}{GW151226}{HL}{GW170104}{HL}{GW170608}{HL}{GW170729}{HLV}{GW170809}{HLV}{GW170814}{HLV}{GW170817}{HLV}{GW170818}{HLV}{GW170823}{HL}}[\textcolor{red}{???}]}

\DeclareRobustCommand{\INSTRUMENTSOTHREEA}[1]{\IfEqCase{#1}{{GW190408A}{HLV}{GW190412A}{HLV}{GW190413A}{HL}{GW190413B}{HLV}{GW190421A}{HL}{GW190424A}{L}{GW190425A}{LV}{GW190426A}{HLV}{GW190503A}{HLV}{GW190512A}{HLV}{GW190513A}{HLV}{GW190514A}{HL}{GW190517A}{HLV}{GW190519A}{HLV}{GW190521A}{HLV}{GW190521B}{HL}{GW190527A}{HL}{GW190602A}{HLV}{GW190620A}{LV}{GW190630A}{LV}{GW190701A}{HLV}{GW190706A}{HLV}{GW190707A}{HL}{GW190708A}{LV}{GW190719A}{HL}{GW190720A}{HLV}{GW190727A}{HLV}{GW190728A}{HLV}{GW190731A}{HL}{GW190803A}{HLV}{GW190814A}{HLV}{GW190828A}{HLV}{GW190828B}{HLV}{GW190909A}{HL}{GW190910A}{LV}{GW190915A}{HLV}{GW190924A}{HLV}{GW190929A}{HLV}{GW190930A}{HL}}[\textcolor{red}{???}]}

\DeclareRobustCommand{\INSTRUMENTSOTHREEB}[1]{\IfEqCase{#1}{{GW191103A}{HL}{GW191105C}{HLV}{GW191109A}{HL}{GW191113B}{HLV}{GW191118N}{LV}{GW191126C}{HL}{GW191127B}{HLV}{GW191129G}{HL}{GW191204A}{HL}{GW191204G}{HL}{GW191215G}{HLV}{GW191216G}{HV}{GW191219E}{HLV}{GW191222A}{HL}{GW191230H}{HLV}{GW200105F}{LV}{GW200112H}{LV}{GW200115A}{HLV}{GW200121A}{HV}{GW200128C}{HL}{GW200129D}{HLV}{GW200201F}{HLV}{GW200202F}{HLV}{GW200208G}{HLV}{GW200208K}{HLV}{GW200209E}{HLV}{GW200210B}{HLV}{GW200214K}{HLV}{GW200216G}{HLV}{GW200219D}{HLV}{GW200219K}{HLV}{GW200220E}{HLV}{GW200220H}{HL}{GW200224H}{HLV}{GW200225B}{HL}{GW200302A}{HV}{GW200306A}{HL}{GW200308G}{HLV}{GW200311H}{HL}{GW200311L}{HLV}{GW200316I}{HLV}{GW200322G}{HLV}}[\textcolor{red}{???}]}

\input{common__pe_macros}
\input{common__pe_macros_gwtc2point1}

\DeclareRobustCommand{\finalspingwtcthreeminus}[1]{\IfEqCase{#1}{{GW200224_222234}{0.07}{GW191129_134029}{0.05}{GW200311_115853}{0.08}{GW191230_180458}{0.12}{GW191222_033537}{0.11}{GW200225_060421}{0.13}{GW200302_015811}{0.15}{GW200128_022011}{0.10}{GW191204_171526}{0.03}{GW200112_155838}{0.06}{GW200105_162426}{0.03}{GW191105_143521}{0.05}{GW191109_010717}{0.19}{GW200209_085452}{0.11}{GW200115_042309}{0.06}{GW191127_050227}{0.29}{GW200216_220804}{0.24}{GW191215_223052}{0.07}{GW200208_130117}{0.13}{GW200219_094415}{0.13}{GW191103_012549}{0.05}{GW200316_215756}{0.04}{GW200202_154313}{0.04}{GW200129_065458}{0.05}{GW191216_213338}{0.04}}[{\red{???}}]}
\DeclareRobustCommand{\finalspingwtcthreemed}[1]{\IfEqCase{#1}{{GW200224_222234}{0.73}{GW191129_134029}{0.69}{GW200311_115853}{0.69}{GW191230_180458}{0.69}{GW191222_033537}{0.67}{GW200225_060421}{0.66}{GW200302_015811}{0.67}{GW200128_022011}{0.75}{GW191204_171526}{0.73}{GW200112_155838}{0.71}{GW200105_162426}{0.44}{GW191105_143521}{0.67}{GW191109_010717}{0.61}{GW200209_085452}{0.67}{GW200115_042309}{0.43}{GW191127_050227}{0.75}{GW200216_220804}{0.70}{GW191215_223052}{0.68}{GW200208_130117}{0.66}{GW200219_094415}{0.66}{GW191103_012549}{0.75}{GW200316_215756}{0.70}{GW200202_154313}{0.69}{GW200129_065458}{0.73}{GW191216_213338}{0.70}}[{\red{???}}]}
\DeclareRobustCommand{\finalspingwtcthreeplus}[1]{\IfEqCase{#1}{{GW200224_222234}{0.07}{GW191129_134029}{0.03}{GW200311_115853}{0.07}{GW191230_180458}{0.11}{GW191222_033537}{0.08}{GW200225_060421}{0.07}{GW200302_015811}{0.13}{GW200128_022011}{0.10}{GW191204_171526}{0.03}{GW200112_155838}{0.06}{GW200105_162426}{0.07}{GW191105_143521}{0.04}{GW191109_010717}{0.18}{GW200209_085452}{0.10}{GW200115_042309}{0.10}{GW191127_050227}{0.13}{GW200216_220804}{0.14}{GW191215_223052}{0.07}{GW200208_130117}{0.09}{GW200219_094415}{0.10}{GW191103_012549}{0.06}{GW200316_215756}{0.04}{GW200202_154313}{0.03}{GW200129_065458}{0.06}{GW191216_213338}{0.03}}[{\red{???}}]}
\DeclareRobustCommand{\finalspingwtcthreetenthpercentile}[1]{\IfEqCase{#1}{{GW200224_222234}{0.68}{GW191129_134029}{0.65}{GW200311_115853}{0.63}{GW191230_180458}{0.60}{GW191222_033537}{0.60}{GW200225_060421}{0.56}{GW200302_015811}{0.56}{GW200128_022011}{0.67}{GW191204_171526}{0.71}{GW200112_155838}{0.66}{GW200105_162426}{0.41}{GW191105_143521}{0.63}{GW191109_010717}{0.47}{GW200209_085452}{0.59}{GW200115_042309}{0.38}{GW191127_050227}{0.56}{GW200216_220804}{0.53}{GW191215_223052}{0.63}{GW200208_130117}{0.57}{GW200219_094415}{0.57}{GW191103_012549}{0.71}{GW200316_215756}{0.67}{GW200202_154313}{0.66}{GW200129_065458}{0.70}{GW191216_213338}{0.67}}[{\red{???}}]}
\DeclareRobustCommand{\finalspingwtcthreenintiethpercentile}[1]{\IfEqCase{#1}{{GW200224_222234}{0.78}{GW191129_134029}{0.71}{GW200311_115853}{0.74}{GW191230_180458}{0.77}{GW191222_033537}{0.74}{GW200225_060421}{0.71}{GW200302_015811}{0.78}{GW200128_022011}{0.83}{GW191204_171526}{0.75}{GW200112_155838}{0.75}{GW200105_162426}{0.49}{GW191105_143521}{0.70}{GW191109_010717}{0.74}{GW200209_085452}{0.74}{GW200115_042309}{0.51}{GW191127_050227}{0.86}{GW200216_220804}{0.82}{GW191215_223052}{0.73}{GW200208_130117}{0.72}{GW200219_094415}{0.74}{GW191103_012549}{0.79}{GW200316_215756}{0.73}{GW200202_154313}{0.71}{GW200129_065458}{0.78}{GW191216_213338}{0.72}}[{\red{???}}]}
\DeclareRobustCommand{\spintwoygwtcthreeminus}[1]{\IfEqCase{#1}{{GW200224_222234}{0.57}{GW191129_134029}{0.46}{GW200311_115853}{0.56}{GW191230_180458}{0.61}{GW191222_033537}{0.54}{GW200225_060421}{0.54}{GW200302_015811}{0.58}{GW200128_022011}{0.61}{GW191204_171526}{0.53}{GW200112_155838}{0.52}{GW200105_162426}{0.51}{GW191105_143521}{0.52}{GW191109_010717}{0.69}{GW200209_085452}{0.60}{GW200115_042309}{0.50}{GW191127_050227}{0.58}{GW200216_220804}{0.60}{GW191215_223052}{0.58}{GW200208_130117}{0.55}{GW200219_094415}{0.57}{GW191103_012549}{0.54}{GW200316_215756}{0.53}{GW200202_154313}{0.49}{GW200129_065458}{0.55}{GW191216_213338}{0.48}}[{\red{???}}]}
\DeclareRobustCommand{\spintwoygwtcthreemed}[1]{\IfEqCase{#1}{{GW200224_222234}{0.00}{GW191129_134029}{0.00}{GW200311_115853}{0.00}{GW191230_180458}{0.00}{GW191222_033537}{0.00}{GW200225_060421}{0.00}{GW200302_015811}{0.00}{GW200128_022011}{0.00}{GW191204_171526}{0.00}{GW200112_155838}{0.00}{GW200105_162426}{0.00}{GW191105_143521}{0.00}{GW191109_010717}{0.01}{GW200209_085452}{0.00}{GW200115_042309}{0.00}{GW191127_050227}{0.00}{GW200216_220804}{0.00}{GW191215_223052}{0.00}{GW200208_130117}{0.00}{GW200219_094415}{0.00}{GW191103_012549}{0.00}{GW200316_215756}{0.00}{GW200202_154313}{0.00}{GW200129_065458}{0.00}{GW191216_213338}{0.00}}[{\red{???}}]}
\DeclareRobustCommand{\spintwoygwtcthreeplus}[1]{\IfEqCase{#1}{{GW200224_222234}{0.55}{GW191129_134029}{0.47}{GW200311_115853}{0.58}{GW191230_180458}{0.60}{GW191222_033537}{0.56}{GW200225_060421}{0.54}{GW200302_015811}{0.57}{GW200128_022011}{0.62}{GW191204_171526}{0.50}{GW200112_155838}{0.53}{GW200105_162426}{0.51}{GW191105_143521}{0.52}{GW191109_010717}{0.69}{GW200209_085452}{0.60}{GW200115_042309}{0.51}{GW191127_050227}{0.58}{GW200216_220804}{0.57}{GW191215_223052}{0.60}{GW200208_130117}{0.55}{GW200219_094415}{0.59}{GW191103_012549}{0.55}{GW200316_215756}{0.54}{GW200202_154313}{0.50}{GW200129_065458}{0.57}{GW191216_213338}{0.47}}[{\red{???}}]}
\DeclareRobustCommand{\spintwoygwtcthreetenthpercentile}[1]{\IfEqCase{#1}{{GW200224_222234}{-0.43}{GW191129_134029}{-0.33}{GW200311_115853}{-0.40}{GW191230_180458}{-0.46}{GW191222_033537}{-0.39}{GW200225_060421}{-0.40}{GW200302_015811}{-0.42}{GW200128_022011}{-0.46}{GW191204_171526}{-0.40}{GW200112_155838}{-0.38}{GW200105_162426}{-0.35}{GW191105_143521}{-0.36}{GW191109_010717}{-0.54}{GW200209_085452}{-0.44}{GW200115_042309}{-0.35}{GW191127_050227}{-0.43}{GW200216_220804}{-0.44}{GW191215_223052}{-0.43}{GW200208_130117}{-0.41}{GW200219_094415}{-0.43}{GW191103_012549}{-0.40}{GW200316_215756}{-0.39}{GW200202_154313}{-0.35}{GW200129_065458}{-0.38}{GW191216_213338}{-0.34}}[{\red{???}}]}
\DeclareRobustCommand{\spintwoygwtcthreenintiethpercentile}[1]{\IfEqCase{#1}{{GW200224_222234}{0.40}{GW191129_134029}{0.34}{GW200311_115853}{0.43}{GW191230_180458}{0.44}{GW191222_033537}{0.41}{GW200225_060421}{0.40}{GW200302_015811}{0.42}{GW200128_022011}{0.46}{GW191204_171526}{0.37}{GW200112_155838}{0.39}{GW200105_162426}{0.36}{GW191105_143521}{0.36}{GW191109_010717}{0.56}{GW200209_085452}{0.45}{GW200115_042309}{0.37}{GW191127_050227}{0.44}{GW200216_220804}{0.43}{GW191215_223052}{0.43}{GW200208_130117}{0.40}{GW200219_094415}{0.43}{GW191103_012549}{0.40}{GW200316_215756}{0.39}{GW200202_154313}{0.35}{GW200129_065458}{0.42}{GW191216_213338}{0.33}}[{\red{???}}]}
\DeclareRobustCommand{\finalmasssourcegwtcthreeminus}[1]{\IfEqCase{#1}{{GW200224_222234}{4.5}{GW191129_134029}{1.1}{GW200311_115853}{3.9}{GW191230_180458}{10}{GW191222_033537}{9.9}{GW200225_060421}{2.8}{GW200302_015811}{6.5}{GW200128_022011}{9.9}{GW191204_171526}{0.92}{GW200112_155838}{4.3}{GW200105_162426}{1.7}{GW191105_143521}{1.2}{GW191109_010717}{15}{GW200209_085452}{8.2}{GW200115_042309}{1.6}{GW191127_050227}{21}{GW200216_220804}{13}{GW191215_223052}{3.8}{GW200208_130117}{6.4}{GW200219_094415}{7.8}{GW191103_012549}{1.7}{GW200316_215756}{1.9}{GW200202_154313}{0.66}{GW200129_065458}{3.3}{GW191216_213338}{0.93}}[{\red{???}}]}
\DeclareRobustCommand{\finalmasssourcegwtcthreemed}[1]{\IfEqCase{#1}{{GW200224_222234}{68.3}{GW191129_134029}{16.7}{GW200311_115853}{59.0}{GW191230_180458}{80}{GW191222_033537}{75.5}{GW200225_060421}{32.1}{GW200302_015811}{55.0}{GW200128_022011}{67.9}{GW191204_171526}{19.14}{GW200112_155838}{60.8}{GW200105_162426}{10.7}{GW191105_143521}{17.6}{GW191109_010717}{107}{GW200209_085452}{58.5}{GW200115_042309}{7.1}{GW191127_050227}{76}{GW200216_220804}{78}{GW191215_223052}{40.7}{GW200208_130117}{62.5}{GW200219_094415}{62.2}{GW191103_012549}{19.0}{GW200316_215756}{20.2}{GW200202_154313}{16.76}{GW200129_065458}{60.3}{GW191216_213338}{18.87}}[{\red{???}}]}
\DeclareRobustCommand{\finalmasssourcegwtcthreeplus}[1]{\IfEqCase{#1}{{GW200224_222234}{6.3}{GW191129_134029}{2.5}{GW200311_115853}{4.8}{GW191230_180458}{16}{GW191222_033537}{15.3}{GW200225_060421}{3.5}{GW200302_015811}{8.9}{GW200128_022011}{14.1}{GW191204_171526}{1.79}{GW200112_155838}{5.3}{GW200105_162426}{2.0}{GW191105_143521}{2.1}{GW191109_010717}{18}{GW200209_085452}{12.2}{GW200115_042309}{2.2}{GW191127_050227}{39}{GW200216_220804}{19}{GW191215_223052}{5.3}{GW200208_130117}{7.3}{GW200219_094415}{11.7}{GW191103_012549}{3.8}{GW200316_215756}{7.5}{GW200202_154313}{1.87}{GW200129_065458}{4.0}{GW191216_213338}{2.84}}[{\red{???}}]}
\DeclareRobustCommand{\finalmasssourcegwtcthreetenthpercentile}[1]{\IfEqCase{#1}{{GW200224_222234}{64.7}{GW191129_134029}{15.7}{GW200311_115853}{56.0}{GW191230_180458}{71}{GW191222_033537}{67.3}{GW200225_060421}{29.9}{GW200302_015811}{49.8}{GW200128_022011}{59.8}{GW191204_171526}{18.38}{GW200112_155838}{57.3}{GW200105_162426}{9.6}{GW191105_143521}{16.7}{GW191109_010717}{96}{GW200209_085452}{51.9}{GW200115_042309}{5.7}{GW191127_050227}{59}{GW200216_220804}{67}{GW191215_223052}{37.6}{GW200208_130117}{57.5}{GW200219_094415}{55.9}{GW191103_012549}{17.5}{GW200316_215756}{18.6}{GW200202_154313}{16.22}{GW200129_065458}{57.6}{GW191216_213338}{18.08}}[{\red{???}}]}
\DeclareRobustCommand{\finalmasssourcegwtcthreenintiethpercentile}[1]{\IfEqCase{#1}{{GW200224_222234}{73.1}{GW191129_134029}{18.5}{GW200311_115853}{62.7}{GW191230_180458}{91}{GW191222_033537}{87.3}{GW200225_060421}{34.7}{GW200302_015811}{61.5}{GW200128_022011}{78.6}{GW191204_171526}{20.40}{GW200112_155838}{64.6}{GW200105_162426}{12.0}{GW191105_143521}{19.1}{GW191109_010717}{121}{GW200209_085452}{67.5}{GW200115_042309}{8.7}{GW191127_050227}{105}{GW200216_220804}{92}{GW191215_223052}{44.8}{GW200208_130117}{68.2}{GW200219_094415}{71.3}{GW191103_012549}{21.4}{GW200316_215756}{24.8}{GW200202_154313}{17.99}{GW200129_065458}{63.4}{GW191216_213338}{20.65}}[{\red{???}}]}
\DeclareRobustCommand{\spinoneygwtcthreeminus}[1]{\IfEqCase{#1}{{GW200224_222234}{0.59}{GW191129_134029}{0.36}{GW200311_115853}{0.56}{GW191230_180458}{0.65}{GW191222_033537}{0.53}{GW200225_060421}{0.65}{GW200302_015811}{0.54}{GW200128_022011}{0.66}{GW191204_171526}{0.48}{GW200112_155838}{0.49}{GW200105_162426}{0.13}{GW191105_143521}{0.41}{GW191109_010717}{0.74}{GW200209_085452}{0.65}{GW200115_042309}{0.35}{GW191127_050227}{0.70}{GW200216_220804}{0.60}{GW191215_223052}{0.63}{GW200208_130117}{0.47}{GW200219_094415}{0.61}{GW191103_012549}{0.51}{GW200316_215756}{0.39}{GW200202_154313}{0.36}{GW200129_065458}{0.58}{GW191216_213338}{0.29}}[{\red{???}}]}
\DeclareRobustCommand{\spinoneygwtcthreemed}[1]{\IfEqCase{#1}{{GW200224_222234}{0.00}{GW191129_134029}{0.00}{GW200311_115853}{0.00}{GW191230_180458}{0.00}{GW191222_033537}{0.00}{GW200225_060421}{0.00}{GW200302_015811}{0.00}{GW200128_022011}{0.00}{GW191204_171526}{0.00}{GW200112_155838}{0.00}{GW200105_162426}{0.00}{GW191105_143521}{0.00}{GW191109_010717}{0.01}{GW200209_085452}{0.00}{GW200115_042309}{0.00}{GW191127_050227}{0.00}{GW200216_220804}{0.00}{GW191215_223052}{0.00}{GW200208_130117}{0.00}{GW200219_094415}{0.00}{GW191103_012549}{0.00}{GW200316_215756}{-0.01}{GW200202_154313}{0.00}{GW200129_065458}{0.00}{GW191216_213338}{0.00}}[{\red{???}}]}
\DeclareRobustCommand{\spinoneygwtcthreeplus}[1]{\IfEqCase{#1}{{GW200224_222234}{0.57}{GW191129_134029}{0.37}{GW200311_115853}{0.58}{GW191230_180458}{0.64}{GW191222_033537}{0.53}{GW200225_060421}{0.64}{GW200302_015811}{0.53}{GW200128_022011}{0.66}{GW191204_171526}{0.49}{GW200112_155838}{0.48}{GW200105_162426}{0.13}{GW191105_143521}{0.41}{GW191109_010717}{0.70}{GW200209_085452}{0.66}{GW200115_042309}{0.33}{GW191127_050227}{0.71}{GW200216_220804}{0.61}{GW191215_223052}{0.63}{GW200208_130117}{0.49}{GW200219_094415}{0.59}{GW191103_012549}{0.53}{GW200316_215756}{0.38}{GW200202_154313}{0.37}{GW200129_065458}{0.72}{GW191216_213338}{0.29}}[{\red{???}}]}
\DeclareRobustCommand{\spinoneygwtcthreetenthpercentile}[1]{\IfEqCase{#1}{{GW200224_222234}{-0.45}{GW191129_134029}{-0.25}{GW200311_115853}{-0.40}{GW191230_180458}{-0.50}{GW191222_033537}{-0.37}{GW200225_060421}{-0.52}{GW200302_015811}{-0.40}{GW200128_022011}{-0.52}{GW191204_171526}{-0.37}{GW200112_155838}{-0.35}{GW200105_162426}{-0.09}{GW191105_143521}{-0.27}{GW191109_010717}{-0.60}{GW200209_085452}{-0.49}{GW200115_042309}{-0.26}{GW191127_050227}{-0.54}{GW200216_220804}{-0.46}{GW191215_223052}{-0.47}{GW200208_130117}{-0.33}{GW200219_094415}{-0.46}{GW191103_012549}{-0.38}{GW200316_215756}{-0.29}{GW200202_154313}{-0.25}{GW200129_065458}{-0.42}{GW191216_213338}{-0.19}}[{\red{???}}]}
\DeclareRobustCommand{\spinoneygwtcthreenintiethpercentile}[1]{\IfEqCase{#1}{{GW200224_222234}{0.43}{GW191129_134029}{0.26}{GW200311_115853}{0.43}{GW191230_180458}{0.48}{GW191222_033537}{0.37}{GW200225_060421}{0.51}{GW200302_015811}{0.39}{GW200128_022011}{0.53}{GW191204_171526}{0.37}{GW200112_155838}{0.35}{GW200105_162426}{0.08}{GW191105_143521}{0.28}{GW191109_010717}{0.59}{GW200209_085452}{0.50}{GW200115_042309}{0.23}{GW191127_050227}{0.55}{GW200216_220804}{0.46}{GW191215_223052}{0.49}{GW200208_130117}{0.36}{GW200219_094415}{0.43}{GW191103_012549}{0.38}{GW200316_215756}{0.26}{GW200202_154313}{0.25}{GW200129_065458}{0.56}{GW191216_213338}{0.21}}[{\red{???}}]}
\DeclareRobustCommand{\costilttwogwtcthreeminus}[1]{\IfEqCase{#1}{{GW200224_222234}{1.01}{GW191129_134029}{1.07}{GW200311_115853}{0.84}{GW191230_180458}{0.79}{GW191222_033537}{0.81}{GW200225_060421}{0.71}{GW200302_015811}{0.98}{GW200128_022011}{0.99}{GW191204_171526}{1.06}{GW200112_155838}{1.02}{GW200105_162426}{0.89}{GW191105_143521}{0.83}{GW191109_010717}{0.66}{GW200209_085452}{0.67}{GW200115_042309}{0.64}{GW191127_050227}{1.06}{GW200216_220804}{1.03}{GW191215_223052}{0.79}{GW200208_130117}{0.76}{GW200219_094415}{0.75}{GW191103_012549}{1.19}{GW200316_215756}{1.05}{GW200202_154313}{0.97}{GW200129_065458}{1.21}{GW191216_213338}{1.16}}[{\red{???}}]}
\DeclareRobustCommand{\costilttwogwtcthreemed}[1]{\IfEqCase{#1}{{GW200224_222234}{0.19}{GW191129_134029}{0.30}{GW200311_115853}{-0.02}{GW191230_180458}{-0.12}{GW191222_033537}{-0.09}{GW200225_060421}{-0.22}{GW200302_015811}{0.12}{GW200128_022011}{0.15}{GW191204_171526}{0.45}{GW200112_155838}{0.25}{GW200105_162426}{0.02}{GW191105_143521}{-0.03}{GW191109_010717}{-0.26}{GW200209_085452}{-0.26}{GW200115_042309}{-0.31}{GW191127_050227}{0.22}{GW200216_220804}{0.18}{GW191215_223052}{-0.09}{GW200208_130117}{-0.15}{GW200219_094415}{-0.17}{GW191103_012549}{0.48}{GW200316_215756}{0.37}{GW200202_154313}{0.22}{GW200129_065458}{0.41}{GW191216_213338}{0.40}}[{\red{???}}]}
\DeclareRobustCommand{\costilttwogwtcthreeplus}[1]{\IfEqCase{#1}{{GW200224_222234}{0.71}{GW191129_134029}{0.63}{GW200311_115853}{0.87}{GW191230_180458}{0.96}{GW191222_033537}{0.95}{GW200225_060421}{1.04}{GW200302_015811}{0.78}{GW200128_022011}{0.74}{GW191204_171526}{0.50}{GW200112_155838}{0.67}{GW200105_162426}{0.84}{GW191105_143521}{0.89}{GW191109_010717}{1.04}{GW200209_085452}{1.05}{GW200115_042309}{1.13}{GW191127_050227}{0.71}{GW200216_220804}{0.75}{GW191215_223052}{0.93}{GW200208_130117}{1.00}{GW200219_094415}{0.99}{GW191103_012549}{0.48}{GW200316_215756}{0.57}{GW200202_154313}{0.68}{GW200129_065458}{0.54}{GW191216_213338}{0.54}}[{\red{???}}]}
\DeclareRobustCommand{\costilttwogwtcthreetenthpercentile}[1]{\IfEqCase{#1}{{GW200224_222234}{-0.67}{GW191129_134029}{-0.56}{GW200311_115853}{-0.75}{GW191230_180458}{-0.81}{GW191222_033537}{-0.80}{GW200225_060421}{-0.85}{GW200302_015811}{-0.72}{GW200128_022011}{-0.69}{GW191204_171526}{-0.38}{GW200112_155838}{-0.61}{GW200105_162426}{-0.74}{GW191105_143521}{-0.73}{GW191109_010717}{-0.85}{GW200209_085452}{-0.86}{GW200115_042309}{-0.90}{GW191127_050227}{-0.69}{GW200216_220804}{-0.71}{GW191215_223052}{-0.78}{GW200208_130117}{-0.82}{GW200219_094415}{-0.84}{GW191103_012549}{-0.50}{GW200316_215756}{-0.46}{GW200202_154313}{-0.55}{GW200129_065458}{-0.61}{GW191216_213338}{-0.54}}[{\red{???}}]}
\DeclareRobustCommand{\costilttwogwtcthreenintiethpercentile}[1]{\IfEqCase{#1}{{GW200224_222234}{0.81}{GW191129_134029}{0.87}{GW200311_115853}{0.72}{GW191230_180458}{0.70}{GW191222_033537}{0.71}{GW200225_060421}{0.65}{GW200302_015811}{0.82}{GW200128_022011}{0.79}{GW191204_171526}{0.89}{GW200112_155838}{0.82}{GW200105_162426}{0.74}{GW191105_143521}{0.74}{GW191109_010717}{0.62}{GW200209_085452}{0.61}{GW200115_042309}{0.65}{GW191127_050227}{0.86}{GW200216_220804}{0.85}{GW191215_223052}{0.70}{GW200208_130117}{0.70}{GW200219_094415}{0.69}{GW191103_012549}{0.91}{GW200316_215756}{0.87}{GW200202_154313}{0.82}{GW200129_065458}{0.90}{GW191216_213338}{0.89}}[{\red{???}}]}
\DeclareRobustCommand{\massonesourcegwtcthreeminus}[1]{\IfEqCase{#1}{{GW200224_222234}{4.4}{GW191129_134029}{2.1}{GW200311_115853}{3.8}{GW191230_180458}{8.6}{GW191222_033537}{8.0}{GW200225_060421}{3.0}{GW200302_015811}{8.4}{GW200128_022011}{7.6}{GW191204_171526}{1.8}{GW200112_155838}{4.5}{GW200105_162426}{1.7}{GW191105_143521}{1.6}{GW191109_010717}{11}{GW200209_085452}{6.3}{GW200115_042309}{2.5}{GW191127_050227}{20}{GW200216_220804}{13}{GW191215_223052}{4.1}{GW200208_130117}{6.2}{GW200219_094415}{6.9}{GW191103_012549}{2.2}{GW200316_215756}{2.9}{GW200202_154313}{1.4}{GW200129_065458}{3.2}{GW191216_213338}{2.2}}[{\red{???}}]}
\DeclareRobustCommand{\massonesourcegwtcthreemed}[1]{\IfEqCase{#1}{{GW200224_222234}{39.8}{GW191129_134029}{10.6}{GW200311_115853}{34.2}{GW191230_180458}{47.7}{GW191222_033537}{45.1}{GW200225_060421}{19.3}{GW200302_015811}{37.0}{GW200128_022011}{40.4}{GW191204_171526}{11.9}{GW200112_155838}{35.6}{GW200105_162426}{9.0}{GW191105_143521}{10.7}{GW191109_010717}{65}{GW200209_085452}{34.6}{GW200115_042309}{5.9}{GW191127_050227}{53}{GW200216_220804}{51}{GW191215_223052}{24.6}{GW200208_130117}{37.8}{GW200219_094415}{37.5}{GW191103_012549}{11.8}{GW200316_215756}{13.1}{GW200202_154313}{10.1}{GW200129_065458}{34.5}{GW191216_213338}{12.1}}[{\red{???}}]}
\DeclareRobustCommand{\massonesourcegwtcthreeplus}[1]{\IfEqCase{#1}{{GW200224_222234}{6.9}{GW191129_134029}{4.1}{GW200311_115853}{6.4}{GW191230_180458}{13.4}{GW191222_033537}{10.9}{GW200225_060421}{5.0}{GW200302_015811}{8.9}{GW200128_022011}{11.1}{GW191204_171526}{3.3}{GW200112_155838}{6.7}{GW200105_162426}{1.7}{GW191105_143521}{3.7}{GW191109_010717}{11}{GW200209_085452}{10.0}{GW200115_042309}{2.0}{GW191127_050227}{47}{GW200216_220804}{22}{GW191215_223052}{7.0}{GW200208_130117}{9.2}{GW200219_094415}{10.1}{GW191103_012549}{6.2}{GW200316_215756}{10.2}{GW200202_154313}{3.5}{GW200129_065458}{9.9}{GW191216_213338}{4.6}}[{\red{???}}]}
\DeclareRobustCommand{\massonesourcegwtcthreetenthpercentile}[1]{\IfEqCase{#1}{{GW200224_222234}{36.1}{GW191129_134029}{8.8}{GW200311_115853}{31.1}{GW191230_180458}{40.6}{GW191222_033537}{38.5}{GW200225_060421}{16.8}{GW200302_015811}{30.3}{GW200128_022011}{34.3}{GW191204_171526}{10.3}{GW200112_155838}{31.9}{GW200105_162426}{8.0}{GW191105_143521}{9.3}{GW191109_010717}{57}{GW200209_085452}{29.5}{GW200115_042309}{3.7}{GW191127_050227}{36}{GW200216_220804}{40}{GW191215_223052}{21.2}{GW200208_130117}{32.7}{GW200219_094415}{31.8}{GW191103_012549}{9.9}{GW200316_215756}{10.6}{GW200202_154313}{8.9}{GW200129_065458}{31.9}{GW191216_213338}{10.1}}[{\red{???}}]}
\DeclareRobustCommand{\massonesourcegwtcthreenintiethpercentile}[1]{\IfEqCase{#1}{{GW200224_222234}{45.0}{GW191129_134029}{13.7}{GW200311_115853}{38.9}{GW191230_180458}{57.6}{GW191222_033537}{53.2}{GW200225_060421}{23.0}{GW200302_015811}{43.9}{GW200128_022011}{48.7}{GW191204_171526}{14.3}{GW200112_155838}{40.8}{GW200105_162426}{10.0}{GW191105_143521}{13.3}{GW191109_010717}{73}{GW200209_085452}{41.9}{GW200115_042309}{7.4}{GW191127_050227}{88}{GW200216_220804}{68}{GW191215_223052}{29.8}{GW200208_130117}{44.9}{GW200219_094415}{45.1}{GW191103_012549}{16.1}{GW200316_215756}{20.0}{GW200202_154313}{12.7}{GW200129_065458}{42.3}{GW191216_213338}{15.2}}[{\red{???}}]}
\DeclareRobustCommand{\spintwoxgwtcthreeminus}[1]{\IfEqCase{#1}{{GW200224_222234}{0.58}{GW191129_134029}{0.47}{GW200311_115853}{0.55}{GW191230_180458}{0.60}{GW191222_033537}{0.55}{GW200225_060421}{0.55}{GW200302_015811}{0.58}{GW200128_022011}{0.61}{GW191204_171526}{0.53}{GW200112_155838}{0.51}{GW200105_162426}{0.52}{GW191105_143521}{0.51}{GW191109_010717}{0.68}{GW200209_085452}{0.58}{GW200115_042309}{0.51}{GW191127_050227}{0.60}{GW200216_220804}{0.60}{GW191215_223052}{0.58}{GW200208_130117}{0.54}{GW200219_094415}{0.60}{GW191103_012549}{0.53}{GW200316_215756}{0.55}{GW200202_154313}{0.50}{GW200129_065458}{0.57}{GW191216_213338}{0.44}}[{\red{???}}]}
\DeclareRobustCommand{\spintwoxgwtcthreemed}[1]{\IfEqCase{#1}{{GW200224_222234}{0.01}{GW191129_134029}{0.00}{GW200311_115853}{0.00}{GW191230_180458}{0.00}{GW191222_033537}{0.00}{GW200225_060421}{0.00}{GW200302_015811}{0.00}{GW200128_022011}{0.00}{GW191204_171526}{0.00}{GW200112_155838}{0.00}{GW200105_162426}{0.00}{GW191105_143521}{0.00}{GW191109_010717}{0.00}{GW200209_085452}{0.00}{GW200115_042309}{0.00}{GW191127_050227}{0.00}{GW200216_220804}{0.00}{GW191215_223052}{0.00}{GW200208_130117}{0.00}{GW200219_094415}{0.00}{GW191103_012549}{0.00}{GW200316_215756}{0.00}{GW200202_154313}{0.00}{GW200129_065458}{0.00}{GW191216_213338}{-0.01}}[{\red{???}}]}
\DeclareRobustCommand{\spintwoxgwtcthreeplus}[1]{\IfEqCase{#1}{{GW200224_222234}{0.58}{GW191129_134029}{0.46}{GW200311_115853}{0.56}{GW191230_180458}{0.60}{GW191222_033537}{0.55}{GW200225_060421}{0.55}{GW200302_015811}{0.58}{GW200128_022011}{0.61}{GW191204_171526}{0.50}{GW200112_155838}{0.53}{GW200105_162426}{0.50}{GW191105_143521}{0.51}{GW191109_010717}{0.65}{GW200209_085452}{0.60}{GW200115_042309}{0.51}{GW191127_050227}{0.60}{GW200216_220804}{0.60}{GW191215_223052}{0.58}{GW200208_130117}{0.57}{GW200219_094415}{0.58}{GW191103_012549}{0.55}{GW200316_215756}{0.53}{GW200202_154313}{0.50}{GW200129_065458}{0.55}{GW191216_213338}{0.47}}[{\red{???}}]}
\DeclareRobustCommand{\spintwoxgwtcthreetenthpercentile}[1]{\IfEqCase{#1}{{GW200224_222234}{-0.41}{GW191129_134029}{-0.33}{GW200311_115853}{-0.41}{GW191230_180458}{-0.45}{GW191222_033537}{-0.41}{GW200225_060421}{-0.40}{GW200302_015811}{-0.42}{GW200128_022011}{-0.46}{GW191204_171526}{-0.39}{GW200112_155838}{-0.37}{GW200105_162426}{-0.36}{GW191105_143521}{-0.35}{GW191109_010717}{-0.53}{GW200209_085452}{-0.44}{GW200115_042309}{-0.38}{GW191127_050227}{-0.46}{GW200216_220804}{-0.45}{GW191215_223052}{-0.42}{GW200208_130117}{-0.40}{GW200219_094415}{-0.47}{GW191103_012549}{-0.40}{GW200316_215756}{-0.41}{GW200202_154313}{-0.36}{GW200129_065458}{-0.41}{GW191216_213338}{-0.33}}[{\red{???}}]}
\DeclareRobustCommand{\spintwoxgwtcthreenintiethpercentile}[1]{\IfEqCase{#1}{{GW200224_222234}{0.44}{GW191129_134029}{0.33}{GW200311_115853}{0.40}{GW191230_180458}{0.45}{GW191222_033537}{0.39}{GW200225_060421}{0.40}{GW200302_015811}{0.43}{GW200128_022011}{0.47}{GW191204_171526}{0.38}{GW200112_155838}{0.39}{GW200105_162426}{0.35}{GW191105_143521}{0.36}{GW191109_010717}{0.51}{GW200209_085452}{0.45}{GW200115_042309}{0.36}{GW191127_050227}{0.45}{GW200216_220804}{0.45}{GW191215_223052}{0.43}{GW200208_130117}{0.42}{GW200219_094415}{0.43}{GW191103_012549}{0.40}{GW200316_215756}{0.39}{GW200202_154313}{0.35}{GW200129_065458}{0.41}{GW191216_213338}{0.31}}[{\red{???}}]}
\DeclareRobustCommand{\phitwogwtcthreeminus}[1]{\IfEqCase{#1}{{GW200224_222234}{2.9}{GW191129_134029}{2.8}{GW200311_115853}{2.8}{GW191230_180458}{2.8}{GW191222_033537}{2.9}{GW200225_060421}{2.8}{GW200302_015811}{2.9}{GW200128_022011}{2.9}{GW191204_171526}{2.8}{GW200112_155838}{2.9}{GW200105_162426}{2.8}{GW191105_143521}{2.8}{GW191109_010717}{2.7}{GW200209_085452}{2.8}{GW200115_042309}{2.8}{GW191127_050227}{2.8}{GW200216_220804}{2.9}{GW191215_223052}{2.8}{GW200208_130117}{2.8}{GW200219_094415}{2.8}{GW191103_012549}{2.8}{GW200316_215756}{2.8}{GW200202_154313}{2.8}{GW200129_065458}{2.7}{GW191216_213338}{2.8}}[{\red{???}}]}
\DeclareRobustCommand{\phitwogwtcthreemed}[1]{\IfEqCase{#1}{{GW200224_222234}{3.2}{GW191129_134029}{3.1}{GW200311_115853}{3.1}{GW191230_180458}{3.1}{GW191222_033537}{3.2}{GW200225_060421}{3.1}{GW200302_015811}{3.1}{GW200128_022011}{3.2}{GW191204_171526}{3.2}{GW200112_155838}{3.2}{GW200105_162426}{3.1}{GW191105_143521}{3.2}{GW191109_010717}{3.0}{GW200209_085452}{3.1}{GW200115_042309}{3.1}{GW191127_050227}{3.1}{GW200216_220804}{3.2}{GW191215_223052}{3.1}{GW200208_130117}{3.1}{GW200219_094415}{3.2}{GW191103_012549}{3.1}{GW200316_215756}{3.1}{GW200202_154313}{3.2}{GW200129_065458}{3.1}{GW191216_213338}{3.2}}[{\red{???}}]}
\DeclareRobustCommand{\phitwogwtcthreeplus}[1]{\IfEqCase{#1}{{GW200224_222234}{2.8}{GW191129_134029}{2.9}{GW200311_115853}{2.9}{GW191230_180458}{2.8}{GW191222_033537}{2.8}{GW200225_060421}{2.9}{GW200302_015811}{2.8}{GW200128_022011}{2.8}{GW191204_171526}{2.8}{GW200112_155838}{2.8}{GW200105_162426}{2.8}{GW191105_143521}{2.8}{GW191109_010717}{2.9}{GW200209_085452}{2.8}{GW200115_042309}{2.8}{GW191127_050227}{2.9}{GW200216_220804}{2.8}{GW191215_223052}{2.8}{GW200208_130117}{2.8}{GW200219_094415}{2.8}{GW191103_012549}{2.8}{GW200316_215756}{2.8}{GW200202_154313}{2.8}{GW200129_065458}{2.9}{GW191216_213338}{2.8}}[{\red{???}}]}
\DeclareRobustCommand{\phitwogwtcthreetenthpercentile}[1]{\IfEqCase{#1}{{GW200224_222234}{0.6}{GW191129_134029}{0.6}{GW200311_115853}{0.6}{GW191230_180458}{0.6}{GW191222_033537}{0.6}{GW200225_060421}{0.7}{GW200302_015811}{0.6}{GW200128_022011}{0.6}{GW191204_171526}{0.6}{GW200112_155838}{0.6}{GW200105_162426}{0.6}{GW191105_143521}{0.6}{GW191109_010717}{0.6}{GW200209_085452}{0.6}{GW200115_042309}{0.6}{GW191127_050227}{0.6}{GW200216_220804}{0.6}{GW191215_223052}{0.6}{GW200208_130117}{0.6}{GW200219_094415}{0.6}{GW191103_012549}{0.7}{GW200316_215756}{0.6}{GW200202_154313}{0.6}{GW200129_065458}{0.6}{GW191216_213338}{0.7}}[{\red{???}}]}
\DeclareRobustCommand{\phitwogwtcthreenintiethpercentile}[1]{\IfEqCase{#1}{{GW200224_222234}{5.7}{GW191129_134029}{5.7}{GW200311_115853}{5.6}{GW191230_180458}{5.6}{GW191222_033537}{5.7}{GW200225_060421}{5.6}{GW200302_015811}{5.7}{GW200128_022011}{5.7}{GW191204_171526}{5.6}{GW200112_155838}{5.6}{GW200105_162426}{5.6}{GW191105_143521}{5.7}{GW191109_010717}{5.6}{GW200209_085452}{5.7}{GW200115_042309}{5.6}{GW191127_050227}{5.6}{GW200216_220804}{5.7}{GW191215_223052}{5.6}{GW200208_130117}{5.7}{GW200219_094415}{5.7}{GW191103_012549}{5.7}{GW200316_215756}{5.6}{GW200202_154313}{5.7}{GW200129_065458}{5.7}{GW191216_213338}{5.6}}[{\red{???}}]}
\DeclareRobustCommand{\chipgwtcthreeminus}[1]{\IfEqCase{#1}{{GW200224_222234}{0.36}{GW191129_134029}{0.19}{GW200311_115853}{0.35}{GW191230_180458}{0.39}{GW191222_033537}{0.32}{GW200225_060421}{0.38}{GW200302_015811}{0.29}{GW200128_022011}{0.40}{GW191204_171526}{0.26}{GW200112_155838}{0.30}{GW200105_162426}{0.07}{GW191105_143521}{0.24}{GW191109_010717}{0.37}{GW200209_085452}{0.38}{GW200115_042309}{0.16}{GW191127_050227}{0.41}{GW200216_220804}{0.35}{GW191215_223052}{0.38}{GW200208_130117}{0.29}{GW200219_094415}{0.35}{GW191103_012549}{0.26}{GW200316_215756}{0.20}{GW200202_154313}{0.22}{GW200129_065458}{0.37}{GW191216_213338}{0.15}}[{\red{???}}]}
\DeclareRobustCommand{\chipgwtcthreemed}[1]{\IfEqCase{#1}{{GW200224_222234}{0.50}{GW191129_134029}{0.26}{GW200311_115853}{0.45}{GW191230_180458}{0.52}{GW191222_033537}{0.41}{GW200225_060421}{0.53}{GW200302_015811}{0.38}{GW200128_022011}{0.57}{GW191204_171526}{0.39}{GW200112_155838}{0.39}{GW200105_162426}{0.09}{GW191105_143521}{0.30}{GW191109_010717}{0.63}{GW200209_085452}{0.52}{GW200115_042309}{0.20}{GW191127_050227}{0.52}{GW200216_220804}{0.45}{GW191215_223052}{0.50}{GW200208_130117}{0.38}{GW200219_094415}{0.48}{GW191103_012549}{0.40}{GW200316_215756}{0.29}{GW200202_154313}{0.28}{GW200129_065458}{0.52}{GW191216_213338}{0.23}}[{\red{???}}]}
\DeclareRobustCommand{\chipgwtcthreeplus}[1]{\IfEqCase{#1}{{GW200224_222234}{0.37}{GW191129_134029}{0.36}{GW200311_115853}{0.40}{GW191230_180458}{0.38}{GW191222_033537}{0.41}{GW200225_060421}{0.34}{GW200302_015811}{0.44}{GW200128_022011}{0.33}{GW191204_171526}{0.35}{GW200112_155838}{0.39}{GW200105_162426}{0.17}{GW191105_143521}{0.45}{GW191109_010717}{0.29}{GW200209_085452}{0.38}{GW200115_042309}{0.34}{GW191127_050227}{0.41}{GW200216_220804}{0.42}{GW191215_223052}{0.38}{GW200208_130117}{0.41}{GW200219_094415}{0.40}{GW191103_012549}{0.41}{GW200316_215756}{0.38}{GW200202_154313}{0.40}{GW200129_065458}{0.42}{GW191216_213338}{0.35}}[{\red{???}}]}
\DeclareRobustCommand{\chipgwtcthreetenthpercentile}[1]{\IfEqCase{#1}{{GW200224_222234}{0.20}{GW191129_134029}{0.10}{GW200311_115853}{0.16}{GW191230_180458}{0.20}{GW191222_033537}{0.15}{GW200225_060421}{0.22}{GW200302_015811}{0.13}{GW200128_022011}{0.24}{GW191204_171526}{0.18}{GW200112_155838}{0.13}{GW200105_162426}{0.03}{GW191105_143521}{0.09}{GW191109_010717}{0.33}{GW200209_085452}{0.20}{GW200115_042309}{0.06}{GW191127_050227}{0.17}{GW200216_220804}{0.15}{GW191215_223052}{0.19}{GW200208_130117}{0.13}{GW200219_094415}{0.18}{GW191103_012549}{0.18}{GW200316_215756}{0.12}{GW200202_154313}{0.09}{GW200129_065458}{0.21}{GW191216_213338}{0.10}}[{\red{???}}]}
\DeclareRobustCommand{\chipgwtcthreenintiethpercentile}[1]{\IfEqCase{#1}{{GW200224_222234}{0.80}{GW191129_134029}{0.54}{GW200311_115853}{0.77}{GW191230_180458}{0.84}{GW191222_033537}{0.75}{GW200225_060421}{0.82}{GW200302_015811}{0.74}{GW200128_022011}{0.85}{GW191204_171526}{0.67}{GW200112_155838}{0.70}{GW200105_162426}{0.19}{GW191105_143521}{0.65}{GW191109_010717}{0.87}{GW200209_085452}{0.84}{GW200115_042309}{0.46}{GW191127_050227}{0.88}{GW200216_220804}{0.80}{GW191215_223052}{0.82}{GW200208_130117}{0.71}{GW200219_094415}{0.80}{GW191103_012549}{0.72}{GW200316_215756}{0.58}{GW200202_154313}{0.59}{GW200129_065458}{0.91}{GW191216_213338}{0.48}}[{\red{???}}]}
\DeclareRobustCommand{\chirpmassdetgwtcthreeminus}[1]{\IfEqCase{#1}{{GW200224_222234}{3.8}{GW191129_134029}{0.05}{GW200311_115853}{2.8}{GW191230_180458}{9.5}{GW191222_033537}{6.5}{GW200225_060421}{1.97}{GW200302_015811}{4.3}{GW200128_022011}{6.6}{GW191204_171526}{0.05}{GW200112_155838}{2.4}{GW200105_162426}{0.01}{GW191105_143521}{0.14}{GW191109_010717}{9.3}{GW200209_085452}{7.3}{GW200115_042309}{0.01}{GW191127_050227}{19}{GW200216_220804}{20}{GW191215_223052}{1.4}{GW200208_130117}{4.8}{GW200219_094415}{6.2}{GW191103_012549}{0.12}{GW200316_215756}{0.12}{GW200202_154313}{0.05}{GW200129_065458}{2.6}{GW191216_213338}{0.05}}[{\red{???}}]}
\DeclareRobustCommand{\chirpmassdetgwtcthreemed}[1]{\IfEqCase{#1}{{GW200224_222234}{41.1}{GW191129_134029}{8.49}{GW200311_115853}{32.7}{GW191230_180458}{62.8}{GW191222_033537}{51.0}{GW200225_060421}{17.65}{GW200302_015811}{30.4}{GW200128_022011}{50.5}{GW191204_171526}{9.69}{GW200112_155838}{33.9}{GW200105_162426}{3.62}{GW191105_143521}{9.58}{GW191109_010717}{60.1}{GW200209_085452}{42.9}{GW200115_042309}{2.58}{GW191127_050227}{48}{GW200216_220804}{56}{GW191215_223052}{24.9}{GW200208_130117}{38.8}{GW200219_094415}{43.7}{GW191103_012549}{9.98}{GW200316_215756}{10.68}{GW200202_154313}{8.15}{GW200129_065458}{32.1}{GW191216_213338}{8.94}}[{\red{???}}]}
\DeclareRobustCommand{\chirpmassdetgwtcthreeplus}[1]{\IfEqCase{#1}{{GW200224_222234}{3.6}{GW191129_134029}{0.06}{GW200311_115853}{2.7}{GW191230_180458}{9.4}{GW191222_033537}{7.2}{GW200225_060421}{0.98}{GW200302_015811}{7.7}{GW200128_022011}{7.5}{GW191204_171526}{0.05}{GW200112_155838}{2.9}{GW200105_162426}{0.01}{GW191105_143521}{0.12}{GW191109_010717}{9.8}{GW200209_085452}{8.7}{GW200115_042309}{0.01}{GW191127_050227}{21}{GW200216_220804}{14}{GW191215_223052}{1.5}{GW200208_130117}{5.2}{GW200219_094415}{6.3}{GW191103_012549}{0.13}{GW200316_215756}{0.12}{GW200202_154313}{0.05}{GW200129_065458}{1.8}{GW191216_213338}{0.05}}[{\red{???}}]}
\DeclareRobustCommand{\chirpmassdetgwtcthreetenthpercentile}[1]{\IfEqCase{#1}{{GW200224_222234}{38.2}{GW191129_134029}{8.45}{GW200311_115853}{30.6}{GW191230_180458}{55.5}{GW191222_033537}{46.1}{GW200225_060421}{15.96}{GW200302_015811}{27.1}{GW200128_022011}{45.4}{GW191204_171526}{9.65}{GW200112_155838}{32.1}{GW200105_162426}{3.61}{GW191105_143521}{9.48}{GW191109_010717}{52.7}{GW200209_085452}{37.3}{GW200115_042309}{2.57}{GW191127_050227}{33}{GW200216_220804}{39}{GW191215_223052}{23.8}{GW200208_130117}{35.0}{GW200219_094415}{39.0}{GW191103_012549}{9.88}{GW200316_215756}{10.59}{GW200202_154313}{8.11}{GW200129_065458}{30.2}{GW191216_213338}{8.90}}[{\red{???}}]}
\DeclareRobustCommand{\chirpmassdetgwtcthreenintiethpercentile}[1]{\IfEqCase{#1}{{GW200224_222234}{43.8}{GW191129_134029}{8.53}{GW200311_115853}{34.6}{GW191230_180458}{69.9}{GW191222_033537}{56.3}{GW200225_060421}{18.43}{GW200302_015811}{36.4}{GW200128_022011}{56.3}{GW191204_171526}{9.74}{GW200112_155838}{36.2}{GW200105_162426}{3.62}{GW191105_143521}{9.67}{GW191109_010717}{67.5}{GW200209_085452}{49.4}{GW200115_042309}{2.59}{GW191127_050227}{65}{GW200216_220804}{67}{GW191215_223052}{26.1}{GW200208_130117}{42.8}{GW200219_094415}{48.5}{GW191103_012549}{10.07}{GW200316_215756}{10.76}{GW200202_154313}{8.19}{GW200129_065458}{33.4}{GW191216_213338}{8.98}}[{\red{???}}]}
\DeclareRobustCommand{\chirpmasssourcegwtcthreeminus}[1]{\IfEqCase{#1}{{GW200224_222234}{2.5}{GW191129_134029}{0.27}{GW200311_115853}{2.0}{GW191230_180458}{4.9}{GW191222_033537}{5.0}{GW200225_060421}{1.4}{GW200302_015811}{2.9}{GW200128_022011}{4.6}{GW191204_171526}{0.25}{GW200112_155838}{2.1}{GW200105_162426}{0.08}{GW191105_143521}{0.45}{GW191109_010717}{7.5}{GW200209_085452}{3.8}{GW200115_042309}{0.07}{GW191127_050227}{9.1}{GW200216_220804}{8.5}{GW191215_223052}{1.5}{GW200208_130117}{3.1}{GW200219_094415}{3.8}{GW191103_012549}{0.57}{GW200316_215756}{0.55}{GW200202_154313}{0.20}{GW200129_065458}{2.3}{GW191216_213338}{0.19}}[{\red{???}}]}
\DeclareRobustCommand{\chirpmasssourcegwtcthreemed}[1]{\IfEqCase{#1}{{GW200224_222234}{31.0}{GW191129_134029}{7.28}{GW200311_115853}{26.6}{GW191230_180458}{35.5}{GW191222_033537}{33.8}{GW200225_060421}{14.2}{GW200302_015811}{23.3}{GW200128_022011}{30.6}{GW191204_171526}{8.53}{GW200112_155838}{27.4}{GW200105_162426}{3.42}{GW191105_143521}{7.82}{GW191109_010717}{47.5}{GW200209_085452}{26.1}{GW200115_042309}{2.43}{GW191127_050227}{29.9}{GW200216_220804}{32.9}{GW191215_223052}{18.1}{GW200208_130117}{27.7}{GW200219_094415}{27.6}{GW191103_012549}{8.34}{GW200316_215756}{8.75}{GW200202_154313}{7.49}{GW200129_065458}{27.2}{GW191216_213338}{8.33}}[{\red{???}}]}
\DeclareRobustCommand{\chirpmasssourcegwtcthreeplus}[1]{\IfEqCase{#1}{{GW200224_222234}{3.1}{GW191129_134029}{0.42}{GW200311_115853}{2.4}{GW191230_180458}{7.5}{GW191222_033537}{7.1}{GW200225_060421}{1.5}{GW200302_015811}{4.6}{GW200128_022011}{6.7}{GW191204_171526}{0.38}{GW200112_155838}{2.6}{GW200105_162426}{0.08}{GW191105_143521}{0.61}{GW191109_010717}{9.6}{GW200209_085452}{5.6}{GW200115_042309}{0.05}{GW191127_050227}{11.7}{GW200216_220804}{9.3}{GW191215_223052}{2.2}{GW200208_130117}{3.6}{GW200219_094415}{5.6}{GW191103_012549}{0.66}{GW200316_215756}{0.62}{GW200202_154313}{0.24}{GW200129_065458}{2.1}{GW191216_213338}{0.22}}[{\red{???}}]}
\DeclareRobustCommand{\chirpmasssourcegwtcthreetenthpercentile}[1]{\IfEqCase{#1}{{GW200224_222234}{29.0}{GW191129_134029}{7.07}{GW200311_115853}{25.0}{GW191230_180458}{31.6}{GW191222_033537}{29.7}{GW200225_060421}{13.1}{GW200302_015811}{21.0}{GW200128_022011}{26.8}{GW191204_171526}{8.32}{GW200112_155838}{25.7}{GW200105_162426}{3.36}{GW191105_143521}{7.46}{GW191109_010717}{41.5}{GW200209_085452}{23.1}{GW200115_042309}{2.38}{GW191127_050227}{22.8}{GW200216_220804}{26.2}{GW191215_223052}{16.9}{GW200208_130117}{25.3}{GW200219_094415}{24.6}{GW191103_012549}{7.88}{GW200316_215756}{8.31}{GW200202_154313}{7.33}{GW200129_065458}{25.4}{GW191216_213338}{8.18}}[{\red{???}}]}
\DeclareRobustCommand{\chirpmasssourcegwtcthreenintiethpercentile}[1]{\IfEqCase{#1}{{GW200224_222234}{33.3}{GW191129_134029}{7.62}{GW200311_115853}{28.4}{GW191230_180458}{41.0}{GW191222_033537}{39.4}{GW200225_060421}{15.3}{GW200302_015811}{26.7}{GW200128_022011}{35.6}{GW191204_171526}{8.83}{GW200112_155838}{29.3}{GW200105_162426}{3.48}{GW191105_143521}{8.30}{GW191109_010717}{54.5}{GW200209_085452}{30.2}{GW200115_042309}{2.47}{GW191127_050227}{38.9}{GW200216_220804}{40.0}{GW191215_223052}{19.7}{GW200208_130117}{30.5}{GW200219_094415}{32.0}{GW191103_012549}{8.88}{GW200316_215756}{9.26}{GW200202_154313}{7.68}{GW200129_065458}{28.8}{GW191216_213338}{8.51}}[{\red{???}}]}
\DeclareRobustCommand{\totalmassdetgwtcthreeminus}[1]{\IfEqCase{#1}{{GW200224_222234}{7.2}{GW191129_134029}{0.65}{GW200311_115853}{5.7}{GW191230_180458}{19}{GW191222_033537}{13}{GW200225_060421}{4.0}{GW200302_015811}{8.1}{GW200128_022011}{14}{GW191204_171526}{0.48}{GW200112_155838}{5.1}{GW200105_162426}{1.5}{GW191105_143521}{0.50}{GW191109_010717}{17}{GW200209_085452}{16}{GW200115_042309}{1.8}{GW191127_050227}{45}{GW200216_220804}{32}{GW191215_223052}{3.7}{GW200208_130117}{10}{GW200219_094415}{12}{GW191103_012549}{0.68}{GW200316_215756}{1.1}{GW200202_154313}{0.34}{GW200129_065458}{3.8}{GW191216_213338}{0.66}}[{\red{???}}]}
\DeclareRobustCommand{\totalmassdetgwtcthreemed}[1]{\IfEqCase{#1}{{GW200224_222234}{95.3}{GW191129_134029}{20.11}{GW200311_115853}{75.9}{GW191230_180458}{147}{GW191222_033537}{119}{GW200225_060421}{41.2}{GW200302_015811}{74.5}{GW200128_022011}{118}{GW191204_171526}{22.73}{GW200112_155838}{79.1}{GW200105_162426}{11.6}{GW191105_143521}{22.38}{GW191109_010717}{140}{GW200209_085452}{100}{GW200115_042309}{7.8}{GW191127_050227}{130}{GW200216_220804}{135}{GW191215_223052}{58.6}{GW200208_130117}{91}{GW200219_094415}{103}{GW191103_012549}{23.47}{GW200316_215756}{25.5}{GW200202_154313}{19.01}{GW200129_065458}{74.6}{GW191216_213338}{21.17}}[{\red{???}}]}
\DeclareRobustCommand{\totalmassdetgwtcthreeplus}[1]{\IfEqCase{#1}{{GW200224_222234}{8.4}{GW191129_134029}{2.94}{GW200311_115853}{6.2}{GW191230_180458}{21}{GW191222_033537}{16}{GW200225_060421}{3.0}{GW200302_015811}{15.7}{GW200128_022011}{18}{GW191204_171526}{1.93}{GW200112_155838}{6.5}{GW200105_162426}{1.6}{GW191105_143521}{2.35}{GW191109_010717}{21}{GW200209_085452}{20}{GW200115_042309}{1.8}{GW191127_050227}{53}{GW200216_220804}{30}{GW191215_223052}{5.1}{GW200208_130117}{11}{GW200219_094415}{14}{GW191103_012549}{4.58}{GW200316_215756}{8.8}{GW200202_154313}{1.99}{GW200129_065458}{4.5}{GW191216_213338}{2.96}}[{\red{???}}]}
\DeclareRobustCommand{\totalmassdetgwtcthreetenthpercentile}[1]{\IfEqCase{#1}{{GW200224_222234}{89.6}{GW191129_134029}{19.51}{GW200311_115853}{71.5}{GW191230_180458}{132}{GW191222_033537}{109}{GW200225_060421}{37.9}{GW200302_015811}{68.1}{GW200128_022011}{107}{GW191204_171526}{22.30}{GW200112_155838}{75.2}{GW200105_162426}{10.7}{GW191105_143521}{21.96}{GW191109_010717}{127}{GW200209_085452}{88}{GW200115_042309}{6.2}{GW191127_050227}{91}{GW200216_220804}{109}{GW191215_223052}{55.7}{GW200208_130117}{84}{GW200219_094415}{93}{GW191103_012549}{22.87}{GW200316_215756}{24.5}{GW200202_154313}{18.71}{GW200129_065458}{71.7}{GW191216_213338}{20.56}}[{\red{???}}]}
\DeclareRobustCommand{\totalmassdetgwtcthreenintiethpercentile}[1]{\IfEqCase{#1}{{GW200224_222234}{101.5}{GW191129_134029}{22.18}{GW200311_115853}{80.4}{GW191230_180458}{163}{GW191222_033537}{131}{GW200225_060421}{43.5}{GW200302_015811}{86.5}{GW200128_022011}{131}{GW191204_171526}{24.06}{GW200112_155838}{84.0}{GW200105_162426}{12.5}{GW191105_143521}{23.91}{GW191109_010717}{156}{GW200209_085452}{115}{GW200115_042309}{9.2}{GW191127_050227}{171}{GW200216_220804}{158}{GW191215_223052}{62.1}{GW200208_130117}{100}{GW200219_094415}{113}{GW191103_012549}{26.45}{GW200316_215756}{31.0}{GW200202_154313}{20.33}{GW200129_065458}{78.0}{GW191216_213338}{23.01}}[{\red{???}}]}
\DeclareRobustCommand{\redshiftgwtcthreeminus}[1]{\IfEqCase{#1}{{GW200224_222234}{0.10}{GW191129_134029}{0.06}{GW200311_115853}{0.07}{GW191230_180458}{0.27}{GW191222_033537}{0.26}{GW200225_060421}{0.10}{GW200302_015811}{0.13}{GW200128_022011}{0.29}{GW191204_171526}{0.05}{GW200112_155838}{0.08}{GW200105_162426}{0.02}{GW191105_143521}{0.09}{GW191109_010717}{0.12}{GW200209_085452}{0.25}{GW200115_042309}{0.02}{GW191127_050227}{0.29}{GW200216_220804}{0.29}{GW191215_223052}{0.15}{GW200208_130117}{0.14}{GW200219_094415}{0.22}{GW191103_012549}{0.09}{GW200316_215756}{0.08}{GW200202_154313}{0.03}{GW200129_065458}{0.07}{GW191216_213338}{0.03}}[{\red{???}}]}
\DeclareRobustCommand{\redshiftgwtcthreemed}[1]{\IfEqCase{#1}{{GW200224_222234}{0.33}{GW191129_134029}{0.17}{GW200311_115853}{0.23}{GW191230_180458}{0.76}{GW191222_033537}{0.51}{GW200225_060421}{0.22}{GW200302_015811}{0.31}{GW200128_022011}{0.66}{GW191204_171526}{0.14}{GW200112_155838}{0.24}{GW200105_162426}{0.06}{GW191105_143521}{0.23}{GW191109_010717}{0.25}{GW200209_085452}{0.64}{GW200115_042309}{0.06}{GW191127_050227}{0.57}{GW200216_220804}{0.63}{GW191215_223052}{0.38}{GW200208_130117}{0.40}{GW200219_094415}{0.57}{GW191103_012549}{0.20}{GW200316_215756}{0.22}{GW200202_154313}{0.09}{GW200129_065458}{0.18}{GW191216_213338}{0.07}}[{\red{???}}]}
\DeclareRobustCommand{\redshiftgwtcthreeplus}[1]{\IfEqCase{#1}{{GW200224_222234}{0.07}{GW191129_134029}{0.04}{GW200311_115853}{0.05}{GW191230_180458}{0.26}{GW191222_033537}{0.23}{GW200225_060421}{0.09}{GW200302_015811}{0.17}{GW200128_022011}{0.26}{GW191204_171526}{0.04}{GW200112_155838}{0.07}{GW200105_162426}{0.02}{GW191105_143521}{0.07}{GW191109_010717}{0.18}{GW200209_085452}{0.25}{GW200115_042309}{0.03}{GW191127_050227}{0.40}{GW200216_220804}{0.37}{GW191215_223052}{0.13}{GW200208_130117}{0.15}{GW200219_094415}{0.22}{GW191103_012549}{0.09}{GW200316_215756}{0.08}{GW200202_154313}{0.03}{GW200129_065458}{0.05}{GW191216_213338}{0.02}}[{\red{???}}]}
\DeclareRobustCommand{\redshiftgwtcthreetenthpercentile}[1]{\IfEqCase{#1}{{GW200224_222234}{0.25}{GW191129_134029}{0.11}{GW200311_115853}{0.17}{GW191230_180458}{0.54}{GW191222_033537}{0.31}{GW200225_060421}{0.15}{GW200302_015811}{0.20}{GW200128_022011}{0.43}{GW191204_171526}{0.10}{GW200112_155838}{0.18}{GW200105_162426}{0.04}{GW191105_143521}{0.15}{GW191109_010717}{0.15}{GW200209_085452}{0.44}{GW200115_042309}{0.05}{GW191127_050227}{0.34}{GW200216_220804}{0.39}{GW191215_223052}{0.26}{GW200208_130117}{0.29}{GW200219_094415}{0.39}{GW191103_012549}{0.12}{GW200316_215756}{0.15}{GW200202_154313}{0.06}{GW200129_065458}{0.12}{GW191216_213338}{0.05}}[{\red{???}}]}
\DeclareRobustCommand{\redshiftgwtcthreenintiethpercentile}[1]{\IfEqCase{#1}{{GW200224_222234}{0.39}{GW191129_134029}{0.20}{GW200311_115853}{0.27}{GW191230_180458}{0.96}{GW191222_033537}{0.69}{GW200225_060421}{0.29}{GW200302_015811}{0.44}{GW200128_022011}{0.86}{GW191204_171526}{0.17}{GW200112_155838}{0.30}{GW200105_162426}{0.08}{GW191105_143521}{0.28}{GW191109_010717}{0.38}{GW200209_085452}{0.83}{GW200115_042309}{0.08}{GW191127_050227}{0.87}{GW200216_220804}{0.91}{GW191215_223052}{0.48}{GW200208_130117}{0.51}{GW200219_094415}{0.75}{GW191103_012549}{0.26}{GW200316_215756}{0.28}{GW200202_154313}{0.11}{GW200129_065458}{0.22}{GW191216_213338}{0.09}}[{\red{???}}]}
\DeclareRobustCommand{\geocenttimegwtcthreeminus}[1]{\IfEqCase{#1}{{GW200224_222234}{0.0179}{GW191129_134029}{0.019}{GW200311_115853}{0.0}{GW191230_180458}{0.0}{GW191222_033537}{0.023}{GW200225_060421}{0.0049}{GW200302_015811}{0.0350}{GW200128_022011}{0.0313}{GW191204_171526}{0.0}{GW200112_155838}{0.0358}{GW200105_162426}{0.0017}{GW191105_143521}{0.023}{GW191109_010717}{0.0311}{GW200209_085452}{0.0304}{GW200115_042309}{0.0400}{GW191127_050227}{0.012}{GW200216_220804}{0.0}{GW191215_223052}{0.0439}{GW200208_130117}{0.0032}{GW200219_094415}{0.026}{GW191103_012549}{0.019}{GW200316_215756}{0.0}{GW200202_154313}{0.1}{GW200129_065458}{0.0}{GW191216_213338}{0.0045}}[{\red{???}}]}
\DeclareRobustCommand{\geocenttimegwtcthreemed}[1]{\IfEqCase{#1}{{GW200224_222234}{1266618172.3978}{GW191129_134029}{1259070047.197}{GW200311_115853}{1267963151.4}{GW191230_180458}{1261764316.4}{GW191222_033537}{1261020955.117}{GW200225_060421}{1266645879.4018}{GW200302_015811}{1267149509.5273}{GW200128_022011}{1264213229.9033}{GW191204_171526}{1259514944.1}{GW200112_155838}{1262879936.1035}{GW200105_162426}{1262276684.0349}{GW191105_143521}{1256999739.933}{GW191109_010717}{1257296855.2191}{GW200209_085452}{1265273710.1831}{GW200115_042309}{1263097407.7623}{GW191127_050227}{1258866165.541}{GW200216_220804}{1265926102.9}{GW191215_223052}{1260484270.3546}{GW200208_130117}{1265202095.9383}{GW200219_094415}{1266140673.195}{GW191103_012549}{1256779567.535}{GW200316_215756}{1268431094.2}{GW200202_154313}{1264693411.6}{GW200129_065458}{1264316116.4}{GW191216_213338}{1260567236.4871}}[{\red{???}}]}
\DeclareRobustCommand{\geocenttimegwtcthreeplus}[1]{\IfEqCase{#1}{{GW200224_222234}{0.0080}{GW191129_134029}{0.020}{GW200311_115853}{0.0}{GW191230_180458}{0.0}{GW191222_033537}{0.017}{GW200225_060421}{0.0124}{GW200302_015811}{0.0070}{GW200128_022011}{0.0100}{GW191204_171526}{0.0}{GW200112_155838}{0.0035}{GW200105_162426}{0.0448}{GW191105_143521}{0.015}{GW191109_010717}{0.0016}{GW200209_085452}{0.0067}{GW200115_042309}{0.0043}{GW191127_050227}{0.024}{GW200216_220804}{0.0}{GW191215_223052}{0.0023}{GW200208_130117}{0.0114}{GW200219_094415}{0.015}{GW191103_012549}{0.013}{GW200316_215756}{0.0}{GW200202_154313}{0.0}{GW200129_065458}{0.0}{GW191216_213338}{0.0044}}[{\red{???}}]}
\DeclareRobustCommand{\geocenttimegwtcthreetenthpercentile}[1]{\IfEqCase{#1}{{GW200224_222234}{1266618172.3803}{GW191129_134029}{1259070047.178}{GW200311_115853}{1267963151.4}{GW191230_180458}{1261764316.4}{GW191222_033537}{1261020955.095}{GW200225_060421}{1266645879.3970}{GW200302_015811}{1267149509.4933}{GW200128_022011}{1264213229.8733}{GW191204_171526}{1259514944.1}{GW200112_155838}{1262879936.0709}{GW200105_162426}{1262276684.0332}{GW191105_143521}{1256999739.910}{GW191109_010717}{1257296855.1896}{GW200209_085452}{1265273710.1818}{GW200115_042309}{1263097407.7501}{GW191127_050227}{1258866165.531}{GW200216_220804}{1265926102.9}{GW191215_223052}{1260484270.3114}{GW200208_130117}{1265202095.9356}{GW200219_094415}{1266140673.170}{GW191103_012549}{1256779567.517}{GW200316_215756}{1268431094.2}{GW200202_154313}{1264693411.6}{GW200129_065458}{1264316116.4}{GW191216_213338}{1260567236.4851}}[{\red{???}}]}
\DeclareRobustCommand{\geocenttimegwtcthreenintiethpercentile}[1]{\IfEqCase{#1}{{GW200224_222234}{1266618172.4058}{GW191129_134029}{1259070047.217}{GW200311_115853}{1267963151.4}{GW191230_180458}{1261764316.4}{GW191222_033537}{1261020955.129}{GW200225_060421}{1266645879.4139}{GW200302_015811}{1267149509.5335}{GW200128_022011}{1264213229.9121}{GW191204_171526}{1259514944.1}{GW200112_155838}{1262879936.1064}{GW200105_162426}{1262276684.0789}{GW191105_143521}{1256999739.945}{GW191109_010717}{1257296855.2207}{GW200209_085452}{1265273710.1888}{GW200115_042309}{1263097407.7666}{GW191127_050227}{1258866165.562}{GW200216_220804}{1265926102.9}{GW191215_223052}{1260484270.3569}{GW200208_130117}{1265202095.9497}{GW200219_094415}{1266140673.209}{GW191103_012549}{1256779567.546}{GW200316_215756}{1268431094.2}{GW200202_154313}{1264693411.6}{GW200129_065458}{1264316116.4}{GW191216_213338}{1260567236.4909}}[{\red{???}}]}
\DeclareRobustCommand{\luminositydistancegwtcthreeminus}[1]{\IfEqCase{#1}{{GW200224_222234}{0.63}{GW191129_134029}{0.33}{GW200311_115853}{0.40}{GW191230_180458}{2.0}{GW191222_033537}{1.7}{GW200225_060421}{0.53}{GW200302_015811}{0.77}{GW200128_022011}{2.0}{GW191204_171526}{0.25}{GW200112_155838}{0.46}{GW200105_162426}{0.11}{GW191105_143521}{0.48}{GW191109_010717}{0.65}{GW200209_085452}{1.8}{GW200115_042309}{0.10}{GW191127_050227}{1.9}{GW200216_220804}{2.0}{GW191215_223052}{0.91}{GW200208_130117}{0.85}{GW200219_094415}{1.5}{GW191103_012549}{0.47}{GW200316_215756}{0.44}{GW200202_154313}{0.16}{GW200129_065458}{0.38}{GW191216_213338}{0.13}}[{\red{???}}]}
\DeclareRobustCommand{\luminositydistancegwtcthreemed}[1]{\IfEqCase{#1}{{GW200224_222234}{1.77}{GW191129_134029}{0.82}{GW200311_115853}{1.17}{GW191230_180458}{4.8}{GW191222_033537}{3.0}{GW200225_060421}{1.15}{GW200302_015811}{1.67}{GW200128_022011}{4.0}{GW191204_171526}{0.66}{GW200112_155838}{1.25}{GW200105_162426}{0.27}{GW191105_143521}{1.15}{GW191109_010717}{1.29}{GW200209_085452}{3.9}{GW200115_042309}{0.29}{GW191127_050227}{3.4}{GW200216_220804}{3.8}{GW191215_223052}{2.10}{GW200208_130117}{2.23}{GW200219_094415}{3.4}{GW191103_012549}{0.99}{GW200316_215756}{1.12}{GW200202_154313}{0.41}{GW200129_065458}{0.90}{GW191216_213338}{0.34}}[{\red{???}}]}
\DeclareRobustCommand{\luminositydistancegwtcthreeplus}[1]{\IfEqCase{#1}{{GW200224_222234}{0.47}{GW191129_134029}{0.25}{GW200311_115853}{0.28}{GW191230_180458}{2.1}{GW191222_033537}{1.7}{GW200225_060421}{0.51}{GW200302_015811}{1.09}{GW200128_022011}{2.1}{GW191204_171526}{0.19}{GW200112_155838}{0.43}{GW200105_162426}{0.12}{GW191105_143521}{0.43}{GW191109_010717}{1.13}{GW200209_085452}{2.0}{GW200115_042309}{0.15}{GW191127_050227}{3.1}{GW200216_220804}{3.0}{GW191215_223052}{0.86}{GW200208_130117}{1.00}{GW200219_094415}{1.7}{GW191103_012549}{0.50}{GW200316_215756}{0.47}{GW200202_154313}{0.15}{GW200129_065458}{0.29}{GW191216_213338}{0.12}}[{\red{???}}]}
\DeclareRobustCommand{\luminositydistancegwtcthreetenthpercentile}[1]{\IfEqCase{#1}{{GW200224_222234}{1.29}{GW191129_134029}{0.55}{GW200311_115853}{0.87}{GW191230_180458}{3.2}{GW191222_033537}{1.6}{GW200225_060421}{0.74}{GW200302_015811}{1.03}{GW200128_022011}{2.4}{GW191204_171526}{0.47}{GW200112_155838}{0.89}{GW200105_162426}{0.18}{GW191105_143521}{0.76}{GW191109_010717}{0.76}{GW200209_085452}{2.5}{GW200115_042309}{0.21}{GW191127_050227}{1.8}{GW200216_220804}{2.2}{GW191215_223052}{1.36}{GW200208_130117}{1.54}{GW200219_094415}{2.2}{GW191103_012549}{0.60}{GW200316_215756}{0.76}{GW200202_154313}{0.28}{GW200129_065458}{0.60}{GW191216_213338}{0.23}}[{\red{???}}]}
\DeclareRobustCommand{\luminositydistancegwtcthreenintiethpercentile}[1]{\IfEqCase{#1}{{GW200224_222234}{2.15}{GW191129_134029}{1.02}{GW200311_115853}{1.39}{GW191230_180458}{6.4}{GW191222_033537}{4.3}{GW200225_060421}{1.55}{GW200302_015811}{2.49}{GW200128_022011}{5.7}{GW191204_171526}{0.82}{GW200112_155838}{1.60}{GW200105_162426}{0.36}{GW191105_143521}{1.49}{GW191109_010717}{2.13}{GW200209_085452}{5.4}{GW200115_042309}{0.40}{GW191127_050227}{5.7}{GW200216_220804}{6.0}{GW191215_223052}{2.78}{GW200208_130117}{2.98}{GW200219_094415}{4.7}{GW191103_012549}{1.38}{GW200316_215756}{1.49}{GW200202_154313}{0.53}{GW200129_065458}{1.13}{GW191216_213338}{0.44}}[{\red{???}}]}
\DeclareRobustCommand{\thetajngwtcthreeminus}[1]{\IfEqCase{#1}{{GW200224_222234}{0.42}{GW191129_134029}{1.5}{GW200311_115853}{0.40}{GW191230_180458}{1.85}{GW191222_033537}{1.3}{GW200225_060421}{1.00}{GW200302_015811}{0.98}{GW200128_022011}{1.1}{GW191204_171526}{2.04}{GW200112_155838}{0.68}{GW200105_162426}{1.2}{GW191105_143521}{0.85}{GW191109_010717}{1.18}{GW200209_085452}{1.5}{GW200115_042309}{0.44}{GW191127_050227}{1.2}{GW200216_220804}{0.69}{GW191215_223052}{0.82}{GW200208_130117}{0.58}{GW200219_094415}{0.92}{GW191103_012549}{1.1}{GW200316_215756}{1.84}{GW200202_154313}{0.59}{GW200129_065458}{0.41}{GW191216_213338}{0.81}}[{\red{???}}]}
\DeclareRobustCommand{\thetajngwtcthreemed}[1]{\IfEqCase{#1}{{GW200224_222234}{0.57}{GW191129_134029}{1.8}{GW200311_115853}{0.55}{GW191230_180458}{2.13}{GW191222_033537}{1.6}{GW200225_060421}{1.31}{GW200302_015811}{1.26}{GW200128_022011}{1.4}{GW191204_171526}{2.29}{GW200112_155838}{0.88}{GW200105_162426}{1.5}{GW191105_143521}{1.07}{GW191109_010717}{1.91}{GW200209_085452}{1.8}{GW200115_042309}{0.62}{GW191127_050227}{1.5}{GW200216_220804}{0.89}{GW191215_223052}{1.12}{GW200208_130117}{2.53}{GW200219_094415}{1.19}{GW191103_012549}{1.4}{GW200316_215756}{2.32}{GW200202_154313}{2.57}{GW200129_065458}{0.66}{GW191216_213338}{2.50}}[{\red{???}}]}
\DeclareRobustCommand{\thetajngwtcthreeplus}[1]{\IfEqCase{#1}{{GW200224_222234}{0.54}{GW191129_134029}{1.1}{GW200311_115853}{0.52}{GW191230_180458}{0.79}{GW191222_033537}{1.2}{GW200225_060421}{1.47}{GW200302_015811}{1.55}{GW200128_022011}{1.5}{GW191204_171526}{0.64}{GW200112_155838}{2.04}{GW200105_162426}{1.3}{GW191105_143521}{1.82}{GW191109_010717}{0.87}{GW200209_085452}{1.1}{GW200115_042309}{1.94}{GW191127_050227}{1.4}{GW200216_220804}{1.87}{GW191215_223052}{1.65}{GW200208_130117}{0.44}{GW200219_094415}{1.59}{GW191103_012549}{1.5}{GW200316_215756}{0.58}{GW200202_154313}{0.42}{GW200129_065458}{0.59}{GW191216_213338}{0.45}}[{\red{???}}]}
\DeclareRobustCommand{\thetajngwtcthreetenthpercentile}[1]{\IfEqCase{#1}{{GW200224_222234}{0.22}{GW191129_134029}{0.4}{GW200311_115853}{0.21}{GW191230_180458}{0.40}{GW191222_033537}{0.4}{GW200225_060421}{0.45}{GW200302_015811}{0.40}{GW200128_022011}{0.3}{GW191204_171526}{0.36}{GW200112_155838}{0.28}{GW200105_162426}{0.5}{GW191105_143521}{0.32}{GW191109_010717}{1.02}{GW200209_085452}{0.4}{GW200115_042309}{0.26}{GW191127_050227}{0.4}{GW200216_220804}{0.30}{GW191215_223052}{0.42}{GW200208_130117}{2.09}{GW200219_094415}{0.40}{GW191103_012549}{0.3}{GW200316_215756}{0.71}{GW200202_154313}{2.10}{GW200129_065458}{0.31}{GW191216_213338}{2.00}}[{\red{???}}]}
\DeclareRobustCommand{\thetajngwtcthreenintiethpercentile}[1]{\IfEqCase{#1}{{GW200224_222234}{1.00}{GW191129_134029}{2.8}{GW200311_115853}{0.97}{GW191230_180458}{2.81}{GW191222_033537}{2.7}{GW200225_060421}{2.64}{GW200302_015811}{2.68}{GW200128_022011}{2.8}{GW191204_171526}{2.84}{GW200112_155838}{2.82}{GW200105_162426}{2.7}{GW191105_143521}{2.77}{GW191109_010717}{2.63}{GW200209_085452}{2.8}{GW200115_042309}{2.12}{GW191127_050227}{2.7}{GW200216_220804}{2.57}{GW191215_223052}{2.61}{GW200208_130117}{2.90}{GW200219_094415}{2.60}{GW191103_012549}{2.8}{GW200316_215756}{2.80}{GW200202_154313}{2.92}{GW200129_065458}{1.11}{GW191216_213338}{2.88}}[{\red{???}}]}
\DeclareRobustCommand{\chieffgwtcthreeminus}[1]{\IfEqCase{#1}{{GW200224_222234}{0.15}{GW191129_134029}{0.08}{GW200311_115853}{0.20}{GW191230_180458}{0.30}{GW191222_033537}{0.25}{GW200225_060421}{0.28}{GW200302_015811}{0.25}{GW200128_022011}{0.25}{GW191204_171526}{0.05}{GW200112_155838}{0.15}{GW200105_162426}{0.18}{GW191105_143521}{0.09}{GW191109_010717}{0.31}{GW200209_085452}{0.30}{GW200115_042309}{0.41}{GW191127_050227}{0.36}{GW200216_220804}{0.36}{GW191215_223052}{0.21}{GW200208_130117}{0.27}{GW200219_094415}{0.29}{GW191103_012549}{0.10}{GW200316_215756}{0.10}{GW200202_154313}{0.06}{GW200129_065458}{0.16}{GW191216_213338}{0.06}}[{\red{???}}]}
\DeclareRobustCommand{\chieffgwtcthreemed}[1]{\IfEqCase{#1}{{GW200224_222234}{0.11}{GW191129_134029}{0.06}{GW200311_115853}{-0.02}{GW191230_180458}{-0.03}{GW191222_033537}{-0.04}{GW200225_060421}{-0.12}{GW200302_015811}{0.03}{GW200128_022011}{0.14}{GW191204_171526}{0.16}{GW200112_155838}{0.06}{GW200105_162426}{0.00}{GW191105_143521}{-0.02}{GW191109_010717}{-0.29}{GW200209_085452}{-0.10}{GW200115_042309}{-0.15}{GW191127_050227}{0.18}{GW200216_220804}{0.10}{GW191215_223052}{-0.03}{GW200208_130117}{-0.07}{GW200219_094415}{-0.08}{GW191103_012549}{0.21}{GW200316_215756}{0.13}{GW200202_154313}{0.04}{GW200129_065458}{0.11}{GW191216_213338}{0.11}}[{\red{???}}]}
\DeclareRobustCommand{\chieffgwtcthreeplus}[1]{\IfEqCase{#1}{{GW200224_222234}{0.15}{GW191129_134029}{0.16}{GW200311_115853}{0.16}{GW191230_180458}{0.26}{GW191222_033537}{0.20}{GW200225_060421}{0.17}{GW200302_015811}{0.26}{GW200128_022011}{0.24}{GW191204_171526}{0.08}{GW200112_155838}{0.15}{GW200105_162426}{0.13}{GW191105_143521}{0.13}{GW191109_010717}{0.42}{GW200209_085452}{0.24}{GW200115_042309}{0.24}{GW191127_050227}{0.34}{GW200216_220804}{0.34}{GW191215_223052}{0.17}{GW200208_130117}{0.22}{GW200219_094415}{0.23}{GW191103_012549}{0.16}{GW200316_215756}{0.27}{GW200202_154313}{0.13}{GW200129_065458}{0.11}{GW191216_213338}{0.13}}[{\red{???}}]}
\DeclareRobustCommand{\chieffgwtcthreetenthpercentile}[1]{\IfEqCase{#1}{{GW200224_222234}{-0.01}{GW191129_134029}{0.00}{GW200311_115853}{-0.17}{GW191230_180458}{-0.26}{GW191222_033537}{-0.23}{GW200225_060421}{-0.34}{GW200302_015811}{-0.15}{GW200128_022011}{-0.05}{GW191204_171526}{0.12}{GW200112_155838}{-0.05}{GW200105_162426}{-0.10}{GW191105_143521}{-0.09}{GW191109_010717}{-0.54}{GW200209_085452}{-0.33}{GW200115_042309}{-0.51}{GW191127_050227}{-0.10}{GW200216_220804}{-0.17}{GW191215_223052}{-0.19}{GW200208_130117}{-0.27}{GW200219_094415}{-0.30}{GW191103_012549}{0.13}{GW200316_215756}{0.04}{GW200202_154313}{-0.01}{GW200129_065458}{0.00}{GW191216_213338}{0.06}}[{\red{???}}]}
\DeclareRobustCommand{\chieffgwtcthreenintiethpercentile}[1]{\IfEqCase{#1}{{GW200224_222234}{0.22}{GW191129_134029}{0.19}{GW200311_115853}{0.10}{GW191230_180458}{0.18}{GW191222_033537}{0.11}{GW200225_060421}{0.02}{GW200302_015811}{0.23}{GW200128_022011}{0.33}{GW191204_171526}{0.22}{GW200112_155838}{0.18}{GW200105_162426}{0.08}{GW191105_143521}{0.07}{GW191109_010717}{0.00}{GW200209_085452}{0.09}{GW200115_042309}{0.04}{GW191127_050227}{0.45}{GW200216_220804}{0.36}{GW191215_223052}{0.10}{GW200208_130117}{0.09}{GW200219_094415}{0.10}{GW191103_012549}{0.33}{GW200316_215756}{0.32}{GW200202_154313}{0.13}{GW200129_065458}{0.20}{GW191216_213338}{0.20}}[{\red{???}}]}
\DeclareRobustCommand{\spinonegwtcthreeminus}[1]{\IfEqCase{#1}{{GW200224_222234}{0.42}{GW191129_134029}{0.22}{GW200311_115853}{0.36}{GW191230_180458}{0.46}{GW191222_033537}{0.34}{GW200225_060421}{0.51}{GW200302_015811}{0.35}{GW200128_022011}{0.53}{GW191204_171526}{0.35}{GW200112_155838}{0.31}{GW200105_162426}{0.07}{GW191105_143521}{0.21}{GW191109_010717}{0.58}{GW200209_085452}{0.46}{GW200115_042309}{0.29}{GW191127_050227}{0.58}{GW200216_220804}{0.43}{GW191215_223052}{0.42}{GW200208_130117}{0.32}{GW200219_094415}{0.43}{GW191103_012549}{0.40}{GW200316_215756}{0.28}{GW200202_154313}{0.20}{GW200129_065458}{0.47}{GW191216_213338}{0.22}}[{\red{???}}]}
\DeclareRobustCommand{\spinonegwtcthreemed}[1]{\IfEqCase{#1}{{GW200224_222234}{0.47}{GW191129_134029}{0.25}{GW200311_115853}{0.39}{GW191230_180458}{0.51}{GW191222_033537}{0.38}{GW200225_060421}{0.59}{GW200302_015811}{0.38}{GW200128_022011}{0.60}{GW191204_171526}{0.40}{GW200112_155838}{0.34}{GW200105_162426}{0.08}{GW191105_143521}{0.23}{GW191109_010717}{0.83}{GW200209_085452}{0.52}{GW200115_042309}{0.32}{GW191127_050227}{0.66}{GW200216_220804}{0.48}{GW191215_223052}{0.47}{GW200208_130117}{0.36}{GW200219_094415}{0.47}{GW191103_012549}{0.46}{GW200316_215756}{0.32}{GW200202_154313}{0.22}{GW200129_065458}{0.53}{GW191216_213338}{0.24}}[{\red{???}}]}
\DeclareRobustCommand{\spinonegwtcthreeplus}[1]{\IfEqCase{#1}{{GW200224_222234}{0.44}{GW191129_134029}{0.37}{GW200311_115853}{0.48}{GW191230_180458}{0.44}{GW191222_033537}{0.50}{GW200225_060421}{0.35}{GW200302_015811}{0.51}{GW200128_022011}{0.36}{GW191204_171526}{0.38}{GW200112_155838}{0.46}{GW200105_162426}{0.30}{GW191105_143521}{0.53}{GW191109_010717}{0.15}{GW200209_085452}{0.43}{GW200115_042309}{0.50}{GW191127_050227}{0.31}{GW200216_220804}{0.46}{GW191215_223052}{0.45}{GW200208_130117}{0.51}{GW200219_094415}{0.46}{GW191103_012549}{0.40}{GW200316_215756}{0.37}{GW200202_154313}{0.45}{GW200129_065458}{0.42}{GW191216_213338}{0.36}}[{\red{???}}]}
\DeclareRobustCommand{\spinonegwtcthreetenthpercentile}[1]{\IfEqCase{#1}{{GW200224_222234}{0.10}{GW191129_134029}{0.05}{GW200311_115853}{0.07}{GW191230_180458}{0.10}{GW191222_033537}{0.07}{GW200225_060421}{0.14}{GW200302_015811}{0.07}{GW200128_022011}{0.14}{GW191204_171526}{0.10}{GW200112_155838}{0.07}{GW200105_162426}{0.01}{GW191105_143521}{0.04}{GW191109_010717}{0.42}{GW200209_085452}{0.10}{GW200115_042309}{0.05}{GW191127_050227}{0.16}{GW200216_220804}{0.10}{GW191215_223052}{0.09}{GW200208_130117}{0.07}{GW200219_094415}{0.09}{GW191103_012549}{0.13}{GW200316_215756}{0.08}{GW200202_154313}{0.04}{GW200129_065458}{0.12}{GW191216_213338}{0.05}}[{\red{???}}]}
\DeclareRobustCommand{\spinonegwtcthreenintiethpercentile}[1]{\IfEqCase{#1}{{GW200224_222234}{0.84}{GW191129_134029}{0.54}{GW200311_115853}{0.79}{GW191230_180458}{0.90}{GW191222_033537}{0.80}{GW200225_060421}{0.89}{GW200302_015811}{0.81}{GW200128_022011}{0.92}{GW191204_171526}{0.70}{GW200112_155838}{0.71}{GW200105_162426}{0.27}{GW191105_143521}{0.64}{GW191109_010717}{0.97}{GW200209_085452}{0.90}{GW200115_042309}{0.73}{GW191127_050227}{0.94}{GW200216_220804}{0.89}{GW191215_223052}{0.85}{GW200208_130117}{0.78}{GW200219_094415}{0.88}{GW191103_012549}{0.78}{GW200316_215756}{0.61}{GW200202_154313}{0.55}{GW200129_065458}{0.93}{GW191216_213338}{0.50}}[{\red{???}}]}
\DeclareRobustCommand{\cosiotagwtcthreeminus}[1]{\IfEqCase{#1}{{GW200224_222234}{0.43}{GW191129_134029}{0.78}{GW200311_115853}{0.37}{GW191230_180458}{0.41}{GW191222_033537}{0.91}{GW200225_060421}{1.15}{GW200302_015811}{1.24}{GW200128_022011}{1.13}{GW191204_171526}{0.33}{GW200112_155838}{1.59}{GW200105_162426}{0.97}{GW191105_143521}{1.44}{GW191109_010717}{0.55}{GW200209_085452}{0.83}{GW200115_042309}{1.64}{GW191127_050227}{1.00}{GW200216_220804}{1.50}{GW191215_223052}{1.28}{GW200208_130117}{0.17}{GW200219_094415}{1.26}{GW191103_012549}{1.17}{GW200316_215756}{0.29}{GW200202_154313}{0.15}{GW200129_065458}{0.50}{GW191216_213338}{0.18}}[{\red{???}}]}
\DeclareRobustCommand{\cosiotagwtcthreemed}[1]{\IfEqCase{#1}{{GW200224_222234}{0.84}{GW191129_134029}{-0.19}{GW200311_115853}{0.86}{GW191230_180458}{-0.57}{GW191222_033537}{-0.05}{GW200225_060421}{0.21}{GW200302_015811}{0.30}{GW200128_022011}{0.16}{GW191204_171526}{-0.65}{GW200112_155838}{0.62}{GW200105_162426}{0.03}{GW191105_143521}{0.47}{GW191109_010717}{-0.40}{GW200209_085452}{-0.14}{GW200115_042309}{0.81}{GW191127_050227}{0.06}{GW200216_220804}{0.57}{GW191215_223052}{0.36}{GW200208_130117}{-0.81}{GW200219_094415}{0.33}{GW191103_012549}{0.20}{GW200316_215756}{-0.68}{GW200202_154313}{-0.84}{GW200129_065458}{0.80}{GW191216_213338}{-0.80}}[{\red{???}}]}
\DeclareRobustCommand{\cosiotagwtcthreeplus}[1]{\IfEqCase{#1}{{GW200224_222234}{0.15}{GW191129_134029}{1.16}{GW200311_115853}{0.13}{GW191230_180458}{1.53}{GW191222_033537}{1.00}{GW200225_060421}{0.74}{GW200302_015811}{0.66}{GW200128_022011}{0.80}{GW191204_171526}{1.62}{GW200112_155838}{0.36}{GW200105_162426}{0.91}{GW191105_143521}{0.50}{GW191109_010717}{1.24}{GW200209_085452}{1.11}{GW200115_042309}{0.18}{GW191127_050227}{0.88}{GW200216_220804}{0.40}{GW191215_223052}{0.60}{GW200208_130117}{0.49}{GW200219_094415}{0.63}{GW191103_012549}{0.78}{GW200316_215756}{1.57}{GW200202_154313}{0.45}{GW200129_065458}{0.19}{GW191216_213338}{0.68}}[{\red{???}}]}
\DeclareRobustCommand{\cosiotagwtcthreetenthpercentile}[1]{\IfEqCase{#1}{{GW200224_222234}{0.52}{GW191129_134029}{-0.94}{GW200311_115853}{0.57}{GW191230_180458}{-0.95}{GW191222_033537}{-0.91}{GW200225_060421}{-0.87}{GW200302_015811}{-0.89}{GW200128_022011}{-0.92}{GW191204_171526}{-0.96}{GW200112_155838}{-0.94}{GW200105_162426}{-0.89}{GW191105_143521}{-0.93}{GW191109_010717}{-0.90}{GW200209_085452}{-0.93}{GW200115_042309}{-0.52}{GW191127_050227}{-0.87}{GW200216_220804}{-0.82}{GW191215_223052}{-0.85}{GW200208_130117}{-0.97}{GW200219_094415}{-0.85}{GW191103_012549}{-0.94}{GW200316_215756}{-0.95}{GW200202_154313}{-0.98}{GW200129_065458}{0.43}{GW191216_213338}{-0.97}}[{\red{???}}]}
\DeclareRobustCommand{\cosiotagwtcthreenintiethpercentile}[1]{\IfEqCase{#1}{{GW200224_222234}{0.98}{GW191129_134029}{0.94}{GW200311_115853}{0.98}{GW191230_180458}{0.93}{GW191222_033537}{0.90}{GW200225_060421}{0.90}{GW200302_015811}{0.92}{GW200128_022011}{0.92}{GW191204_171526}{0.94}{GW200112_155838}{0.96}{GW200105_162426}{0.89}{GW191105_143521}{0.95}{GW191109_010717}{0.68}{GW200209_085452}{0.93}{GW200115_042309}{0.97}{GW191127_050227}{0.88}{GW200216_220804}{0.94}{GW191215_223052}{0.91}{GW200208_130117}{-0.47}{GW200219_094415}{0.92}{GW191103_012549}{0.94}{GW200316_215756}{0.75}{GW200202_154313}{-0.50}{GW200129_065458}{0.97}{GW191216_213338}{-0.41}}[{\red{???}}]}
\DeclareRobustCommand{\radiatedenergygwtcthreeminus}[1]{\IfEqCase{#1}{{GW200224_222234}{0.67}{GW191129_134029}{0.116}{GW200311_115853}{0.57}{GW191230_180458}{1.0}{GW191222_033537}{1.00}{GW200225_060421}{0.39}{GW200302_015811}{0.79}{GW200128_022011}{0.96}{GW191204_171526}{0.106}{GW200112_155838}{0.53}{GW200105_162426}{0.026}{GW191105_143521}{0.118}{GW191109_010717}{1.3}{GW200209_085452}{0.70}{GW200115_042309}{0.029}{GW191127_050227}{2.1}{GW200216_220804}{2.0}{GW191215_223052}{0.32}{GW200208_130117}{0.77}{GW200219_094415}{0.81}{GW191103_012549}{0.17}{GW200316_215756}{0.22}{GW200202_154313}{0.094}{GW200129_065458}{0.86}{GW191216_213338}{0.130}}[{\red{???}}]}
\DeclareRobustCommand{\radiatedenergygwtcthreemed}[1]{\IfEqCase{#1}{{GW200224_222234}{3.61}{GW191129_134029}{0.778}{GW200311_115853}{2.87}{GW191230_180458}{3.8}{GW191222_033537}{3.57}{GW200225_060421}{1.43}{GW200302_015811}{2.35}{GW200128_022011}{3.61}{GW191204_171526}{0.991}{GW200112_155838}{3.08}{GW200105_162426}{0.204}{GW191105_143521}{0.818}{GW191109_010717}{4.3}{GW200209_085452}{2.68}{GW200115_042309}{0.146}{GW191127_050227}{3.0}{GW200216_220804}{3.5}{GW191215_223052}{1.88}{GW200208_130117}{2.83}{GW200219_094415}{2.84}{GW191103_012549}{0.98}{GW200316_215756}{0.94}{GW200202_154313}{0.814}{GW200129_065458}{3.18}{GW191216_213338}{0.921}}[{\red{???}}]}
\DeclareRobustCommand{\radiatedenergygwtcthreeplus}[1]{\IfEqCase{#1}{{GW200224_222234}{0.67}{GW191129_134029}{0.072}{GW200311_115853}{0.52}{GW191230_180458}{1.2}{GW191222_033537}{1.04}{GW200225_060421}{0.27}{GW200302_015811}{1.10}{GW200128_022011}{1.26}{GW191204_171526}{0.069}{GW200112_155838}{0.59}{GW200105_162426}{0.032}{GW191105_143521}{0.088}{GW191109_010717}{2.3}{GW200209_085452}{0.83}{GW200115_042309}{0.045}{GW191127_050227}{2.6}{GW200216_220804}{2.0}{GW191215_223052}{0.37}{GW200208_130117}{0.77}{GW200219_094415}{0.91}{GW191103_012549}{0.13}{GW200316_215756}{0.11}{GW200202_154313}{0.047}{GW200129_065458}{0.42}{GW191216_213338}{0.057}}[{\red{???}}]}
\DeclareRobustCommand{\radiatedenergygwtcthreetenthpercentile}[1]{\IfEqCase{#1}{{GW200224_222234}{3.09}{GW191129_134029}{0.690}{GW200311_115853}{2.44}{GW191230_180458}{3.0}{GW191222_033537}{2.84}{GW200225_060421}{1.12}{GW200302_015811}{1.72}{GW200128_022011}{2.85}{GW191204_171526}{0.911}{GW200112_155838}{2.67}{GW200105_162426}{0.187}{GW191105_143521}{0.731}{GW191109_010717}{3.3}{GW200209_085452}{2.14}{GW200115_042309}{0.124}{GW191127_050227}{1.3}{GW200216_220804}{1.9}{GW191215_223052}{1.63}{GW200208_130117}{2.24}{GW200219_094415}{2.23}{GW191103_012549}{0.85}{GW200316_215756}{0.79}{GW200202_154313}{0.748}{GW200129_065458}{2.58}{GW191216_213338}{0.829}}[{\red{???}}]}
\DeclareRobustCommand{\radiatedenergygwtcthreenintiethpercentile}[1]{\IfEqCase{#1}{{GW200224_222234}{4.11}{GW191129_134029}{0.835}{GW200311_115853}{3.27}{GW191230_180458}{4.7}{GW191222_033537}{4.39}{GW200225_060421}{1.64}{GW200302_015811}{3.18}{GW200128_022011}{4.56}{GW191204_171526}{1.045}{GW200112_155838}{3.52}{GW200105_162426}{0.225}{GW191105_143521}{0.887}{GW191109_010717}{5.8}{GW200209_085452}{3.30}{GW200115_042309}{0.183}{GW191127_050227}{5.0}{GW200216_220804}{5.0}{GW191215_223052}{2.15}{GW200208_130117}{3.42}{GW200219_094415}{3.53}{GW191103_012549}{1.08}{GW200316_215756}{1.03}{GW200202_154313}{0.851}{GW200129_065458}{3.53}{GW191216_213338}{0.967}}[{\red{???}}]}
\DeclareRobustCommand{\costhetajngwtcthreeminus}[1]{\IfEqCase{#1}{{GW200224_222234}{0.40}{GW191129_134029}{0.78}{GW200311_115853}{0.37}{GW191230_180458}{0.45}{GW191222_033537}{0.91}{GW200225_060421}{1.19}{GW200302_015811}{1.25}{GW200128_022011}{1.19}{GW191204_171526}{0.32}{GW200112_155838}{1.61}{GW200105_162426}{0.98}{GW191105_143521}{1.45}{GW191109_010717}{0.60}{GW200209_085452}{0.75}{GW200115_042309}{1.65}{GW191127_050227}{1.07}{GW200216_220804}{1.56}{GW191215_223052}{1.37}{GW200208_130117}{0.17}{GW200219_094415}{1.30}{GW191103_012549}{1.16}{GW200316_215756}{0.29}{GW200202_154313}{0.15}{GW200129_065458}{0.47}{GW191216_213338}{0.18}}[{\red{???}}]}
\DeclareRobustCommand{\costhetajngwtcthreemed}[1]{\IfEqCase{#1}{{GW200224_222234}{0.84}{GW191129_134029}{-0.19}{GW200311_115853}{0.85}{GW191230_180458}{-0.53}{GW191222_033537}{-0.05}{GW200225_060421}{0.25}{GW200302_015811}{0.30}{GW200128_022011}{0.22}{GW191204_171526}{-0.66}{GW200112_155838}{0.64}{GW200105_162426}{0.03}{GW191105_143521}{0.48}{GW191109_010717}{-0.33}{GW200209_085452}{-0.22}{GW200115_042309}{0.81}{GW191127_050227}{0.11}{GW200216_220804}{0.63}{GW191215_223052}{0.44}{GW200208_130117}{-0.82}{GW200219_094415}{0.37}{GW191103_012549}{0.19}{GW200316_215756}{-0.68}{GW200202_154313}{-0.84}{GW200129_065458}{0.79}{GW191216_213338}{-0.80}}[{\red{???}}]}
\DeclareRobustCommand{\costhetajngwtcthreeplus}[1]{\IfEqCase{#1}{{GW200224_222234}{0.15}{GW191129_134029}{1.16}{GW200311_115853}{0.14}{GW191230_180458}{1.49}{GW191222_033537}{1.01}{GW200225_060421}{0.70}{GW200302_015811}{0.66}{GW200128_022011}{0.75}{GW191204_171526}{1.63}{GW200112_155838}{0.34}{GW200105_162426}{0.92}{GW191105_143521}{0.50}{GW191109_010717}{1.07}{GW200209_085452}{1.19}{GW200115_042309}{0.17}{GW191127_050227}{0.85}{GW200216_220804}{0.35}{GW191215_223052}{0.52}{GW200208_130117}{0.44}{GW200219_094415}{0.59}{GW191103_012549}{0.78}{GW200316_215756}{1.57}{GW200202_154313}{0.45}{GW200129_065458}{0.18}{GW191216_213338}{0.68}}[{\red{???}}]}
\DeclareRobustCommand{\costhetajngwtcthreetenthpercentile}[1]{\IfEqCase{#1}{{GW200224_222234}{0.54}{GW191129_134029}{-0.94}{GW200311_115853}{0.57}{GW191230_180458}{-0.95}{GW191222_033537}{-0.92}{GW200225_060421}{-0.88}{GW200302_015811}{-0.89}{GW200128_022011}{-0.94}{GW191204_171526}{-0.96}{GW200112_155838}{-0.95}{GW200105_162426}{-0.90}{GW191105_143521}{-0.93}{GW191109_010717}{-0.87}{GW200209_085452}{-0.94}{GW200115_042309}{-0.52}{GW191127_050227}{-0.92}{GW200216_220804}{-0.84}{GW191215_223052}{-0.86}{GW200208_130117}{-0.97}{GW200219_094415}{-0.85}{GW191103_012549}{-0.94}{GW200316_215756}{-0.94}{GW200202_154313}{-0.98}{GW200129_065458}{0.45}{GW191216_213338}{-0.97}}[{\red{???}}]}
\DeclareRobustCommand{\costhetajngwtcthreenintiethpercentile}[1]{\IfEqCase{#1}{{GW200224_222234}{0.98}{GW191129_134029}{0.94}{GW200311_115853}{0.98}{GW191230_180458}{0.92}{GW191222_033537}{0.91}{GW200225_060421}{0.90}{GW200302_015811}{0.92}{GW200128_022011}{0.94}{GW191204_171526}{0.94}{GW200112_155838}{0.96}{GW200105_162426}{0.90}{GW191105_143521}{0.95}{GW191109_010717}{0.52}{GW200209_085452}{0.93}{GW200115_042309}{0.97}{GW191127_050227}{0.91}{GW200216_220804}{0.95}{GW191215_223052}{0.91}{GW200208_130117}{-0.50}{GW200219_094415}{0.92}{GW191103_012549}{0.94}{GW200316_215756}{0.76}{GW200202_154313}{-0.50}{GW200129_065458}{0.95}{GW191216_213338}{-0.41}}[{\red{???}}]}
\DeclareRobustCommand{\totalmasssourcegwtcthreeminus}[1]{\IfEqCase{#1}{{GW200224_222234}{5.0}{GW191129_134029}{1.1}{GW200311_115853}{4.2}{GW191230_180458}{11}{GW191222_033537}{11}{GW200225_060421}{3.0}{GW200302_015811}{6.7}{GW200128_022011}{11}{GW191204_171526}{0.93}{GW200112_155838}{4.6}{GW200105_162426}{1.4}{GW191105_143521}{1.3}{GW191109_010717}{16}{GW200209_085452}{8.6}{GW200115_042309}{1.6}{GW191127_050227}{22}{GW200216_220804}{14}{GW191215_223052}{4.0}{GW200208_130117}{6.8}{GW200219_094415}{8.2}{GW191103_012549}{1.8}{GW200316_215756}{2.0}{GW200202_154313}{0.67}{GW200129_065458}{3.6}{GW191216_213338}{0.93}}[{\red{???}}]}
\DeclareRobustCommand{\totalmasssourcegwtcthreemed}[1]{\IfEqCase{#1}{{GW200224_222234}{71.9}{GW191129_134029}{17.5}{GW200311_115853}{61.9}{GW191230_180458}{83}{GW191222_033537}{79}{GW200225_060421}{33.5}{GW200302_015811}{57.3}{GW200128_022011}{72}{GW191204_171526}{20.14}{GW200112_155838}{63.9}{GW200105_162426}{11.0}{GW191105_143521}{18.5}{GW191109_010717}{112}{GW200209_085452}{61.1}{GW200115_042309}{7.3}{GW191127_050227}{80}{GW200216_220804}{81}{GW191215_223052}{42.6}{GW200208_130117}{65.4}{GW200219_094415}{65.0}{GW191103_012549}{20.0}{GW200316_215756}{21.2}{GW200202_154313}{17.58}{GW200129_065458}{63.4}{GW191216_213338}{19.80}}[{\red{???}}]}
\DeclareRobustCommand{\totalmasssourcegwtcthreeplus}[1]{\IfEqCase{#1}{{GW200224_222234}{6.8}{GW191129_134029}{2.4}{GW200311_115853}{5.3}{GW191230_180458}{17}{GW191222_033537}{16}{GW200225_060421}{3.6}{GW200302_015811}{9.6}{GW200128_022011}{15}{GW191204_171526}{1.70}{GW200112_155838}{5.7}{GW200105_162426}{1.5}{GW191105_143521}{2.1}{GW191109_010717}{20}{GW200209_085452}{12.8}{GW200115_042309}{1.7}{GW191127_050227}{39}{GW200216_220804}{20}{GW191215_223052}{5.4}{GW200208_130117}{7.8}{GW200219_094415}{12.6}{GW191103_012549}{3.7}{GW200316_215756}{7.2}{GW200202_154313}{1.78}{GW200129_065458}{4.3}{GW191216_213338}{2.72}}[{\red{???}}]}
\DeclareRobustCommand{\totalmasssourcegwtcthreetenthpercentile}[1]{\IfEqCase{#1}{{GW200224_222234}{67.9}{GW191129_134029}{16.5}{GW200311_115853}{58.5}{GW191230_180458}{75}{GW191222_033537}{70}{GW200225_060421}{31.1}{GW200302_015811}{51.9}{GW200128_022011}{63}{GW191204_171526}{19.37}{GW200112_155838}{60.1}{GW200105_162426}{10.1}{GW191105_143521}{17.4}{GW191109_010717}{99}{GW200209_085452}{54.2}{GW200115_042309}{5.8}{GW191127_050227}{61}{GW200216_220804}{70}{GW191215_223052}{39.4}{GW200208_130117}{60.0}{GW200219_094415}{58.3}{GW191103_012549}{18.5}{GW200316_215756}{19.5}{GW200202_154313}{17.04}{GW200129_065458}{60.5}{GW191216_213338}{19.02}}[{\red{???}}]}
\DeclareRobustCommand{\totalmasssourcegwtcthreenintiethpercentile}[1]{\IfEqCase{#1}{{GW200224_222234}{77.1}{GW191129_134029}{19.2}{GW200311_115853}{65.9}{GW191230_180458}{96}{GW191222_033537}{92}{GW200225_060421}{36.2}{GW200302_015811}{64.3}{GW200128_022011}{83}{GW191204_171526}{21.36}{GW200112_155838}{68.1}{GW200105_162426}{11.8}{GW191105_143521}{19.9}{GW191109_010717}{126}{GW200209_085452}{70.6}{GW200115_042309}{8.6}{GW191127_050227}{108}{GW200216_220804}{96}{GW191215_223052}{46.8}{GW200208_130117}{71.4}{GW200219_094415}{74.7}{GW191103_012549}{22.4}{GW200316_215756}{25.6}{GW200202_154313}{18.75}{GW200129_065458}{66.7}{GW191216_213338}{21.49}}[{\red{???}}]}
\DeclareRobustCommand{\phijlgwtcthreeminus}[1]{\IfEqCase{#1}{{GW200224_222234}{2.6}{GW191129_134029}{2.8}{GW200311_115853}{2.3}{GW191230_180458}{3.0}{GW191222_033537}{2.8}{GW200225_060421}{2.8}{GW200302_015811}{2.8}{GW200128_022011}{2.9}{GW191204_171526}{2.8}{GW200112_155838}{2.8}{GW200105_162426}{2.8}{GW191105_143521}{2.8}{GW191109_010717}{3.1}{GW200209_085452}{2.8}{GW200115_042309}{2.7}{GW191127_050227}{2.7}{GW200216_220804}{3.0}{GW191215_223052}{2.9}{GW200208_130117}{2.8}{GW200219_094415}{2.8}{GW191103_012549}{2.8}{GW200316_215756}{3.0}{GW200202_154313}{2.8}{GW200129_065458}{2.0}{GW191216_213338}{3.0}}[{\red{???}}]}
\DeclareRobustCommand{\phijlgwtcthreemed}[1]{\IfEqCase{#1}{{GW200224_222234}{3.0}{GW191129_134029}{3.1}{GW200311_115853}{2.7}{GW191230_180458}{3.4}{GW191222_033537}{3.1}{GW200225_060421}{3.1}{GW200302_015811}{3.1}{GW200128_022011}{3.2}{GW191204_171526}{3.2}{GW200112_155838}{3.1}{GW200105_162426}{3.2}{GW191105_143521}{3.1}{GW191109_010717}{3.5}{GW200209_085452}{3.1}{GW200115_042309}{3.0}{GW191127_050227}{3.1}{GW200216_220804}{3.4}{GW191215_223052}{3.2}{GW200208_130117}{3.2}{GW200219_094415}{3.1}{GW191103_012549}{3.1}{GW200316_215756}{3.3}{GW200202_154313}{3.1}{GW200129_065458}{2.7}{GW191216_213338}{3.3}}[{\red{???}}]}
\DeclareRobustCommand{\phijlgwtcthreeplus}[1]{\IfEqCase{#1}{{GW200224_222234}{2.8}{GW191129_134029}{2.8}{GW200311_115853}{3.3}{GW191230_180458}{2.6}{GW191222_033537}{2.8}{GW200225_060421}{2.9}{GW200302_015811}{2.8}{GW200128_022011}{2.8}{GW191204_171526}{2.8}{GW200112_155838}{2.9}{GW200105_162426}{2.8}{GW191105_143521}{2.8}{GW191109_010717}{2.4}{GW200209_085452}{2.8}{GW200115_042309}{3.0}{GW191127_050227}{2.8}{GW200216_220804}{2.5}{GW191215_223052}{2.8}{GW200208_130117}{2.7}{GW200219_094415}{2.8}{GW191103_012549}{2.9}{GW200316_215756}{2.6}{GW200202_154313}{2.9}{GW200129_065458}{3.1}{GW191216_213338}{2.7}}[{\red{???}}]}
\DeclareRobustCommand{\phijlgwtcthreetenthpercentile}[1]{\IfEqCase{#1}{{GW200224_222234}{0.8}{GW191129_134029}{0.6}{GW200311_115853}{0.6}{GW191230_180458}{0.7}{GW191222_033537}{0.7}{GW200225_060421}{0.6}{GW200302_015811}{0.6}{GW200128_022011}{0.6}{GW191204_171526}{0.6}{GW200112_155838}{0.6}{GW200105_162426}{0.7}{GW191105_143521}{0.6}{GW191109_010717}{0.8}{GW200209_085452}{0.7}{GW200115_042309}{0.7}{GW191127_050227}{0.7}{GW200216_220804}{0.8}{GW191215_223052}{0.6}{GW200208_130117}{0.7}{GW200219_094415}{0.8}{GW191103_012549}{0.7}{GW200316_215756}{0.7}{GW200202_154313}{0.6}{GW200129_065458}{1.0}{GW191216_213338}{0.6}}[{\red{???}}]}
\DeclareRobustCommand{\phijlgwtcthreenintiethpercentile}[1]{\IfEqCase{#1}{{GW200224_222234}{5.1}{GW191129_134029}{5.6}{GW200311_115853}{5.6}{GW191230_180458}{5.7}{GW191222_033537}{5.7}{GW200225_060421}{5.7}{GW200302_015811}{5.6}{GW200128_022011}{5.7}{GW191204_171526}{5.7}{GW200112_155838}{5.7}{GW200105_162426}{5.7}{GW191105_143521}{5.6}{GW191109_010717}{5.5}{GW200209_085452}{5.6}{GW200115_042309}{5.6}{GW191127_050227}{5.6}{GW200216_220804}{5.7}{GW191215_223052}{5.7}{GW200208_130117}{5.6}{GW200219_094415}{5.6}{GW191103_012549}{5.7}{GW200316_215756}{5.6}{GW200202_154313}{5.7}{GW200129_065458}{5.4}{GW191216_213338}{5.7}}[{\red{???}}]}
\DeclareRobustCommand{\masstwodetgwtcthreeminus}[1]{\IfEqCase{#1}{{GW200224_222234}{9.7}{GW191129_134029}{1.9}{GW200311_115853}{7.3}{GW191230_180458}{20}{GW191222_033537}{15}{GW200225_060421}{4.6}{GW200302_015811}{8.1}{GW200128_022011}{13}{GW191204_171526}{1.8}{GW200112_155838}{7.4}{GW200105_162426}{0.25}{GW191105_143521}{2.2}{GW191109_010717}{17}{GW200209_085452}{13}{GW200115_042309}{0.30}{GW191127_050227}{25}{GW200216_220804}{31}{GW191215_223052}{5.1}{GW200208_130117}{11.2}{GW200219_094415}{14.1}{GW191103_012549}{2.9}{GW200316_215756}{3.5}{GW200202_154313}{1.9}{GW200129_065458}{10.8}{GW191216_213338}{2.0}}[{\red{???}}]}
\DeclareRobustCommand{\masstwodetgwtcthreemed}[1]{\IfEqCase{#1}{{GW200224_222234}{43.0}{GW191129_134029}{7.8}{GW200311_115853}{34.0}{GW191230_180458}{64}{GW191222_033537}{52}{GW200225_060421}{17.2}{GW200302_015811}{26.5}{GW200128_022011}{52}{GW191204_171526}{9.3}{GW200112_155838}{35.2}{GW200105_162426}{2.02}{GW191105_143521}{9.4}{GW191109_010717}{60}{GW200209_085452}{44}{GW200115_042309}{1.53}{GW191127_050227}{38}{GW200216_220804}{51}{GW191215_223052}{24.5}{GW200208_130117}{38.5}{GW200219_094415}{44.4}{GW191103_012549}{9.4}{GW200316_215756}{9.5}{GW200202_154313}{8.0}{GW200129_065458}{34.1}{GW191216_213338}{8.2}}[{\red{???}}]}
\DeclareRobustCommand{\masstwodetgwtcthreeplus}[1]{\IfEqCase{#1}{{GW200224_222234}{5.8}{GW191129_134029}{1.7}{GW200311_115853}{4.7}{GW191230_180458}{14}{GW191222_033537}{11}{GW200225_060421}{3.0}{GW200302_015811}{12.5}{GW200128_022011}{10}{GW191204_171526}{1.5}{GW200112_155838}{5.1}{GW200105_162426}{0.35}{GW191105_143521}{1.4}{GW191109_010717}{16}{GW200209_085452}{11}{GW200115_042309}{0.91}{GW191127_050227}{31}{GW200216_220804}{23}{GW191215_223052}{4.1}{GW200208_130117}{8.6}{GW200219_094415}{9.3}{GW191103_012549}{1.8}{GW200316_215756}{2.3}{GW200202_154313}{1.2}{GW200129_065458}{3.3}{GW191216_213338}{1.7}}[{\red{???}}]}
\DeclareRobustCommand{\masstwodetgwtcthreetenthpercentile}[1]{\IfEqCase{#1}{{GW200224_222234}{35.7}{GW191129_134029}{6.2}{GW200311_115853}{28.6}{GW191230_180458}{48}{GW191222_033537}{41}{GW200225_060421}{13.7}{GW200302_015811}{19.8}{GW200128_022011}{41}{GW191204_171526}{7.8}{GW200112_155838}{29.5}{GW200105_162426}{1.87}{GW191105_143521}{7.6}{GW191109_010717}{46}{GW200209_085452}{34}{GW200115_042309}{1.29}{GW191127_050227}{17}{GW200216_220804}{24}{GW191215_223052}{20.5}{GW200208_130117}{29.6}{GW200219_094415}{33.7}{GW191103_012549}{7.2}{GW200316_215756}{6.8}{GW200202_154313}{6.5}{GW200129_065458}{25.3}{GW191216_213338}{6.7}}[{\red{???}}]}
\DeclareRobustCommand{\masstwodetgwtcthreenintiethpercentile}[1]{\IfEqCase{#1}{{GW200224_222234}{47.8}{GW191129_134029}{9.3}{GW200311_115853}{37.7}{GW191230_180458}{75}{GW191222_033537}{61}{GW200225_060421}{19.8}{GW200302_015811}{36.3}{GW200128_022011}{60}{GW191204_171526}{10.6}{GW200112_155838}{39.3}{GW200105_162426}{2.20}{GW191105_143521}{10.6}{GW191109_010717}{72}{GW200209_085452}{52}{GW200115_042309}{2.25}{GW191127_050227}{63}{GW200216_220804}{70}{GW191215_223052}{28.0}{GW200208_130117}{45.5}{GW200219_094415}{51.8}{GW191103_012549}{11.0}{GW200316_215756}{11.5}{GW200202_154313}{9.1}{GW200129_065458}{36.8}{GW191216_213338}{9.8}}[{\red{???}}]}
\DeclareRobustCommand{\ragwtcthreeminus}[1]{\IfEqCase{#1}{{GW200224_222234}{0.048}{GW191129_134029}{2.77}{GW200311_115853}{0.029}{GW191230_180458}{0.29}{GW191222_033537}{3.0}{GW200225_060421}{0.30}{GW200302_015811}{3.13}{GW200128_022011}{3.0}{GW191204_171526}{0.53}{GW200112_155838}{2.9}{GW200105_162426}{1.3}{GW191105_143521}{0.36}{GW191109_010717}{1.4}{GW200209_085452}{0.92}{GW200115_042309}{0.10}{GW191127_050227}{1.1}{GW200216_220804}{1.49}{GW191215_223052}{0.76}{GW200208_130117}{0.039}{GW200219_094415}{0.12}{GW191103_012549}{1.85}{GW200316_215756}{0.35}{GW200202_154313}{0.117}{GW200129_065458}{0.15}{GW191216_213338}{3.518}}[{\red{???}}]}
\DeclareRobustCommand{\ragwtcthreemed}[1]{\IfEqCase{#1}{{GW200224_222234}{3.050}{GW191129_134029}{5.59}{GW200311_115853}{0.038}{GW191230_180458}{1.07}{GW191222_033537}{3.6}{GW200225_060421}{1.92}{GW200302_015811}{3.83}{GW200128_022011}{3.8}{GW191204_171526}{1.27}{GW200112_155838}{3.5}{GW200105_162426}{2.0}{GW191105_143521}{0.44}{GW191109_010717}{3.7}{GW200209_085452}{2.51}{GW200115_042309}{0.74}{GW191127_050227}{1.2}{GW200216_220804}{5.31}{GW191215_223052}{2.63}{GW200208_130117}{2.438}{GW200219_094415}{0.39}{GW191103_012549}{4.35}{GW200316_215756}{1.51}{GW200202_154313}{2.523}{GW200129_065458}{5.56}{GW191216_213338}{5.557}}[{\red{???}}]}
\DeclareRobustCommand{\ragwtcthreeplus}[1]{\IfEqCase{#1}{{GW200224_222234}{0.041}{GW191129_134029}{0.42}{GW200311_115853}{0.041}{GW191230_180458}{3.73}{GW191222_033537}{1.8}{GW200225_060421}{3.25}{GW200302_015811}{0.96}{GW200128_022011}{1.1}{GW191204_171526}{1.91}{GW200112_155838}{1.7}{GW200105_162426}{3.0}{GW191105_143521}{5.68}{GW191109_010717}{1.2}{GW200209_085452}{0.63}{GW200115_042309}{4.06}{GW191127_050227}{4.9}{GW200216_220804}{0.28}{GW191215_223052}{3.28}{GW200208_130117}{0.040}{GW200219_094415}{2.85}{GW191103_012549}{0.39}{GW200316_215756}{2.01}{GW200202_154313}{0.058}{GW200129_065458}{0.47}{GW191216_213338}{0.066}}[{\red{???}}]}
\DeclareRobustCommand{\ragwtcthreetenthpercentile}[1]{\IfEqCase{#1}{{GW200224_222234}{3.015}{GW191129_134029}{3.01}{GW200311_115853}{0.014}{GW191230_180458}{0.84}{GW191222_033537}{1.1}{GW200225_060421}{1.71}{GW200302_015811}{0.89}{GW200128_022011}{0.9}{GW191204_171526}{0.82}{GW200112_155838}{0.6}{GW200105_162426}{0.8}{GW191105_143521}{0.12}{GW191109_010717}{2.4}{GW200209_085452}{2.22}{GW200115_042309}{0.66}{GW191127_050227}{0.1}{GW200216_220804}{3.96}{GW191215_223052}{1.93}{GW200208_130117}{2.409}{GW200219_094415}{0.30}{GW191103_012549}{2.56}{GW200316_215756}{1.21}{GW200202_154313}{2.425}{GW200129_065458}{5.45}{GW191216_213338}{5.260}}[{\red{???}}]}
\DeclareRobustCommand{\ragwtcthreenintiethpercentile}[1]{\IfEqCase{#1}{{GW200224_222234}{3.082}{GW191129_134029}{5.92}{GW200311_115853}{0.068}{GW191230_180458}{4.65}{GW191222_033537}{4.7}{GW200225_060421}{2.42}{GW200302_015811}{4.59}{GW200128_022011}{4.5}{GW191204_171526}{2.05}{GW200112_155838}{5.1}{GW200105_162426}{4.9}{GW191105_143521}{6.07}{GW191109_010717}{4.6}{GW200209_085452}{2.91}{GW200115_042309}{4.65}{GW191127_050227}{5.0}{GW200216_220804}{5.54}{GW191215_223052}{5.77}{GW200208_130117}{2.468}{GW200219_094415}{3.06}{GW191103_012549}{4.71}{GW200316_215756}{3.42}{GW200202_154313}{2.569}{GW200129_065458}{5.59}{GW191216_213338}{5.615}}[{\red{???}}]}
\DeclareRobustCommand{\finalmassdetgwtcthreeminus}[1]{\IfEqCase{#1}{{GW200224_222234}{6.4}{GW191129_134029}{0.67}{GW200311_115853}{5.1}{GW191230_180458}{17}{GW191222_033537}{12}{GW200225_060421}{3.6}{GW200302_015811}{7.6}{GW200128_022011}{12}{GW191204_171526}{0.50}{GW200112_155838}{4.6}{GW200105_162426}{1.8}{GW191105_143521}{0.47}{GW191109_010717}{15}{GW200209_085452}{14}{GW200115_042309}{1.7}{GW191127_050227}{43}{GW200216_220804}{30}{GW191215_223052}{3.3}{GW200208_130117}{9.1}{GW200219_094415}{11}{GW191103_012549}{0.66}{GW200316_215756}{1.1}{GW200202_154313}{0.35}{GW200129_065458}{3.4}{GW191216_213338}{0.70}}[{\red{???}}]}
\DeclareRobustCommand{\finalmassdetgwtcthreemed}[1]{\IfEqCase{#1}{{GW200224_222234}{90.5}{GW191129_134029}{19.20}{GW200311_115853}{72.4}{GW191230_180458}{140}{GW191222_033537}{114}{GW200225_060421}{39.4}{GW200302_015811}{71.6}{GW200128_022011}{112}{GW191204_171526}{21.60}{GW200112_155838}{75.3}{GW200105_162426}{11.4}{GW191105_143521}{21.36}{GW191109_010717}{135}{GW200209_085452}{96}{GW200115_042309}{7.6}{GW191127_050227}{124}{GW200216_220804}{129}{GW191215_223052}{55.9}{GW200208_130117}{87.5}{GW200219_094415}{98}{GW191103_012549}{22.27}{GW200316_215756}{24.4}{GW200202_154313}{18.12}{GW200129_065458}{70.9}{GW191216_213338}{20.18}}[{\red{???}}]}
\DeclareRobustCommand{\finalmassdetgwtcthreeplus}[1]{\IfEqCase{#1}{{GW200224_222234}{7.6}{GW191129_134029}{3.08}{GW200311_115853}{5.6}{GW191230_180458}{20}{GW191222_033537}{14}{GW200225_060421}{2.9}{GW200302_015811}{14.1}{GW200128_022011}{16}{GW191204_171526}{2.04}{GW200112_155838}{5.8}{GW200105_162426}{2.1}{GW191105_143521}{2.48}{GW191109_010717}{19}{GW200209_085452}{19}{GW200115_042309}{2.3}{GW191127_050227}{52}{GW200216_220804}{27}{GW191215_223052}{5.0}{GW200208_130117}{10.3}{GW200219_094415}{13}{GW191103_012549}{4.79}{GW200316_215756}{9.0}{GW200202_154313}{2.09}{GW200129_065458}{4.2}{GW191216_213338}{3.10}}[{\red{???}}]}
\DeclareRobustCommand{\finalmassdetgwtcthreetenthpercentile}[1]{\IfEqCase{#1}{{GW200224_222234}{85.4}{GW191129_134029}{18.57}{GW200311_115853}{68.4}{GW191230_180458}{127}{GW191222_033537}{105}{GW200225_060421}{36.5}{GW200302_015811}{65.5}{GW200128_022011}{102}{GW191204_171526}{21.14}{GW200112_155838}{71.8}{GW200105_162426}{10.2}{GW191105_143521}{20.95}{GW191109_010717}{123}{GW200209_085452}{84}{GW200115_042309}{6.1}{GW191127_050227}{88}{GW200216_220804}{106}{GW191215_223052}{53.3}{GW200208_130117}{80.3}{GW200219_094415}{89}{GW191103_012549}{21.68}{GW200316_215756}{23.3}{GW200202_154313}{17.80}{GW200129_065458}{68.3}{GW191216_213338}{19.53}}[{\red{???}}]}
\DeclareRobustCommand{\finalmassdetgwtcthreenintiethpercentile}[1]{\IfEqCase{#1}{{GW200224_222234}{96.2}{GW191129_134029}{21.37}{GW200311_115853}{76.5}{GW191230_180458}{155}{GW191222_033537}{125}{GW200225_060421}{41.6}{GW200302_015811}{82.3}{GW200128_022011}{124}{GW191204_171526}{23.02}{GW200112_155838}{79.7}{GW200105_162426}{12.7}{GW191105_143521}{22.99}{GW191109_010717}{149}{GW200209_085452}{110}{GW200115_042309}{9.3}{GW191127_050227}{164}{GW200216_220804}{150}{GW191215_223052}{59.4}{GW200208_130117}{95.3}{GW200219_094415}{108}{GW191103_012549}{25.37}{GW200316_215756}{30.0}{GW200202_154313}{19.51}{GW200129_065458}{74.2}{GW191216_213338}{22.11}}[{\red{???}}]}
\DeclareRobustCommand{\spinonexgwtcthreeminus}[1]{\IfEqCase{#1}{{GW200224_222234}{0.59}{GW191129_134029}{0.36}{GW200311_115853}{0.55}{GW191230_180458}{0.64}{GW191222_033537}{0.53}{GW200225_060421}{0.65}{GW200302_015811}{0.55}{GW200128_022011}{0.68}{GW191204_171526}{0.50}{GW200112_155838}{0.51}{GW200105_162426}{0.14}{GW191105_143521}{0.41}{GW191109_010717}{0.70}{GW200209_085452}{0.64}{GW200115_042309}{0.33}{GW191127_050227}{0.70}{GW200216_220804}{0.60}{GW191215_223052}{0.64}{GW200208_130117}{0.47}{GW200219_094415}{0.61}{GW191103_012549}{0.53}{GW200316_215756}{0.40}{GW200202_154313}{0.37}{GW200129_065458}{0.81}{GW191216_213338}{0.29}}[{\red{???}}]}
\DeclareRobustCommand{\spinonexgwtcthreemed}[1]{\IfEqCase{#1}{{GW200224_222234}{0.01}{GW191129_134029}{0.00}{GW200311_115853}{0.00}{GW191230_180458}{0.00}{GW191222_033537}{0.00}{GW200225_060421}{0.00}{GW200302_015811}{0.00}{GW200128_022011}{0.00}{GW191204_171526}{0.00}{GW200112_155838}{0.00}{GW200105_162426}{0.00}{GW191105_143521}{0.00}{GW191109_010717}{0.00}{GW200209_085452}{0.00}{GW200115_042309}{0.00}{GW191127_050227}{0.00}{GW200216_220804}{0.00}{GW191215_223052}{0.00}{GW200208_130117}{0.00}{GW200219_094415}{0.00}{GW191103_012549}{0.00}{GW200316_215756}{0.00}{GW200202_154313}{0.00}{GW200129_065458}{-0.02}{GW191216_213338}{0.00}}[{\red{???}}]}
\DeclareRobustCommand{\spinonexgwtcthreeplus}[1]{\IfEqCase{#1}{{GW200224_222234}{0.61}{GW191129_134029}{0.37}{GW200311_115853}{0.54}{GW191230_180458}{0.65}{GW191222_033537}{0.54}{GW200225_060421}{0.65}{GW200302_015811}{0.55}{GW200128_022011}{0.67}{GW191204_171526}{0.48}{GW200112_155838}{0.48}{GW200105_162426}{0.13}{GW191105_143521}{0.42}{GW191109_010717}{0.68}{GW200209_085452}{0.65}{GW200115_042309}{0.33}{GW191127_050227}{0.67}{GW200216_220804}{0.59}{GW191215_223052}{0.63}{GW200208_130117}{0.52}{GW200219_094415}{0.61}{GW191103_012549}{0.51}{GW200316_215756}{0.38}{GW200202_154313}{0.37}{GW200129_065458}{0.65}{GW191216_213338}{0.29}}[{\red{???}}]}
\DeclareRobustCommand{\spinonexgwtcthreetenthpercentile}[1]{\IfEqCase{#1}{{GW200224_222234}{-0.43}{GW191129_134029}{-0.25}{GW200311_115853}{-0.42}{GW191230_180458}{-0.49}{GW191222_033537}{-0.39}{GW200225_060421}{-0.52}{GW200302_015811}{-0.41}{GW200128_022011}{-0.53}{GW191204_171526}{-0.39}{GW200112_155838}{-0.37}{GW200105_162426}{-0.09}{GW191105_143521}{-0.27}{GW191109_010717}{-0.56}{GW200209_085452}{-0.48}{GW200115_042309}{-0.24}{GW191127_050227}{-0.54}{GW200216_220804}{-0.46}{GW191215_223052}{-0.50}{GW200208_130117}{-0.34}{GW200219_094415}{-0.45}{GW191103_012549}{-0.39}{GW200316_215756}{-0.27}{GW200202_154313}{-0.25}{GW200129_065458}{-0.72}{GW191216_213338}{-0.21}}[{\red{???}}]}
\DeclareRobustCommand{\spinonexgwtcthreenintiethpercentile}[1]{\IfEqCase{#1}{{GW200224_222234}{0.49}{GW191129_134029}{0.26}{GW200311_115853}{0.38}{GW191230_180458}{0.49}{GW191222_033537}{0.38}{GW200225_060421}{0.52}{GW200302_015811}{0.40}{GW200128_022011}{0.52}{GW191204_171526}{0.36}{GW200112_155838}{0.35}{GW200105_162426}{0.08}{GW191105_143521}{0.29}{GW191109_010717}{0.57}{GW200209_085452}{0.49}{GW200115_042309}{0.25}{GW191127_050227}{0.52}{GW200216_220804}{0.44}{GW191215_223052}{0.48}{GW200208_130117}{0.37}{GW200219_094415}{0.45}{GW191103_012549}{0.38}{GW200316_215756}{0.27}{GW200202_154313}{0.24}{GW200129_065458}{0.46}{GW191216_213338}{0.20}}[{\red{???}}]}
\DeclareRobustCommand{\loglikelihoodgwtcthreeminus}[1]{\IfEqCase{#1}{{GW200224_222234}{5.0}{GW191129_134029}{5.3}{GW200311_115853}{4.7}{GW191230_180458}{4.4}{GW191222_033537}{4.3}{GW200225_060421}{5.0}{GW200302_015811}{4.6}{GW200128_022011}{4.5}{GW191204_171526}{5.1}{GW200112_155838}{4.9}{GW200105_162426}{5.5}{GW191105_143521}{5.9}{GW191109_010717}{6.7}{GW200209_085452}{4.7}{GW200115_042309}{6.4}{GW191127_050227}{5.3}{GW200216_220804}{4.2}{GW191215_223052}{5.0}{GW200208_130117}{5.2}{GW200219_094415}{5.1}{GW191103_012549}{5.1}{GW200316_215756}{6.6}{GW200202_154313}{5.7}{GW200129_065458}{21.5}{GW191216_213338}{6.7}}[{\red{???}}]}
\DeclareRobustCommand{\loglikelihoodgwtcthreemed}[1]{\IfEqCase{#1}{{GW200224_222234}{188.2}{GW191129_134029}{73.6}{GW200311_115853}{148.0}{GW191230_180458}{46.2}{GW191222_033537}{68.9}{GW200225_060421}{66.4}{GW200302_015811}{47.0}{GW200128_022011}{48.7}{GW191204_171526}{138.2}{GW200112_155838}{183.8}{GW200105_162426}{82.7}{GW191105_143521}{34.7}{GW191109_010717}{138.1}{GW200209_085452}{37.1}{GW200115_042309}{47.8}{GW191127_050227}{32.1}{GW200216_220804}{25.0}{GW191215_223052}{51.9}{GW200208_130117}{49.4}{GW200219_094415}{48.6}{GW191103_012549}{27.5}{GW200316_215756}{40.3}{GW200202_154313}{44.9}{GW200129_065458}{343.3}{GW191216_213338}{156.6}}[{\red{???}}]}
\DeclareRobustCommand{\loglikelihoodgwtcthreeplus}[1]{\IfEqCase{#1}{{GW200224_222234}{3.3}{GW191129_134029}{5.1}{GW200311_115853}{4.6}{GW191230_180458}{2.8}{GW191222_033537}{2.8}{GW200225_060421}{3.8}{GW200302_015811}{4.0}{GW200128_022011}{3.5}{GW191204_171526}{4.4}{GW200112_155838}{3.4}{GW200105_162426}{4.0}{GW191105_143521}{5.4}{GW191109_010717}{7.4}{GW200209_085452}{3.2}{GW200115_042309}{8.1}{GW191127_050227}{4.9}{GW200216_220804}{3.1}{GW191215_223052}{4.2}{GW200208_130117}{2.9}{GW200219_094415}{3.4}{GW191103_012549}{4.8}{GW200316_215756}{4.8}{GW200202_154313}{6.9}{GW200129_065458}{6.3}{GW191216_213338}{8.0}}[{\red{???}}]}
\DeclareRobustCommand{\loglikelihoodgwtcthreetenthpercentile}[1]{\IfEqCase{#1}{{GW200224_222234}{184.6}{GW191129_134029}{69.5}{GW200311_115853}{144.4}{GW191230_180458}{43.0}{GW191222_033537}{65.7}{GW200225_060421}{62.7}{GW200302_015811}{43.7}{GW200128_022011}{45.3}{GW191204_171526}{134.4}{GW200112_155838}{180.1}{GW200105_162426}{78.6}{GW191105_143521}{30.2}{GW191109_010717}{132.9}{GW200209_085452}{33.6}{GW200115_042309}{42.9}{GW191127_050227}{28.0}{GW200216_220804}{21.9}{GW191215_223052}{48.2}{GW200208_130117}{45.7}{GW200219_094415}{44.8}{GW191103_012549}{23.6}{GW200316_215756}{35.4}{GW200202_154313}{40.4}{GW200129_065458}{330.2}{GW191216_213338}{151.5}}[{\red{???}}]}
\DeclareRobustCommand{\loglikelihoodgwtcthreenintiethpercentile}[1]{\IfEqCase{#1}{{GW200224_222234}{190.9}{GW191129_134029}{77.9}{GW200311_115853}{151.8}{GW191230_180458}{48.5}{GW191222_033537}{71.1}{GW200225_060421}{69.4}{GW200302_015811}{50.1}{GW200128_022011}{51.4}{GW191204_171526}{141.8}{GW200112_155838}{186.6}{GW200105_162426}{85.8}{GW191105_143521}{39.1}{GW191109_010717}{144.1}{GW200209_085452}{39.7}{GW200115_042309}{54.8}{GW191127_050227}{35.9}{GW200216_220804}{27.5}{GW191215_223052}{55.2}{GW200208_130117}{51.8}{GW200219_094415}{51.4}{GW191103_012549}{31.6}{GW200316_215756}{44.2}{GW200202_154313}{51.1}{GW200129_065458}{348.4}{GW191216_213338}{163.5}}[{\red{???}}]}
\DeclareRobustCommand{\tilttwogwtcthreeminus}[1]{\IfEqCase{#1}{{GW200224_222234}{0.92}{GW191129_134029}{0.89}{GW200311_115853}{1.0}{GW191230_180458}{1.1}{GW191222_033537}{1.1}{GW200225_060421}{1.18}{GW200302_015811}{1.0}{GW200128_022011}{0.95}{GW191204_171526}{0.77}{GW200112_155838}{0.91}{GW200105_162426}{1.0}{GW191105_143521}{1.07}{GW191109_010717}{1.16}{GW200209_085452}{1.17}{GW200115_042309}{1.27}{GW191127_050227}{0.98}{GW200216_220804}{1.0}{GW191215_223052}{1.09}{GW200208_130117}{1.17}{GW200219_094415}{1.1}{GW191103_012549}{0.77}{GW200316_215756}{0.83}{GW200202_154313}{0.90}{GW200129_065458}{0.84}{GW191216_213338}{0.83}}[{\red{???}}]}
\DeclareRobustCommand{\tilttwogwtcthreemed}[1]{\IfEqCase{#1}{{GW200224_222234}{1.38}{GW191129_134029}{1.27}{GW200311_115853}{1.6}{GW191230_180458}{1.7}{GW191222_033537}{1.7}{GW200225_060421}{1.79}{GW200302_015811}{1.4}{GW200128_022011}{1.42}{GW191204_171526}{1.11}{GW200112_155838}{1.32}{GW200105_162426}{1.5}{GW191105_143521}{1.60}{GW191109_010717}{1.84}{GW200209_085452}{1.84}{GW200115_042309}{1.89}{GW191127_050227}{1.35}{GW200216_220804}{1.4}{GW191215_223052}{1.66}{GW200208_130117}{1.72}{GW200219_094415}{1.7}{GW191103_012549}{1.07}{GW200316_215756}{1.19}{GW200202_154313}{1.34}{GW200129_065458}{1.15}{GW191216_213338}{1.16}}[{\red{???}}]}
\DeclareRobustCommand{\tilttwogwtcthreeplus}[1]{\IfEqCase{#1}{{GW200224_222234}{1.16}{GW191129_134029}{1.19}{GW200311_115853}{1.0}{GW191230_180458}{1.0}{GW191222_033537}{1.0}{GW200225_060421}{0.97}{GW200302_015811}{1.1}{GW200128_022011}{1.15}{GW191204_171526}{1.12}{GW200112_155838}{1.14}{GW200105_162426}{1.1}{GW191105_143521}{0.99}{GW191109_010717}{0.92}{GW200209_085452}{0.93}{GW200115_042309}{0.93}{GW191127_050227}{1.22}{GW200216_220804}{1.2}{GW191215_223052}{0.99}{GW200208_130117}{0.99}{GW200219_094415}{1.0}{GW191103_012549}{1.29}{GW200316_215756}{1.13}{GW200202_154313}{1.07}{GW200129_065458}{1.36}{GW191216_213338}{1.28}}[{\red{???}}]}
\DeclareRobustCommand{\tilttwogwtcthreetenthpercentile}[1]{\IfEqCase{#1}{{GW200224_222234}{0.63}{GW191129_134029}{0.52}{GW200311_115853}{0.8}{GW191230_180458}{0.8}{GW191222_033537}{0.8}{GW200225_060421}{0.87}{GW200302_015811}{0.6}{GW200128_022011}{0.66}{GW191204_171526}{0.47}{GW200112_155838}{0.60}{GW200105_162426}{0.7}{GW191105_143521}{0.74}{GW191109_010717}{0.91}{GW200209_085452}{0.92}{GW200115_042309}{0.86}{GW191127_050227}{0.53}{GW200216_220804}{0.6}{GW191215_223052}{0.79}{GW200208_130117}{0.79}{GW200219_094415}{0.8}{GW191103_012549}{0.43}{GW200316_215756}{0.51}{GW200202_154313}{0.62}{GW200129_065458}{0.45}{GW191216_213338}{0.47}}[{\red{???}}]}
\DeclareRobustCommand{\tilttwogwtcthreenintiethpercentile}[1]{\IfEqCase{#1}{{GW200224_222234}{2.30}{GW191129_134029}{2.17}{GW200311_115853}{2.4}{GW191230_180458}{2.5}{GW191222_033537}{2.5}{GW200225_060421}{2.59}{GW200302_015811}{2.4}{GW200128_022011}{2.34}{GW191204_171526}{1.96}{GW200112_155838}{2.23}{GW200105_162426}{2.4}{GW191105_143521}{2.39}{GW191109_010717}{2.59}{GW200209_085452}{2.60}{GW200115_042309}{2.69}{GW191127_050227}{2.34}{GW200216_220804}{2.4}{GW191215_223052}{2.46}{GW200208_130117}{2.54}{GW200219_094415}{2.6}{GW191103_012549}{2.09}{GW200316_215756}{2.05}{GW200202_154313}{2.16}{GW200129_065458}{2.23}{GW191216_213338}{2.15}}[{\red{???}}]}
\DeclareRobustCommand{\tiltonegwtcthreeminus}[1]{\IfEqCase{#1}{{GW200224_222234}{0.84}{GW191129_134029}{0.87}{GW200311_115853}{1.02}{GW191230_180458}{1.0}{GW191222_033537}{1.10}{GW200225_060421}{0.90}{GW200302_015811}{0.97}{GW200128_022011}{0.79}{GW191204_171526}{0.72}{GW200112_155838}{0.95}{GW200105_162426}{1.1}{GW191105_143521}{1.05}{GW191109_010717}{0.90}{GW200209_085452}{1.01}{GW200115_042309}{1.37}{GW191127_050227}{0.84}{GW200216_220804}{0.92}{GW191215_223052}{0.96}{GW200208_130117}{1.15}{GW200219_094415}{1.06}{GW191103_012549}{0.68}{GW200316_215756}{0.80}{GW200202_154313}{0.94}{GW200129_065458}{0.92}{GW191216_213338}{0.73}}[{\red{???}}]}
\DeclareRobustCommand{\tiltonegwtcthreemed}[1]{\IfEqCase{#1}{{GW200224_222234}{1.27}{GW191129_134029}{1.29}{GW200311_115853}{1.66}{GW191230_180458}{1.6}{GW191222_033537}{1.73}{GW200225_060421}{1.88}{GW200302_015811}{1.51}{GW200128_022011}{1.20}{GW191204_171526}{1.13}{GW200112_155838}{1.43}{GW200105_162426}{1.6}{GW191105_143521}{1.64}{GW191109_010717}{2.22}{GW200209_085452}{1.76}{GW200115_042309}{2.24}{GW191127_050227}{1.19}{GW200216_220804}{1.33}{GW191215_223052}{1.63}{GW200208_130117}{1.82}{GW200219_094415}{1.80}{GW191103_012549}{1.00}{GW200316_215756}{1.08}{GW200202_154313}{1.39}{GW200129_065458}{1.45}{GW191216_213338}{1.01}}[{\red{???}}]}
\DeclareRobustCommand{\tiltonegwtcthreeplus}[1]{\IfEqCase{#1}{{GW200224_222234}{1.03}{GW191129_134029}{1.01}{GW200311_115853}{0.97}{GW191230_180458}{1.0}{GW191222_033537}{0.94}{GW200225_060421}{0.80}{GW200302_015811}{1.06}{GW200128_022011}{1.03}{GW191204_171526}{0.91}{GW200112_155838}{1.02}{GW200105_162426}{1.2}{GW191105_143521}{0.95}{GW191109_010717}{0.67}{GW200209_085452}{0.91}{GW200115_042309}{0.66}{GW191127_050227}{1.06}{GW200216_220804}{1.14}{GW191215_223052}{0.83}{GW200208_130117}{0.93}{GW200219_094415}{0.92}{GW191103_012549}{0.98}{GW200316_215756}{1.07}{GW200202_154313}{1.06}{GW200129_065458}{1.04}{GW191216_213338}{1.20}}[{\red{???}}]}
\DeclareRobustCommand{\tiltonegwtcthreetenthpercentile}[1]{\IfEqCase{#1}{{GW200224_222234}{0.60}{GW191129_134029}{0.59}{GW200311_115853}{0.89}{GW191230_180458}{0.8}{GW191222_033537}{0.88}{GW200225_060421}{1.24}{GW200302_015811}{0.75}{GW200128_022011}{0.59}{GW191204_171526}{0.56}{GW200112_155838}{0.67}{GW200105_162426}{0.6}{GW191105_143521}{0.82}{GW191109_010717}{1.55}{GW200209_085452}{0.98}{GW200115_042309}{1.15}{GW191127_050227}{0.50}{GW200216_220804}{0.57}{GW191215_223052}{0.92}{GW200208_130117}{0.91}{GW200219_094415}{0.98}{GW191103_012549}{0.44}{GW200316_215756}{0.40}{GW200202_154313}{0.64}{GW200129_065458}{0.70}{GW191216_213338}{0.39}}[{\red{???}}]}
\DeclareRobustCommand{\tiltonegwtcthreenintiethpercentile}[1]{\IfEqCase{#1}{{GW200224_222234}{2.03}{GW191129_134029}{2.03}{GW200311_115853}{2.42}{GW191230_180458}{2.4}{GW191222_033537}{2.49}{GW200225_060421}{2.51}{GW200302_015811}{2.35}{GW200128_022011}{1.96}{GW191204_171526}{1.76}{GW200112_155838}{2.22}{GW200105_162426}{2.5}{GW191105_143521}{2.38}{GW191109_010717}{2.77}{GW200209_085452}{2.49}{GW200115_042309}{2.79}{GW191127_050227}{1.96}{GW200216_220804}{2.23}{GW191215_223052}{2.27}{GW200208_130117}{2.58}{GW200219_094415}{2.56}{GW191103_012549}{1.69}{GW200316_215756}{1.86}{GW200202_154313}{2.20}{GW200129_065458}{2.18}{GW191216_213338}{1.89}}[{\red{???}}]}
\DeclareRobustCommand{\psigwtcthreeminus}[1]{\IfEqCase{#1}{{GW200224_222234}{1.6}{GW191129_134029}{1.9}{GW200311_115853}{1.5}{GW191230_180458}{1.7}{GW191222_033537}{0.99}{GW200225_060421}{1.5}{GW200302_015811}{1.3}{GW200128_022011}{1.4}{GW191204_171526}{1.5}{GW200112_155838}{1.5}{GW200105_162426}{2.3}{GW191105_143521}{1.4}{GW191109_010717}{1.9}{GW200209_085452}{1.1}{GW200115_042309}{2.1}{GW191127_050227}{1.4}{GW200216_220804}{1.4}{GW191215_223052}{1.4}{GW200208_130117}{1.1}{GW200219_094415}{1.4}{GW191103_012549}{1.5}{GW200316_215756}{1.5}{GW200202_154313}{1.3}{GW200129_065458}{0.99}{GW191216_213338}{1.2}}[{\red{???}}]}
\DeclareRobustCommand{\psigwtcthreemed}[1]{\IfEqCase{#1}{{GW200224_222234}{1.7}{GW191129_134029}{2.1}{GW200311_115853}{1.8}{GW191230_180458}{1.9}{GW191222_033537}{1.11}{GW200225_060421}{1.7}{GW200302_015811}{1.4}{GW200128_022011}{1.6}{GW191204_171526}{1.6}{GW200112_155838}{1.6}{GW200105_162426}{2.4}{GW191105_143521}{1.5}{GW191109_010717}{2.0}{GW200209_085452}{1.3}{GW200115_042309}{2.3}{GW191127_050227}{1.6}{GW200216_220804}{1.6}{GW191215_223052}{1.6}{GW200208_130117}{1.3}{GW200219_094415}{1.7}{GW191103_012549}{1.6}{GW200316_215756}{1.6}{GW200202_154313}{1.5}{GW200129_065458}{1.18}{GW191216_213338}{1.7}}[{\red{???}}]}
\DeclareRobustCommand{\psigwtcthreeplus}[1]{\IfEqCase{#1}{{GW200224_222234}{1.4}{GW191129_134029}{3.6}{GW200311_115853}{1.1}{GW191230_180458}{3.6}{GW191222_033537}{1.88}{GW200225_060421}{1.3}{GW200302_015811}{1.5}{GW200128_022011}{1.4}{GW191204_171526}{1.3}{GW200112_155838}{1.4}{GW200105_162426}{3.5}{GW191105_143521}{1.5}{GW191109_010717}{1.1}{GW200209_085452}{1.6}{GW200115_042309}{3.5}{GW191127_050227}{1.4}{GW200216_220804}{1.4}{GW191215_223052}{1.4}{GW200208_130117}{1.6}{GW200219_094415}{1.2}{GW191103_012549}{1.4}{GW200316_215756}{1.4}{GW200202_154313}{1.5}{GW200129_065458}{1.65}{GW191216_213338}{3.7}}[{\red{???}}]}
\DeclareRobustCommand{\psigwtcthreetenthpercentile}[1]{\IfEqCase{#1}{{GW200224_222234}{0.2}{GW191129_134029}{0.4}{GW200311_115853}{0.5}{GW191230_180458}{0.4}{GW191222_033537}{0.24}{GW200225_060421}{0.3}{GW200302_015811}{0.3}{GW200128_022011}{0.3}{GW191204_171526}{0.4}{GW200112_155838}{0.3}{GW200105_162426}{0.3}{GW191105_143521}{0.2}{GW191109_010717}{0.2}{GW200209_085452}{0.3}{GW200115_042309}{0.4}{GW191127_050227}{0.3}{GW200216_220804}{0.3}{GW191215_223052}{0.3}{GW200208_130117}{0.4}{GW200219_094415}{0.5}{GW191103_012549}{0.3}{GW200316_215756}{0.2}{GW200202_154313}{0.3}{GW200129_065458}{0.40}{GW191216_213338}{0.7}}[{\red{???}}]}
\DeclareRobustCommand{\psigwtcthreenintiethpercentile}[1]{\IfEqCase{#1}{{GW200224_222234}{3.0}{GW191129_134029}{5.0}{GW200311_115853}{2.7}{GW191230_180458}{4.9}{GW191222_033537}{2.83}{GW200225_060421}{2.8}{GW200302_015811}{2.8}{GW200128_022011}{2.8}{GW191204_171526}{2.8}{GW200112_155838}{2.8}{GW200105_162426}{5.2}{GW191105_143521}{2.9}{GW191109_010717}{3.0}{GW200209_085452}{2.8}{GW200115_042309}{5.2}{GW191127_050227}{2.8}{GW200216_220804}{2.8}{GW191215_223052}{2.9}{GW200208_130117}{2.7}{GW200219_094415}{2.7}{GW191103_012549}{2.8}{GW200316_215756}{3.0}{GW200202_154313}{2.8}{GW200129_065458}{2.46}{GW191216_213338}{4.8}}[{\red{???}}]}
\DeclareRobustCommand{\decgwtcthreeminus}[1]{\IfEqCase{#1}{{GW200224_222234}{0.076}{GW191129_134029}{0.62}{GW200311_115853}{0.093}{GW191230_180458}{0.52}{GW191222_033537}{0.54}{GW200225_060421}{0.73}{GW200302_015811}{0.55}{GW200128_022011}{0.73}{GW191204_171526}{0.21}{GW200112_155838}{0.57}{GW200105_162426}{0.82}{GW191105_143521}{0.14}{GW191109_010717}{0.20}{GW200209_085452}{1.22}{GW200115_042309}{0.62}{GW191127_050227}{1.86}{GW200216_220804}{0.59}{GW191215_223052}{0.87}{GW200208_130117}{0.063}{GW200219_094415}{0.19}{GW191103_012549}{1.06}{GW200316_215756}{1.433}{GW200202_154313}{0.11}{GW200129_065458}{0.53}{GW191216_213338}{1.09}}[{\red{???}}]}
\DeclareRobustCommand{\decgwtcthreemed}[1]{\IfEqCase{#1}{{GW200224_222234}{-0.168}{GW191129_134029}{-0.58}{GW200311_115853}{-0.133}{GW191230_180458}{-0.62}{GW191222_033537}{-0.68}{GW200225_060421}{0.94}{GW200302_015811}{-0.51}{GW200128_022011}{-0.38}{GW191204_171526}{-0.54}{GW200112_155838}{-0.21}{GW200105_162426}{-0.05}{GW191105_143521}{-0.61}{GW191109_010717}{-0.59}{GW200209_085452}{0.88}{GW200115_042309}{-0.03}{GW191127_050227}{1.00}{GW200216_220804}{0.75}{GW191215_223052}{-0.24}{GW200208_130117}{-0.597}{GW200219_094415}{-0.46}{GW191103_012549}{0.65}{GW200316_215756}{0.819}{GW200202_154313}{0.38}{GW200129_065458}{0.09}{GW191216_213338}{0.44}}[{\red{???}}]}
\DeclareRobustCommand{\decgwtcthreeplus}[1]{\IfEqCase{#1}{{GW200224_222234}{0.095}{GW191129_134029}{1.30}{GW200311_115853}{0.079}{GW191230_180458}{0.98}{GW191222_033537}{1.56}{GW200225_060421}{0.53}{GW200302_015811}{1.65}{GW200128_022011}{1.35}{GW191204_171526}{0.90}{GW200112_155838}{1.12}{GW200105_162426}{0.86}{GW191105_143521}{1.86}{GW191109_010717}{0.99}{GW200209_085452}{0.48}{GW200115_042309}{0.52}{GW191127_050227}{0.50}{GW200216_220804}{0.37}{GW191215_223052}{0.93}{GW200208_130117}{0.074}{GW200219_094415}{1.39}{GW191103_012549}{0.73}{GW200316_215756}{0.060}{GW200202_154313}{0.13}{GW200129_065458}{0.39}{GW191216_213338}{0.47}}[{\red{???}}]}
\DeclareRobustCommand{\decgwtcthreetenthpercentile}[1]{\IfEqCase{#1}{{GW200224_222234}{-0.227}{GW191129_134029}{-1.06}{GW200311_115853}{-0.205}{GW191230_180458}{-1.07}{GW191222_033537}{-1.15}{GW200225_060421}{0.40}{GW200302_015811}{-1.03}{GW200128_022011}{-1.01}{GW191204_171526}{-0.74}{GW200112_155838}{-0.66}{GW200105_162426}{-0.73}{GW191105_143521}{-0.73}{GW191109_010717}{-0.77}{GW200209_085452}{-0.19}{GW200115_042309}{-0.53}{GW191127_050227}{-0.79}{GW200216_220804}{0.25}{GW191215_223052}{-0.96}{GW200208_130117}{-0.645}{GW200219_094415}{-0.60}{GW191103_012549}{-0.29}{GW200316_215756}{-0.432}{GW200202_154313}{0.30}{GW200129_065458}{0.00}{GW191216_213338}{-0.20}}[{\red{???}}]}
\DeclareRobustCommand{\decgwtcthreenintiethpercentile}[1]{\IfEqCase{#1}{{GW200224_222234}{-0.099}{GW191129_134029}{0.63}{GW200311_115853}{-0.071}{GW191230_180458}{0.07}{GW191222_033537}{0.72}{GW200225_060421}{1.39}{GW200302_015811}{0.96}{GW200128_022011}{0.86}{GW191204_171526}{0.08}{GW200112_155838}{0.80}{GW200105_162426}{0.71}{GW191105_143521}{0.92}{GW191109_010717}{0.14}{GW200209_085452}{1.33}{GW200115_042309}{0.15}{GW191127_050227}{1.46}{GW200216_220804}{1.05}{GW191215_223052}{0.49}{GW200208_130117}{-0.544}{GW200219_094415}{0.72}{GW191103_012549}{1.32}{GW200316_215756}{0.866}{GW200202_154313}{0.48}{GW200129_065458}{0.40}{GW191216_213338}{0.79}}[{\red{???}}]}
\DeclareRobustCommand{\symmetricmassratiogwtcthreeminus}[1]{\IfEqCase{#1}{{GW200224_222234}{0.017}{GW191129_134029}{0.048}{GW200311_115853}{0.019}{GW191230_180458}{0.030}{GW191222_033537}{0.028}{GW200225_060421}{0.030}{GW200302_015811}{0.041}{GW200128_022011}{0.024}{GW191204_171526}{0.031}{GW200112_155838}{0.019}{GW200105_162426}{0.028}{GW191105_143521}{0.037}{GW191109_010717}{0.023}{GW200209_085452}{0.025}{GW200115_042309}{0.046}{GW191127_050227}{0.126}{GW200216_220804}{0.093}{GW191215_223052}{0.029}{GW200208_130117}{0.032}{GW200219_094415}{0.032}{GW191103_012549}{0.061}{GW200316_215756}{0.089}{GW200202_154313}{0.037}{GW200129_065458}{0.035}{GW191216_213338}{0.046}}[{\red{???}}]}
\DeclareRobustCommand{\symmetricmassratiogwtcthreemed}[1]{\IfEqCase{#1}{{GW200224_222234}{0.248}{GW191129_134029}{0.237}{GW200311_115853}{0.248}{GW191230_180458}{0.246}{GW191222_033537}{0.247}{GW200225_060421}{0.244}{GW200302_015811}{0.229}{GW200128_022011}{0.247}{GW191204_171526}{0.242}{GW200112_155838}{0.247}{GW200105_162426}{0.144}{GW191105_143521}{0.243}{GW191109_010717}{0.244}{GW200209_085452}{0.247}{GW200115_042309}{0.157}{GW191127_050227}{0.217}{GW200216_220804}{0.235}{GW191215_223052}{0.244}{GW200208_130117}{0.244}{GW200219_094415}{0.246}{GW191103_012549}{0.240}{GW200316_215756}{0.234}{GW200202_154313}{0.244}{GW200129_065458}{0.248}{GW191216_213338}{0.238}}[{\red{???}}]}
\DeclareRobustCommand{\symmetricmassratiogwtcthreeplus}[1]{\IfEqCase{#1}{{GW200224_222234}{0.002}{GW191129_134029}{0.012}{GW200311_115853}{0.002}{GW191230_180458}{0.004}{GW191222_033537}{0.003}{GW200225_060421}{0.006}{GW200302_015811}{0.021}{GW200128_022011}{0.003}{GW191204_171526}{0.008}{GW200112_155838}{0.003}{GW200105_162426}{0.036}{GW191105_143521}{0.006}{GW191109_010717}{0.006}{GW200209_085452}{0.003}{GW200115_042309}{0.083}{GW191127_050227}{0.033}{GW200216_220804}{0.015}{GW191215_223052}{0.006}{GW200208_130117}{0.006}{GW200219_094415}{0.004}{GW191103_012549}{0.009}{GW200316_215756}{0.016}{GW200202_154313}{0.006}{GW200129_065458}{0.002}{GW191216_213338}{0.012}}[{\red{???}}]}
\DeclareRobustCommand{\symmetricmassratiogwtcthreetenthpercentile}[1]{\IfEqCase{#1}{{GW200224_222234}{0.236}{GW191129_134029}{0.202}{GW200311_115853}{0.235}{GW191230_180458}{0.226}{GW191222_033537}{0.228}{GW200225_060421}{0.223}{GW200302_015811}{0.197}{GW200128_022011}{0.230}{GW191204_171526}{0.220}{GW200112_155838}{0.234}{GW200105_162426}{0.128}{GW191105_143521}{0.217}{GW191109_010717}{0.227}{GW200209_085452}{0.230}{GW200115_042309}{0.121}{GW191127_050227}{0.115}{GW200216_220804}{0.161}{GW191215_223052}{0.224}{GW200208_130117}{0.221}{GW200219_094415}{0.224}{GW191103_012549}{0.198}{GW200316_215756}{0.171}{GW200202_154313}{0.218}{GW200129_065458}{0.222}{GW191216_213338}{0.207}}[{\red{???}}]}
\DeclareRobustCommand{\symmetricmassratiogwtcthreenintiethpercentile}[1]{\IfEqCase{#1}{{GW200224_222234}{0.250}{GW191129_134029}{0.249}{GW200311_115853}{0.250}{GW191230_180458}{0.250}{GW191222_033537}{0.250}{GW200225_060421}{0.250}{GW200302_015811}{0.248}{GW200128_022011}{0.250}{GW191204_171526}{0.249}{GW200112_155838}{0.250}{GW200105_162426}{0.163}{GW191105_143521}{0.250}{GW191109_010717}{0.249}{GW200209_085452}{0.250}{GW200115_042309}{0.231}{GW191127_050227}{0.249}{GW200216_220804}{0.250}{GW191215_223052}{0.250}{GW200208_130117}{0.250}{GW200219_094415}{0.250}{GW191103_012549}{0.250}{GW200316_215756}{0.249}{GW200202_154313}{0.250}{GW200129_065458}{0.250}{GW191216_213338}{0.249}}[{\red{???}}]}
\DeclareRobustCommand{\masstwosourcegwtcthreeminus}[1]{\IfEqCase{#1}{{GW200224_222234}{7.0}{GW191129_134029}{1.7}{GW200311_115853}{5.9}{GW191230_180458}{10.2}{GW191222_033537}{10.5}{GW200225_060421}{3.5}{GW200302_015811}{5.8}{GW200128_022011}{8.0}{GW191204_171526}{1.6}{GW200112_155838}{5.9}{GW200105_162426}{0.24}{GW191105_143521}{1.9}{GW191109_010717}{13}{GW200209_085452}{7.0}{GW200115_042309}{0.28}{GW191127_050227}{14}{GW200216_220804}{16}{GW191215_223052}{3.8}{GW200208_130117}{7.4}{GW200219_094415}{8.4}{GW191103_012549}{2.4}{GW200316_215756}{2.9}{GW200202_154313}{1.7}{GW200129_065458}{9.3}{GW191216_213338}{1.9}}[{\red{???}}]}
\DeclareRobustCommand{\masstwosourcegwtcthreemed}[1]{\IfEqCase{#1}{{GW200224_222234}{32.4}{GW191129_134029}{6.7}{GW200311_115853}{27.7}{GW191230_180458}{36.0}{GW191222_033537}{34.7}{GW200225_060421}{14.0}{GW200302_015811}{20.3}{GW200128_022011}{31.1}{GW191204_171526}{8.2}{GW200112_155838}{28.3}{GW200105_162426}{1.91}{GW191105_143521}{7.7}{GW191109_010717}{47}{GW200209_085452}{26.6}{GW200115_042309}{1.44}{GW191127_050227}{24}{GW200216_220804}{30}{GW191215_223052}{17.8}{GW200208_130117}{27.4}{GW200219_094415}{27.9}{GW191103_012549}{7.9}{GW200316_215756}{7.8}{GW200202_154313}{7.3}{GW200129_065458}{28.9}{GW191216_213338}{7.7}}[{\red{???}}]}
\DeclareRobustCommand{\masstwosourcegwtcthreeplus}[1]{\IfEqCase{#1}{{GW200224_222234}{4.7}{GW191129_134029}{1.5}{GW200311_115853}{4.1}{GW191230_180458}{9.7}{GW191222_033537}{9.3}{GW200225_060421}{2.8}{GW200302_015811}{8.0}{GW200128_022011}{8.7}{GW191204_171526}{1.4}{GW200112_155838}{4.4}{GW200105_162426}{0.33}{GW191105_143521}{1.4}{GW191109_010717}{15}{GW200209_085452}{7.0}{GW200115_042309}{0.85}{GW191127_050227}{17}{GW200216_220804}{14}{GW191215_223052}{3.7}{GW200208_130117}{6.1}{GW200219_094415}{7.4}{GW191103_012549}{1.7}{GW200316_215756}{1.9}{GW200202_154313}{1.1}{GW200129_065458}{3.4}{GW191216_213338}{1.6}}[{\red{???}}]}
\DeclareRobustCommand{\masstwosourcegwtcthreetenthpercentile}[1]{\IfEqCase{#1}{{GW200224_222234}{27.1}{GW191129_134029}{5.3}{GW200311_115853}{23.2}{GW191230_180458}{28.0}{GW191222_033537}{26.6}{GW200225_060421}{11.2}{GW200302_015811}{15.5}{GW200128_022011}{24.8}{GW191204_171526}{6.9}{GW200112_155838}{23.7}{GW200105_162426}{1.77}{GW191105_143521}{6.2}{GW191109_010717}{36}{GW200209_085452}{21.3}{GW200115_042309}{1.21}{GW191127_050227}{12}{GW200216_220804}{16}{GW191215_223052}{14.7}{GW200208_130117}{21.5}{GW200219_094415}{21.6}{GW191103_012549}{6.0}{GW200316_215756}{5.6}{GW200202_154313}{6.0}{GW200129_065458}{21.3}{GW191216_213338}{6.3}}[{\red{???}}]}
\DeclareRobustCommand{\masstwosourcegwtcthreenintiethpercentile}[1]{\IfEqCase{#1}{{GW200224_222234}{36.1}{GW191129_134029}{8.0}{GW200311_115853}{31.0}{GW191230_180458}{43.4}{GW191222_033537}{42.1}{GW200225_060421}{16.3}{GW200302_015811}{26.6}{GW200128_022011}{37.8}{GW191204_171526}{9.4}{GW200112_155838}{31.8}{GW200105_162426}{2.08}{GW191105_143521}{8.8}{GW191109_010717}{58}{GW200209_085452}{32.0}{GW200115_042309}{2.12}{GW191127_050227}{37}{GW200216_220804}{41}{GW191215_223052}{20.7}{GW200208_130117}{32.3}{GW200219_094415}{33.8}{GW191103_012549}{9.3}{GW200316_215756}{9.4}{GW200202_154313}{8.3}{GW200129_065458}{31.7}{GW191216_213338}{9.1}}[{\red{???}}]}
\DeclareRobustCommand{\iotagwtcthreeminus}[1]{\IfEqCase{#1}{{GW200224_222234}{0.42}{GW191129_134029}{1.5}{GW200311_115853}{0.39}{GW191230_180458}{1.92}{GW191222_033537}{1.3}{GW200225_060421}{1.0}{GW200302_015811}{0.99}{GW200128_022011}{1.1}{GW191204_171526}{2.03}{GW200112_155838}{0.69}{GW200105_162426}{1.2}{GW191105_143521}{0.86}{GW191109_010717}{1.41}{GW200209_085452}{1.5}{GW200115_042309}{0.46}{GW191127_050227}{1.2}{GW200216_220804}{0.73}{GW191215_223052}{0.91}{GW200208_130117}{0.62}{GW200219_094415}{0.96}{GW191103_012549}{1.1}{GW200316_215756}{1.84}{GW200202_154313}{0.59}{GW200129_065458}{0.50}{GW191216_213338}{0.80}}[{\red{???}}]}
\DeclareRobustCommand{\iotagwtcthreemed}[1]{\IfEqCase{#1}{{GW200224_222234}{0.58}{GW191129_134029}{1.8}{GW200311_115853}{0.54}{GW191230_180458}{2.17}{GW191222_033537}{1.6}{GW200225_060421}{1.4}{GW200302_015811}{1.27}{GW200128_022011}{1.4}{GW191204_171526}{2.28}{GW200112_155838}{0.90}{GW200105_162426}{1.5}{GW191105_143521}{1.08}{GW191109_010717}{1.98}{GW200209_085452}{1.7}{GW200115_042309}{0.63}{GW191127_050227}{1.5}{GW200216_220804}{0.96}{GW191215_223052}{1.21}{GW200208_130117}{2.52}{GW200219_094415}{1.23}{GW191103_012549}{1.4}{GW200316_215756}{2.32}{GW200202_154313}{2.57}{GW200129_065458}{0.65}{GW191216_213338}{2.50}}[{\red{???}}]}
\DeclareRobustCommand{\iotagwtcthreeplus}[1]{\IfEqCase{#1}{{GW200224_222234}{0.57}{GW191129_134029}{1.1}{GW200311_115853}{0.51}{GW191230_180458}{0.76}{GW191222_033537}{1.2}{GW200225_060421}{1.4}{GW200302_015811}{1.55}{GW200128_022011}{1.5}{GW191204_171526}{0.66}{GW200112_155838}{2.00}{GW200105_162426}{1.3}{GW191105_143521}{1.80}{GW191109_010717}{0.85}{GW200209_085452}{1.2}{GW200115_042309}{1.92}{GW191127_050227}{1.3}{GW200216_220804}{1.79}{GW191215_223052}{1.55}{GW200208_130117}{0.45}{GW200219_094415}{1.53}{GW191103_012549}{1.5}{GW200316_215756}{0.59}{GW200202_154313}{0.42}{GW200129_065458}{0.62}{GW191216_213338}{0.46}}[{\red{???}}]}
\DeclareRobustCommand{\iotagwtcthreetenthpercentile}[1]{\IfEqCase{#1}{{GW200224_222234}{0.22}{GW191129_134029}{0.4}{GW200311_115853}{0.21}{GW191230_180458}{0.38}{GW191222_033537}{0.4}{GW200225_060421}{0.5}{GW200302_015811}{0.40}{GW200128_022011}{0.4}{GW191204_171526}{0.36}{GW200112_155838}{0.29}{GW200105_162426}{0.5}{GW191105_143521}{0.32}{GW191109_010717}{0.83}{GW200209_085452}{0.4}{GW200115_042309}{0.25}{GW191127_050227}{0.5}{GW200216_220804}{0.34}{GW191215_223052}{0.43}{GW200208_130117}{2.05}{GW200219_094415}{0.40}{GW191103_012549}{0.3}{GW200316_215756}{0.72}{GW200202_154313}{2.09}{GW200129_065458}{0.23}{GW191216_213338}{2.00}}[{\red{???}}]}
\DeclareRobustCommand{\iotagwtcthreenintiethpercentile}[1]{\IfEqCase{#1}{{GW200224_222234}{1.03}{GW191129_134029}{2.8}{GW200311_115853}{0.96}{GW191230_180458}{2.83}{GW191222_033537}{2.7}{GW200225_060421}{2.6}{GW200302_015811}{2.67}{GW200128_022011}{2.7}{GW191204_171526}{2.85}{GW200112_155838}{2.81}{GW200105_162426}{2.7}{GW191105_143521}{2.76}{GW191109_010717}{2.69}{GW200209_085452}{2.8}{GW200115_042309}{2.12}{GW191127_050227}{2.6}{GW200216_220804}{2.53}{GW191215_223052}{2.59}{GW200208_130117}{2.89}{GW200219_094415}{2.58}{GW191103_012549}{2.8}{GW200316_215756}{2.81}{GW200202_154313}{2.92}{GW200129_065458}{1.12}{GW191216_213338}{2.88}}[{\red{???}}]}
\DeclareRobustCommand{\costiltonegwtcthreeminus}[1]{\IfEqCase{#1}{{GW200224_222234}{0.96}{GW191129_134029}{0.95}{GW200311_115853}{0.78}{GW191230_180458}{0.83}{GW191222_033537}{0.73}{GW200225_060421}{0.59}{GW200302_015811}{0.90}{GW200128_022011}{0.98}{GW191204_171526}{0.88}{GW200112_155838}{0.91}{GW200105_162426}{0.93}{GW191105_143521}{0.78}{GW191109_010717}{0.36}{GW200209_085452}{0.70}{GW200115_042309}{0.35}{GW191127_050227}{1.00}{GW200216_220804}{1.02}{GW191215_223052}{0.72}{GW200208_130117}{0.68}{GW200219_094415}{0.69}{GW191103_012549}{0.94}{GW200316_215756}{1.02}{GW200202_154313}{0.95}{GW200129_065458}{0.92}{GW191216_213338}{1.13}}[{\red{???}}]}
\DeclareRobustCommand{\costiltonegwtcthreemed}[1]{\IfEqCase{#1}{{GW200224_222234}{0.30}{GW191129_134029}{0.27}{GW200311_115853}{-0.09}{GW191230_180458}{-0.04}{GW191222_033537}{-0.16}{GW200225_060421}{-0.30}{GW200302_015811}{0.06}{GW200128_022011}{0.36}{GW191204_171526}{0.43}{GW200112_155838}{0.14}{GW200105_162426}{0.01}{GW191105_143521}{-0.07}{GW191109_010717}{-0.61}{GW200209_085452}{-0.19}{GW200115_042309}{-0.62}{GW191127_050227}{0.37}{GW200216_220804}{0.24}{GW191215_223052}{-0.06}{GW200208_130117}{-0.24}{GW200219_094415}{-0.23}{GW191103_012549}{0.54}{GW200316_215756}{0.47}{GW200202_154313}{0.18}{GW200129_065458}{0.13}{GW191216_213338}{0.54}}[{\red{???}}]}
\DeclareRobustCommand{\costiltonegwtcthreeplus}[1]{\IfEqCase{#1}{{GW200224_222234}{0.61}{GW191129_134029}{0.64}{GW200311_115853}{0.89}{GW191230_180458}{0.86}{GW191222_033537}{0.97}{GW200225_060421}{0.86}{GW200302_015811}{0.80}{GW200128_022011}{0.56}{GW191204_171526}{0.49}{GW200112_155838}{0.75}{GW200105_162426}{0.90}{GW191105_143521}{0.90}{GW191109_010717}{0.85}{GW200209_085452}{0.92}{GW200115_042309}{1.26}{GW191127_050227}{0.56}{GW200216_220804}{0.68}{GW191215_223052}{0.85}{GW200208_130117}{1.03}{GW200219_094415}{0.96}{GW191103_012549}{0.41}{GW200316_215756}{0.49}{GW200202_154313}{0.72}{GW200129_065458}{0.74}{GW191216_213338}{0.43}}[{\red{???}}]}
\DeclareRobustCommand{\costiltonegwtcthreetenthpercentile}[1]{\IfEqCase{#1}{{GW200224_222234}{-0.44}{GW191129_134029}{-0.45}{GW200311_115853}{-0.75}{GW191230_180458}{-0.73}{GW191222_033537}{-0.80}{GW200225_060421}{-0.81}{GW200302_015811}{-0.70}{GW200128_022011}{-0.38}{GW191204_171526}{-0.19}{GW200112_155838}{-0.60}{GW200105_162426}{-0.82}{GW191105_143521}{-0.72}{GW191109_010717}{-0.93}{GW200209_085452}{-0.79}{GW200115_042309}{-0.94}{GW191127_050227}{-0.38}{GW200216_220804}{-0.62}{GW191215_223052}{-0.64}{GW200208_130117}{-0.85}{GW200219_094415}{-0.83}{GW191103_012549}{-0.12}{GW200316_215756}{-0.29}{GW200202_154313}{-0.59}{GW200129_065458}{-0.57}{GW191216_213338}{-0.32}}[{\red{???}}]}
\DeclareRobustCommand{\costiltonegwtcthreenintiethpercentile}[1]{\IfEqCase{#1}{{GW200224_222234}{0.82}{GW191129_134029}{0.83}{GW200311_115853}{0.63}{GW191230_180458}{0.69}{GW191222_033537}{0.64}{GW200225_060421}{0.32}{GW200302_015811}{0.73}{GW200128_022011}{0.83}{GW191204_171526}{0.85}{GW200112_155838}{0.78}{GW200105_162426}{0.82}{GW191105_143521}{0.68}{GW191109_010717}{0.02}{GW200209_085452}{0.55}{GW200115_042309}{0.41}{GW191127_050227}{0.88}{GW200216_220804}{0.84}{GW191215_223052}{0.61}{GW200208_130117}{0.61}{GW200219_094415}{0.56}{GW191103_012549}{0.90}{GW200316_215756}{0.92}{GW200202_154313}{0.80}{GW200129_065458}{0.76}{GW191216_213338}{0.92}}[{\red{???}}]}
\DeclareRobustCommand{\phionetwogwtcthreeminus}[1]{\IfEqCase{#1}{{GW200224_222234}{2.7}{GW191129_134029}{2.8}{GW200311_115853}{2.8}{GW191230_180458}{2.9}{GW191222_033537}{2.8}{GW200225_060421}{2.8}{GW200302_015811}{2.8}{GW200128_022011}{3.0}{GW191204_171526}{2.8}{GW200112_155838}{2.7}{GW200105_162426}{2.8}{GW191105_143521}{2.8}{GW191109_010717}{3.5}{GW200209_085452}{2.8}{GW200115_042309}{2.8}{GW191127_050227}{2.8}{GW200216_220804}{2.9}{GW191215_223052}{2.9}{GW200208_130117}{2.7}{GW200219_094415}{2.9}{GW191103_012549}{2.8}{GW200316_215756}{2.8}{GW200202_154313}{2.9}{GW200129_065458}{2.6}{GW191216_213338}{2.8}}[{\red{???}}]}
\DeclareRobustCommand{\phionetwogwtcthreemed}[1]{\IfEqCase{#1}{{GW200224_222234}{3.1}{GW191129_134029}{3.1}{GW200311_115853}{3.1}{GW191230_180458}{3.1}{GW191222_033537}{3.2}{GW200225_060421}{3.2}{GW200302_015811}{3.1}{GW200128_022011}{3.3}{GW191204_171526}{3.1}{GW200112_155838}{3.1}{GW200105_162426}{3.2}{GW191105_143521}{3.2}{GW191109_010717}{3.7}{GW200209_085452}{3.1}{GW200115_042309}{3.2}{GW191127_050227}{3.2}{GW200216_220804}{3.2}{GW191215_223052}{3.2}{GW200208_130117}{3.0}{GW200219_094415}{3.2}{GW191103_012549}{3.1}{GW200316_215756}{3.1}{GW200202_154313}{3.2}{GW200129_065458}{3.0}{GW191216_213338}{3.2}}[{\red{???}}]}
\DeclareRobustCommand{\phionetwogwtcthreeplus}[1]{\IfEqCase{#1}{{GW200224_222234}{2.8}{GW191129_134029}{2.8}{GW200311_115853}{2.8}{GW191230_180458}{2.9}{GW191222_033537}{2.8}{GW200225_060421}{2.8}{GW200302_015811}{2.8}{GW200128_022011}{2.7}{GW191204_171526}{2.8}{GW200112_155838}{2.9}{GW200105_162426}{2.8}{GW191105_143521}{2.8}{GW191109_010717}{2.3}{GW200209_085452}{2.9}{GW200115_042309}{2.8}{GW191127_050227}{2.8}{GW200216_220804}{2.8}{GW191215_223052}{2.8}{GW200208_130117}{2.9}{GW200219_094415}{2.8}{GW191103_012549}{2.9}{GW200316_215756}{2.8}{GW200202_154313}{2.8}{GW200129_065458}{2.9}{GW191216_213338}{2.8}}[{\red{???}}]}
\DeclareRobustCommand{\phionetwogwtcthreetenthpercentile}[1]{\IfEqCase{#1}{{GW200224_222234}{0.7}{GW191129_134029}{0.7}{GW200311_115853}{0.7}{GW191230_180458}{0.6}{GW191222_033537}{0.7}{GW200225_060421}{0.7}{GW200302_015811}{0.7}{GW200128_022011}{0.6}{GW191204_171526}{0.7}{GW200112_155838}{0.7}{GW200105_162426}{0.7}{GW191105_143521}{0.7}{GW191109_010717}{0.5}{GW200209_085452}{0.5}{GW200115_042309}{0.7}{GW191127_050227}{0.6}{GW200216_220804}{0.6}{GW191215_223052}{0.6}{GW200208_130117}{0.6}{GW200219_094415}{0.6}{GW191103_012549}{0.6}{GW200316_215756}{0.6}{GW200202_154313}{0.7}{GW200129_065458}{0.6}{GW191216_213338}{0.7}}[{\red{???}}]}
\DeclareRobustCommand{\phionetwogwtcthreenintiethpercentile}[1]{\IfEqCase{#1}{{GW200224_222234}{5.6}{GW191129_134029}{5.6}{GW200311_115853}{5.6}{GW191230_180458}{5.7}{GW191222_033537}{5.6}{GW200225_060421}{5.6}{GW200302_015811}{5.6}{GW200128_022011}{5.7}{GW191204_171526}{5.6}{GW200112_155838}{5.6}{GW200105_162426}{5.6}{GW191105_143521}{5.6}{GW191109_010717}{5.8}{GW200209_085452}{5.7}{GW200115_042309}{5.6}{GW191127_050227}{5.7}{GW200216_220804}{5.6}{GW191215_223052}{5.6}{GW200208_130117}{5.6}{GW200219_094415}{5.7}{GW191103_012549}{5.6}{GW200316_215756}{5.6}{GW200202_154313}{5.6}{GW200129_065458}{5.6}{GW191216_213338}{5.6}}[{\red{???}}]}
\DeclareRobustCommand{\massratiogwtcthreeminus}[1]{\IfEqCase{#1}{{GW200224_222234}{0.26}{GW191129_134029}{0.29}{GW200311_115853}{0.27}{GW191230_180458}{0.32}{GW191222_033537}{0.32}{GW200225_060421}{0.28}{GW200302_015811}{0.21}{GW200128_022011}{0.29}{GW191204_171526}{0.26}{GW200112_155838}{0.26}{GW200105_162426}{0.056}{GW191105_143521}{0.31}{GW191109_010717}{0.24}{GW200209_085452}{0.30}{GW200115_042309}{0.097}{GW191127_050227}{0.35}{GW200216_220804}{0.40}{GW191215_223052}{0.27}{GW200208_130117}{0.29}{GW200219_094415}{0.32}{GW191103_012549}{0.37}{GW200316_215756}{0.38}{GW200202_154313}{0.31}{GW200129_065458}{0.41}{GW191216_213338}{0.29}}[{\red{???}}]}
\DeclareRobustCommand{\massratiogwtcthreemed}[1]{\IfEqCase{#1}{{GW200224_222234}{0.82}{GW191129_134029}{0.63}{GW200311_115853}{0.82}{GW191230_180458}{0.78}{GW191222_033537}{0.79}{GW200225_060421}{0.73}{GW200302_015811}{0.55}{GW200128_022011}{0.79}{GW191204_171526}{0.69}{GW200112_155838}{0.81}{GW200105_162426}{0.211}{GW191105_143521}{0.72}{GW191109_010717}{0.73}{GW200209_085452}{0.79}{GW200115_042309}{0.243}{GW191127_050227}{0.47}{GW200216_220804}{0.61}{GW191215_223052}{0.73}{GW200208_130117}{0.73}{GW200219_094415}{0.77}{GW191103_012549}{0.67}{GW200316_215756}{0.59}{GW200202_154313}{0.72}{GW200129_065458}{0.85}{GW191216_213338}{0.63}}[{\red{???}}]}
\DeclareRobustCommand{\massratiogwtcthreeplus}[1]{\IfEqCase{#1}{{GW200224_222234}{0.16}{GW191129_134029}{0.31}{GW200311_115853}{0.16}{GW191230_180458}{0.20}{GW191222_033537}{0.18}{GW200225_060421}{0.23}{GW200302_015811}{0.36}{GW200128_022011}{0.18}{GW191204_171526}{0.25}{GW200112_155838}{0.17}{GW200105_162426}{0.095}{GW191105_143521}{0.24}{GW191109_010717}{0.21}{GW200209_085452}{0.19}{GW200115_042309}{0.432}{GW191127_050227}{0.47}{GW200216_220804}{0.35}{GW191215_223052}{0.24}{GW200208_130117}{0.23}{GW200219_094415}{0.21}{GW191103_012549}{0.29}{GW200316_215756}{0.34}{GW200202_154313}{0.24}{GW200129_065458}{0.12}{GW191216_213338}{0.31}}[{\red{???}}]}
\DeclareRobustCommand{\massratiogwtcthreetenthpercentile}[1]{\IfEqCase{#1}{{GW200224_222234}{0.62}{GW191129_134029}{0.39}{GW200311_115853}{0.61}{GW191230_180458}{0.52}{GW191222_033537}{0.54}{GW200225_060421}{0.50}{GW200302_015811}{0.37}{GW200128_022011}{0.56}{GW191204_171526}{0.48}{GW200112_155838}{0.60}{GW200105_162426}{0.177}{GW191105_143521}{0.47}{GW191109_010717}{0.54}{GW200209_085452}{0.56}{GW200115_042309}{0.164}{GW191127_050227}{0.15}{GW200216_220804}{0.25}{GW191215_223052}{0.51}{GW200208_130117}{0.49}{GW200219_094415}{0.51}{GW191103_012549}{0.37}{GW200316_215756}{0.28}{GW200202_154313}{0.48}{GW200129_065458}{0.50}{GW191216_213338}{0.41}}[{\red{???}}]}
\DeclareRobustCommand{\massratiogwtcthreenintiethpercentile}[1]{\IfEqCase{#1}{{GW200224_222234}{0.96}{GW191129_134029}{0.91}{GW200311_115853}{0.96}{GW191230_180458}{0.95}{GW191222_033537}{0.96}{GW200225_060421}{0.93}{GW200302_015811}{0.84}{GW200128_022011}{0.96}{GW191204_171526}{0.91}{GW200112_155838}{0.96}{GW200105_162426}{0.259}{GW191105_143521}{0.93}{GW191109_010717}{0.91}{GW200209_085452}{0.96}{GW200115_042309}{0.571}{GW191127_050227}{0.88}{GW200216_220804}{0.92}{GW191215_223052}{0.94}{GW200208_130117}{0.94}{GW200219_094415}{0.95}{GW191103_012549}{0.92}{GW200316_215756}{0.89}{GW200202_154313}{0.94}{GW200129_065458}{0.96}{GW191216_213338}{0.90}}[{\red{???}}]}
\DeclareRobustCommand{\comovingdistgwtcthreeminus}[1]{\IfEqCase{#1}{{GW200224_222234}{400}{GW191129_134029}{260}{GW200311_115853}{280}{GW191230_180458}{850}{GW191222_033537}{920}{GW200225_060421}{390}{GW200302_015811}{510}{GW200128_022011}{970}{GW191204_171526}{200}{GW200112_155838}{330}{GW200105_162426}{100}{GW191105_143521}{350}{GW191109_010717}{470}{GW200209_085452}{840}{GW200115_042309}{92}{GW191127_050227}{1000}{GW200216_220804}{1000}{GW191215_223052}{560}{GW200208_130117}{500}{GW200219_094415}{750}{GW191103_012549}{360}{GW200316_215756}{320}{GW200202_154313}{140}{GW200129_065458}{290}{GW191216_213338}{120}}[{\red{???}}]}
\DeclareRobustCommand{\comovingdistgwtcthreemed}[1]{\IfEqCase{#1}{{GW200224_222234}{1340}{GW191129_134029}{700}{GW200311_115853}{950}{GW191230_180458}{2740}{GW191222_033537}{1980}{GW200225_060421}{940}{GW200302_015811}{1270}{GW200128_022011}{2440}{GW191204_171526}{580}{GW200112_155838}{1010}{GW200105_162426}{260}{GW191105_143521}{940}{GW191109_010717}{1040}{GW200209_085452}{2390}{GW200115_042309}{274}{GW191127_050227}{2200}{GW200216_220804}{2350}{GW191215_223052}{1520}{GW200208_130117}{1590}{GW200219_094415}{2180}{GW191103_012549}{830}{GW200316_215756}{920}{GW200202_154313}{380}{GW200129_065458}{760}{GW191216_213338}{320}}[{\red{???}}]}
\DeclareRobustCommand{\comovingdistgwtcthreeplus}[1]{\IfEqCase{#1}{{GW200224_222234}{270}{GW191129_134029}{180}{GW200311_115853}{180}{GW191230_180458}{690}{GW191222_033537}{710}{GW200225_060421}{330}{GW200302_015811}{590}{GW200128_022011}{750}{GW191204_171526}{140}{GW200112_155838}{270}{GW200105_162426}{100}{GW191105_143521}{280}{GW191109_010717}{660}{GW200209_085452}{720}{GW200115_042309}{126}{GW191127_050227}{1100}{GW200216_220804}{1040}{GW191215_223052}{440}{GW200208_130117}{500}{GW200219_094415}{680}{GW191103_012549}{340}{GW200316_215756}{310}{GW200202_154313}{120}{GW200129_065458}{200}{GW191216_213338}{100}}[{\red{???}}]}
\DeclareRobustCommand{\comovingdistgwtcthreetenthpercentile}[1]{\IfEqCase{#1}{{GW200224_222234}{1030}{GW191129_134029}{490}{GW200311_115853}{740}{GW191230_180458}{2080}{GW191222_033537}{1260}{GW200225_060421}{640}{GW200302_015811}{860}{GW200128_022011}{1690}{GW191204_171526}{420}{GW200112_155838}{750}{GW200105_162426}{170}{GW191105_143521}{660}{GW191109_010717}{660}{GW200209_085452}{1750}{GW200115_042309}{202}{GW191127_050227}{1400}{GW200216_220804}{1560}{GW191215_223052}{1080}{GW200208_130117}{1190}{GW200219_094415}{1550}{GW191103_012549}{530}{GW200316_215756}{660}{GW200202_154313}{270}{GW200129_065458}{540}{GW191216_213338}{220}}[{\red{???}}]}
\DeclareRobustCommand{\comovingdistgwtcthreenintiethpercentile}[1]{\IfEqCase{#1}{{GW200224_222234}{1550}{GW191129_134029}{850}{GW200311_115853}{1100}{GW191230_180458}{3280}{GW191222_033537}{2550}{GW200225_060421}{1200}{GW200302_015811}{1730}{GW200128_022011}{3040}{GW191204_171526}{700}{GW200112_155838}{1230}{GW200105_162426}{340}{GW191105_143521}{1160}{GW191109_010717}{1540}{GW200209_085452}{2950}{GW200115_042309}{365}{GW191127_050227}{3100}{GW200216_220804}{3160}{GW191215_223052}{1880}{GW200208_130117}{1970}{GW200219_094415}{2710}{GW191103_012549}{1090}{GW200316_215756}{1160}{GW200202_154313}{480}{GW200129_065458}{930}{GW191216_213338}{400}}[{\red{???}}]}
\DeclareRobustCommand{\phasegwtcthreeminus}[1]{\IfEqCase{#1}{{GW200224_222234}{2.8}{GW191129_134029}{2.9}{GW200311_115853}{2.7}{GW191230_180458}{2.8}{GW191222_033537}{2.8}{GW200225_060421}{2.4}{GW200302_015811}{2.2}{GW200128_022011}{3.1}{GW191204_171526}{2.7}{GW200112_155838}{2.9}{GW200105_162426}{3.1}{GW191105_143521}{2.8}{GW191109_010717}{1.4}{GW200209_085452}{2.9}{GW200115_042309}{2.8}{GW191127_050227}{2.8}{GW200216_220804}{2.6}{GW191215_223052}{2.4}{GW200208_130117}{3.5}{GW200219_094415}{3.5}{GW191103_012549}{2.6}{GW200316_215756}{2.4}{GW200202_154313}{2.6}{GW200129_065458}{3.0}{GW191216_213338}{2.1}}[{\red{???}}]}
\DeclareRobustCommand{\phasegwtcthreemed}[1]{\IfEqCase{#1}{{GW200224_222234}{3.0}{GW191129_134029}{3.4}{GW200311_115853}{3.1}{GW191230_180458}{3.1}{GW191222_033537}{3.1}{GW200225_060421}{2.8}{GW200302_015811}{2.5}{GW200128_022011}{3.4}{GW191204_171526}{3.1}{GW200112_155838}{3.4}{GW200105_162426}{4.2}{GW191105_143521}{3.1}{GW191109_010717}{1.6}{GW200209_085452}{3.2}{GW200115_042309}{3.2}{GW191127_050227}{3.0}{GW200216_220804}{3.0}{GW191215_223052}{2.6}{GW200208_130117}{3.8}{GW200219_094415}{3.8}{GW191103_012549}{3.0}{GW200316_215756}{2.9}{GW200202_154313}{3.0}{GW200129_065458}{3.5}{GW191216_213338}{2.2}}[{\red{???}}]}
\DeclareRobustCommand{\phasegwtcthreeplus}[1]{\IfEqCase{#1}{{GW200224_222234}{3.0}{GW191129_134029}{2.5}{GW200311_115853}{2.8}{GW191230_180458}{2.9}{GW191222_033537}{3.0}{GW200225_060421}{3.2}{GW200302_015811}{3.4}{GW200128_022011}{2.7}{GW191204_171526}{2.8}{GW200112_155838}{2.5}{GW200105_162426}{1.4}{GW191105_143521}{2.9}{GW191109_010717}{4.3}{GW200209_085452}{2.8}{GW200115_042309}{2.8}{GW191127_050227}{2.9}{GW200216_220804}{3.0}{GW191215_223052}{3.4}{GW200208_130117}{2.2}{GW200219_094415}{2.2}{GW191103_012549}{3.0}{GW200316_215756}{2.8}{GW200202_154313}{2.9}{GW200129_065458}{2.4}{GW191216_213338}{3.9}}[{\red{???}}]}
\DeclareRobustCommand{\phasegwtcthreetenthpercentile}[1]{\IfEqCase{#1}{{GW200224_222234}{0.4}{GW191129_134029}{0.9}{GW200311_115853}{0.8}{GW191230_180458}{0.6}{GW191222_033537}{0.5}{GW200225_060421}{0.7}{GW200302_015811}{0.7}{GW200128_022011}{0.6}{GW191204_171526}{0.8}{GW200112_155838}{1.0}{GW200105_162426}{1.8}{GW191105_143521}{0.6}{GW191109_010717}{0.3}{GW200209_085452}{0.5}{GW200115_042309}{0.6}{GW191127_050227}{0.5}{GW200216_220804}{0.7}{GW191215_223052}{0.4}{GW200208_130117}{0.7}{GW200219_094415}{0.8}{GW191103_012549}{0.7}{GW200316_215756}{1.0}{GW200202_154313}{0.7}{GW200129_065458}{0.9}{GW191216_213338}{0.3}}[{\red{???}}]}
\DeclareRobustCommand{\phasegwtcthreenintiethpercentile}[1]{\IfEqCase{#1}{{GW200224_222234}{5.8}{GW191129_134029}{5.5}{GW200311_115853}{5.5}{GW191230_180458}{5.6}{GW191222_033537}{5.8}{GW200225_060421}{5.5}{GW200302_015811}{5.5}{GW200128_022011}{5.8}{GW191204_171526}{5.4}{GW200112_155838}{5.5}{GW200105_162426}{5.3}{GW191105_143521}{5.7}{GW191109_010717}{5.0}{GW200209_085452}{5.7}{GW200115_042309}{5.7}{GW191127_050227}{5.7}{GW200216_220804}{5.6}{GW191215_223052}{5.7}{GW200208_130117}{5.7}{GW200219_094415}{5.7}{GW191103_012549}{5.6}{GW200316_215756}{5.2}{GW200202_154313}{5.5}{GW200129_065458}{5.6}{GW191216_213338}{6.0}}[{\red{???}}]}
\DeclareRobustCommand{\phionegwtcthreeminus}[1]{\IfEqCase{#1}{{GW200224_222234}{2.9}{GW191129_134029}{2.9}{GW200311_115853}{2.7}{GW191230_180458}{2.8}{GW191222_033537}{2.8}{GW200225_060421}{2.9}{GW200302_015811}{2.8}{GW200128_022011}{2.8}{GW191204_171526}{2.8}{GW200112_155838}{2.8}{GW200105_162426}{2.9}{GW191105_143521}{2.8}{GW191109_010717}{2.7}{GW200209_085452}{2.8}{GW200115_042309}{3.0}{GW191127_050227}{2.8}{GW200216_220804}{2.9}{GW191215_223052}{2.8}{GW200208_130117}{2.8}{GW200219_094415}{2.9}{GW191103_012549}{2.8}{GW200316_215756}{3.0}{GW200202_154313}{2.9}{GW200129_065458}{2.8}{GW191216_213338}{2.7}}[{\red{???}}]}
\DeclareRobustCommand{\phionegwtcthreemed}[1]{\IfEqCase{#1}{{GW200224_222234}{3.1}{GW191129_134029}{3.2}{GW200311_115853}{3.1}{GW191230_180458}{3.2}{GW191222_033537}{3.1}{GW200225_060421}{3.2}{GW200302_015811}{3.2}{GW200128_022011}{3.1}{GW191204_171526}{3.1}{GW200112_155838}{3.1}{GW200105_162426}{3.2}{GW191105_143521}{3.1}{GW191109_010717}{3.1}{GW200209_085452}{3.2}{GW200115_042309}{3.3}{GW191127_050227}{3.1}{GW200216_220804}{3.2}{GW191215_223052}{3.1}{GW200208_130117}{3.0}{GW200219_094415}{3.2}{GW191103_012549}{3.1}{GW200316_215756}{3.3}{GW200202_154313}{3.2}{GW200129_065458}{3.1}{GW191216_213338}{3.0}}[{\red{???}}]}
\DeclareRobustCommand{\phionegwtcthreeplus}[1]{\IfEqCase{#1}{{GW200224_222234}{2.9}{GW191129_134029}{2.8}{GW200311_115853}{2.9}{GW191230_180458}{2.8}{GW191222_033537}{2.9}{GW200225_060421}{2.8}{GW200302_015811}{2.8}{GW200128_022011}{2.8}{GW191204_171526}{2.9}{GW200112_155838}{2.9}{GW200105_162426}{2.8}{GW191105_143521}{2.8}{GW191109_010717}{2.9}{GW200209_085452}{2.8}{GW200115_042309}{2.7}{GW191127_050227}{2.9}{GW200216_220804}{2.8}{GW191215_223052}{2.9}{GW200208_130117}{2.9}{GW200219_094415}{2.8}{GW191103_012549}{2.9}{GW200316_215756}{2.7}{GW200202_154313}{2.8}{GW200129_065458}{2.9}{GW191216_213338}{3.0}}[{\red{???}}]}
\DeclareRobustCommand{\phionegwtcthreetenthpercentile}[1]{\IfEqCase{#1}{{GW200224_222234}{0.6}{GW191129_134029}{0.6}{GW200311_115853}{0.7}{GW191230_180458}{0.6}{GW191222_033537}{0.6}{GW200225_060421}{0.6}{GW200302_015811}{0.6}{GW200128_022011}{0.7}{GW191204_171526}{0.7}{GW200112_155838}{0.6}{GW200105_162426}{0.6}{GW191105_143521}{0.6}{GW191109_010717}{0.6}{GW200209_085452}{0.6}{GW200115_042309}{0.6}{GW191127_050227}{0.7}{GW200216_220804}{0.7}{GW191215_223052}{0.6}{GW200208_130117}{0.6}{GW200219_094415}{0.6}{GW191103_012549}{0.7}{GW200316_215756}{0.6}{GW200202_154313}{0.6}{GW200129_065458}{0.6}{GW191216_213338}{0.6}}[{\red{???}}]}
\DeclareRobustCommand{\phionegwtcthreenintiethpercentile}[1]{\IfEqCase{#1}{{GW200224_222234}{5.7}{GW191129_134029}{5.7}{GW200311_115853}{5.6}{GW191230_180458}{5.6}{GW191222_033537}{5.6}{GW200225_060421}{5.7}{GW200302_015811}{5.7}{GW200128_022011}{5.6}{GW191204_171526}{5.6}{GW200112_155838}{5.6}{GW200105_162426}{5.6}{GW191105_143521}{5.7}{GW191109_010717}{5.6}{GW200209_085452}{5.7}{GW200115_042309}{5.7}{GW191127_050227}{5.7}{GW200216_220804}{5.6}{GW191215_223052}{5.7}{GW200208_130117}{5.7}{GW200219_094415}{5.7}{GW191103_012549}{5.6}{GW200316_215756}{5.7}{GW200202_154313}{5.7}{GW200129_065458}{5.7}{GW191216_213338}{5.6}}[{\red{???}}]}
\DeclareRobustCommand{\spintwogwtcthreeminus}[1]{\IfEqCase{#1}{{GW200224_222234}{0.39}{GW191129_134029}{0.31}{GW200311_115853}{0.37}{GW191230_180458}{0.44}{GW191222_033537}{0.38}{GW200225_060421}{0.39}{GW200302_015811}{0.41}{GW200128_022011}{0.45}{GW191204_171526}{0.40}{GW200112_155838}{0.34}{GW200105_162426}{0.30}{GW191105_143521}{0.31}{GW191109_010717}{0.58}{GW200209_085452}{0.44}{GW200115_042309}{0.39}{GW191127_050227}{0.49}{GW200216_220804}{0.46}{GW191215_223052}{0.40}{GW200208_130117}{0.39}{GW200219_094415}{0.43}{GW191103_012549}{0.44}{GW200316_215756}{0.39}{GW200202_154313}{0.30}{GW200129_065458}{0.42}{GW191216_213338}{0.32}}[{\red{???}}]}
\DeclareRobustCommand{\spintwogwtcthreemed}[1]{\IfEqCase{#1}{{GW200224_222234}{0.44}{GW191129_134029}{0.35}{GW200311_115853}{0.41}{GW191230_180458}{0.49}{GW191222_033537}{0.41}{GW200225_060421}{0.42}{GW200302_015811}{0.45}{GW200128_022011}{0.50}{GW191204_171526}{0.46}{GW200112_155838}{0.39}{GW200105_162426}{0.33}{GW191105_143521}{0.34}{GW191109_010717}{0.65}{GW200209_085452}{0.49}{GW200115_042309}{0.44}{GW191127_050227}{0.54}{GW200216_220804}{0.51}{GW191215_223052}{0.44}{GW200208_130117}{0.43}{GW200219_094415}{0.48}{GW191103_012549}{0.50}{GW200316_215756}{0.44}{GW200202_154313}{0.33}{GW200129_065458}{0.49}{GW191216_213338}{0.36}}[{\red{???}}]}
\DeclareRobustCommand{\spintwogwtcthreeplus}[1]{\IfEqCase{#1}{{GW200224_222234}{0.48}{GW191129_134029}{0.51}{GW200311_115853}{0.51}{GW191230_180458}{0.45}{GW191222_033537}{0.50}{GW200225_060421}{0.49}{GW200302_015811}{0.48}{GW200128_022011}{0.44}{GW191204_171526}{0.41}{GW200112_155838}{0.50}{GW200105_162426}{0.56}{GW191105_143521}{0.54}{GW191109_010717}{0.32}{GW200209_085452}{0.45}{GW200115_042309}{0.48}{GW191127_050227}{0.41}{GW200216_220804}{0.44}{GW191215_223052}{0.49}{GW200208_130117}{0.49}{GW200219_094415}{0.46}{GW191103_012549}{0.44}{GW200316_215756}{0.47}{GW200202_154313}{0.53}{GW200129_065458}{0.44}{GW191216_213338}{0.50}}[{\red{???}}]}
\DeclareRobustCommand{\spintwogwtcthreetenthpercentile}[1]{\IfEqCase{#1}{{GW200224_222234}{0.09}{GW191129_134029}{0.07}{GW200311_115853}{0.08}{GW191230_180458}{0.10}{GW191222_033537}{0.07}{GW200225_060421}{0.08}{GW200302_015811}{0.09}{GW200128_022011}{0.10}{GW191204_171526}{0.11}{GW200112_155838}{0.08}{GW200105_162426}{0.06}{GW191105_143521}{0.06}{GW191109_010717}{0.14}{GW200209_085452}{0.10}{GW200115_042309}{0.09}{GW191127_050227}{0.11}{GW200216_220804}{0.10}{GW191215_223052}{0.09}{GW200208_130117}{0.08}{GW200219_094415}{0.10}{GW191103_012549}{0.10}{GW200316_215756}{0.10}{GW200202_154313}{0.06}{GW200129_065458}{0.13}{GW191216_213338}{0.08}}[{\red{???}}]}
\DeclareRobustCommand{\spintwogwtcthreenintiethpercentile}[1]{\IfEqCase{#1}{{GW200224_222234}{0.85}{GW191129_134029}{0.76}{GW200311_115853}{0.84}{GW191230_180458}{0.89}{GW191222_033537}{0.84}{GW200225_060421}{0.85}{GW200302_015811}{0.87}{GW200128_022011}{0.89}{GW191204_171526}{0.79}{GW200112_155838}{0.81}{GW200105_162426}{0.79}{GW191105_143521}{0.78}{GW191109_010717}{0.94}{GW200209_085452}{0.90}{GW200115_042309}{0.84}{GW191127_050227}{0.91}{GW200216_220804}{0.90}{GW191215_223052}{0.86}{GW200208_130117}{0.85}{GW200219_094415}{0.88}{GW191103_012549}{0.88}{GW200316_215756}{0.83}{GW200202_154313}{0.77}{GW200129_065458}{0.86}{GW191216_213338}{0.76}}[{\red{???}}]}
\DeclareRobustCommand{\spinonezgwtcthreeminus}[1]{\IfEqCase{#1}{{GW200224_222234}{0.31}{GW191129_134029}{0.20}{GW200311_115853}{0.41}{GW191230_180458}{0.49}{GW191222_033537}{0.45}{GW200225_060421}{0.48}{GW200302_015811}{0.36}{GW200128_022011}{0.38}{GW191204_171526}{0.27}{GW200112_155838}{0.31}{GW200105_162426}{0.21}{GW191105_143521}{0.30}{GW191109_010717}{0.41}{GW200209_085452}{0.52}{GW200115_042309}{0.55}{GW191127_050227}{0.45}{GW200216_220804}{0.44}{GW191215_223052}{0.39}{GW200208_130117}{0.48}{GW200219_094415}{0.50}{GW191103_012549}{0.32}{GW200316_215756}{0.24}{GW200202_154313}{0.23}{GW200129_065458}{0.35}{GW191216_213338}{0.20}}[{\red{???}}]}
\DeclareRobustCommand{\spinonezgwtcthreemed}[1]{\IfEqCase{#1}{{GW200224_222234}{0.11}{GW191129_134029}{0.05}{GW200311_115853}{-0.02}{GW191230_180458}{-0.01}{GW191222_033537}{-0.03}{GW200225_060421}{-0.13}{GW200302_015811}{0.01}{GW200128_022011}{0.17}{GW191204_171526}{0.16}{GW200112_155838}{0.03}{GW200105_162426}{0.00}{GW191105_143521}{-0.01}{GW191109_010717}{-0.44}{GW200209_085452}{-0.06}{GW200115_042309}{-0.15}{GW191127_050227}{0.18}{GW200216_220804}{0.07}{GW191215_223052}{-0.02}{GW200208_130117}{-0.05}{GW200219_094415}{-0.07}{GW191103_012549}{0.24}{GW200316_215756}{0.12}{GW200202_154313}{0.02}{GW200129_065458}{0.05}{GW191216_213338}{0.11}}[{\red{???}}]}
\DeclareRobustCommand{\spinonezgwtcthreeplus}[1]{\IfEqCase{#1}{{GW200224_222234}{0.42}{GW191129_134029}{0.24}{GW200311_115853}{0.31}{GW191230_180458}{0.44}{GW191222_033537}{0.33}{GW200225_060421}{0.32}{GW200302_015811}{0.43}{GW200128_022011}{0.47}{GW191204_171526}{0.21}{GW200112_155838}{0.36}{GW200105_162426}{0.16}{GW191105_143521}{0.21}{GW191109_010717}{0.60}{GW200209_085452}{0.37}{GW200115_042309}{0.25}{GW191127_050227}{0.51}{GW200216_220804}{0.53}{GW191215_223052}{0.29}{GW200208_130117}{0.31}{GW200219_094415}{0.38}{GW191103_012549}{0.29}{GW200316_215756}{0.34}{GW200202_154313}{0.23}{GW200129_065458}{0.41}{GW191216_213338}{0.26}}[{\red{???}}]}
\DeclareRobustCommand{\spinonezgwtcthreetenthpercentile}[1]{\IfEqCase{#1}{{GW200224_222234}{-0.12}{GW191129_134029}{-0.08}{GW200311_115853}{-0.32}{GW191230_180458}{-0.37}{GW191222_033537}{-0.37}{GW200225_060421}{-0.52}{GW200302_015811}{-0.25}{GW200128_022011}{-0.12}{GW191204_171526}{-0.04}{GW200112_155838}{-0.18}{GW200105_162426}{-0.11}{GW191105_143521}{-0.22}{GW191109_010717}{-0.79}{GW200209_085452}{-0.46}{GW200115_042309}{-0.62}{GW191127_050227}{-0.15}{GW200216_220804}{-0.24}{GW191215_223052}{-0.30}{GW200208_130117}{-0.42}{GW200219_094415}{-0.45}{GW191103_012549}{-0.02}{GW200316_215756}{-0.05}{GW200202_154313}{-0.12}{GW200129_065458}{-0.22}{GW191216_213338}{-0.03}}[{\red{???}}]}
\DeclareRobustCommand{\spinonezgwtcthreenintiethpercentile}[1]{\IfEqCase{#1}{{GW200224_222234}{0.43}{GW191129_134029}{0.23}{GW200311_115853}{0.20}{GW191230_180458}{0.31}{GW191222_033537}{0.20}{GW200225_060421}{0.11}{GW200302_015811}{0.33}{GW200128_022011}{0.54}{GW191204_171526}{0.33}{GW200112_155838}{0.29}{GW200105_162426}{0.09}{GW191105_143521}{0.14}{GW191109_010717}{0.01}{GW200209_085452}{0.21}{GW200115_042309}{0.05}{GW191127_050227}{0.60}{GW200216_220804}{0.49}{GW191215_223052}{0.20}{GW200208_130117}{0.16}{GW200219_094415}{0.20}{GW191103_012549}{0.46}{GW200316_215756}{0.39}{GW200202_154313}{0.20}{GW200129_065458}{0.36}{GW191216_213338}{0.31}}[{\red{???}}]}
\DeclareRobustCommand{\spintwozgwtcthreeminus}[1]{\IfEqCase{#1}{{GW200224_222234}{0.44}{GW191129_134029}{0.30}{GW200311_115853}{0.45}{GW191230_180458}{0.56}{GW191222_033537}{0.50}{GW200225_060421}{0.55}{GW200302_015811}{0.47}{GW200128_022011}{0.47}{GW191204_171526}{0.35}{GW200112_155838}{0.35}{GW200105_162426}{0.42}{GW191105_143521}{0.34}{GW191109_010717}{0.58}{GW200209_085452}{0.58}{GW200115_042309}{0.59}{GW191127_050227}{0.52}{GW200216_220804}{0.52}{GW191215_223052}{0.50}{GW200208_130117}{0.52}{GW200219_094415}{0.57}{GW191103_012549}{0.42}{GW200316_215756}{0.35}{GW200202_154313}{0.28}{GW200129_065458}{0.50}{GW191216_213338}{0.37}}[{\red{???}}]}
\DeclareRobustCommand{\spintwozgwtcthreemed}[1]{\IfEqCase{#1}{{GW200224_222234}{0.05}{GW191129_134029}{0.07}{GW200311_115853}{0.00}{GW191230_180458}{-0.03}{GW191222_033537}{-0.02}{GW200225_060421}{-0.05}{GW200302_015811}{0.03}{GW200128_022011}{0.04}{GW191204_171526}{0.17}{GW200112_155838}{0.06}{GW200105_162426}{0.00}{GW191105_143521}{0.00}{GW191109_010717}{-0.10}{GW200209_085452}{-0.08}{GW200115_042309}{-0.08}{GW191127_050227}{0.06}{GW200216_220804}{0.05}{GW191215_223052}{-0.02}{GW200208_130117}{-0.03}{GW200219_094415}{-0.04}{GW191103_012549}{0.18}{GW200316_215756}{0.12}{GW200202_154313}{0.05}{GW200129_065458}{0.15}{GW191216_213338}{0.11}}[{\red{???}}]}
\DeclareRobustCommand{\spintwozgwtcthreeplus}[1]{\IfEqCase{#1}{{GW200224_222234}{0.45}{GW191129_134029}{0.50}{GW200311_115853}{0.41}{GW191230_180458}{0.46}{GW191222_033537}{0.42}{GW200225_060421}{0.40}{GW200302_015811}{0.54}{GW200128_022011}{0.52}{GW191204_171526}{0.43}{GW200112_155838}{0.46}{GW200105_162426}{0.38}{GW191105_143521}{0.42}{GW191109_010717}{0.62}{GW200209_085452}{0.41}{GW200115_042309}{0.42}{GW191127_050227}{0.64}{GW200216_220804}{0.61}{GW191215_223052}{0.41}{GW200208_130117}{0.42}{GW200219_094415}{0.47}{GW191103_012549}{0.52}{GW200316_215756}{0.49}{GW200202_154313}{0.42}{GW200129_065458}{0.49}{GW191216_213338}{0.41}}[{\red{???}}]}
\DeclareRobustCommand{\spintwozgwtcthreetenthpercentile}[1]{\IfEqCase{#1}{{GW200224_222234}{-0.25}{GW191129_134029}{-0.13}{GW200311_115853}{-0.33}{GW191230_180458}{-0.46}{GW191222_033537}{-0.39}{GW200225_060421}{-0.47}{GW200302_015811}{-0.31}{GW200128_022011}{-0.30}{GW191204_171526}{-0.09}{GW200112_155838}{-0.19}{GW200105_162426}{-0.26}{GW191105_143521}{-0.24}{GW191109_010717}{-0.56}{GW200209_085452}{-0.53}{GW200115_042309}{-0.57}{GW191127_050227}{-0.31}{GW200216_220804}{-0.32}{GW191215_223052}{-0.39}{GW200208_130117}{-0.44}{GW200219_094415}{-0.48}{GW191103_012549}{-0.13}{GW200316_215756}{-0.12}{GW200202_154313}{-0.13}{GW200129_065458}{-0.21}{GW191216_213338}{-0.14}}[{\red{???}}]}
\DeclareRobustCommand{\spintwozgwtcthreenintiethpercentile}[1]{\IfEqCase{#1}{{GW200224_222234}{0.40}{GW191129_134029}{0.44}{GW200311_115853}{0.29}{GW191230_180458}{0.31}{GW191222_033537}{0.28}{GW200225_060421}{0.22}{GW200302_015811}{0.44}{GW200128_022011}{0.45}{GW191204_171526}{0.51}{GW200112_155838}{0.40}{GW200105_162426}{0.26}{GW191105_143521}{0.29}{GW191109_010717}{0.34}{GW200209_085452}{0.21}{GW200115_042309}{0.22}{GW191127_050227}{0.58}{GW200216_220804}{0.53}{GW191215_223052}{0.27}{GW200208_130117}{0.27}{GW200219_094415}{0.29}{GW191103_012549}{0.60}{GW200316_215756}{0.50}{GW200202_154313}{0.35}{GW200129_065458}{0.53}{GW191216_213338}{0.43}}[{\red{???}}]}
\DeclareRobustCommand{\massonedetgwtcthreeminus}[1]{\IfEqCase{#1}{{GW200224_222234}{5.4}{GW191129_134029}{2.3}{GW200311_115853}{4.6}{GW191230_180458}{13}{GW191222_033537}{9.5}{GW200225_060421}{3.2}{GW200302_015811}{9.7}{GW200128_022011}{9.9}{GW191204_171526}{2.0}{GW200112_155838}{5.2}{GW200105_162426}{1.8}{GW191105_143521}{1.8}{GW191109_010717}{8.9}{GW200209_085452}{9.9}{GW200115_042309}{2.7}{GW191127_050227}{37}{GW200216_220804}{21}{GW191215_223052}{4.8}{GW200208_130117}{8.4}{GW200219_094415}{8.9}{GW191103_012549}{2.3}{GW200316_215756}{3.4}{GW200202_154313}{1.5}{GW200129_065458}{3.3}{GW191216_213338}{2.4}}[{\red{???}}]}
\DeclareRobustCommand{\massonedetgwtcthreemed}[1]{\IfEqCase{#1}{{GW200224_222234}{52.5}{GW191129_134029}{12.3}{GW200311_115853}{41.8}{GW191230_180458}{83}{GW191222_033537}{67.2}{GW200225_060421}{23.6}{GW200302_015811}{48.5}{GW200128_022011}{66.1}{GW191204_171526}{13.4}{GW200112_155838}{44.0}{GW200105_162426}{9.6}{GW191105_143521}{13.0}{GW191109_010717}{81.2}{GW200209_085452}{56.5}{GW200115_042309}{6.3}{GW191127_050227}{86}{GW200216_220804}{84}{GW191215_223052}{33.7}{GW200208_130117}{53.0}{GW200219_094415}{58.6}{GW191103_012549}{14.0}{GW200316_215756}{16.0}{GW200202_154313}{11.0}{GW200129_065458}{40.2}{GW191216_213338}{13.0}}[{\red{???}}]}
\DeclareRobustCommand{\massonedetgwtcthreeplus}[1]{\IfEqCase{#1}{{GW200224_222234}{9.1}{GW191129_134029}{4.9}{GW200311_115853}{8.2}{GW191230_180458}{19}{GW191222_033537}{14.7}{GW200225_060421}{5.6}{GW200302_015811}{10.8}{GW200128_022011}{16.6}{GW191204_171526}{3.8}{GW200112_155838}{8.2}{GW200105_162426}{1.9}{GW191105_143521}{4.5}{GW191109_010717}{12.9}{GW200209_085452}{15.4}{GW200115_042309}{2.1}{GW191127_050227}{60}{GW200216_220804}{28}{GW191215_223052}{9.4}{GW200208_130117}{12.2}{GW200219_094415}{13.5}{GW191103_012549}{7.4}{GW200316_215756}{12.3}{GW200202_154313}{3.8}{GW200129_065458}{12.2}{GW191216_213338}{5.0}}[{\red{???}}]}
\DeclareRobustCommand{\massonedetgwtcthreetenthpercentile}[1]{\IfEqCase{#1}{{GW200224_222234}{48.1}{GW191129_134029}{10.2}{GW200311_115853}{38.1}{GW191230_180458}{73}{GW191222_033537}{59.4}{GW200225_060421}{20.8}{GW200302_015811}{40.7}{GW200128_022011}{58.0}{GW191204_171526}{11.7}{GW200112_155838}{39.8}{GW200105_162426}{8.5}{GW191105_143521}{11.4}{GW191109_010717}{74.0}{GW200209_085452}{48.5}{GW200115_042309}{3.9}{GW191127_050227}{55}{GW200216_220804}{68}{GW191215_223052}{29.6}{GW200208_130117}{46.1}{GW200219_094415}{51.1}{GW191103_012549}{11.9}{GW200316_215756}{13.0}{GW200202_154313}{9.7}{GW200129_065458}{37.5}{GW191216_213338}{10.8}}[{\red{???}}]}
\DeclareRobustCommand{\massonedetgwtcthreenintiethpercentile}[1]{\IfEqCase{#1}{{GW200224_222234}{59.3}{GW191129_134029}{15.9}{GW200311_115853}{47.8}{GW191230_180458}{98}{GW191222_033537}{77.9}{GW200225_060421}{27.9}{GW200302_015811}{56.6}{GW200128_022011}{78.3}{GW191204_171526}{16.2}{GW200112_155838}{50.3}{GW200105_162426}{10.6}{GW191105_143521}{16.3}{GW191109_010717}{90.3}{GW200209_085452}{67.7}{GW200115_042309}{7.9}{GW191127_050227}{133}{GW200216_220804}{104}{GW191215_223052}{40.5}{GW200208_130117}{62.2}{GW200219_094415}{69.0}{GW191103_012549}{19.3}{GW200316_215756}{24.2}{GW200202_154313}{13.8}{GW200129_065458}{50.4}{GW191216_213338}{16.3}}[{\red{???}}]}
\DeclareRobustCommand{\chieffinfinityonlyprecavggwtcthreeminus}[1]{\IfEqCase{#1}{{GW200224_222234}{0.15}{GW191129_134029}{0.08}{GW200311_115853}{0.20}{GW191230_180458}{0.30}{GW191222_033537}{0.25}{GW200225_060421}{0.28}{GW200302_015811}{0.25}{GW200128_022011}{0.25}{GW191204_171526}{0.05}{GW200112_155838}{0.15}{GW200105_162426}{0.18}{GW191105_143521}{0.09}{GW191109_010717}{0.31}{GW200209_085452}{0.30}{GW200115_042309}{0.41}{GW191127_050227}{0.36}{GW200216_220804}{0.36}{GW191215_223052}{0.21}{GW200208_130117}{0.27}{GW200219_094415}{0.29}{GW191103_012549}{0.10}{GW200316_215756}{0.10}{GW200202_154313}{0.06}{GW200129_065458}{0.16}{GW191216_213338}{0.06}}[{\red{???}}]}
\DeclareRobustCommand{\chieffinfinityonlyprecavggwtcthreemed}[1]{\IfEqCase{#1}{{GW200224_222234}{0.11}{GW191129_134029}{0.06}{GW200311_115853}{-0.02}{GW191230_180458}{-0.03}{GW191222_033537}{-0.04}{GW200225_060421}{-0.12}{GW200302_015811}{0.03}{GW200128_022011}{0.14}{GW191204_171526}{0.16}{GW200112_155838}{0.06}{GW200105_162426}{0.00}{GW191105_143521}{-0.02}{GW191109_010717}{-0.29}{GW200209_085452}{-0.10}{GW200115_042309}{-0.15}{GW191127_050227}{0.18}{GW200216_220804}{0.10}{GW191215_223052}{-0.03}{GW200208_130117}{-0.07}{GW200219_094415}{-0.08}{GW191103_012549}{0.21}{GW200316_215756}{0.13}{GW200202_154313}{0.04}{GW200129_065458}{0.11}{GW191216_213338}{0.11}}[{\red{???}}]}
\DeclareRobustCommand{\chieffinfinityonlyprecavggwtcthreeplus}[1]{\IfEqCase{#1}{{GW200224_222234}{0.15}{GW191129_134029}{0.16}{GW200311_115853}{0.16}{GW191230_180458}{0.26}{GW191222_033537}{0.20}{GW200225_060421}{0.17}{GW200302_015811}{0.26}{GW200128_022011}{0.24}{GW191204_171526}{0.08}{GW200112_155838}{0.15}{GW200105_162426}{0.13}{GW191105_143521}{0.13}{GW191109_010717}{0.42}{GW200209_085452}{0.24}{GW200115_042309}{0.24}{GW191127_050227}{0.34}{GW200216_220804}{0.34}{GW191215_223052}{0.17}{GW200208_130117}{0.22}{GW200219_094415}{0.23}{GW191103_012549}{0.16}{GW200316_215756}{0.27}{GW200202_154313}{0.13}{GW200129_065458}{0.11}{GW191216_213338}{0.13}}[{\red{???}}]}
\DeclareRobustCommand{\chieffinfinityonlyprecavggwtcthreetenthpercentile}[1]{\IfEqCase{#1}{{GW200224_222234}{-0.01}{GW191129_134029}{0.00}{GW200311_115853}{-0.17}{GW191230_180458}{-0.26}{GW191222_033537}{-0.23}{GW200225_060421}{-0.34}{GW200302_015811}{-0.15}{GW200128_022011}{-0.05}{GW191204_171526}{0.12}{GW200112_155838}{-0.05}{GW200105_162426}{-0.10}{GW191105_143521}{-0.09}{GW191109_010717}{-0.54}{GW200209_085452}{-0.33}{GW200115_042309}{-0.51}{GW191127_050227}{-0.10}{GW200216_220804}{-0.17}{GW191215_223052}{-0.19}{GW200208_130117}{-0.27}{GW200219_094415}{-0.30}{GW191103_012549}{0.13}{GW200316_215756}{0.04}{GW200202_154313}{-0.01}{GW200129_065458}{0.00}{GW191216_213338}{0.06}}[{\red{???}}]}
\DeclareRobustCommand{\chieffinfinityonlyprecavggwtcthreenintiethpercentile}[1]{\IfEqCase{#1}{{GW200224_222234}{0.22}{GW191129_134029}{0.19}{GW200311_115853}{0.10}{GW191230_180458}{0.18}{GW191222_033537}{0.11}{GW200225_060421}{0.02}{GW200302_015811}{0.23}{GW200128_022011}{0.33}{GW191204_171526}{0.22}{GW200112_155838}{0.18}{GW200105_162426}{0.08}{GW191105_143521}{0.07}{GW191109_010717}{0.00}{GW200209_085452}{0.09}{GW200115_042309}{0.04}{GW191127_050227}{0.45}{GW200216_220804}{0.36}{GW191215_223052}{0.10}{GW200208_130117}{0.09}{GW200219_094415}{0.10}{GW191103_012549}{0.33}{GW200316_215756}{0.32}{GW200202_154313}{0.13}{GW200129_065458}{0.20}{GW191216_213338}{0.20}}[{\red{???}}]}
\DeclareRobustCommand{\chipinfinityonlyprecavggwtcthreeminus}[1]{\IfEqCase{#1}{{GW200224_222234}{0.37}{GW191129_134029}{0.19}{GW200311_115853}{0.35}{GW191230_180458}{0.39}{GW191222_033537}{0.32}{GW200225_060421}{0.38}{GW200302_015811}{0.29}{GW200128_022011}{0.40}{GW191204_171526}{0.26}{GW200112_155838}{0.30}{GW200105_162426}{0.07}{GW191105_143521}{0.24}{GW191109_010717}{0.38}{GW200209_085452}{0.38}{GW200115_042309}{0.16}{GW191127_050227}{0.41}{GW200216_220804}{0.35}{GW191215_223052}{0.38}{GW200208_130117}{0.29}{GW200219_094415}{0.35}{GW191103_012549}{0.27}{GW200316_215756}{0.20}{GW200202_154313}{0.22}{GW200129_065458}{0.39}{GW191216_213338}{0.15}}[{\red{???}}]}
\DeclareRobustCommand{\chipinfinityonlyprecavggwtcthreemed}[1]{\IfEqCase{#1}{{GW200224_222234}{0.50}{GW191129_134029}{0.26}{GW200311_115853}{0.45}{GW191230_180458}{0.52}{GW191222_033537}{0.41}{GW200225_060421}{0.53}{GW200302_015811}{0.38}{GW200128_022011}{0.56}{GW191204_171526}{0.39}{GW200112_155838}{0.40}{GW200105_162426}{0.09}{GW191105_143521}{0.30}{GW191109_010717}{0.63}{GW200209_085452}{0.52}{GW200115_042309}{0.20}{GW191127_050227}{0.52}{GW200216_220804}{0.45}{GW191215_223052}{0.50}{GW200208_130117}{0.38}{GW200219_094415}{0.47}{GW191103_012549}{0.40}{GW200316_215756}{0.29}{GW200202_154313}{0.28}{GW200129_065458}{0.54}{GW191216_213338}{0.23}}[{\red{???}}]}
\DeclareRobustCommand{\chipinfinityonlyprecavggwtcthreeplus}[1]{\IfEqCase{#1}{{GW200224_222234}{0.37}{GW191129_134029}{0.36}{GW200311_115853}{0.39}{GW191230_180458}{0.38}{GW191222_033537}{0.42}{GW200225_060421}{0.35}{GW200302_015811}{0.44}{GW200128_022011}{0.34}{GW191204_171526}{0.35}{GW200112_155838}{0.38}{GW200105_162426}{0.17}{GW191105_143521}{0.45}{GW191109_010717}{0.28}{GW200209_085452}{0.38}{GW200115_042309}{0.34}{GW191127_050227}{0.40}{GW200216_220804}{0.42}{GW191215_223052}{0.38}{GW200208_130117}{0.42}{GW200219_094415}{0.40}{GW191103_012549}{0.41}{GW200316_215756}{0.38}{GW200202_154313}{0.40}{GW200129_065458}{0.39}{GW191216_213338}{0.35}}[{\red{???}}]}
\DeclareRobustCommand{\chipinfinityonlyprecavggwtcthreetenthpercentile}[1]{\IfEqCase{#1}{{GW200224_222234}{0.20}{GW191129_134029}{0.10}{GW200311_115853}{0.16}{GW191230_180458}{0.20}{GW191222_033537}{0.14}{GW200225_060421}{0.22}{GW200302_015811}{0.13}{GW200128_022011}{0.23}{GW191204_171526}{0.18}{GW200112_155838}{0.14}{GW200105_162426}{0.03}{GW191105_143521}{0.09}{GW191109_010717}{0.33}{GW200209_085452}{0.20}{GW200115_042309}{0.06}{GW191127_050227}{0.17}{GW200216_220804}{0.15}{GW191215_223052}{0.19}{GW200208_130117}{0.13}{GW200219_094415}{0.18}{GW191103_012549}{0.18}{GW200316_215756}{0.12}{GW200202_154313}{0.09}{GW200129_065458}{0.21}{GW191216_213338}{0.10}}[{\red{???}}]}
\DeclareRobustCommand{\chipinfinityonlyprecavggwtcthreenintiethpercentile}[1]{\IfEqCase{#1}{{GW200224_222234}{0.80}{GW191129_134029}{0.54}{GW200311_115853}{0.77}{GW191230_180458}{0.84}{GW191222_033537}{0.75}{GW200225_060421}{0.82}{GW200302_015811}{0.74}{GW200128_022011}{0.84}{GW191204_171526}{0.67}{GW200112_155838}{0.70}{GW200105_162426}{0.19}{GW191105_143521}{0.65}{GW191109_010717}{0.87}{GW200209_085452}{0.84}{GW200115_042309}{0.46}{GW191127_050227}{0.88}{GW200216_220804}{0.80}{GW191215_223052}{0.82}{GW200208_130117}{0.71}{GW200219_094415}{0.80}{GW191103_012549}{0.72}{GW200316_215756}{0.58}{GW200202_154313}{0.59}{GW200129_065458}{0.91}{GW191216_213338}{0.48}}[{\red{???}}]}
\DeclareRobustCommand{\logpriorgwtcthreeminus}[1]{\IfEqCase{#1}{{GW200224_222234}{10.9}{GW200311_115853}{10.8}{GW191230_180458}{10.8}{GW191222_033537}{9.0}{GW200225_060421}{9.3}{GW200302_015811}{9.0}{GW200128_022011}{9.2}{GW200112_155838}{9.0}{GW191105_143521}{11.1}{GW191109_010717}{9.4}{GW200209_085452}{11.0}{GW191127_050227}{11.0}{GW200216_220804}{10.5}{GW191215_223052}{10.9}{GW200208_130117}{10.9}{GW200219_094415}{10.8}{GW191103_012549}{9.0}{GW200202_154313}{10.9}{GW200129_065458}{11.1}}[{\red{???}}]}
\DeclareRobustCommand{\logpriorgwtcthreemed}[1]{\IfEqCase{#1}{{GW200224_222234}{127.6}{GW200311_115853}{132.7}{GW191230_180458}{129.7}{GW191222_033537}{90.7}{GW200225_060421}{100.3}{GW200302_015811}{82.6}{GW200128_022011}{93.6}{GW200112_155838}{72.1}{GW191105_143521}{129.4}{GW191109_010717}{90.9}{GW200209_085452}{132.4}{GW191127_050227}{129.5}{GW200216_220804}{131.3}{GW191215_223052}{131.5}{GW200208_130117}{128.4}{GW200219_094415}{132.2}{GW191103_012549}{100.7}{GW200202_154313}{135.3}{GW200129_065458}{132.8}}[{\red{???}}]}
\DeclareRobustCommand{\logpriorgwtcthreeplus}[1]{\IfEqCase{#1}{{GW200224_222234}{8.7}{GW200311_115853}{8.6}{GW191230_180458}{8.8}{GW191222_033537}{6.9}{GW200225_060421}{6.9}{GW200302_015811}{6.9}{GW200128_022011}{6.9}{GW200112_155838}{6.8}{GW191105_143521}{8.8}{GW191109_010717}{7.2}{GW200209_085452}{8.7}{GW191127_050227}{8.9}{GW200216_220804}{8.7}{GW191215_223052}{8.8}{GW200208_130117}{8.7}{GW200219_094415}{8.8}{GW191103_012549}{6.9}{GW200202_154313}{8.7}{GW200129_065458}{8.8}}[{\red{???}}]}
\DeclareRobustCommand{\logpriorgwtcthreetenthpercentile}[1]{\IfEqCase{#1}{{GW200224_222234}{119.3}{GW200311_115853}{124.6}{GW191230_180458}{121.4}{GW191222_033537}{83.9}{GW200225_060421}{93.4}{GW200302_015811}{75.8}{GW200128_022011}{86.7}{GW200112_155838}{65.4}{GW191105_143521}{121.1}{GW191109_010717}{83.8}{GW200209_085452}{124.1}{GW191127_050227}{121.2}{GW200216_220804}{123.3}{GW191215_223052}{123.1}{GW200208_130117}{120.2}{GW200219_094415}{124.0}{GW191103_012549}{93.9}{GW200202_154313}{127.1}{GW200129_065458}{124.4}}[{\red{???}}]}
\DeclareRobustCommand{\logpriorgwtcthreenintiethpercentile}[1]{\IfEqCase{#1}{{GW200224_222234}{134.5}{GW200311_115853}{139.6}{GW191230_180458}{136.7}{GW191222_033537}{96.3}{GW200225_060421}{105.9}{GW200302_015811}{88.2}{GW200128_022011}{99.2}{GW200112_155838}{77.6}{GW191105_143521}{136.5}{GW191109_010717}{96.6}{GW200209_085452}{139.4}{GW191127_050227}{136.7}{GW200216_220804}{138.2}{GW191215_223052}{138.6}{GW200208_130117}{135.4}{GW200219_094415}{139.2}{GW191103_012549}{106.3}{GW200202_154313}{142.2}{GW200129_065458}{139.9}}[{\red{???}}]}
\DeclareRobustCommand{\networkmatchedfiltersnrgwtcthreeminus}[1]{\IfEqCase{#1}{{GW200224_222234}{0.2}{GW191129_134029}{0.3}{GW200311_115853}{0.2}{GW191230_180458}{0.4}{GW191222_033537}{0.3}{GW200225_060421}{0.4}{GW200302_015811}{0.4}{GW200128_022011}{0.4}{GW191204_171526}{0.2}{GW200112_155838}{0.2}{GW200105_162426}{0.4}{GW191105_143521}{0.5}{GW191109_010717}{0.5}{GW200209_085452}{0.5}{GW200115_042309}{0.5}{GW191127_050227}{0.6}{GW200216_220804}{0.6}{GW191215_223052}{0.4}{GW200208_130117}{0.5}{GW200219_094415}{0.5}{GW191103_012549}{0.5}{GW200316_215756}{0.7}{GW200202_154313}{0.4}{GW200129_065458}{0.2}{GW191216_213338}{0.2}}[{\red{???}}]}
\DeclareRobustCommand{\networkmatchedfiltersnrgwtcthreemed}[1]{\IfEqCase{#1}{{GW200224_222234}{20.0}{GW191129_134029}{13.1}{GW200311_115853}{17.8}{GW191230_180458}{10.4}{GW191222_033537}{12.5}{GW200225_060421}{12.5}{GW200302_015811}{10.8}{GW200128_022011}{10.6}{GW191204_171526}{17.5}{GW200112_155838}{19.8}{GW200105_162426}{13.7}{GW191105_143521}{9.7}{GW191109_010717}{17.2}{GW200209_085452}{9.6}{GW200115_042309}{11.3}{GW191127_050227}{9.1}{GW200216_220804}{8.1}{GW191215_223052}{11.2}{GW200208_130117}{10.8}{GW200219_094415}{10.7}{GW191103_012549}{8.9}{GW200316_215756}{10.3}{GW200202_154313}{10.8}{GW200129_065458}{26.8}{GW191216_213338}{18.6}}[{\red{???}}]}
\DeclareRobustCommand{\networkmatchedfiltersnrgwtcthreeplus}[1]{\IfEqCase{#1}{{GW200224_222234}{0.2}{GW191129_134029}{0.2}{GW200311_115853}{0.2}{GW191230_180458}{0.3}{GW191222_033537}{0.2}{GW200225_060421}{0.3}{GW200302_015811}{0.3}{GW200128_022011}{0.3}{GW191204_171526}{0.2}{GW200112_155838}{0.1}{GW200105_162426}{0.2}{GW191105_143521}{0.3}{GW191109_010717}{0.5}{GW200209_085452}{0.4}{GW200115_042309}{0.3}{GW191127_050227}{0.5}{GW200216_220804}{0.3}{GW191215_223052}{0.3}{GW200208_130117}{0.3}{GW200219_094415}{0.3}{GW191103_012549}{0.3}{GW200316_215756}{0.4}{GW200202_154313}{0.2}{GW200129_065458}{0.2}{GW191216_213338}{0.2}}[{\red{???}}]}
\DeclareRobustCommand{\networkmatchedfiltersnrgwtcthreetenthpercentile}[1]{\IfEqCase{#1}{{GW200224_222234}{19.8}{GW191129_134029}{12.9}{GW200311_115853}{17.7}{GW191230_180458}{10.1}{GW191222_033537}{12.2}{GW200225_060421}{12.2}{GW200302_015811}{10.5}{GW200128_022011}{10.3}{GW191204_171526}{17.3}{GW200112_155838}{19.6}{GW200105_162426}{13.4}{GW191105_143521}{9.3}{GW191109_010717}{16.9}{GW200209_085452}{9.2}{GW200115_042309}{10.9}{GW191127_050227}{8.7}{GW200216_220804}{7.7}{GW191215_223052}{10.9}{GW200208_130117}{10.5}{GW200219_094415}{10.3}{GW191103_012549}{8.5}{GW200316_215756}{9.8}{GW200202_154313}{10.5}{GW200129_065458}{26.6}{GW191216_213338}{18.4}}[{\red{???}}]}
\DeclareRobustCommand{\networkmatchedfiltersnrgwtcthreenintiethpercentile}[1]{\IfEqCase{#1}{{GW200224_222234}{20.1}{GW191129_134029}{13.3}{GW200311_115853}{18.0}{GW191230_180458}{10.7}{GW191222_033537}{12.6}{GW200225_060421}{12.7}{GW200302_015811}{11.1}{GW200128_022011}{10.9}{GW191204_171526}{17.6}{GW200112_155838}{19.9}{GW200105_162426}{13.9}{GW191105_143521}{9.9}{GW191109_010717}{17.7}{GW200209_085452}{9.9}{GW200115_042309}{11.5}{GW191127_050227}{9.5}{GW200216_220804}{8.4}{GW191215_223052}{11.5}{GW200208_130117}{11.0}{GW200219_094415}{10.9}{GW191103_012549}{9.1}{GW200316_215756}{10.6}{GW200202_154313}{11.0}{GW200129_065458}{26.9}{GW191216_213338}{18.7}}[{\red{???}}]}
\DeclareRobustCommand{\networkoptimalsnrgwtcthreeminus}[1]{\IfEqCase{#1}{{GW200224_222234}{1.7}{GW191129_134029}{1.7}{GW200311_115853}{1.7}{GW191230_180458}{1.7}{GW191222_033537}{1.7}{GW200225_060421}{1.7}{GW200302_015811}{1.7}{GW200128_022011}{1.7}{GW191204_171526}{1.7}{GW200112_155838}{1.7}{GW200105_162426}{1.7}{GW191105_143521}{1.8}{GW191109_010717}{1.7}{GW200209_085452}{1.7}{GW200115_042309}{1.7}{GW191127_050227}{1.8}{GW200216_220804}{1.7}{GW191215_223052}{1.7}{GW200208_130117}{1.7}{GW200219_094415}{1.8}{GW191103_012549}{1.8}{GW200316_215756}{1.8}{GW200202_154313}{1.7}{GW200129_065458}{1.6}{GW191216_213338}{1.7}}[{\red{???}}]}
\DeclareRobustCommand{\networkoptimalsnrgwtcthreemed}[1]{\IfEqCase{#1}{{GW200224_222234}{19.8}{GW191129_134029}{12.9}{GW200311_115853}{17.6}{GW191230_180458}{10.2}{GW191222_033537}{12.1}{GW200225_060421}{12.2}{GW200302_015811}{10.5}{GW200128_022011}{10.4}{GW191204_171526}{17.2}{GW200112_155838}{19.6}{GW200105_162426}{13.3}{GW191105_143521}{9.2}{GW191109_010717}{17.0}{GW200209_085452}{9.3}{GW200115_042309}{10.9}{GW191127_050227}{8.6}{GW200216_220804}{7.6}{GW191215_223052}{10.9}{GW200208_130117}{10.4}{GW200219_094415}{10.3}{GW191103_012549}{8.4}{GW200316_215756}{9.8}{GW200202_154313}{10.4}{GW200129_065458}{26.6}{GW191216_213338}{18.3}}[{\red{???}}]}
\DeclareRobustCommand{\networkoptimalsnrgwtcthreeplus}[1]{\IfEqCase{#1}{{GW200224_222234}{1.7}{GW191129_134029}{1.7}{GW200311_115853}{1.7}{GW191230_180458}{1.7}{GW191222_033537}{1.7}{GW200225_060421}{1.7}{GW200302_015811}{1.7}{GW200128_022011}{1.6}{GW191204_171526}{1.7}{GW200112_155838}{1.7}{GW200105_162426}{1.7}{GW191105_143521}{1.7}{GW191109_010717}{1.7}{GW200209_085452}{1.7}{GW200115_042309}{1.7}{GW191127_050227}{1.8}{GW200216_220804}{1.8}{GW191215_223052}{1.7}{GW200208_130117}{1.7}{GW200219_094415}{1.7}{GW191103_012549}{1.7}{GW200316_215756}{1.7}{GW200202_154313}{1.7}{GW200129_065458}{1.7}{GW191216_213338}{1.7}}[{\red{???}}]}
\DeclareRobustCommand{\networkoptimalsnrgwtcthreetenthpercentile}[1]{\IfEqCase{#1}{{GW200224_222234}{18.5}{GW191129_134029}{11.6}{GW200311_115853}{16.3}{GW191230_180458}{8.8}{GW191222_033537}{10.8}{GW200225_060421}{10.8}{GW200302_015811}{9.2}{GW200128_022011}{9.0}{GW191204_171526}{15.9}{GW200112_155838}{18.3}{GW200105_162426}{12.0}{GW191105_143521}{7.8}{GW191109_010717}{15.7}{GW200209_085452}{8.0}{GW200115_042309}{9.5}{GW191127_050227}{7.2}{GW200216_220804}{6.2}{GW191215_223052}{9.6}{GW200208_130117}{9.0}{GW200219_094415}{8.9}{GW191103_012549}{7.0}{GW200316_215756}{8.5}{GW200202_154313}{9.1}{GW200129_065458}{25.3}{GW191216_213338}{17.0}}[{\red{???}}]}
\DeclareRobustCommand{\networkoptimalsnrgwtcthreenintiethpercentile}[1]{\IfEqCase{#1}{{GW200224_222234}{21.1}{GW191129_134029}{14.2}{GW200311_115853}{18.9}{GW191230_180458}{11.5}{GW191222_033537}{13.4}{GW200225_060421}{13.5}{GW200302_015811}{11.8}{GW200128_022011}{11.7}{GW191204_171526}{18.5}{GW200112_155838}{20.9}{GW200105_162426}{14.6}{GW191105_143521}{10.5}{GW191109_010717}{18.4}{GW200209_085452}{10.7}{GW200115_042309}{12.2}{GW191127_050227}{10.0}{GW200216_220804}{8.9}{GW191215_223052}{12.2}{GW200208_130117}{11.7}{GW200219_094415}{11.6}{GW191103_012549}{9.7}{GW200316_215756}{11.2}{GW200202_154313}{11.8}{GW200129_065458}{27.9}{GW191216_213338}{19.6}}[{\red{???}}]}
\DeclareRobustCommand{\PEpercentBNSgwtcthree}[1]{\IfEqCase{#1}{{GW200224_222234}{0}{GW191129_134029}{0}{GW200311_115853}{0}{GW191230_180458}{0}{GW191222_033537}{0}{GW200225_060421}{0}{GW200302_015811}{0}{GW200128_022011}{0}{GW191204_171526}{0}{GW200112_155838}{0}{GW200105_162426}{0}{GW191105_143521}{0}{GW191109_010717}{0}{GW200209_085452}{0}{GW200115_042309}{1}{GW191127_050227}{0}{GW200216_220804}{0}{GW191215_223052}{0}{GW200208_130117}{0}{GW200219_094415}{0}{GW191103_012549}{0}{GW200316_215756}{0}{GW200202_154313}{0}{GW200129_065458}{0}{GW191216_213338}{0}}[{\red{???}}]}
\DeclareRobustCommand{\PEpercentNSBHgwtcthree}[1]{\IfEqCase{#1}{{GW200224_222234}{0}{GW191129_134029}{0}{GW200311_115853}{0}{GW191230_180458}{0}{GW191222_033537}{0}{GW200225_060421}{0}{GW200302_015811}{0}{GW200128_022011}{0}{GW191204_171526}{0}{GW200112_155838}{0}{GW200105_162426}{99}{GW191105_143521}{0}{GW191109_010717}{0}{GW200209_085452}{0}{GW200115_042309}{99}{GW191127_050227}{0}{GW200216_220804}{0}{GW191215_223052}{0}{GW200208_130117}{0}{GW200219_094415}{0}{GW191103_012549}{0}{GW200316_215756}{0}{GW200202_154313}{0}{GW200129_065458}{0}{GW191216_213338}{0}}[{\red{???}}]}
\DeclareRobustCommand{\PEpercentBBHgwtcthree}[1]{\IfEqCase{#1}{{GW200224_222234}{100}{GW191129_134029}{100}{GW200311_115853}{100}{GW191230_180458}{100}{GW191222_033537}{100}{GW200225_060421}{100}{GW200302_015811}{100}{GW200128_022011}{100}{GW191204_171526}{100}{GW200112_155838}{100}{GW200105_162426}{1}{GW191105_143521}{100}{GW191109_010717}{100}{GW200209_085452}{100}{GW200115_042309}{0}{GW191127_050227}{100}{GW200216_220804}{100}{GW191215_223052}{100}{GW200208_130117}{100}{GW200219_094415}{100}{GW191103_012549}{100}{GW200316_215756}{100}{GW200202_154313}{100}{GW200129_065458}{100}{GW191216_213338}{100}}[{\red{???}}]}
\DeclareRobustCommand{\PEpercentMassGapgwtcthree}[1]{\IfEqCase{#1}{{GW200224_222234}{0}{GW191129_134029}{0}{GW200311_115853}{0}{GW191230_180458}{0}{GW191222_033537}{0}{GW200225_060421}{0}{GW200302_015811}{0}{GW200128_022011}{0}{GW191204_171526}{0}{GW200112_155838}{0}{GW200105_162426}{0}{GW191105_143521}{0}{GW191109_010717}{0}{GW200209_085452}{0}{GW200115_042309}{0}{GW191127_050227}{0}{GW200216_220804}{0}{GW191215_223052}{0}{GW200208_130117}{0}{GW200219_094415}{0}{GW191103_012549}{0}{GW200316_215756}{0}{GW200202_154313}{0}{GW200129_065458}{0}{GW191216_213338}{0}}[{\red{???}}]}
\DeclareRobustCommand{\percentmassonelessthanthreegwtcthree}[1]{\IfEqCase{#1}{{GW200224_222234}{0}{GW191129_134029}{0}{GW200311_115853}{0}{GW191230_180458}{0}{GW191222_033537}{0}{GW200225_060421}{0}{GW200302_015811}{0}{GW200128_022011}{0}{GW191204_171526}{0}{GW200112_155838}{0}{GW200105_162426}{0}{GW191105_143521}{0}{GW191109_010717}{0}{GW200209_085452}{0}{GW200115_042309}{1}{GW191127_050227}{0}{GW200216_220804}{0}{GW191215_223052}{0}{GW200208_130117}{0}{GW200219_094415}{0}{GW191103_012549}{0}{GW200316_215756}{0}{GW200202_154313}{0}{GW200129_065458}{0}{GW191216_213338}{0}}[{\red{???}}]}
\DeclareRobustCommand{\percentmasstwolessthanthreegwtcthree}[1]{\IfEqCase{#1}{{GW200224_222234}{0}{GW191129_134029}{0}{GW200311_115853}{0}{GW191230_180458}{0}{GW191222_033537}{0}{GW200225_060421}{0}{GW200302_015811}{0}{GW200128_022011}{0}{GW191204_171526}{0}{GW200112_155838}{0}{GW200105_162426}{99}{GW191105_143521}{0}{GW191109_010717}{0}{GW200209_085452}{0}{GW200115_042309}{100}{GW191127_050227}{0}{GW200216_220804}{0}{GW191215_223052}{0}{GW200208_130117}{0}{GW200219_094415}{0}{GW191103_012549}{0}{GW200316_215756}{0}{GW200202_154313}{0}{GW200129_065458}{0}{GW191216_213338}{0}}[{\red{???}}]}
\DeclareRobustCommand{\percentmassonelessthanfivegwtcthree}[1]{\IfEqCase{#1}{{GW200224_222234}{0}{GW191129_134029}{0}{GW200311_115853}{0}{GW191230_180458}{0}{GW191222_033537}{0}{GW200225_060421}{0}{GW200302_015811}{0}{GW200128_022011}{0}{GW191204_171526}{0}{GW200112_155838}{0}{GW200105_162426}{1}{GW191105_143521}{0}{GW191109_010717}{0}{GW200209_085452}{0}{GW200115_042309}{29}{GW191127_050227}{0}{GW200216_220804}{0}{GW191215_223052}{0}{GW200208_130117}{0}{GW200219_094415}{0}{GW191103_012549}{0}{GW200316_215756}{0}{GW200202_154313}{0}{GW200129_065458}{0}{GW191216_213338}{0}}[{\red{???}}]}
\DeclareRobustCommand{\percentmasstwolessthanfivegwtcthree}[1]{\IfEqCase{#1}{{GW200224_222234}{0}{GW191129_134029}{5}{GW200311_115853}{0}{GW191230_180458}{0}{GW191222_033537}{0}{GW200225_060421}{0}{GW200302_015811}{0}{GW200128_022011}{0}{GW191204_171526}{0}{GW200112_155838}{0}{GW200105_162426}{100}{GW191105_143521}{1}{GW191109_010717}{0}{GW200209_085452}{0}{GW200115_042309}{100}{GW191127_050227}{0}{GW200216_220804}{0}{GW191215_223052}{0}{GW200208_130117}{0}{GW200219_094415}{0}{GW191103_012549}{2}{GW200316_215756}{5}{GW200202_154313}{1}{GW200129_065458}{0}{GW191216_213338}{2}}[{\red{???}}]}
\DeclareRobustCommand{\percentmassonemorethansixtyfivegwtcthree}[1]{\IfEqCase{#1}{{GW200224_222234}{0}{GW191129_134029}{0}{GW200311_115853}{0}{GW191230_180458}{2}{GW191222_033537}{0}{GW200225_060421}{0}{GW200302_015811}{0}{GW200128_022011}{0}{GW191204_171526}{0}{GW200112_155838}{0}{GW200105_162426}{0}{GW191105_143521}{0}{GW191109_010717}{51}{GW200209_085452}{0}{GW200115_042309}{0}{GW191127_050227}{30}{GW200216_220804}{13}{GW191215_223052}{0}{GW200208_130117}{0}{GW200219_094415}{0}{GW191103_012549}{0}{GW200316_215756}{0}{GW200202_154313}{0}{GW200129_065458}{0}{GW191216_213338}{0}}[{\red{???}}]}
\DeclareRobustCommand{\percentmasstwomorethansixtyfivegwtcthree}[1]{\IfEqCase{#1}{{GW200224_222234}{0}{GW191129_134029}{0}{GW200311_115853}{0}{GW191230_180458}{0}{GW191222_033537}{0}{GW200225_060421}{0}{GW200302_015811}{0}{GW200128_022011}{0}{GW191204_171526}{0}{GW200112_155838}{0}{GW200105_162426}{0}{GW191105_143521}{0}{GW191109_010717}{2}{GW200209_085452}{0}{GW200115_042309}{0}{GW191127_050227}{0}{GW200216_220804}{0}{GW191215_223052}{0}{GW200208_130117}{0}{GW200219_094415}{0}{GW191103_012549}{0}{GW200316_215756}{0}{GW200202_154313}{0}{GW200129_065458}{0}{GW191216_213338}{0}}[{\red{???}}]}
\DeclareRobustCommand{\percentmassonemorethanonetwentygwtcthree}[1]{\IfEqCase{#1}{{GW200224_222234}{0}{GW191129_134029}{0}{GW200311_115853}{0}{GW191230_180458}{0}{GW191222_033537}{0}{GW200225_060421}{0}{GW200302_015811}{0}{GW200128_022011}{0}{GW191204_171526}{0}{GW200112_155838}{0}{GW200105_162426}{0}{GW191105_143521}{0}{GW191109_010717}{0}{GW200209_085452}{0}{GW200115_042309}{0}{GW191127_050227}{1}{GW200216_220804}{0}{GW191215_223052}{0}{GW200208_130117}{0}{GW200219_094415}{0}{GW191103_012549}{0}{GW200316_215756}{0}{GW200202_154313}{0}{GW200129_065458}{0}{GW191216_213338}{0}}[{\red{???}}]}
\DeclareRobustCommand{\percentmasstwomorethanonetwentygwtcthree}[1]{\IfEqCase{#1}{{GW200224_222234}{0}{GW191129_134029}{0}{GW200311_115853}{0}{GW191230_180458}{0}{GW191222_033537}{0}{GW200225_060421}{0}{GW200302_015811}{0}{GW200128_022011}{0}{GW191204_171526}{0}{GW200112_155838}{0}{GW200105_162426}{0}{GW191105_143521}{0}{GW191109_010717}{0}{GW200209_085452}{0}{GW200115_042309}{0}{GW191127_050227}{0}{GW200216_220804}{0}{GW191215_223052}{0}{GW200208_130117}{0}{GW200219_094415}{0}{GW191103_012549}{0}{GW200316_215756}{0}{GW200202_154313}{0}{GW200129_065458}{0}{GW191216_213338}{0}}[{\red{???}}]}
\DeclareRobustCommand{\percentmassfinalmorethanonehundredgwtcthree}[1]{\IfEqCase{#1}{{GW200224_222234}{0}{GW191129_134029}{0}{GW200311_115853}{0}{GW191230_180458}{2}{GW191222_033537}{0}{GW200225_060421}{0}{GW200302_015811}{0}{GW200128_022011}{0}{GW191204_171526}{0}{GW200112_155838}{0}{GW200105_162426}{0}{GW191105_143521}{0}{GW191109_010717}{78}{GW200209_085452}{0}{GW200115_042309}{0}{GW191127_050227}{14}{GW200216_220804}{4}{GW191215_223052}{0}{GW200208_130117}{0}{GW200219_094415}{0}{GW191103_012549}{0}{GW200316_215756}{0}{GW200202_154313}{0}{GW200129_065458}{0}{GW191216_213338}{0}}[{\red{???}}]}
\DeclareRobustCommand{\percentchieffmorethanzerogwtcthree}[1]{\IfEqCase{#1}{{GW200224_222234}{87}{GW191129_134029}{91}{GW200311_115853}{42}{GW191230_180458}{43}{GW191222_033537}{37}{GW200225_060421}{15}{GW200302_015811}{60}{GW200128_022011}{83}{GW191204_171526}{100}{GW200112_155838}{77}{GW200105_162426}{53}{GW191105_143521}{37}{GW191109_010717}{10}{GW200209_085452}{26}{GW200115_042309}{18}{GW191127_050227}{79}{GW200216_220804}{69}{GW191215_223052}{38}{GW200208_130117}{30}{GW200219_094415}{29}{GW191103_012549}{100}{GW200316_215756}{98}{GW200202_154313}{86}{GW200129_065458}{89}{GW191216_213338}{100}}[{\red{???}}]}
\DeclareRobustCommand{\percentchiefflessthanzerogwtcthree}[1]{\IfEqCase{#1}{{GW200224_222234}{13}{GW191129_134029}{9}{GW200311_115853}{58}{GW191230_180458}{57}{GW191222_033537}{63}{GW200225_060421}{85}{GW200302_015811}{40}{GW200128_022011}{17}{GW191204_171526}{0}{GW200112_155838}{23}{GW200105_162426}{47}{GW191105_143521}{63}{GW191109_010717}{90}{GW200209_085452}{74}{GW200115_042309}{82}{GW191127_050227}{21}{GW200216_220804}{31}{GW191215_223052}{62}{GW200208_130117}{70}{GW200219_094415}{71}{GW191103_012549}{0}{GW200316_215756}{2}{GW200202_154313}{14}{GW200129_065458}{11}{GW191216_213338}{0}}[{\red{???}}]}
\DeclareRobustCommand{\percentchionemorethanpointeightgwtcthree}[1]{\IfEqCase{#1}{{GW200224_222234}{14}{GW191129_134029}{1}{GW200311_115853}{10}{GW191230_180458}{21}{GW191222_033537}{10}{GW200225_060421}{21}{GW200302_015811}{11}{GW200128_022011}{26}{GW191204_171526}{4}{GW200112_155838}{5}{GW200105_162426}{0}{GW191105_143521}{4}{GW191109_010717}{55}{GW200209_085452}{21}{GW200115_042309}{6}{GW191127_050227}{33}{GW200216_220804}{19}{GW191215_223052}{15}{GW200208_130117}{9}{GW200219_094415}{17}{GW191103_012549}{9}{GW200316_215756}{2}{GW200202_154313}{2}{GW200129_065458}{26}{GW191216_213338}{1}}[{\red{???}}]}
\DeclareRobustCommand{\percentchitwomorethanpointeightgwtcthree}[1]{\IfEqCase{#1}{{GW200224_222234}{14}{GW191129_134029}{8}{GW200311_115853}{13}{GW191230_180458}{19}{GW191222_033537}{13}{GW200225_060421}{14}{GW200302_015811}{16}{GW200128_022011}{19}{GW191204_171526}{9}{GW200112_155838}{11}{GW200105_162426}{10}{GW191105_143521}{9}{GW191109_010717}{33}{GW200209_085452}{19}{GW200115_042309}{13}{GW191127_050227}{22}{GW200216_220804}{21}{GW191215_223052}{16}{GW200208_130117}{14}{GW200219_094415}{18}{GW191103_012549}{17}{GW200316_215756}{13}{GW200202_154313}{8}{GW200129_065458}{15}{GW191216_213338}{8}}[{\red{???}}]}
\DeclareRobustCommand{\percentanychimorethanpointeightgwtcthree}[1]{\IfEqCase{#1}{{GW200224_222234}{26}{GW191129_134029}{9}{GW200311_115853}{21}{GW191230_180458}{36}{GW191222_033537}{21}{GW200225_060421}{33}{GW200302_015811}{25}{GW200128_022011}{41}{GW191204_171526}{13}{GW200112_155838}{15}{GW200105_162426}{10}{GW191105_143521}{12}{GW191109_010717}{72}{GW200209_085452}{37}{GW200115_042309}{18}{GW191127_050227}{49}{GW200216_220804}{36}{GW191215_223052}{29}{GW200208_130117}{22}{GW200219_094415}{32}{GW191103_012549}{24}{GW200316_215756}{14}{GW200202_154313}{10}{GW200129_065458}{36}{GW191216_213338}{8}}[{\red{???}}]}
\newcommand{\chirpmassdetgwtcthreeuncert}[1]{\ensuremath{ \chirpmassdetgwtcthreemed{#1}_{-\chirpmassdetgwtcthreeminus{#1}}^{+\chirpmassdetgwtcthreeplus{#1}}  } }

\newcommand{\networkmatchedfiltersnrgwtcthreeuncert}[1]{\ensuremath{ \networkmatchedfiltersnrgwtcthreemed{#1}_{-\networkmatchedfiltersnrgwtcthreeminus{#1}}^{+\networkmatchedfiltersnrgwtcthreeplus{#1}}  } }

\newcommand{\imrppnrtMCobsCOMPACTOneSevenZeroEightOneSevenHighSpinCat}{\macro{\ensuremath{1.1976_{-0.0002}^{+0.0004}}}}

\newcommand{\imrppnrtPEMATCHSNRCOMPACTOneSevenZeroEightOneSevenHighSpinCat}{\macro{\ensuremath{32.7_{-0.1}^{+0.1}}}}

\newcommand{\loglikelihoodfourtwofiveminus}[1]{\IfEqCase{#1}{{AlignedSpinInspiralTidalHS}{5.5}{AlignedSpinInspiralTidalLS}{5.4}{AlignedSpinTidalHS}{6.5}{AlignedSpinTidalLS}{6.3}{IMRPhenomDNRTidal-HS}{5.4}{IMRPhenomDNRTidal-LS}{5.5}{IMRPhenomPv2NRTidal-HS}{5.7}{IMRPhenomPv2NRTidal-LS}{5.5}{SEOBNRv4TsurrogateHS}{5.2}{SEOBNRv4TsurrogateLS}{5.4}{SEOBNRv4TsurrogatehighspinRIFT}{18.6}{SEOBNRv4TsurrogatelowspinRIFT}{5.0}{TEOBResumS-HS}{15.8}{TEOBResumS-LS}{5.1}{TaylorF2-HS}{5.5}{TaylorF2-LS}{5.4}{PrecessingSpinIMRTidalHS}{5.7}{PrecessingSpinIMRTidalLS}{5.5}{PublicationSamples}{5.6}}}
\newcommand{\loglikelihoodfourtwofivemed}[1]{\IfEqCase{#1}{{AlignedSpinInspiralTidalHS}{-240965.7}{AlignedSpinInspiralTidalLS}{-240964.4}{AlignedSpinTidalHS}{-1045826.3}{AlignedSpinTidalLS}{-1045825.4}{IMRPhenomDNRTidal-HS}{-1045828.5}{IMRPhenomDNRTidal-LS}{-1045827.1}{IMRPhenomPv2NRTidal-HS}{-500483.9}{IMRPhenomPv2NRTidal-LS}{-500482.8}{SEOBNRv4TsurrogateHS}{-1045828.0}{SEOBNRv4TsurrogateLS}{-1045827.1}{SEOBNRv4TsurrogatehighspinRIFT}{66.9}{SEOBNRv4TsurrogatelowspinRIFT}{68.5}{TEOBResumS-HS}{67.1}{TEOBResumS-LS}{68.8}{TaylorF2-HS}{-240965.7}{TaylorF2-LS}{-240964.4}{PrecessingSpinIMRTidalHS}{-500483.9}{PrecessingSpinIMRTidalLS}{-500482.8}{PublicationSamples}{-500483.9}}}
\newcommand{\loglikelihoodfourtwofiveplus}[1]{\IfEqCase{#1}{{AlignedSpinInspiralTidalHS}{3.8}{AlignedSpinInspiralTidalLS}{3.2}{AlignedSpinTidalHS}{1045896.4}{AlignedSpinTidalLS}{1045896.5}{IMRPhenomDNRTidal-HS}{4.5}{IMRPhenomDNRTidal-LS}{3.7}{IMRPhenomPv2NRTidal-HS}{4.5}{IMRPhenomPv2NRTidal-LS}{3.8}{SEOBNRv4TsurrogateHS}{3.9}{SEOBNRv4TsurrogateLS}{3.5}{SEOBNRv4TsurrogatehighspinRIFT}{4.3}{SEOBNRv4TsurrogatelowspinRIFT}{3.4}{TEOBResumS-HS}{4.4}{TEOBResumS-LS}{3.3}{TaylorF2-HS}{3.8}{TaylorF2-LS}{3.2}{PrecessingSpinIMRTidalHS}{4.5}{PrecessingSpinIMRTidalLS}{3.8}{PublicationSamples}{4.5}}}
\newcommand{\chiefffourtwofiveminus}[1]{\IfEqCase{#1}{{AlignedSpinInspiralTidalHS}{0.04}{AlignedSpinInspiralTidalLS}{0.01}{AlignedSpinTidalHS}{0.03}{AlignedSpinTidalLS}{0.01}{IMRPhenomDNRTidal-HS}{0.03}{IMRPhenomDNRTidal-LS}{0.01}{IMRPhenomPv2NRTidal-HS}{0.05}{IMRPhenomPv2NRTidal-LS}{0.01}{SEOBNRv4TsurrogateHS}{0.03}{SEOBNRv4TsurrogateLS}{0.01}{SEOBNRv4TsurrogatehighspinRIFT}{0.03}{SEOBNRv4TsurrogatelowspinRIFT}{0.01}{TEOBResumS-HS}{0.03}{TEOBResumS-LS}{0.01}{TaylorF2-HS}{0.04}{TaylorF2-LS}{0.01}{PrecessingSpinIMRTidalHS}{0.05}{PrecessingSpinIMRTidalLS}{0.01}{PublicationSamples}{0.05}}}
\newcommand{\chiefffourtwofivemed}[1]{\IfEqCase{#1}{{AlignedSpinInspiralTidalHS}{0.05}{AlignedSpinInspiralTidalLS}{0.01}{AlignedSpinTidalHS}{0.04}{AlignedSpinTidalLS}{0.01}{IMRPhenomDNRTidal-HS}{0.04}{IMRPhenomDNRTidal-LS}{0.01}{IMRPhenomPv2NRTidal-HS}{0.06}{IMRPhenomPv2NRTidal-LS}{0.01}{SEOBNRv4TsurrogateHS}{0.04}{SEOBNRv4TsurrogateLS}{0.01}{SEOBNRv4TsurrogatehighspinRIFT}{0.03}{SEOBNRv4TsurrogatelowspinRIFT}{0.01}{TEOBResumS-HS}{0.04}{TEOBResumS-LS}{0.01}{TaylorF2-HS}{0.05}{TaylorF2-LS}{0.01}{PrecessingSpinIMRTidalHS}{0.06}{PrecessingSpinIMRTidalLS}{0.01}{PublicationSamples}{0.06}}}
\newcommand{\chiefffourtwofiveplus}[1]{\IfEqCase{#1}{{AlignedSpinInspiralTidalHS}{0.09}{AlignedSpinInspiralTidalLS}{0.01}{AlignedSpinTidalHS}{0.08}{AlignedSpinTidalLS}{0.01}{IMRPhenomDNRTidal-HS}{0.10}{IMRPhenomDNRTidal-LS}{0.01}{IMRPhenomPv2NRTidal-HS}{0.11}{IMRPhenomPv2NRTidal-LS}{0.01}{SEOBNRv4TsurrogateHS}{0.07}{SEOBNRv4TsurrogateLS}{0.01}{SEOBNRv4TsurrogatehighspinRIFT}{0.07}{SEOBNRv4TsurrogatelowspinRIFT}{0.02}{TEOBResumS-HS}{0.06}{TEOBResumS-LS}{0.02}{TaylorF2-HS}{0.09}{TaylorF2-LS}{0.01}{PrecessingSpinIMRTidalHS}{0.11}{PrecessingSpinIMRTidalLS}{0.01}{PublicationSamples}{0.11}}}
\newcommand{\totalmasssourcefourtwofiveminus}[1]{\IfEqCase{#1}{{AlignedSpinInspiralTidalHS}{0.09}{AlignedSpinInspiralTidalLS}{0.05}{AlignedSpinTidalHS}{0.08}{AlignedSpinTidalLS}{0.05}{IMRPhenomDNRTidal-HS}{0.09}{IMRPhenomDNRTidal-LS}{0.05}{IMRPhenomPv2NRTidal-HS}{0.1}{IMRPhenomPv2NRTidal-LS}{0.05}{SEOBNRv4TsurrogateHS}{0.07}{SEOBNRv4TsurrogateLS}{0.05}{SEOBNRv4TsurrogatehighspinRIFT}{0.07}{SEOBNRv4TsurrogatelowspinRIFT}{0.05}{TEOBResumS-HS}{0.08}{TEOBResumS-LS}{0.05}{TaylorF2-HS}{0.09}{TaylorF2-LS}{0.05}{PrecessingSpinIMRTidalHS}{0.1}{PrecessingSpinIMRTidalLS}{0.05}{PublicationSamples}{0.1}}}
\newcommand{\totalmasssourcefourtwofivemed}[1]{\IfEqCase{#1}{{AlignedSpinInspiralTidalHS}{3.37}{AlignedSpinInspiralTidalLS}{3.31}{AlignedSpinTidalHS}{3.35}{AlignedSpinTidalLS}{3.31}{IMRPhenomDNRTidal-HS}{3.36}{IMRPhenomDNRTidal-LS}{3.31}{IMRPhenomPv2NRTidal-HS}{3.4}{IMRPhenomPv2NRTidal-LS}{3.31}{SEOBNRv4TsurrogateHS}{3.35}{SEOBNRv4TsurrogateLS}{3.31}{SEOBNRv4TsurrogatehighspinRIFT}{3.34}{SEOBNRv4TsurrogatelowspinRIFT}{3.31}{TEOBResumS-HS}{3.35}{TEOBResumS-LS}{3.31}{TaylorF2-HS}{3.37}{TaylorF2-LS}{3.31}{PrecessingSpinIMRTidalHS}{3.4}{PrecessingSpinIMRTidalLS}{3.31}{PublicationSamples}{3.4}}}
\newcommand{\totalmasssourcefourtwofiveplus}[1]{\IfEqCase{#1}{{AlignedSpinInspiralTidalHS}{0.2}{AlignedSpinInspiralTidalLS}{0.06}{AlignedSpinTidalHS}{0.3}{AlignedSpinTidalLS}{0.06}{IMRPhenomDNRTidal-HS}{0.4}{IMRPhenomDNRTidal-LS}{0.06}{IMRPhenomPv2NRTidal-HS}{0.3}{IMRPhenomPv2NRTidal-LS}{0.06}{SEOBNRv4TsurrogateHS}{0.2}{SEOBNRv4TsurrogateLS}{0.06}{SEOBNRv4TsurrogatehighspinRIFT}{0.2}{SEOBNRv4TsurrogatelowspinRIFT}{0.06}{TEOBResumS-HS}{0.2}{TEOBResumS-LS}{0.06}{TaylorF2-HS}{0.2}{TaylorF2-LS}{0.06}{PrecessingSpinIMRTidalHS}{0.3}{PrecessingSpinIMRTidalLS}{0.06}{PublicationSamples}{0.3}}}
\newcommand{\chipfourtwofiveminus}[1]{\IfEqCase{#1}{{AlignedSpinInspiralTidalHS}{0.00}{AlignedSpinInspiralTidalLS}{0.00}{AlignedSpinTidalHS}{0.00}{AlignedSpinTidalLS}{0.00}{IMRPhenomDNRTidal-HS}{0.00}{IMRPhenomDNRTidal-LS}{0.00}{IMRPhenomPv2NRTidal-HS}{0.27}{IMRPhenomPv2NRTidal-LS}{0.02}{SEOBNRv4TsurrogateHS}{0.00}{SEOBNRv4TsurrogateLS}{0.00}{SEOBNRv4TsurrogatehighspinRIFT}{0.00}{SEOBNRv4TsurrogatelowspinRIFT}{0.00}{TEOBResumS-HS}{0.00}{TEOBResumS-LS}{0.00}{TaylorF2-HS}{0.00}{TaylorF2-LS}{0.00}{PrecessingSpinIMRTidalHS}{0.27}{PrecessingSpinIMRTidalLS}{0.02}{PublicationSamples}{0.27}}}
\newcommand{\chipfourtwofivemed}[1]{\IfEqCase{#1}{{AlignedSpinInspiralTidalHS}{0.00}{AlignedSpinInspiralTidalLS}{0.00}{AlignedSpinTidalHS}{0.00}{AlignedSpinTidalLS}{0.00}{IMRPhenomDNRTidal-HS}{0.00}{IMRPhenomDNRTidal-LS}{0.00}{IMRPhenomPv2NRTidal-HS}{0.34}{IMRPhenomPv2NRTidal-LS}{0.03}{SEOBNRv4TsurrogateHS}{0.00}{SEOBNRv4TsurrogateLS}{0.00}{SEOBNRv4TsurrogatehighspinRIFT}{0.00}{SEOBNRv4TsurrogatelowspinRIFT}{0.00}{TEOBResumS-HS}{0.00}{TEOBResumS-LS}{0.00}{TaylorF2-HS}{0.00}{TaylorF2-LS}{0.00}{PrecessingSpinIMRTidalHS}{0.34}{PrecessingSpinIMRTidalLS}{0.03}{PublicationSamples}{0.34}}}
\newcommand{\chipfourtwofiveplus}[1]{\IfEqCase{#1}{{AlignedSpinInspiralTidalHS}{0.00}{AlignedSpinInspiralTidalLS}{0.00}{AlignedSpinTidalHS}{0.00}{AlignedSpinTidalLS}{0.00}{IMRPhenomDNRTidal-HS}{0.00}{IMRPhenomDNRTidal-LS}{0.00}{IMRPhenomPv2NRTidal-HS}{0.43}{IMRPhenomPv2NRTidal-LS}{0.02}{SEOBNRv4TsurrogateHS}{0.00}{SEOBNRv4TsurrogateLS}{0.00}{SEOBNRv4TsurrogatehighspinRIFT}{0.00}{SEOBNRv4TsurrogatelowspinRIFT}{0.00}{TEOBResumS-HS}{0.00}{TEOBResumS-LS}{0.00}{TaylorF2-HS}{0.00}{TaylorF2-LS}{0.00}{PrecessingSpinIMRTidalHS}{0.43}{PrecessingSpinIMRTidalLS}{0.02}{PublicationSamples}{0.43}}}
\newcommand{\spinoneyfourtwofiveminus}[1]{\IfEqCase{#1}{{AlignedSpinInspiralTidalHS}{0.00}{AlignedSpinInspiralTidalLS}{0.00}{AlignedSpinTidalHS}{0.00}{AlignedSpinTidalLS}{0.00}{IMRPhenomDNRTidal-HS}{0.00}{IMRPhenomDNRTidal-LS}{0.00}{IMRPhenomPv2NRTidal-HS}{0.49}{IMRPhenomPv2NRTidal-LS}{0.03}{SEOBNRv4TsurrogateHS}{0.00}{SEOBNRv4TsurrogateLS}{0.00}{SEOBNRv4TsurrogatehighspinRIFT}{0.00}{SEOBNRv4TsurrogatelowspinRIFT}{0.00}{TEOBResumS-HS}{0.00}{TEOBResumS-LS}{0.00}{TaylorF2-HS}{0.00}{TaylorF2-LS}{0.00}{PrecessingSpinIMRTidalHS}{0.48}{PrecessingSpinIMRTidalLS}{0.03}{PublicationSamples}{0.48}}}
\newcommand{\spinoneyfourtwofivemed}[1]{\IfEqCase{#1}{{AlignedSpinInspiralTidalHS}{0.00}{AlignedSpinInspiralTidalLS}{0.00}{AlignedSpinTidalHS}{0.00}{AlignedSpinTidalLS}{0.00}{IMRPhenomDNRTidal-HS}{0.00}{IMRPhenomDNRTidal-LS}{0.00}{IMRPhenomPv2NRTidal-HS}{0.003}{IMRPhenomPv2NRTidal-LS}{0.00}{SEOBNRv4TsurrogateHS}{0.00}{SEOBNRv4TsurrogateLS}{0.00}{SEOBNRv4TsurrogatehighspinRIFT}{0.00}{SEOBNRv4TsurrogatelowspinRIFT}{0.00}{TEOBResumS-HS}{0.00}{TEOBResumS-LS}{0.00}{TaylorF2-HS}{0.00}{TaylorF2-LS}{0.00}{PrecessingSpinIMRTidalHS}{0.003}{PrecessingSpinIMRTidalLS}{0.00}{PublicationSamples}{0.003}}}
\newcommand{\spinoneyfourtwofiveplus}[1]{\IfEqCase{#1}{{AlignedSpinInspiralTidalHS}{0.00}{AlignedSpinInspiralTidalLS}{0.00}{AlignedSpinTidalHS}{0.00}{AlignedSpinTidalLS}{0.00}{IMRPhenomDNRTidal-HS}{0.00}{IMRPhenomDNRTidal-LS}{0.00}{IMRPhenomPv2NRTidal-HS}{0.48}{IMRPhenomPv2NRTidal-LS}{0.03}{SEOBNRv4TsurrogateHS}{0.00}{SEOBNRv4TsurrogateLS}{0.00}{SEOBNRv4TsurrogatehighspinRIFT}{0.00}{SEOBNRv4TsurrogatelowspinRIFT}{0.00}{TEOBResumS-HS}{0.00}{TEOBResumS-LS}{0.00}{TaylorF2-HS}{0.00}{TaylorF2-LS}{0.00}{PrecessingSpinIMRTidalHS}{0.48}{PrecessingSpinIMRTidalLS}{0.03}{PublicationSamples}{0.48}}}
\newcommand{\phitwofourtwofiveminus}[1]{\IfEqCase{#1}{{AlignedSpinInspiralTidalHS}{0.00}{AlignedSpinInspiralTidalLS}{0.00}{AlignedSpinTidalHS}{0.00}{AlignedSpinTidalLS}{0.00}{IMRPhenomDNRTidal-HS}{0.00}{IMRPhenomDNRTidal-LS}{0.00}{IMRPhenomPv2NRTidal-HS}{2.83}{IMRPhenomPv2NRTidal-LS}{2.82}{SEOBNRv4TsurrogateHS}{0.00}{SEOBNRv4TsurrogateLS}{0.00}{SEOBNRv4TsurrogatehighspinRIFT}{0.00}{SEOBNRv4TsurrogatelowspinRIFT}{0.00}{TEOBResumS-HS}{0.00}{TEOBResumS-LS}{0.00}{TaylorF2-HS}{0.00}{TaylorF2-LS}{0.00}{PrecessingSpinIMRTidalHS}{2.84}{PrecessingSpinIMRTidalLS}{2.82}{PublicationSamples}{2.84}}}
\newcommand{\phitwofourtwofivemed}[1]{\IfEqCase{#1}{{AlignedSpinInspiralTidalHS}{0.00}{AlignedSpinInspiralTidalLS}{0.00}{AlignedSpinTidalHS}{0.00}{AlignedSpinTidalLS}{0.00}{IMRPhenomDNRTidal-HS}{0.00}{IMRPhenomDNRTidal-LS}{0.00}{IMRPhenomPv2NRTidal-HS}{3.14}{IMRPhenomPv2NRTidal-LS}{3.13}{SEOBNRv4TsurrogateHS}{0.00}{SEOBNRv4TsurrogateLS}{0.00}{SEOBNRv4TsurrogatehighspinRIFT}{0.00}{SEOBNRv4TsurrogatelowspinRIFT}{0.00}{TEOBResumS-HS}{0.00}{TEOBResumS-LS}{0.00}{TaylorF2-HS}{0.00}{TaylorF2-LS}{0.00}{PrecessingSpinIMRTidalHS}{3.15}{PrecessingSpinIMRTidalLS}{3.13}{PublicationSamples}{3.15}}}
\newcommand{\phitwofourtwofiveplus}[1]{\IfEqCase{#1}{{AlignedSpinInspiralTidalHS}{0.00}{AlignedSpinInspiralTidalLS}{0.00}{AlignedSpinTidalHS}{0.00}{AlignedSpinTidalLS}{0.00}{IMRPhenomDNRTidal-HS}{0.00}{IMRPhenomDNRTidal-LS}{0.00}{IMRPhenomPv2NRTidal-HS}{2.84}{IMRPhenomPv2NRTidal-LS}{2.83}{SEOBNRv4TsurrogateHS}{0.00}{SEOBNRv4TsurrogateLS}{0.00}{SEOBNRv4TsurrogatehighspinRIFT}{0.00}{SEOBNRv4TsurrogatelowspinRIFT}{0.00}{TEOBResumS-HS}{0.00}{TEOBResumS-LS}{0.00}{TaylorF2-HS}{0.00}{TaylorF2-LS}{0.00}{PrecessingSpinIMRTidalHS}{2.83}{PrecessingSpinIMRTidalLS}{2.83}{PublicationSamples}{2.83}}}
\newcommand{\phionetwofourtwofiveminus}[1]{\IfEqCase{#1}{{AlignedSpinInspiralTidalHS}{0.00}{AlignedSpinInspiralTidalLS}{0.00}{AlignedSpinTidalHS}{0.00}{AlignedSpinTidalLS}{0.00}{IMRPhenomDNRTidal-HS}{0.00}{IMRPhenomDNRTidal-LS}{0.00}{IMRPhenomPv2NRTidal-HS}{2.88}{IMRPhenomPv2NRTidal-LS}{2.74}{SEOBNRv4TsurrogateHS}{0.00}{SEOBNRv4TsurrogateLS}{0.00}{SEOBNRv4TsurrogatehighspinRIFT}{0.00}{SEOBNRv4TsurrogatelowspinRIFT}{0.00}{TEOBResumS-HS}{0.00}{TEOBResumS-LS}{0.00}{TaylorF2-HS}{0.00}{TaylorF2-LS}{0.00}{PrecessingSpinIMRTidalHS}{2.87}{PrecessingSpinIMRTidalLS}{2.75}{PublicationSamples}{2.88}}}
\newcommand{\phionetwofourtwofivemed}[1]{\IfEqCase{#1}{{AlignedSpinInspiralTidalHS}{0.00}{AlignedSpinInspiralTidalLS}{0.00}{AlignedSpinTidalHS}{0.00}{AlignedSpinTidalLS}{0.00}{IMRPhenomDNRTidal-HS}{0.00}{IMRPhenomDNRTidal-LS}{0.00}{IMRPhenomPv2NRTidal-HS}{3.19}{IMRPhenomPv2NRTidal-LS}{3.05}{SEOBNRv4TsurrogateHS}{0.00}{SEOBNRv4TsurrogateLS}{0.00}{SEOBNRv4TsurrogatehighspinRIFT}{0.00}{SEOBNRv4TsurrogatelowspinRIFT}{0.00}{TEOBResumS-HS}{0.00}{TEOBResumS-LS}{0.00}{TaylorF2-HS}{0.00}{TaylorF2-LS}{0.00}{PrecessingSpinIMRTidalHS}{3.18}{PrecessingSpinIMRTidalLS}{3.05}{PublicationSamples}{3.18}}}
\newcommand{\phionetwofourtwofiveplus}[1]{\IfEqCase{#1}{{AlignedSpinInspiralTidalHS}{0.00}{AlignedSpinInspiralTidalLS}{0.00}{AlignedSpinTidalHS}{0.00}{AlignedSpinTidalLS}{0.00}{IMRPhenomDNRTidal-HS}{0.00}{IMRPhenomDNRTidal-LS}{0.00}{IMRPhenomPv2NRTidal-HS}{2.75}{IMRPhenomPv2NRTidal-LS}{2.88}{SEOBNRv4TsurrogateHS}{0.00}{SEOBNRv4TsurrogateLS}{0.00}{SEOBNRv4TsurrogatehighspinRIFT}{0.00}{SEOBNRv4TsurrogatelowspinRIFT}{0.00}{TEOBResumS-HS}{0.00}{TEOBResumS-LS}{0.00}{TaylorF2-HS}{0.00}{TaylorF2-LS}{0.00}{PrecessingSpinIMRTidalHS}{2.76}{PrecessingSpinIMRTidalLS}{2.88}{PublicationSamples}{2.76}}}
\newcommand{\rafourtwofiveminus}[1]{\IfEqCase{#1}{{AlignedSpinInspiralTidalHS}{1.04004}{AlignedSpinInspiralTidalLS}{1.10804}{AlignedSpinTidalHS}{1.04466}{AlignedSpinTidalLS}{1.05437}{IMRPhenomDNRTidal-HS}{1.04415}{IMRPhenomDNRTidal-LS}{1.34502}{IMRPhenomPv2NRTidal-HS}{1.14841}{IMRPhenomPv2NRTidal-LS}{1.49247}{SEOBNRv4TsurrogateHS}{1.04813}{SEOBNRv4TsurrogateLS}{0.99881}{SEOBNRv4TsurrogatehighspinRIFT}{1.03845}{SEOBNRv4TsurrogatelowspinRIFT}{1.07158}{TEOBResumS-HS}{1.04161}{TEOBResumS-LS}{1.06268}{TaylorF2-HS}{1.03744}{TaylorF2-LS}{1.11687}{PrecessingSpinIMRTidalHS}{1.14713}{PrecessingSpinIMRTidalLS}{1.36500}{PublicationSamples}{1.14187}}}
\newcommand{\rafourtwofivemed}[1]{\IfEqCase{#1}{{AlignedSpinInspiralTidalHS}{1.47807}{AlignedSpinInspiralTidalLS}{1.57610}{AlignedSpinTidalHS}{1.52327}{AlignedSpinTidalLS}{1.53218}{IMRPhenomDNRTidal-HS}{1.52282}{IMRPhenomDNRTidal-LS}{1.86790}{IMRPhenomPv2NRTidal-HS}{1.62902}{IMRPhenomPv2NRTidal-LS}{1.99502}{SEOBNRv4TsurrogateHS}{1.52665}{SEOBNRv4TsurrogateLS}{1.46873}{SEOBNRv4TsurrogatehighspinRIFT}{1.50830}{SEOBNRv4TsurrogatelowspinRIFT}{1.53881}{TEOBResumS-HS}{1.51000}{TEOBResumS-LS}{1.50223}{TaylorF2-HS}{1.47697}{TaylorF2-LS}{1.57735}{PrecessingSpinIMRTidalHS}{1.62833}{PrecessingSpinIMRTidalLS}{1.86958}{PublicationSamples}{1.62336}}}
\newcommand{\rafourtwofiveplus}[1]{\IfEqCase{#1}{{AlignedSpinInspiralTidalHS}{3.24632}{AlignedSpinInspiralTidalLS}{3.13801}{AlignedSpinTidalHS}{3.25020}{AlignedSpinTidalLS}{3.18422}{IMRPhenomDNRTidal-HS}{3.29588}{IMRPhenomDNRTidal-LS}{2.86054}{IMRPhenomPv2NRTidal-HS}{3.13029}{IMRPhenomPv2NRTidal-LS}{2.74917}{SEOBNRv4TsurrogateHS}{3.18219}{SEOBNRv4TsurrogateLS}{3.22628}{SEOBNRv4TsurrogatehighspinRIFT}{3.24313}{SEOBNRv4TsurrogatelowspinRIFT}{3.18050}{TEOBResumS-HS}{3.24215}{TEOBResumS-LS}{3.23570}{TaylorF2-HS}{3.25577}{TaylorF2-LS}{3.13908}{PrecessingSpinIMRTidalHS}{3.12911}{PrecessingSpinIMRTidalLS}{2.87696}{PublicationSamples}{3.13407}}}
\newcommand{\phijlfourtwofiveminus}[1]{\IfEqCase{#1}{{AlignedSpinInspiralTidalHS}{0.97}{AlignedSpinInspiralTidalLS}{0.88}{AlignedSpinTidalHS}{0.78}{AlignedSpinTidalLS}{0.77}{IMRPhenomDNRTidal-HS}{0.74}{IMRPhenomDNRTidal-LS}{0.78}{IMRPhenomPv2NRTidal-HS}{2.89}{IMRPhenomPv2NRTidal-LS}{2.58}{SEOBNRv4TsurrogateHS}{0.89}{SEOBNRv4TsurrogateLS}{0.85}{SEOBNRv4TsurrogatehighspinRIFT}{0.00}{SEOBNRv4TsurrogatelowspinRIFT}{0.00}{TEOBResumS-HS}{0.00}{TEOBResumS-LS}{0.00}{TaylorF2-HS}{0.97}{TaylorF2-LS}{0.88}{PrecessingSpinIMRTidalHS}{2.88}{PrecessingSpinIMRTidalLS}{2.57}{PublicationSamples}{2.89}}}
\newcommand{\phijlfourtwofivemed}[1]{\IfEqCase{#1}{{AlignedSpinInspiralTidalHS}{1.22}{AlignedSpinInspiralTidalLS}{1.11}{AlignedSpinTidalHS}{0.78}{AlignedSpinTidalLS}{0.77}{IMRPhenomDNRTidal-HS}{0.96}{IMRPhenomDNRTidal-LS}{1.02}{IMRPhenomPv2NRTidal-HS}{3.23}{IMRPhenomPv2NRTidal-LS}{2.87}{SEOBNRv4TsurrogateHS}{1.13}{SEOBNRv4TsurrogateLS}{1.09}{SEOBNRv4TsurrogatehighspinRIFT}{0.00}{SEOBNRv4TsurrogatelowspinRIFT}{0.00}{TEOBResumS-HS}{0.00}{TEOBResumS-LS}{0.00}{TaylorF2-HS}{1.22}{TaylorF2-LS}{1.11}{PrecessingSpinIMRTidalHS}{3.23}{PrecessingSpinIMRTidalLS}{2.86}{PublicationSamples}{3.23}}}
\newcommand{\phijlfourtwofiveplus}[1]{\IfEqCase{#1}{{AlignedSpinInspiralTidalHS}{1.64}{AlignedSpinInspiralTidalLS}{1.75}{AlignedSpinTidalHS}{2.37}{AlignedSpinTidalLS}{2.37}{IMRPhenomDNRTidal-HS}{1.88}{IMRPhenomDNRTidal-LS}{1.83}{IMRPhenomPv2NRTidal-HS}{2.75}{IMRPhenomPv2NRTidal-LS}{3.05}{SEOBNRv4TsurrogateHS}{1.72}{SEOBNRv4TsurrogateLS}{1.76}{SEOBNRv4TsurrogatehighspinRIFT}{3.14}{SEOBNRv4TsurrogatelowspinRIFT}{3.14}{TEOBResumS-HS}{3.14}{TEOBResumS-LS}{3.14}{TaylorF2-HS}{1.63}{TaylorF2-LS}{1.75}{PrecessingSpinIMRTidalHS}{2.76}{PrecessingSpinIMRTidalLS}{3.07}{PublicationSamples}{2.75}}}
\newcommand{\tilttwofourtwofiveminus}[1]{\IfEqCase{#1}{{AlignedSpinInspiralTidalHS}{0.00}{AlignedSpinInspiralTidalLS}{0.00}{AlignedSpinTidalHS}{0.00}{AlignedSpinTidalLS}{0.00}{IMRPhenomDNRTidal-HS}{0.00}{IMRPhenomDNRTidal-LS}{0.00}{IMRPhenomPv2NRTidal-HS}{0.87}{IMRPhenomPv2NRTidal-LS}{0.78}{SEOBNRv4TsurrogateHS}{0.00}{SEOBNRv4TsurrogateLS}{0.00}{SEOBNRv4TsurrogatehighspinRIFT}{0.00}{SEOBNRv4TsurrogatelowspinRIFT}{0.00}{TEOBResumS-HS}{0.00}{TEOBResumS-LS}{0.00}{TaylorF2-HS}{0.00}{TaylorF2-LS}{0.00}{PrecessingSpinIMRTidalHS}{0.87}{PrecessingSpinIMRTidalLS}{0.79}{PublicationSamples}{0.87}}}
\newcommand{\tilttwofourtwofivemed}[1]{\IfEqCase{#1}{{AlignedSpinInspiralTidalHS}{0.00}{AlignedSpinInspiralTidalLS}{0.00}{AlignedSpinTidalHS}{0.00}{AlignedSpinTidalLS}{0.00}{IMRPhenomDNRTidal-HS}{0.00}{IMRPhenomDNRTidal-LS}{0.00}{IMRPhenomPv2NRTidal-HS}{1.41}{IMRPhenomPv2NRTidal-LS}{1.09}{SEOBNRv4TsurrogateHS}{0.00}{SEOBNRv4TsurrogateLS}{0.00}{SEOBNRv4TsurrogatehighspinRIFT}{0.00}{SEOBNRv4TsurrogatelowspinRIFT}{0.00}{TEOBResumS-HS}{0.00}{TEOBResumS-LS}{0.00}{TaylorF2-HS}{0.00}{TaylorF2-LS}{0.00}{PrecessingSpinIMRTidalHS}{1.41}{PrecessingSpinIMRTidalLS}{1.09}{PublicationSamples}{1.41}}}
\newcommand{\tilttwofourtwofiveplus}[1]{\IfEqCase{#1}{{AlignedSpinInspiralTidalHS}{3.14}{AlignedSpinInspiralTidalLS}{3.14}{AlignedSpinTidalHS}{3.14}{AlignedSpinTidalLS}{3.14}{IMRPhenomDNRTidal-HS}{3.14}{IMRPhenomDNRTidal-LS}{3.14}{IMRPhenomPv2NRTidal-HS}{0.94}{IMRPhenomPv2NRTidal-LS}{1.21}{SEOBNRv4TsurrogateHS}{3.14}{SEOBNRv4TsurrogateLS}{3.14}{SEOBNRv4TsurrogatehighspinRIFT}{3.14}{SEOBNRv4TsurrogatelowspinRIFT}{3.14}{TEOBResumS-HS}{3.14}{TEOBResumS-LS}{3.14}{TaylorF2-HS}{3.14}{TaylorF2-LS}{3.14}{PrecessingSpinIMRTidalHS}{0.94}{PrecessingSpinIMRTidalLS}{1.20}{PublicationSamples}{0.94}}}
\newcommand{\costhetajnfourtwofiveminus}[1]{\IfEqCase{#1}{{AlignedSpinInspiralTidalHS}{1.30}{AlignedSpinInspiralTidalLS}{1.40}{AlignedSpinTidalHS}{1.44}{AlignedSpinTidalLS}{1.44}{IMRPhenomDNRTidal-HS}{1.53}{IMRPhenomDNRTidal-LS}{1.48}{IMRPhenomPv2NRTidal-HS}{1.43}{IMRPhenomPv2NRTidal-LS}{1.44}{SEOBNRv4TsurrogateHS}{1.38}{SEOBNRv4TsurrogateLS}{1.42}{SEOBNRv4TsurrogatehighspinRIFT}{1.41}{SEOBNRv4TsurrogatelowspinRIFT}{1.42}{TEOBResumS-HS}{1.42}{TEOBResumS-LS}{1.40}{TaylorF2-HS}{1.30}{TaylorF2-LS}{1.40}{PrecessingSpinIMRTidalHS}{1.43}{PrecessingSpinIMRTidalLS}{1.44}{PublicationSamples}{1.43}}}
\newcommand{\costhetajnfourtwofivemed}[1]{\IfEqCase{#1}{{AlignedSpinInspiralTidalHS}{0.34}{AlignedSpinInspiralTidalLS}{0.44}{AlignedSpinTidalHS}{0.49}{AlignedSpinTidalLS}{0.49}{IMRPhenomDNRTidal-HS}{0.58}{IMRPhenomDNRTidal-LS}{0.53}{IMRPhenomPv2NRTidal-HS}{0.47}{IMRPhenomPv2NRTidal-LS}{0.48}{SEOBNRv4TsurrogateHS}{0.43}{SEOBNRv4TsurrogateLS}{0.46}{SEOBNRv4TsurrogatehighspinRIFT}{0.45}{SEOBNRv4TsurrogatelowspinRIFT}{0.46}{TEOBResumS-HS}{0.46}{TEOBResumS-LS}{0.44}{TaylorF2-HS}{0.34}{TaylorF2-LS}{0.44}{PrecessingSpinIMRTidalHS}{0.47}{PrecessingSpinIMRTidalLS}{0.48}{PublicationSamples}{0.47}}}
\newcommand{\costhetajnfourtwofiveplus}[1]{\IfEqCase{#1}{{AlignedSpinInspiralTidalHS}{0.62}{AlignedSpinInspiralTidalLS}{0.53}{AlignedSpinTidalHS}{0.49}{AlignedSpinTidalLS}{0.49}{IMRPhenomDNRTidal-HS}{0.40}{IMRPhenomDNRTidal-LS}{0.45}{IMRPhenomPv2NRTidal-HS}{0.50}{IMRPhenomPv2NRTidal-LS}{0.49}{SEOBNRv4TsurrogateHS}{0.54}{SEOBNRv4TsurrogateLS}{0.51}{SEOBNRv4TsurrogatehighspinRIFT}{0.52}{SEOBNRv4TsurrogatelowspinRIFT}{0.51}{TEOBResumS-HS}{0.51}{TEOBResumS-LS}{0.54}{TaylorF2-HS}{0.63}{TaylorF2-LS}{0.53}{PrecessingSpinIMRTidalHS}{0.50}{PrecessingSpinIMRTidalLS}{0.49}{PublicationSamples}{0.50}}}
\newcommand{\spintwofourtwofiveminus}[1]{\IfEqCase{#1}{{AlignedSpinInspiralTidalHS}{0.07}{AlignedSpinInspiralTidalLS}{0.01}{AlignedSpinTidalHS}{0.07}{AlignedSpinTidalLS}{0.01}{IMRPhenomDNRTidal-HS}{0.10}{IMRPhenomDNRTidal-LS}{0.01}{IMRPhenomPv2NRTidal-HS}{0.25}{IMRPhenomPv2NRTidal-LS}{0.02}{SEOBNRv4TsurrogateHS}{0.06}{SEOBNRv4TsurrogateLS}{0.01}{SEOBNRv4TsurrogatehighspinRIFT}{0.06}{SEOBNRv4TsurrogatelowspinRIFT}{0.01}{TEOBResumS-HS}{0.06}{TEOBResumS-LS}{0.01}{TaylorF2-HS}{0.07}{TaylorF2-LS}{0.01}{PrecessingSpinIMRTidalHS}{0.25}{PrecessingSpinIMRTidalLS}{0.02}{PublicationSamples}{0.25}}}
\newcommand{\spintwofourtwofivemed}[1]{\IfEqCase{#1}{{AlignedSpinInspiralTidalHS}{0.08}{AlignedSpinInspiralTidalLS}{0.01}{AlignedSpinTidalHS}{0.07}{AlignedSpinTidalLS}{0.01}{IMRPhenomDNRTidal-HS}{0.11}{IMRPhenomDNRTidal-LS}{0.01}{IMRPhenomPv2NRTidal-HS}{0.28}{IMRPhenomPv2NRTidal-LS}{0.03}{SEOBNRv4TsurrogateHS}{0.06}{SEOBNRv4TsurrogateLS}{0.01}{SEOBNRv4TsurrogatehighspinRIFT}{0.06}{SEOBNRv4TsurrogatelowspinRIFT}{0.01}{TEOBResumS-HS}{0.06}{TEOBResumS-LS}{0.01}{TaylorF2-HS}{0.08}{TaylorF2-LS}{0.01}{PrecessingSpinIMRTidalHS}{0.28}{PrecessingSpinIMRTidalLS}{0.03}{PublicationSamples}{0.28}}}
\newcommand{\spintwofourtwofiveplus}[1]{\IfEqCase{#1}{{AlignedSpinInspiralTidalHS}{0.26}{AlignedSpinInspiralTidalLS}{0.03}{AlignedSpinTidalHS}{0.27}{AlignedSpinTidalLS}{0.03}{IMRPhenomDNRTidal-HS}{0.38}{IMRPhenomDNRTidal-LS}{0.03}{IMRPhenomPv2NRTidal-HS}{0.51}{IMRPhenomPv2NRTidal-LS}{0.02}{SEOBNRv4TsurrogateHS}{0.19}{SEOBNRv4TsurrogateLS}{0.03}{SEOBNRv4TsurrogatehighspinRIFT}{0.19}{SEOBNRv4TsurrogatelowspinRIFT}{0.03}{TEOBResumS-HS}{0.19}{TEOBResumS-LS}{0.03}{TaylorF2-HS}{0.26}{TaylorF2-LS}{0.03}{PrecessingSpinIMRTidalHS}{0.51}{PrecessingSpinIMRTidalLS}{0.02}{PublicationSamples}{0.51}}}
\newcommand{\massonedetfourtwofiveminus}[1]{\IfEqCase{#1}{{AlignedSpinInspiralTidalHS}{0.3}{AlignedSpinInspiralTidalLS}{0.09}{AlignedSpinTidalHS}{0.2}{AlignedSpinTidalLS}{0.09}{IMRPhenomDNRTidal-HS}{0.3}{IMRPhenomDNRTidal-LS}{0.09}{IMRPhenomPv2NRTidal-HS}{0.4}{IMRPhenomPv2NRTidal-LS}{0.09}{SEOBNRv4TsurrogateHS}{0.2}{SEOBNRv4TsurrogateLS}{0.09}{SEOBNRv4TsurrogatehighspinRIFT}{0.2}{SEOBNRv4TsurrogatelowspinRIFT}{0.09}{TEOBResumS-HS}{0.2}{TEOBResumS-LS}{0.09}{TaylorF2-HS}{0.3}{TaylorF2-LS}{0.09}{PrecessingSpinIMRTidalHS}{0.4}{PrecessingSpinIMRTidalLS}{0.09}{PublicationSamples}{0.4}}}
\newcommand{\massonedetfourtwofivemed}[1]{\IfEqCase{#1}{{AlignedSpinInspiralTidalHS}{2.0}{AlignedSpinInspiralTidalLS}{1.81}{AlignedSpinTidalHS}{2.0}{AlignedSpinTidalLS}{1.81}{IMRPhenomDNRTidal-HS}{2.0}{IMRPhenomDNRTidal-LS}{1.81}{IMRPhenomPv2NRTidal-HS}{2.1}{IMRPhenomPv2NRTidal-LS}{1.80}{SEOBNRv4TsurrogateHS}{2.0}{SEOBNRv4TsurrogateLS}{1.80}{SEOBNRv4TsurrogatehighspinRIFT}{1.9}{SEOBNRv4TsurrogatelowspinRIFT}{1.81}{TEOBResumS-HS}{2.0}{TEOBResumS-LS}{1.81}{TaylorF2-HS}{2.0}{TaylorF2-LS}{1.81}{PrecessingSpinIMRTidalHS}{2.1}{PrecessingSpinIMRTidalLS}{1.80}{PublicationSamples}{2.1}}}
\newcommand{\massonedetfourtwofiveplus}[1]{\IfEqCase{#1}{{AlignedSpinInspiralTidalHS}{0.5}{AlignedSpinInspiralTidalLS}{0.2}{AlignedSpinTidalHS}{0.6}{AlignedSpinTidalLS}{0.2}{IMRPhenomDNRTidal-HS}{0.7}{IMRPhenomDNRTidal-LS}{0.2}{IMRPhenomPv2NRTidal-HS}{0.6}{IMRPhenomPv2NRTidal-LS}{0.2}{SEOBNRv4TsurrogateHS}{0.5}{SEOBNRv4TsurrogateLS}{0.2}{SEOBNRv4TsurrogatehighspinRIFT}{0.5}{SEOBNRv4TsurrogatelowspinRIFT}{0.2}{TEOBResumS-HS}{0.5}{TEOBResumS-LS}{0.2}{TaylorF2-HS}{0.5}{TaylorF2-LS}{0.2}{PrecessingSpinIMRTidalHS}{0.6}{PrecessingSpinIMRTidalLS}{0.2}{PublicationSamples}{0.6}}}
\newcommand{\spintwoxfourtwofiveminus}[1]{\IfEqCase{#1}{{AlignedSpinInspiralTidalHS}{0.00}{AlignedSpinInspiralTidalLS}{0.00}{AlignedSpinTidalHS}{0.00}{AlignedSpinTidalLS}{0.00}{IMRPhenomDNRTidal-HS}{0.00}{IMRPhenomDNRTidal-LS}{0.00}{IMRPhenomPv2NRTidal-HS}{0.47}{IMRPhenomPv2NRTidal-LS}{0.03}{SEOBNRv4TsurrogateHS}{0.00}{SEOBNRv4TsurrogateLS}{0.00}{SEOBNRv4TsurrogatehighspinRIFT}{0.00}{SEOBNRv4TsurrogatelowspinRIFT}{0.00}{TEOBResumS-HS}{0.00}{TEOBResumS-LS}{0.00}{TaylorF2-HS}{0.00}{TaylorF2-LS}{0.00}{PrecessingSpinIMRTidalHS}{0.47}{PrecessingSpinIMRTidalLS}{0.03}{PublicationSamples}{0.47}}}
\newcommand{\spintwoxfourtwofivemed}[1]{\IfEqCase{#1}{{AlignedSpinInspiralTidalHS}{0.00}{AlignedSpinInspiralTidalLS}{0.00}{AlignedSpinTidalHS}{0.00}{AlignedSpinTidalLS}{0.00}{IMRPhenomDNRTidal-HS}{0.00}{IMRPhenomDNRTidal-LS}{0.00}{IMRPhenomPv2NRTidal-HS}{0.0007}{IMRPhenomPv2NRTidal-LS}{0.00}{SEOBNRv4TsurrogateHS}{0.00}{SEOBNRv4TsurrogateLS}{0.00}{SEOBNRv4TsurrogatehighspinRIFT}{0.00}{SEOBNRv4TsurrogatelowspinRIFT}{0.00}{TEOBResumS-HS}{0.00}{TEOBResumS-LS}{0.00}{TaylorF2-HS}{0.00}{TaylorF2-LS}{0.00}{PrecessingSpinIMRTidalHS}{0.0006}{PrecessingSpinIMRTidalLS}{0.00}{PublicationSamples}{0.0007}}}
\newcommand{\spintwoxfourtwofiveplus}[1]{\IfEqCase{#1}{{AlignedSpinInspiralTidalHS}{0.00}{AlignedSpinInspiralTidalLS}{0.00}{AlignedSpinTidalHS}{0.00}{AlignedSpinTidalLS}{0.00}{IMRPhenomDNRTidal-HS}{0.00}{IMRPhenomDNRTidal-LS}{0.00}{IMRPhenomPv2NRTidal-HS}{0.48}{IMRPhenomPv2NRTidal-LS}{0.03}{SEOBNRv4TsurrogateHS}{0.00}{SEOBNRv4TsurrogateLS}{0.00}{SEOBNRv4TsurrogatehighspinRIFT}{0.00}{SEOBNRv4TsurrogatelowspinRIFT}{0.00}{TEOBResumS-HS}{0.00}{TEOBResumS-LS}{0.00}{TaylorF2-HS}{0.00}{TaylorF2-LS}{0.00}{PrecessingSpinIMRTidalHS}{0.47}{PrecessingSpinIMRTidalLS}{0.03}{PublicationSamples}{0.47}}}
\newcommand{\massratiofourtwofiveminus}[1]{\IfEqCase{#1}{{AlignedSpinInspiralTidalHS}{0.24}{AlignedSpinInspiralTidalLS}{0.15}{AlignedSpinTidalHS}{0.30}{AlignedSpinTidalLS}{0.15}{IMRPhenomDNRTidal-HS}{0.31}{IMRPhenomDNRTidal-LS}{0.15}{IMRPhenomPv2NRTidal-HS}{0.25}{IMRPhenomPv2NRTidal-LS}{0.15}{SEOBNRv4TsurrogateHS}{0.29}{SEOBNRv4TsurrogateLS}{0.15}{SEOBNRv4TsurrogatehighspinRIFT}{0.27}{SEOBNRv4TsurrogatelowspinRIFT}{0.15}{TEOBResumS-HS}{0.27}{TEOBResumS-LS}{0.15}{TaylorF2-HS}{0.24}{TaylorF2-LS}{0.15}{PrecessingSpinIMRTidalHS}{0.25}{PrecessingSpinIMRTidalLS}{0.15}{PublicationSamples}{0.25}}}
\newcommand{\massratiofourtwofivemed}[1]{\IfEqCase{#1}{{AlignedSpinInspiralTidalHS}{0.70}{AlignedSpinInspiralTidalLS}{0.89}{AlignedSpinTidalHS}{0.74}{AlignedSpinTidalLS}{0.89}{IMRPhenomDNRTidal-HS}{0.72}{IMRPhenomDNRTidal-LS}{0.89}{IMRPhenomPv2NRTidal-HS}{0.67}{IMRPhenomPv2NRTidal-LS}{0.90}{SEOBNRv4TsurrogateHS}{0.77}{SEOBNRv4TsurrogateLS}{0.90}{SEOBNRv4TsurrogatehighspinRIFT}{0.78}{SEOBNRv4TsurrogatelowspinRIFT}{0.89}{TEOBResumS-HS}{0.75}{TEOBResumS-LS}{0.89}{TaylorF2-HS}{0.70}{TaylorF2-LS}{0.89}{PrecessingSpinIMRTidalHS}{0.67}{PrecessingSpinIMRTidalLS}{0.90}{PublicationSamples}{0.67}}}
\newcommand{\massratiofourtwofiveplus}[1]{\IfEqCase{#1}{{AlignedSpinInspiralTidalHS}{0.26}{AlignedSpinInspiralTidalLS}{0.10}{AlignedSpinTidalHS}{0.22}{AlignedSpinTidalLS}{0.09}{IMRPhenomDNRTidal-HS}{0.25}{IMRPhenomDNRTidal-LS}{0.09}{IMRPhenomPv2NRTidal-HS}{0.29}{IMRPhenomPv2NRTidal-LS}{0.09}{SEOBNRv4TsurrogateHS}{0.20}{SEOBNRv4TsurrogateLS}{0.09}{SEOBNRv4TsurrogatehighspinRIFT}{0.19}{SEOBNRv4TsurrogatelowspinRIFT}{0.10}{TEOBResumS-HS}{0.22}{TEOBResumS-LS}{0.10}{TaylorF2-HS}{0.26}{TaylorF2-LS}{0.10}{PrecessingSpinIMRTidalHS}{0.29}{PrecessingSpinIMRTidalLS}{0.09}{PublicationSamples}{0.29}}}
\newcommand{\spinonefourtwofiveminus}[1]{\IfEqCase{#1}{{AlignedSpinInspiralTidalHS}{0.06}{AlignedSpinInspiralTidalLS}{0.01}{AlignedSpinTidalHS}{0.06}{AlignedSpinTidalLS}{0.01}{IMRPhenomDNRTidal-HS}{0.08}{IMRPhenomDNRTidal-LS}{0.01}{IMRPhenomPv2NRTidal-HS}{0.25}{IMRPhenomPv2NRTidal-LS}{0.03}{SEOBNRv4TsurrogateHS}{0.05}{SEOBNRv4TsurrogateLS}{0.01}{SEOBNRv4TsurrogatehighspinRIFT}{0.05}{SEOBNRv4TsurrogatelowspinRIFT}{0.01}{TEOBResumS-HS}{0.05}{TEOBResumS-LS}{0.01}{TaylorF2-HS}{0.06}{TaylorF2-LS}{0.01}{PrecessingSpinIMRTidalHS}{0.25}{PrecessingSpinIMRTidalLS}{0.03}{PublicationSamples}{0.25}}}
\newcommand{\spinonefourtwofivemed}[1]{\IfEqCase{#1}{{AlignedSpinInspiralTidalHS}{0.06}{AlignedSpinInspiralTidalLS}{0.01}{AlignedSpinTidalHS}{0.06}{AlignedSpinTidalLS}{0.01}{IMRPhenomDNRTidal-HS}{0.09}{IMRPhenomDNRTidal-LS}{0.01}{IMRPhenomPv2NRTidal-HS}{0.27}{IMRPhenomPv2NRTidal-LS}{0.03}{SEOBNRv4TsurrogateHS}{0.06}{SEOBNRv4TsurrogateLS}{0.01}{SEOBNRv4TsurrogatehighspinRIFT}{0.06}{SEOBNRv4TsurrogatelowspinRIFT}{0.01}{TEOBResumS-HS}{0.06}{TEOBResumS-LS}{0.01}{TaylorF2-HS}{0.06}{TaylorF2-LS}{0.01}{PrecessingSpinIMRTidalHS}{0.27}{PrecessingSpinIMRTidalLS}{0.03}{PublicationSamples}{0.27}}}
\newcommand{\spinonefourtwofiveplus}[1]{\IfEqCase{#1}{{AlignedSpinInspiralTidalHS}{0.17}{AlignedSpinInspiralTidalLS}{0.03}{AlignedSpinTidalHS}{0.19}{AlignedSpinTidalLS}{0.03}{IMRPhenomDNRTidal-HS}{0.25}{IMRPhenomDNRTidal-LS}{0.03}{IMRPhenomPv2NRTidal-HS}{0.51}{IMRPhenomPv2NRTidal-LS}{0.02}{SEOBNRv4TsurrogateHS}{0.15}{SEOBNRv4TsurrogateLS}{0.03}{SEOBNRv4TsurrogatehighspinRIFT}{0.15}{SEOBNRv4TsurrogatelowspinRIFT}{0.03}{TEOBResumS-HS}{0.14}{TEOBResumS-LS}{0.03}{TaylorF2-HS}{0.17}{TaylorF2-LS}{0.03}{PrecessingSpinIMRTidalHS}{0.51}{PrecessingSpinIMRTidalLS}{0.02}{PublicationSamples}{0.51}}}
\newcommand{\costiltonefourtwofiveminus}[1]{\IfEqCase{#1}{{AlignedSpinInspiralTidalHS}{2.00}{AlignedSpinInspiralTidalLS}{2.00}{AlignedSpinTidalHS}{2.00}{AlignedSpinTidalLS}{2.00}{IMRPhenomDNRTidal-HS}{2.00}{IMRPhenomDNRTidal-LS}{2.00}{IMRPhenomPv2NRTidal-HS}{0.65}{IMRPhenomPv2NRTidal-LS}{1.10}{SEOBNRv4TsurrogateHS}{2.00}{SEOBNRv4TsurrogateLS}{2.00}{SEOBNRv4TsurrogatehighspinRIFT}{2.00}{SEOBNRv4TsurrogatelowspinRIFT}{2.00}{TEOBResumS-HS}{2.00}{TEOBResumS-LS}{2.00}{TaylorF2-HS}{2.00}{TaylorF2-LS}{2.00}{PrecessingSpinIMRTidalHS}{0.65}{PrecessingSpinIMRTidalLS}{1.10}{PublicationSamples}{0.65}}}
\newcommand{\costiltonefourtwofivemed}[1]{\IfEqCase{#1}{{AlignedSpinInspiralTidalHS}{1.00}{AlignedSpinInspiralTidalLS}{1.00}{AlignedSpinTidalHS}{1.00}{AlignedSpinTidalLS}{1.00}{IMRPhenomDNRTidal-HS}{1.00}{IMRPhenomDNRTidal-LS}{1.00}{IMRPhenomPv2NRTidal-HS}{0.26}{IMRPhenomPv2NRTidal-LS}{0.51}{SEOBNRv4TsurrogateHS}{1.00}{SEOBNRv4TsurrogateLS}{1.00}{SEOBNRv4TsurrogatehighspinRIFT}{1.00}{SEOBNRv4TsurrogatelowspinRIFT}{1.00}{TEOBResumS-HS}{1.00}{TEOBResumS-LS}{1.00}{TaylorF2-HS}{1.00}{TaylorF2-LS}{1.00}{PrecessingSpinIMRTidalHS}{0.26}{PrecessingSpinIMRTidalLS}{0.51}{PublicationSamples}{0.26}}}
\newcommand{\costiltonefourtwofiveplus}[1]{\IfEqCase{#1}{{AlignedSpinInspiralTidalHS}{0.00}{AlignedSpinInspiralTidalLS}{0.00}{AlignedSpinTidalHS}{0.00}{AlignedSpinTidalLS}{0.00}{IMRPhenomDNRTidal-HS}{0.00}{IMRPhenomDNRTidal-LS}{0.00}{IMRPhenomPv2NRTidal-HS}{0.61}{IMRPhenomPv2NRTidal-LS}{0.45}{SEOBNRv4TsurrogateHS}{0.00}{SEOBNRv4TsurrogateLS}{0.00}{SEOBNRv4TsurrogatehighspinRIFT}{0.00}{SEOBNRv4TsurrogatelowspinRIFT}{0.00}{TEOBResumS-HS}{0.00}{TEOBResumS-LS}{0.00}{TaylorF2-HS}{0.00}{TaylorF2-LS}{0.00}{PrecessingSpinIMRTidalHS}{0.61}{PrecessingSpinIMRTidalLS}{0.44}{PublicationSamples}{0.61}}}
\newcommand{\phasefourtwofiveminus}[1]{\IfEqCase{#1}{{AlignedSpinInspiralTidalHS}{3.10}{AlignedSpinInspiralTidalLS}{2.74}{AlignedSpinTidalHS}{2.81}{AlignedSpinTidalLS}{2.76}{IMRPhenomDNRTidal-HS}{2.83}{IMRPhenomDNRTidal-LS}{2.62}{IMRPhenomPv2NRTidal-HS}{2.82}{IMRPhenomPv2NRTidal-LS}{2.85}{SEOBNRv4TsurrogateHS}{2.74}{SEOBNRv4TsurrogateLS}{2.81}{SEOBNRv4TsurrogatehighspinRIFT}{2.78}{SEOBNRv4TsurrogatelowspinRIFT}{2.79}{TEOBResumS-HS}{2.78}{TEOBResumS-LS}{2.83}{TaylorF2-HS}{3.13}{TaylorF2-LS}{2.74}{PrecessingSpinIMRTidalHS}{2.82}{PrecessingSpinIMRTidalLS}{2.86}{PublicationSamples}{2.82}}}
\newcommand{\phasefourtwofivemed}[1]{\IfEqCase{#1}{{AlignedSpinInspiralTidalHS}{3.44}{AlignedSpinInspiralTidalLS}{3.01}{AlignedSpinTidalHS}{3.13}{AlignedSpinTidalLS}{3.07}{IMRPhenomDNRTidal-HS}{3.13}{IMRPhenomDNRTidal-LS}{2.88}{IMRPhenomPv2NRTidal-HS}{3.13}{IMRPhenomPv2NRTidal-LS}{3.20}{SEOBNRv4TsurrogateHS}{3.11}{SEOBNRv4TsurrogateLS}{3.16}{SEOBNRv4TsurrogatehighspinRIFT}{3.12}{SEOBNRv4TsurrogatelowspinRIFT}{3.12}{TEOBResumS-HS}{3.09}{TEOBResumS-LS}{3.14}{TaylorF2-HS}{3.46}{TaylorF2-LS}{3.00}{PrecessingSpinIMRTidalHS}{3.12}{PrecessingSpinIMRTidalLS}{3.20}{PublicationSamples}{3.13}}}
\newcommand{\phasefourtwofiveplus}[1]{\IfEqCase{#1}{{AlignedSpinInspiralTidalHS}{2.49}{AlignedSpinInspiralTidalLS}{2.93}{AlignedSpinTidalHS}{2.82}{AlignedSpinTidalLS}{2.88}{IMRPhenomDNRTidal-HS}{2.82}{IMRPhenomDNRTidal-LS}{3.04}{IMRPhenomPv2NRTidal-HS}{2.86}{IMRPhenomPv2NRTidal-LS}{2.74}{SEOBNRv4TsurrogateHS}{2.81}{SEOBNRv4TsurrogateLS}{2.74}{SEOBNRv4TsurrogatehighspinRIFT}{2.81}{SEOBNRv4TsurrogatelowspinRIFT}{2.79}{TEOBResumS-HS}{2.87}{TEOBResumS-LS}{2.86}{TaylorF2-HS}{2.47}{TaylorF2-LS}{2.94}{PrecessingSpinIMRTidalHS}{2.87}{PrecessingSpinIMRTidalLS}{2.73}{PublicationSamples}{2.86}}}
\newcommand{\masstwodetfourtwofiveminus}[1]{\IfEqCase{#1}{{AlignedSpinInspiralTidalHS}{0.3}{AlignedSpinInspiralTidalLS}{0.1}{AlignedSpinTidalHS}{0.3}{AlignedSpinTidalLS}{0.1}{IMRPhenomDNRTidal-HS}{0.3}{IMRPhenomDNRTidal-LS}{0.1}{IMRPhenomPv2NRTidal-HS}{0.3}{IMRPhenomPv2NRTidal-LS}{0.1}{SEOBNRv4TsurrogateHS}{0.3}{SEOBNRv4TsurrogateLS}{0.1}{SEOBNRv4TsurrogatehighspinRIFT}{0.3}{SEOBNRv4TsurrogatelowspinRIFT}{0.1}{TEOBResumS-HS}{0.3}{TEOBResumS-LS}{0.1}{TaylorF2-HS}{0.3}{TaylorF2-LS}{0.1}{PrecessingSpinIMRTidalHS}{0.3}{PrecessingSpinIMRTidalLS}{0.1}{PublicationSamples}{0.3}}}
\newcommand{\masstwodetfourtwofivemed}[1]{\IfEqCase{#1}{{AlignedSpinInspiralTidalHS}{1.4}{AlignedSpinInspiralTidalLS}{1.61}{AlignedSpinTidalHS}{1.5}{AlignedSpinTidalLS}{1.61}{IMRPhenomDNRTidal-HS}{1.5}{IMRPhenomDNRTidal-LS}{1.62}{IMRPhenomPv2NRTidal-HS}{1.4}{IMRPhenomPv2NRTidal-LS}{1.62}{SEOBNRv4TsurrogateHS}{1.5}{SEOBNRv4TsurrogateLS}{1.62}{SEOBNRv4TsurrogatehighspinRIFT}{1.5}{SEOBNRv4TsurrogatelowspinRIFT}{1.61}{TEOBResumS-HS}{1.5}{TEOBResumS-LS}{1.61}{TaylorF2-HS}{1.4}{TaylorF2-LS}{1.61}{PrecessingSpinIMRTidalHS}{1.4}{PrecessingSpinIMRTidalLS}{1.62}{PublicationSamples}{1.4}}}
\newcommand{\masstwodetfourtwofiveplus}[1]{\IfEqCase{#1}{{AlignedSpinInspiralTidalHS}{0.2}{AlignedSpinInspiralTidalLS}{0.09}{AlignedSpinTidalHS}{0.2}{AlignedSpinTidalLS}{0.08}{IMRPhenomDNRTidal-HS}{0.2}{IMRPhenomDNRTidal-LS}{0.08}{IMRPhenomPv2NRTidal-HS}{0.3}{IMRPhenomPv2NRTidal-LS}{0.08}{SEOBNRv4TsurrogateHS}{0.2}{SEOBNRv4TsurrogateLS}{0.08}{SEOBNRv4TsurrogatehighspinRIFT}{0.2}{SEOBNRv4TsurrogatelowspinRIFT}{0.08}{TEOBResumS-HS}{0.2}{TEOBResumS-LS}{0.09}{TaylorF2-HS}{0.2}{TaylorF2-LS}{0.09}{PrecessingSpinIMRTidalHS}{0.3}{PrecessingSpinIMRTidalLS}{0.08}{PublicationSamples}{0.3}}}
\newcommand{\masstwosourcefourtwofiveminus}[1]{\IfEqCase{#1}{{AlignedSpinInspiralTidalHS}{0.3}{AlignedSpinInspiralTidalLS}{0.1}{AlignedSpinTidalHS}{0.3}{AlignedSpinTidalLS}{0.1}{IMRPhenomDNRTidal-HS}{0.3}{IMRPhenomDNRTidal-LS}{0.1}{IMRPhenomPv2NRTidal-HS}{0.3}{IMRPhenomPv2NRTidal-LS}{0.1}{SEOBNRv4TsurrogateHS}{0.3}{SEOBNRv4TsurrogateLS}{0.1}{SEOBNRv4TsurrogatehighspinRIFT}{0.3}{SEOBNRv4TsurrogatelowspinRIFT}{0.1}{TEOBResumS-HS}{0.3}{TEOBResumS-LS}{0.1}{TaylorF2-HS}{0.3}{TaylorF2-LS}{0.1}{PrecessingSpinIMRTidalHS}{0.3}{PrecessingSpinIMRTidalLS}{0.1}{PublicationSamples}{0.3}}}
\newcommand{\masstwosourcefourtwofivemed}[1]{\IfEqCase{#1}{{AlignedSpinInspiralTidalHS}{1.4}{AlignedSpinInspiralTidalLS}{1.56}{AlignedSpinTidalHS}{1.4}{AlignedSpinTidalLS}{1.56}{IMRPhenomDNRTidal-HS}{1.4}{IMRPhenomDNRTidal-LS}{1.56}{IMRPhenomPv2NRTidal-HS}{1.4}{IMRPhenomPv2NRTidal-LS}{1.57}{SEOBNRv4TsurrogateHS}{1.4}{SEOBNRv4TsurrogateLS}{1.56}{SEOBNRv4TsurrogatehighspinRIFT}{1.5}{SEOBNRv4TsurrogatelowspinRIFT}{1.56}{TEOBResumS-HS}{1.4}{TEOBResumS-LS}{1.56}{TaylorF2-HS}{1.4}{TaylorF2-LS}{1.56}{PrecessingSpinIMRTidalHS}{1.4}{PrecessingSpinIMRTidalLS}{1.57}{PublicationSamples}{1.4}}}
\newcommand{\masstwosourcefourtwofiveplus}[1]{\IfEqCase{#1}{{AlignedSpinInspiralTidalHS}{0.2}{AlignedSpinInspiralTidalLS}{0.09}{AlignedSpinTidalHS}{0.2}{AlignedSpinTidalLS}{0.08}{IMRPhenomDNRTidal-HS}{0.2}{IMRPhenomDNRTidal-LS}{0.08}{IMRPhenomPv2NRTidal-HS}{0.3}{IMRPhenomPv2NRTidal-LS}{0.08}{SEOBNRv4TsurrogateHS}{0.2}{SEOBNRv4TsurrogateLS}{0.08}{SEOBNRv4TsurrogatehighspinRIFT}{0.2}{SEOBNRv4TsurrogatelowspinRIFT}{0.08}{TEOBResumS-HS}{0.2}{TEOBResumS-LS}{0.09}{TaylorF2-HS}{0.2}{TaylorF2-LS}{0.09}{PrecessingSpinIMRTidalHS}{0.3}{PrecessingSpinIMRTidalLS}{0.08}{PublicationSamples}{0.3}}}
\newcommand{\decfourtwofiveminus}[1]{\IfEqCase{#1}{{AlignedSpinInspiralTidalHS}{0.92098}{AlignedSpinInspiralTidalLS}{0.94910}{AlignedSpinTidalHS}{0.88455}{AlignedSpinTidalLS}{0.88762}{IMRPhenomDNRTidal-HS}{0.92222}{IMRPhenomDNRTidal-LS}{0.90469}{IMRPhenomPv2NRTidal-HS}{0.90104}{IMRPhenomPv2NRTidal-LS}{0.97042}{SEOBNRv4TsurrogateHS}{0.87704}{SEOBNRv4TsurrogateLS}{0.89335}{SEOBNRv4TsurrogatehighspinRIFT}{0.89796}{SEOBNRv4TsurrogatelowspinRIFT}{0.90269}{TEOBResumS-HS}{0.88789}{TEOBResumS-LS}{0.89751}{TaylorF2-HS}{0.92183}{TaylorF2-LS}{0.95570}{PrecessingSpinIMRTidalHS}{0.89977}{PrecessingSpinIMRTidalLS}{0.96778}{PublicationSamples}{0.89807}}}
\newcommand{\decfourtwofivemed}[1]{\IfEqCase{#1}{{AlignedSpinInspiralTidalHS}{-0.14949}{AlignedSpinInspiralTidalLS}{-0.10865}{AlignedSpinTidalHS}{-0.15883}{AlignedSpinTidalLS}{-0.13405}{IMRPhenomDNRTidal-HS}{-0.18418}{IMRPhenomDNRTidal-LS}{-0.06597}{IMRPhenomPv2NRTidal-HS}{-0.12984}{IMRPhenomPv2NRTidal-LS}{-0.05685}{SEOBNRv4TsurrogateHS}{-0.14893}{SEOBNRv4TsurrogateLS}{-0.17562}{SEOBNRv4TsurrogatehighspinRIFT}{-0.15053}{SEOBNRv4TsurrogatelowspinRIFT}{-0.12883}{TEOBResumS-HS}{-0.15241}{TEOBResumS-LS}{-0.16207}{TaylorF2-HS}{-0.14824}{TaylorF2-LS}{-0.10463}{PrecessingSpinIMRTidalHS}{-0.13006}{PrecessingSpinIMRTidalLS}{-0.06120}{PublicationSamples}{-0.13133}}}
\newcommand{\decfourtwofiveplus}[1]{\IfEqCase{#1}{{AlignedSpinInspiralTidalHS}{0.99909}{AlignedSpinInspiralTidalLS}{0.93177}{AlignedSpinTidalHS}{0.97810}{AlignedSpinTidalLS}{0.98373}{IMRPhenomDNRTidal-HS}{0.95052}{IMRPhenomDNRTidal-LS}{0.93229}{IMRPhenomPv2NRTidal-HS}{0.96946}{IMRPhenomPv2NRTidal-LS}{0.91574}{SEOBNRv4TsurrogateHS}{1.00436}{SEOBNRv4TsurrogateLS}{1.00055}{SEOBNRv4TsurrogatehighspinRIFT}{0.99086}{SEOBNRv4TsurrogatelowspinRIFT}{0.97101}{TEOBResumS-HS}{0.98792}{TEOBResumS-LS}{1.00135}{TaylorF2-HS}{0.99948}{TaylorF2-LS}{0.92354}{PrecessingSpinIMRTidalHS}{0.96811}{PrecessingSpinIMRTidalLS}{0.91709}{PublicationSamples}{0.96897}}}
\newcommand{\psifourtwofiveminus}[1]{\IfEqCase{#1}{{AlignedSpinInspiralTidalHS}{1.40}{AlignedSpinInspiralTidalLS}{1.41}{AlignedSpinTidalHS}{1.63}{AlignedSpinTidalLS}{1.69}{IMRPhenomDNRTidal-HS}{1.36}{IMRPhenomDNRTidal-LS}{1.41}{IMRPhenomPv2NRTidal-HS}{1.46}{IMRPhenomPv2NRTidal-LS}{1.40}{SEOBNRv4TsurrogateHS}{1.36}{SEOBNRv4TsurrogateLS}{1.43}{SEOBNRv4TsurrogatehighspinRIFT}{2.83}{SEOBNRv4TsurrogatelowspinRIFT}{2.86}{TEOBResumS-HS}{2.86}{TEOBResumS-LS}{2.86}{TaylorF2-HS}{1.41}{TaylorF2-LS}{1.41}{PrecessingSpinIMRTidalHS}{1.46}{PrecessingSpinIMRTidalLS}{1.39}{PublicationSamples}{1.46}}}
\newcommand{\psifourtwofivemed}[1]{\IfEqCase{#1}{{AlignedSpinInspiralTidalHS}{1.55}{AlignedSpinInspiralTidalLS}{1.56}{AlignedSpinTidalHS}{1.80}{AlignedSpinTidalLS}{1.86}{IMRPhenomDNRTidal-HS}{1.50}{IMRPhenomDNRTidal-LS}{1.56}{IMRPhenomPv2NRTidal-HS}{1.61}{IMRPhenomPv2NRTidal-LS}{1.54}{SEOBNRv4TsurrogateHS}{1.50}{SEOBNRv4TsurrogateLS}{1.56}{SEOBNRv4TsurrogatehighspinRIFT}{3.13}{SEOBNRv4TsurrogatelowspinRIFT}{3.14}{TEOBResumS-HS}{3.16}{TEOBResumS-LS}{3.16}{TaylorF2-HS}{1.56}{TaylorF2-LS}{1.55}{PrecessingSpinIMRTidalHS}{1.62}{PrecessingSpinIMRTidalLS}{1.54}{PublicationSamples}{1.61}}}
\newcommand{\psifourtwofiveplus}[1]{\IfEqCase{#1}{{AlignedSpinInspiralTidalHS}{1.42}{AlignedSpinInspiralTidalLS}{1.44}{AlignedSpinTidalHS}{3.54}{AlignedSpinTidalLS}{3.48}{IMRPhenomDNRTidal-HS}{1.50}{IMRPhenomDNRTidal-LS}{1.43}{IMRPhenomPv2NRTidal-HS}{1.38}{IMRPhenomPv2NRTidal-LS}{1.43}{SEOBNRv4TsurrogateHS}{1.47}{SEOBNRv4TsurrogateLS}{1.42}{SEOBNRv4TsurrogatehighspinRIFT}{2.85}{SEOBNRv4TsurrogatelowspinRIFT}{2.83}{TEOBResumS-HS}{2.82}{TEOBResumS-LS}{2.84}{TaylorF2-HS}{1.41}{TaylorF2-LS}{1.44}{PrecessingSpinIMRTidalHS}{1.38}{PrecessingSpinIMRTidalLS}{1.43}{PublicationSamples}{1.38}}}
\newcommand{\networkoptimalsnrfourtwofiveminus}[1]{\IfEqCase{#1}{{AlignedSpinInspiralTidalHS}{1.7}{AlignedSpinInspiralTidalLS}{1.7}{IMRPhenomDNRTidal-HS}{1.7}{IMRPhenomDNRTidal-LS}{1.7}{IMRPhenomPv2NRTidal-HS}{1.7}{IMRPhenomPv2NRTidal-LS}{1.7}{SEOBNRv4TsurrogateHS}{1.7}{SEOBNRv4TsurrogateLS}{1.7}{TaylorF2-HS}{1.7}{TaylorF2-LS}{1.7}{PrecessingSpinIMRTidalHS}{1.7}{PrecessingSpinIMRTidalLS}{1.7}{PublicationSamples}{1.7}}}
\newcommand{\networkoptimalsnrfourtwofivemed}[1]{\IfEqCase{#1}{{AlignedSpinInspiralTidalHS}{12.1}{AlignedSpinInspiralTidalLS}{12.2}{IMRPhenomDNRTidal-HS}{12.0}{IMRPhenomDNRTidal-LS}{12.1}{IMRPhenomPv2NRTidal-HS}{12.0}{IMRPhenomPv2NRTidal-LS}{12.1}{SEOBNRv4TsurrogateHS}{12.0}{SEOBNRv4TsurrogateLS}{12.1}{TaylorF2-HS}{12.1}{TaylorF2-LS}{12.2}{PrecessingSpinIMRTidalHS}{12.0}{PrecessingSpinIMRTidalLS}{12.1}{PublicationSamples}{12.0}}}
\newcommand{\networkoptimalsnrfourtwofiveplus}[1]{\IfEqCase{#1}{{AlignedSpinInspiralTidalHS}{1.7}{AlignedSpinInspiralTidalLS}{1.7}{IMRPhenomDNRTidal-HS}{1.7}{IMRPhenomDNRTidal-LS}{1.7}{IMRPhenomPv2NRTidal-HS}{1.7}{IMRPhenomPv2NRTidal-LS}{1.7}{SEOBNRv4TsurrogateHS}{1.7}{SEOBNRv4TsurrogateLS}{1.7}{TaylorF2-HS}{1.7}{TaylorF2-LS}{1.7}{PrecessingSpinIMRTidalHS}{1.7}{PrecessingSpinIMRTidalLS}{1.7}{PublicationSamples}{1.7}}}
\newcommand{\thetajnfourtwofiveminus}[1]{\IfEqCase{#1}{{AlignedSpinInspiralTidalHS}{0.97}{AlignedSpinInspiralTidalLS}{0.88}{AlignedSpinTidalHS}{0.83}{AlignedSpinTidalLS}{0.83}{IMRPhenomDNRTidal-HS}{0.74}{IMRPhenomDNRTidal-LS}{0.78}{IMRPhenomPv2NRTidal-HS}{0.85}{IMRPhenomPv2NRTidal-LS}{0.84}{SEOBNRv4TsurrogateHS}{0.89}{SEOBNRv4TsurrogateLS}{0.85}{SEOBNRv4TsurrogatehighspinRIFT}{0.86}{SEOBNRv4TsurrogatelowspinRIFT}{0.86}{TEOBResumS-HS}{0.86}{TEOBResumS-LS}{0.88}{TaylorF2-HS}{0.97}{TaylorF2-LS}{0.88}{PrecessingSpinIMRTidalHS}{0.85}{PrecessingSpinIMRTidalLS}{0.84}{PublicationSamples}{0.85}}}
\newcommand{\thetajnfourtwofivemed}[1]{\IfEqCase{#1}{{AlignedSpinInspiralTidalHS}{1.22}{AlignedSpinInspiralTidalLS}{1.11}{AlignedSpinTidalHS}{1.06}{AlignedSpinTidalLS}{1.06}{IMRPhenomDNRTidal-HS}{0.96}{IMRPhenomDNRTidal-LS}{1.02}{IMRPhenomPv2NRTidal-HS}{1.08}{IMRPhenomPv2NRTidal-LS}{1.07}{SEOBNRv4TsurrogateHS}{1.13}{SEOBNRv4TsurrogateLS}{1.09}{SEOBNRv4TsurrogatehighspinRIFT}{1.10}{SEOBNRv4TsurrogatelowspinRIFT}{1.09}{TEOBResumS-HS}{1.09}{TEOBResumS-LS}{1.12}{TaylorF2-HS}{1.22}{TaylorF2-LS}{1.11}{PrecessingSpinIMRTidalHS}{1.08}{PrecessingSpinIMRTidalLS}{1.07}{PublicationSamples}{1.08}}}
\newcommand{\thetajnfourtwofiveplus}[1]{\IfEqCase{#1}{{AlignedSpinInspiralTidalHS}{1.64}{AlignedSpinInspiralTidalLS}{1.75}{AlignedSpinTidalHS}{1.79}{AlignedSpinTidalLS}{1.78}{IMRPhenomDNRTidal-HS}{1.88}{IMRPhenomDNRTidal-LS}{1.83}{IMRPhenomPv2NRTidal-HS}{1.77}{IMRPhenomPv2NRTidal-LS}{1.78}{SEOBNRv4TsurrogateHS}{1.72}{SEOBNRv4TsurrogateLS}{1.76}{SEOBNRv4TsurrogatehighspinRIFT}{1.76}{SEOBNRv4TsurrogatelowspinRIFT}{1.75}{TEOBResumS-HS}{1.77}{TEOBResumS-LS}{1.74}{TaylorF2-HS}{1.63}{TaylorF2-LS}{1.75}{PrecessingSpinIMRTidalHS}{1.77}{PrecessingSpinIMRTidalLS}{1.78}{PublicationSamples}{1.77}}}
\newcommand{\totalmassdetfourtwofiveminus}[1]{\IfEqCase{#1}{{AlignedSpinInspiralTidalHS}{0.06}{AlignedSpinInspiralTidalLS}{0.007}{AlignedSpinTidalHS}{0.04}{AlignedSpinTidalLS}{0.007}{IMRPhenomDNRTidal-HS}{0.06}{IMRPhenomDNRTidal-LS}{0.006}{IMRPhenomPv2NRTidal-HS}{0.08}{IMRPhenomPv2NRTidal-LS}{0.006}{SEOBNRv4TsurrogateHS}{0.03}{SEOBNRv4TsurrogateLS}{0.006}{SEOBNRv4TsurrogatehighspinRIFT}{0.03}{SEOBNRv4TsurrogatelowspinRIFT}{0.007}{TEOBResumS-HS}{0.04}{TEOBResumS-LS}{0.007}{TaylorF2-HS}{0.06}{TaylorF2-LS}{0.007}{PrecessingSpinIMRTidalHS}{0.08}{PrecessingSpinIMRTidalLS}{0.006}{PublicationSamples}{0.08}}}
\newcommand{\totalmassdetfourtwofivemed}[1]{\IfEqCase{#1}{{AlignedSpinInspiralTidalHS}{3.48}{AlignedSpinInspiralTidalLS}{3.42}{AlignedSpinTidalHS}{3.46}{AlignedSpinTidalLS}{3.42}{IMRPhenomDNRTidal-HS}{3.47}{IMRPhenomDNRTidal-LS}{3.42}{IMRPhenomPv2NRTidal-HS}{3.50}{IMRPhenomPv2NRTidal-LS}{3.42}{SEOBNRv4TsurrogateHS}{3.45}{SEOBNRv4TsurrogateLS}{3.42}{SEOBNRv4TsurrogatehighspinRIFT}{3.45}{SEOBNRv4TsurrogatelowspinRIFT}{3.42}{TEOBResumS-HS}{3.46}{TEOBResumS-LS}{3.42}{TaylorF2-HS}{3.48}{TaylorF2-LS}{3.42}{PrecessingSpinIMRTidalHS}{3.50}{PrecessingSpinIMRTidalLS}{3.42}{PublicationSamples}{3.50}}}
\newcommand{\totalmassdetfourtwofiveplus}[1]{\IfEqCase{#1}{{AlignedSpinInspiralTidalHS}{0.3}{AlignedSpinInspiralTidalLS}{0.04}{AlignedSpinTidalHS}{0.3}{AlignedSpinTidalLS}{0.04}{IMRPhenomDNRTidal-HS}{0.4}{IMRPhenomDNRTidal-LS}{0.04}{IMRPhenomPv2NRTidal-HS}{0.3}{IMRPhenomPv2NRTidal-LS}{0.04}{SEOBNRv4TsurrogateHS}{0.2}{SEOBNRv4TsurrogateLS}{0.04}{SEOBNRv4TsurrogatehighspinRIFT}{0.2}{SEOBNRv4TsurrogatelowspinRIFT}{0.04}{TEOBResumS-HS}{0.2}{TEOBResumS-LS}{0.04}{TaylorF2-HS}{0.3}{TaylorF2-LS}{0.04}{PrecessingSpinIMRTidalHS}{0.3}{PrecessingSpinIMRTidalLS}{0.04}{PublicationSamples}{0.3}}}
\newcommand{\redshiftfourtwofiveminus}[1]{\IfEqCase{#1}{{AlignedSpinInspiralTidalHS}{0.02}{AlignedSpinInspiralTidalLS}{0.02}{AlignedSpinTidalHS}{0.02}{AlignedSpinTidalLS}{0.02}{IMRPhenomDNRTidal-HS}{0.02}{IMRPhenomDNRTidal-LS}{0.02}{IMRPhenomPv2NRTidal-HS}{0.02}{IMRPhenomPv2NRTidal-LS}{0.02}{SEOBNRv4TsurrogateHS}{0.02}{SEOBNRv4TsurrogateLS}{0.02}{SEOBNRv4TsurrogatehighspinRIFT}{0.02}{SEOBNRv4TsurrogatelowspinRIFT}{0.02}{TEOBResumS-HS}{0.02}{TEOBResumS-LS}{0.02}{TaylorF2-HS}{0.02}{TaylorF2-LS}{0.02}{PrecessingSpinIMRTidalHS}{0.02}{PrecessingSpinIMRTidalLS}{0.02}{PublicationSamples}{0.02}}}
\newcommand{\redshiftfourtwofivemed}[1]{\IfEqCase{#1}{{AlignedSpinInspiralTidalHS}{0.04}{AlignedSpinInspiralTidalLS}{0.04}{AlignedSpinTidalHS}{0.04}{AlignedSpinTidalLS}{0.03}{IMRPhenomDNRTidal-HS}{0.04}{IMRPhenomDNRTidal-LS}{0.03}{IMRPhenomPv2NRTidal-HS}{0.03}{IMRPhenomPv2NRTidal-LS}{0.03}{SEOBNRv4TsurrogateHS}{0.03}{SEOBNRv4TsurrogateLS}{0.03}{SEOBNRv4TsurrogatehighspinRIFT}{0.04}{SEOBNRv4TsurrogatelowspinRIFT}{0.03}{TEOBResumS-HS}{0.04}{TEOBResumS-LS}{0.03}{TaylorF2-HS}{0.04}{TaylorF2-LS}{0.04}{PrecessingSpinIMRTidalHS}{0.03}{PrecessingSpinIMRTidalLS}{0.03}{PublicationSamples}{0.03}}}
\newcommand{\redshiftfourtwofiveplus}[1]{\IfEqCase{#1}{{AlignedSpinInspiralTidalHS}{0.02}{AlignedSpinInspiralTidalLS}{0.01}{AlignedSpinTidalHS}{0.02}{AlignedSpinTidalLS}{0.01}{IMRPhenomDNRTidal-HS}{0.01}{IMRPhenomDNRTidal-LS}{0.01}{IMRPhenomPv2NRTidal-HS}{0.01}{IMRPhenomPv2NRTidal-LS}{0.01}{SEOBNRv4TsurrogateHS}{0.01}{SEOBNRv4TsurrogateLS}{0.01}{SEOBNRv4TsurrogatehighspinRIFT}{0.02}{SEOBNRv4TsurrogatelowspinRIFT}{0.02}{TEOBResumS-HS}{0.02}{TEOBResumS-LS}{0.01}{TaylorF2-HS}{0.02}{TaylorF2-LS}{0.02}{PrecessingSpinIMRTidalHS}{0.01}{PrecessingSpinIMRTidalLS}{0.01}{PublicationSamples}{0.01}}}
\newcommand{\iotafourtwofiveminus}[1]{\IfEqCase{#1}{{AlignedSpinInspiralTidalHS}{0.97}{AlignedSpinInspiralTidalLS}{0.88}{AlignedSpinTidalHS}{0.83}{AlignedSpinTidalLS}{0.83}{IMRPhenomDNRTidal-HS}{0.74}{IMRPhenomDNRTidal-LS}{0.78}{IMRPhenomPv2NRTidal-HS}{0.85}{IMRPhenomPv2NRTidal-LS}{0.84}{SEOBNRv4TsurrogateHS}{0.89}{SEOBNRv4TsurrogateLS}{0.85}{SEOBNRv4TsurrogatehighspinRIFT}{0.86}{SEOBNRv4TsurrogatelowspinRIFT}{0.86}{TEOBResumS-HS}{0.86}{TEOBResumS-LS}{0.88}{TaylorF2-HS}{0.97}{TaylorF2-LS}{0.88}{PrecessingSpinIMRTidalHS}{0.85}{PrecessingSpinIMRTidalLS}{0.84}{PublicationSamples}{0.85}}}
\newcommand{\iotafourtwofivemed}[1]{\IfEqCase{#1}{{AlignedSpinInspiralTidalHS}{1.22}{AlignedSpinInspiralTidalLS}{1.11}{AlignedSpinTidalHS}{1.06}{AlignedSpinTidalLS}{1.06}{IMRPhenomDNRTidal-HS}{0.96}{IMRPhenomDNRTidal-LS}{1.02}{IMRPhenomPv2NRTidal-HS}{1.09}{IMRPhenomPv2NRTidal-LS}{1.07}{SEOBNRv4TsurrogateHS}{1.13}{SEOBNRv4TsurrogateLS}{1.09}{SEOBNRv4TsurrogatehighspinRIFT}{1.10}{SEOBNRv4TsurrogatelowspinRIFT}{1.09}{TEOBResumS-HS}{1.09}{TEOBResumS-LS}{1.12}{TaylorF2-HS}{1.22}{TaylorF2-LS}{1.11}{PrecessingSpinIMRTidalHS}{1.09}{PrecessingSpinIMRTidalLS}{1.07}{PublicationSamples}{1.09}}}
\newcommand{\iotafourtwofiveplus}[1]{\IfEqCase{#1}{{AlignedSpinInspiralTidalHS}{1.64}{AlignedSpinInspiralTidalLS}{1.75}{AlignedSpinTidalHS}{1.79}{AlignedSpinTidalLS}{1.78}{IMRPhenomDNRTidal-HS}{1.88}{IMRPhenomDNRTidal-LS}{1.83}{IMRPhenomPv2NRTidal-HS}{1.77}{IMRPhenomPv2NRTidal-LS}{1.78}{SEOBNRv4TsurrogateHS}{1.72}{SEOBNRv4TsurrogateLS}{1.76}{SEOBNRv4TsurrogatehighspinRIFT}{1.76}{SEOBNRv4TsurrogatelowspinRIFT}{1.75}{TEOBResumS-HS}{1.77}{TEOBResumS-LS}{1.74}{TaylorF2-HS}{1.63}{TaylorF2-LS}{1.75}{PrecessingSpinIMRTidalHS}{1.77}{PrecessingSpinIMRTidalLS}{1.78}{PublicationSamples}{1.77}}}
\newcommand{\spinonexfourtwofiveminus}[1]{\IfEqCase{#1}{{AlignedSpinInspiralTidalHS}{0.00}{AlignedSpinInspiralTidalLS}{0.00}{AlignedSpinTidalHS}{0.00}{AlignedSpinTidalLS}{0.00}{IMRPhenomDNRTidal-HS}{0.00}{IMRPhenomDNRTidal-LS}{0.00}{IMRPhenomPv2NRTidal-HS}{0.50}{IMRPhenomPv2NRTidal-LS}{0.03}{SEOBNRv4TsurrogateHS}{0.00}{SEOBNRv4TsurrogateLS}{0.00}{SEOBNRv4TsurrogatehighspinRIFT}{0.00}{SEOBNRv4TsurrogatelowspinRIFT}{0.00}{TEOBResumS-HS}{0.00}{TEOBResumS-LS}{0.00}{TaylorF2-HS}{0.00}{TaylorF2-LS}{0.00}{PrecessingSpinIMRTidalHS}{0.50}{PrecessingSpinIMRTidalLS}{0.03}{PublicationSamples}{0.50}}}
\newcommand{\spinonexfourtwofivemed}[1]{\IfEqCase{#1}{{AlignedSpinInspiralTidalHS}{0.00}{AlignedSpinInspiralTidalLS}{0.00}{AlignedSpinTidalHS}{0.00}{AlignedSpinTidalLS}{0.00}{IMRPhenomDNRTidal-HS}{0.00}{IMRPhenomDNRTidal-LS}{0.00}{IMRPhenomPv2NRTidal-HS}{0.00}{IMRPhenomPv2NRTidal-LS}{0.00009}{SEOBNRv4TsurrogateHS}{0.00}{SEOBNRv4TsurrogateLS}{0.00}{SEOBNRv4TsurrogatehighspinRIFT}{0.00}{SEOBNRv4TsurrogatelowspinRIFT}{0.00}{TEOBResumS-HS}{0.00}{TEOBResumS-LS}{0.00}{TaylorF2-HS}{0.00}{TaylorF2-LS}{0.00}{PrecessingSpinIMRTidalHS}{0.00}{PrecessingSpinIMRTidalLS}{0.0001}{PublicationSamples}{0.00}}}
\newcommand{\spinonexfourtwofiveplus}[1]{\IfEqCase{#1}{{AlignedSpinInspiralTidalHS}{0.00}{AlignedSpinInspiralTidalLS}{0.00}{AlignedSpinTidalHS}{0.00}{AlignedSpinTidalLS}{0.00}{IMRPhenomDNRTidal-HS}{0.00}{IMRPhenomDNRTidal-LS}{0.00}{IMRPhenomPv2NRTidal-HS}{0.47}{IMRPhenomPv2NRTidal-LS}{0.03}{SEOBNRv4TsurrogateHS}{0.00}{SEOBNRv4TsurrogateLS}{0.00}{SEOBNRv4TsurrogatehighspinRIFT}{0.00}{SEOBNRv4TsurrogatelowspinRIFT}{0.00}{TEOBResumS-HS}{0.00}{TEOBResumS-LS}{0.00}{TaylorF2-HS}{0.00}{TaylorF2-LS}{0.00}{PrecessingSpinIMRTidalHS}{0.47}{PrecessingSpinIMRTidalLS}{0.03}{PublicationSamples}{0.47}}}
\newcommand{\chirpmassdetfourtwofiveminus}[1]{\IfEqCase{#1}{{AlignedSpinInspiralTidalHS}{0.0005}{AlignedSpinInspiralTidalLS}{0.0003}{AlignedSpinTidalHS}{0.0005}{AlignedSpinTidalLS}{0.0003}{IMRPhenomDNRTidal-HS}{0.0005}{IMRPhenomDNRTidal-LS}{0.0003}{IMRPhenomPv2NRTidal-HS}{0.0006}{IMRPhenomPv2NRTidal-LS}{0.0003}{SEOBNRv4TsurrogateHS}{0.0005}{SEOBNRv4TsurrogateLS}{0.0003}{SEOBNRv4TsurrogatehighspinRIFT}{0.0005}{SEOBNRv4TsurrogatelowspinRIFT}{0.0004}{TEOBResumS-HS}{0.0005}{TEOBResumS-LS}{0.0003}{TaylorF2-HS}{0.0005}{TaylorF2-LS}{0.0003}{PrecessingSpinIMRTidalHS}{0.0006}{PrecessingSpinIMRTidalLS}{0.0003}{PublicationSamples}{0.0006}}}
\newcommand{\chirpmassdetfourtwofivemed}[1]{\IfEqCase{#1}{{AlignedSpinInspiralTidalHS}{1.49}{AlignedSpinInspiralTidalLS}{1.49}{AlignedSpinTidalHS}{1.49}{AlignedSpinTidalLS}{1.49}{IMRPhenomDNRTidal-HS}{1.49}{IMRPhenomDNRTidal-LS}{1.49}{IMRPhenomPv2NRTidal-HS}{1.4873}{IMRPhenomPv2NRTidal-LS}{1.49}{SEOBNRv4TsurrogateHS}{1.49}{SEOBNRv4TsurrogateLS}{1.49}{SEOBNRv4TsurrogatehighspinRIFT}{1.49}{SEOBNRv4TsurrogatelowspinRIFT}{1.49}{TEOBResumS-HS}{1.49}{TEOBResumS-LS}{1.49}{TaylorF2-HS}{1.49}{TaylorF2-LS}{1.49}{PrecessingSpinIMRTidalHS}{1.49}{PrecessingSpinIMRTidalLS}{1.49}{PublicationSamples}{1.49}}}
\newcommand{\chirpmassdetfourtwofiveplus}[1]{\IfEqCase{#1}{{AlignedSpinInspiralTidalHS}{0.0007}{AlignedSpinInspiralTidalLS}{0.0003}{AlignedSpinTidalHS}{0.0007}{AlignedSpinTidalLS}{0.0003}{IMRPhenomDNRTidal-HS}{0.0008}{IMRPhenomDNRTidal-LS}{0.0004}{IMRPhenomPv2NRTidal-HS}{0.0008}{IMRPhenomPv2NRTidal-LS}{0.0003}{SEOBNRv4TsurrogateHS}{0.0006}{SEOBNRv4TsurrogateLS}{0.0003}{SEOBNRv4TsurrogatehighspinRIFT}{0.0006}{SEOBNRv4TsurrogatelowspinRIFT}{0.0003}{TEOBResumS-HS}{0.0005}{TEOBResumS-LS}{0.0003}{TaylorF2-HS}{0.0007}{TaylorF2-LS}{0.0003}{PrecessingSpinIMRTidalHS}{0.0008}{PrecessingSpinIMRTidalLS}{0.0003}{PublicationSamples}{0.0008}}}
\newcommand{\cosiotafourtwofiveminus}[1]{\IfEqCase{#1}{{AlignedSpinInspiralTidalHS}{1.30}{AlignedSpinInspiralTidalLS}{1.40}{AlignedSpinTidalHS}{1.44}{AlignedSpinTidalLS}{1.44}{IMRPhenomDNRTidal-HS}{1.53}{IMRPhenomDNRTidal-LS}{1.48}{IMRPhenomPv2NRTidal-HS}{1.42}{IMRPhenomPv2NRTidal-LS}{1.44}{SEOBNRv4TsurrogateHS}{1.38}{SEOBNRv4TsurrogateLS}{1.42}{SEOBNRv4TsurrogatehighspinRIFT}{1.41}{SEOBNRv4TsurrogatelowspinRIFT}{1.42}{TEOBResumS-HS}{1.42}{TEOBResumS-LS}{1.40}{TaylorF2-HS}{1.30}{TaylorF2-LS}{1.40}{PrecessingSpinIMRTidalHS}{1.42}{PrecessingSpinIMRTidalLS}{1.44}{PublicationSamples}{1.42}}}
\newcommand{\cosiotafourtwofivemed}[1]{\IfEqCase{#1}{{AlignedSpinInspiralTidalHS}{0.34}{AlignedSpinInspiralTidalLS}{0.44}{AlignedSpinTidalHS}{0.49}{AlignedSpinTidalLS}{0.49}{IMRPhenomDNRTidal-HS}{0.58}{IMRPhenomDNRTidal-LS}{0.53}{IMRPhenomPv2NRTidal-HS}{0.46}{IMRPhenomPv2NRTidal-LS}{0.48}{SEOBNRv4TsurrogateHS}{0.43}{SEOBNRv4TsurrogateLS}{0.46}{SEOBNRv4TsurrogatehighspinRIFT}{0.45}{SEOBNRv4TsurrogatelowspinRIFT}{0.46}{TEOBResumS-HS}{0.46}{TEOBResumS-LS}{0.44}{TaylorF2-HS}{0.34}{TaylorF2-LS}{0.44}{PrecessingSpinIMRTidalHS}{0.46}{PrecessingSpinIMRTidalLS}{0.48}{PublicationSamples}{0.46}}}
\newcommand{\cosiotafourtwofiveplus}[1]{\IfEqCase{#1}{{AlignedSpinInspiralTidalHS}{0.62}{AlignedSpinInspiralTidalLS}{0.53}{AlignedSpinTidalHS}{0.49}{AlignedSpinTidalLS}{0.49}{IMRPhenomDNRTidal-HS}{0.40}{IMRPhenomDNRTidal-LS}{0.45}{IMRPhenomPv2NRTidal-HS}{0.51}{IMRPhenomPv2NRTidal-LS}{0.49}{SEOBNRv4TsurrogateHS}{0.54}{SEOBNRv4TsurrogateLS}{0.51}{SEOBNRv4TsurrogatehighspinRIFT}{0.52}{SEOBNRv4TsurrogatelowspinRIFT}{0.51}{TEOBResumS-HS}{0.51}{TEOBResumS-LS}{0.54}{TaylorF2-HS}{0.63}{TaylorF2-LS}{0.53}{PrecessingSpinIMRTidalHS}{0.51}{PrecessingSpinIMRTidalLS}{0.49}{PublicationSamples}{0.51}}}
\newcommand{\comovingdistfourtwofiveminus}[1]{\IfEqCase{#1}{{AlignedSpinInspiralTidalHS}{72}{AlignedSpinInspiralTidalLS}{71}{AlignedSpinTidalHS}{69}{AlignedSpinTidalLS}{70}{IMRPhenomDNRTidal-HS}{69}{IMRPhenomDNRTidal-LS}{69}{IMRPhenomPv2NRTidal-HS}{67}{IMRPhenomPv2NRTidal-LS}{68}{SEOBNRv4TsurrogateHS}{68}{SEOBNRv4TsurrogateLS}{69}{SEOBNRv4TsurrogatehighspinRIFT}{70}{SEOBNRv4TsurrogatelowspinRIFT}{68}{TEOBResumS-HS}{70}{TEOBResumS-LS}{69}{TaylorF2-HS}{72}{TaylorF2-LS}{71}{PrecessingSpinIMRTidalHS}{67}{PrecessingSpinIMRTidalLS}{68}{PublicationSamples}{67}}}
\newcommand{\comovingdistfourtwofivemed}[1]{\IfEqCase{#1}{{AlignedSpinInspiralTidalHS}{156}{AlignedSpinInspiralTidalLS}{157}{AlignedSpinTidalHS}{155}{AlignedSpinTidalLS}{153}{IMRPhenomDNRTidal-HS}{155}{IMRPhenomDNRTidal-LS}{153}{IMRPhenomPv2NRTidal-HS}{151}{IMRPhenomPv2NRTidal-LS}{151}{SEOBNRv4TsurrogateHS}{153}{SEOBNRv4TsurrogateLS}{153}{SEOBNRv4TsurrogatehighspinRIFT}{157}{SEOBNRv4TsurrogatelowspinRIFT}{152}{TEOBResumS-HS}{157}{TEOBResumS-LS}{153}{TaylorF2-HS}{156}{TaylorF2-LS}{157}{PrecessingSpinIMRTidalHS}{151}{PrecessingSpinIMRTidalLS}{151}{PublicationSamples}{151}}}
\newcommand{\comovingdistfourtwofiveplus}[1]{\IfEqCase{#1}{{AlignedSpinInspiralTidalHS}{67}{AlignedSpinInspiralTidalLS}{65}{AlignedSpinTidalHS}{65}{AlignedSpinTidalLS}{63}{IMRPhenomDNRTidal-HS}{63}{IMRPhenomDNRTidal-LS}{64}{IMRPhenomPv2NRTidal-HS}{64}{IMRPhenomPv2NRTidal-LS}{64}{SEOBNRv4TsurrogateHS}{64}{SEOBNRv4TsurrogateLS}{64}{SEOBNRv4TsurrogatehighspinRIFT}{70}{SEOBNRv4TsurrogatelowspinRIFT}{65}{TEOBResumS-HS}{68}{TEOBResumS-LS}{64}{TaylorF2-HS}{67}{TaylorF2-LS}{65}{PrecessingSpinIMRTidalHS}{64}{PrecessingSpinIMRTidalLS}{64}{PublicationSamples}{64}}}
\newcommand{\logpriorfourtwofiveminus}[1]{\IfEqCase{#1}{{AlignedSpinInspiralTidalHS}{8.6}{AlignedSpinInspiralTidalLS}{8.5}{IMRPhenomDNRTidal-HS}{8.6}{IMRPhenomDNRTidal-LS}{8.6}{IMRPhenomPv2NRTidal-HS}{8.6}{IMRPhenomPv2NRTidal-LS}{8.4}{SEOBNRv4TsurrogateHS}{8.4}{SEOBNRv4TsurrogateLS}{8.8}{TaylorF2-HS}{8.6}{TaylorF2-LS}{8.5}{PrecessingSpinIMRTidalHS}{8.6}{PrecessingSpinIMRTidalLS}{8.4}{PublicationSamples}{8.6}}}
\newcommand{\logpriorfourtwofivemed}[1]{\IfEqCase{#1}{{AlignedSpinInspiralTidalHS}{102.5}{AlignedSpinInspiralTidalLS}{106.7}{IMRPhenomDNRTidal-HS}{94.5}{IMRPhenomDNRTidal-LS}{99.2}{IMRPhenomPv2NRTidal-HS}{98.4}{IMRPhenomPv2NRTidal-LS}{97.8}{SEOBNRv4TsurrogateHS}{95.6}{SEOBNRv4TsurrogateLS}{99.0}{TaylorF2-HS}{102.5}{TaylorF2-LS}{106.7}{PrecessingSpinIMRTidalHS}{98.4}{PrecessingSpinIMRTidalLS}{97.8}{PublicationSamples}{98.4}}}
\newcommand{\logpriorfourtwofiveplus}[1]{\IfEqCase{#1}{{AlignedSpinInspiralTidalHS}{6.8}{AlignedSpinInspiralTidalLS}{6.9}{IMRPhenomDNRTidal-HS}{6.8}{IMRPhenomDNRTidal-LS}{6.9}{IMRPhenomPv2NRTidal-HS}{6.7}{IMRPhenomPv2NRTidal-LS}{6.7}{SEOBNRv4TsurrogateHS}{6.9}{SEOBNRv4TsurrogateLS}{6.9}{TaylorF2-HS}{6.8}{TaylorF2-LS}{6.9}{PrecessingSpinIMRTidalHS}{6.7}{PrecessingSpinIMRTidalLS}{6.7}{PublicationSamples}{6.7}}}
\newcommand{\tiltonefourtwofiveminus}[1]{\IfEqCase{#1}{{AlignedSpinInspiralTidalHS}{0.00}{AlignedSpinInspiralTidalLS}{0.00}{AlignedSpinTidalHS}{0.00}{AlignedSpinTidalLS}{0.00}{IMRPhenomDNRTidal-HS}{0.00}{IMRPhenomDNRTidal-LS}{0.00}{IMRPhenomPv2NRTidal-HS}{0.80}{IMRPhenomPv2NRTidal-LS}{0.74}{SEOBNRv4TsurrogateHS}{0.00}{SEOBNRv4TsurrogateLS}{0.00}{SEOBNRv4TsurrogatehighspinRIFT}{0.00}{SEOBNRv4TsurrogatelowspinRIFT}{0.00}{TEOBResumS-HS}{0.00}{TEOBResumS-LS}{0.00}{TaylorF2-HS}{0.00}{TaylorF2-LS}{0.00}{PrecessingSpinIMRTidalHS}{0.80}{PrecessingSpinIMRTidalLS}{0.74}{PublicationSamples}{0.79}}}
\newcommand{\tiltonefourtwofivemed}[1]{\IfEqCase{#1}{{AlignedSpinInspiralTidalHS}{0.00}{AlignedSpinInspiralTidalLS}{0.00}{AlignedSpinTidalHS}{0.00}{AlignedSpinTidalLS}{0.00}{IMRPhenomDNRTidal-HS}{0.00}{IMRPhenomDNRTidal-LS}{0.00}{IMRPhenomPv2NRTidal-HS}{1.31}{IMRPhenomPv2NRTidal-LS}{1.03}{SEOBNRv4TsurrogateHS}{0.00}{SEOBNRv4TsurrogateLS}{0.00}{SEOBNRv4TsurrogatehighspinRIFT}{0.00}{SEOBNRv4TsurrogatelowspinRIFT}{0.00}{TEOBResumS-HS}{0.00}{TEOBResumS-LS}{0.00}{TaylorF2-HS}{0.00}{TaylorF2-LS}{0.00}{PrecessingSpinIMRTidalHS}{1.31}{PrecessingSpinIMRTidalLS}{1.03}{PublicationSamples}{1.31}}}
\newcommand{\tiltonefourtwofiveplus}[1]{\IfEqCase{#1}{{AlignedSpinInspiralTidalHS}{3.14}{AlignedSpinInspiralTidalLS}{3.14}{AlignedSpinTidalHS}{3.14}{AlignedSpinTidalLS}{3.14}{IMRPhenomDNRTidal-HS}{3.14}{IMRPhenomDNRTidal-LS}{3.14}{IMRPhenomPv2NRTidal-HS}{0.66}{IMRPhenomPv2NRTidal-LS}{1.16}{SEOBNRv4TsurrogateHS}{3.14}{SEOBNRv4TsurrogateLS}{3.14}{SEOBNRv4TsurrogatehighspinRIFT}{3.14}{SEOBNRv4TsurrogatelowspinRIFT}{3.14}{TEOBResumS-HS}{3.14}{TEOBResumS-LS}{3.14}{TaylorF2-HS}{3.14}{TaylorF2-LS}{3.14}{PrecessingSpinIMRTidalHS}{0.66}{PrecessingSpinIMRTidalLS}{1.17}{PublicationSamples}{0.66}}}
\newcommand{\spintwoyfourtwofiveminus}[1]{\IfEqCase{#1}{{AlignedSpinInspiralTidalHS}{0.00}{AlignedSpinInspiralTidalLS}{0.00}{AlignedSpinTidalHS}{0.00}{AlignedSpinTidalLS}{0.00}{IMRPhenomDNRTidal-HS}{0.00}{IMRPhenomDNRTidal-LS}{0.00}{IMRPhenomPv2NRTidal-HS}{0.48}{IMRPhenomPv2NRTidal-LS}{0.03}{SEOBNRv4TsurrogateHS}{0.00}{SEOBNRv4TsurrogateLS}{0.00}{SEOBNRv4TsurrogatehighspinRIFT}{0.00}{SEOBNRv4TsurrogatelowspinRIFT}{0.00}{TEOBResumS-HS}{0.00}{TEOBResumS-LS}{0.00}{TaylorF2-HS}{0.00}{TaylorF2-LS}{0.00}{PrecessingSpinIMRTidalHS}{0.48}{PrecessingSpinIMRTidalLS}{0.03}{PublicationSamples}{0.48}}}
\newcommand{\spintwoyfourtwofivemed}[1]{\IfEqCase{#1}{{AlignedSpinInspiralTidalHS}{0.00}{AlignedSpinInspiralTidalLS}{0.00}{AlignedSpinTidalHS}{0.00}{AlignedSpinTidalLS}{0.00}{IMRPhenomDNRTidal-HS}{0.00}{IMRPhenomDNRTidal-LS}{0.00}{IMRPhenomPv2NRTidal-HS}{0.00002}{IMRPhenomPv2NRTidal-LS}{0.00003}{SEOBNRv4TsurrogateHS}{0.00}{SEOBNRv4TsurrogateLS}{0.00}{SEOBNRv4TsurrogatehighspinRIFT}{0.00}{SEOBNRv4TsurrogatelowspinRIFT}{0.00}{TEOBResumS-HS}{0.00}{TEOBResumS-LS}{0.00}{TaylorF2-HS}{0.00}{TaylorF2-LS}{0.00}{PrecessingSpinIMRTidalHS}{0.00}{PrecessingSpinIMRTidalLS}{0.00002}{PublicationSamples}{0.00}}}
\newcommand{\spintwoyfourtwofiveplus}[1]{\IfEqCase{#1}{{AlignedSpinInspiralTidalHS}{0.00}{AlignedSpinInspiralTidalLS}{0.00}{AlignedSpinTidalHS}{0.00}{AlignedSpinTidalLS}{0.00}{IMRPhenomDNRTidal-HS}{0.00}{IMRPhenomDNRTidal-LS}{0.00}{IMRPhenomPv2NRTidal-HS}{0.48}{IMRPhenomPv2NRTidal-LS}{0.03}{SEOBNRv4TsurrogateHS}{0.00}{SEOBNRv4TsurrogateLS}{0.00}{SEOBNRv4TsurrogatehighspinRIFT}{0.00}{SEOBNRv4TsurrogatelowspinRIFT}{0.00}{TEOBResumS-HS}{0.00}{TEOBResumS-LS}{0.00}{TaylorF2-HS}{0.00}{TaylorF2-LS}{0.00}{PrecessingSpinIMRTidalHS}{0.48}{PrecessingSpinIMRTidalLS}{0.03}{PublicationSamples}{0.48}}}
\newcommand{\spintwozfourtwofiveminus}[1]{\IfEqCase{#1}{{AlignedSpinInspiralTidalHS}{0.18}{AlignedSpinInspiralTidalLS}{0.02}{AlignedSpinTidalHS}{0.24}{AlignedSpinTidalLS}{0.02}{IMRPhenomDNRTidal-HS}{0.39}{IMRPhenomDNRTidal-LS}{0.02}{IMRPhenomPv2NRTidal-HS}{0.18}{IMRPhenomPv2NRTidal-LS}{0.02}{SEOBNRv4TsurrogateHS}{0.16}{SEOBNRv4TsurrogateLS}{0.02}{SEOBNRv4TsurrogatehighspinRIFT}{0.18}{SEOBNRv4TsurrogatelowspinRIFT}{0.02}{TEOBResumS-HS}{0.18}{TEOBResumS-LS}{0.02}{TaylorF2-HS}{0.18}{TaylorF2-LS}{0.02}{PrecessingSpinIMRTidalHS}{0.18}{PrecessingSpinIMRTidalLS}{0.02}{PublicationSamples}{0.18}}}
\newcommand{\spintwozfourtwofivemed}[1]{\IfEqCase{#1}{{AlignedSpinInspiralTidalHS}{0.04}{AlignedSpinInspiralTidalLS}{0.008}{AlignedSpinTidalHS}{0.03}{AlignedSpinTidalLS}{0.009}{IMRPhenomDNRTidal-HS}{0.03}{IMRPhenomDNRTidal-LS}{0.009}{IMRPhenomPv2NRTidal-HS}{0.03}{IMRPhenomPv2NRTidal-LS}{0.009}{SEOBNRv4TsurrogateHS}{0.02}{SEOBNRv4TsurrogateLS}{0.009}{SEOBNRv4TsurrogatehighspinRIFT}{0.03}{SEOBNRv4TsurrogatelowspinRIFT}{0.01}{TEOBResumS-HS}{0.03}{TEOBResumS-LS}{0.009}{TaylorF2-HS}{0.04}{TaylorF2-LS}{0.008}{PrecessingSpinIMRTidalHS}{0.03}{PrecessingSpinIMRTidalLS}{0.009}{PublicationSamples}{0.03}}}
\newcommand{\spintwozfourtwofiveplus}[1]{\IfEqCase{#1}{{AlignedSpinInspiralTidalHS}{0.30}{AlignedSpinInspiralTidalLS}{0.03}{AlignedSpinTidalHS}{0.26}{AlignedSpinTidalLS}{0.03}{IMRPhenomDNRTidal-HS}{0.37}{IMRPhenomDNRTidal-LS}{0.03}{IMRPhenomPv2NRTidal-HS}{0.30}{IMRPhenomPv2NRTidal-LS}{0.03}{SEOBNRv4TsurrogateHS}{0.20}{SEOBNRv4TsurrogateLS}{0.03}{SEOBNRv4TsurrogatehighspinRIFT}{0.21}{SEOBNRv4TsurrogatelowspinRIFT}{0.03}{TEOBResumS-HS}{0.21}{TEOBResumS-LS}{0.03}{TaylorF2-HS}{0.30}{TaylorF2-LS}{0.03}{PrecessingSpinIMRTidalHS}{0.30}{PrecessingSpinIMRTidalLS}{0.03}{PublicationSamples}{0.30}}}
\newcommand{\massonesourcefourtwofiveminus}[1]{\IfEqCase{#1}{{AlignedSpinInspiralTidalHS}{0.3}{AlignedSpinInspiralTidalLS}{0.10}{AlignedSpinTidalHS}{0.2}{AlignedSpinTidalLS}{0.09}{IMRPhenomDNRTidal-HS}{0.3}{IMRPhenomDNRTidal-LS}{0.09}{IMRPhenomPv2NRTidal-HS}{0.3}{IMRPhenomPv2NRTidal-LS}{0.09}{SEOBNRv4TsurrogateHS}{0.2}{SEOBNRv4TsurrogateLS}{0.09}{SEOBNRv4TsurrogatehighspinRIFT}{0.2}{SEOBNRv4TsurrogatelowspinRIFT}{0.09}{TEOBResumS-HS}{0.2}{TEOBResumS-LS}{0.09}{TaylorF2-HS}{0.3}{TaylorF2-LS}{0.10}{PrecessingSpinIMRTidalHS}{0.3}{PrecessingSpinIMRTidalLS}{0.09}{PublicationSamples}{0.3}}}
\newcommand{\massonesourcefourtwofivemed}[1]{\IfEqCase{#1}{{AlignedSpinInspiralTidalHS}{2.0}{AlignedSpinInspiralTidalLS}{1.75}{AlignedSpinTidalHS}{1.9}{AlignedSpinTidalLS}{1.75}{IMRPhenomDNRTidal-HS}{2.0}{IMRPhenomDNRTidal-LS}{1.75}{IMRPhenomPv2NRTidal-HS}{2.0}{IMRPhenomPv2NRTidal-LS}{1.74}{SEOBNRv4TsurrogateHS}{1.9}{SEOBNRv4TsurrogateLS}{1.74}{SEOBNRv4TsurrogatehighspinRIFT}{1.9}{SEOBNRv4TsurrogatelowspinRIFT}{1.75}{TEOBResumS-HS}{1.9}{TEOBResumS-LS}{1.75}{TaylorF2-HS}{2.0}{TaylorF2-LS}{1.75}{PrecessingSpinIMRTidalHS}{2.0}{PrecessingSpinIMRTidalLS}{1.74}{PublicationSamples}{2.0}}}
\newcommand{\massonesourcefourtwofiveplus}[1]{\IfEqCase{#1}{{AlignedSpinInspiralTidalHS}{0.5}{AlignedSpinInspiralTidalLS}{0.2}{AlignedSpinTidalHS}{0.6}{AlignedSpinTidalLS}{0.2}{IMRPhenomDNRTidal-HS}{0.7}{IMRPhenomDNRTidal-LS}{0.2}{IMRPhenomPv2NRTidal-HS}{0.6}{IMRPhenomPv2NRTidal-LS}{0.2}{SEOBNRv4TsurrogateHS}{0.5}{SEOBNRv4TsurrogateLS}{0.2}{SEOBNRv4TsurrogatehighspinRIFT}{0.5}{SEOBNRv4TsurrogatelowspinRIFT}{0.2}{TEOBResumS-HS}{0.5}{TEOBResumS-LS}{0.2}{TaylorF2-HS}{0.5}{TaylorF2-LS}{0.2}{PrecessingSpinIMRTidalHS}{0.6}{PrecessingSpinIMRTidalLS}{0.2}{PublicationSamples}{0.6}}}
\newcommand{\geocenttimefourtwofiveminus}[1]{\IfEqCase{#1}{{AlignedSpinInspiralTidalHS}{0.007}{AlignedSpinInspiralTidalLS}{0.008}{AlignedSpinTidalHS}{0.03}{AlignedSpinTidalLS}{0.02}{IMRPhenomDNRTidal-HS}{0.008}{IMRPhenomDNRTidal-LS}{0.01}{IMRPhenomPv2NRTidal-HS}{0.009}{IMRPhenomPv2NRTidal-LS}{0.01}{SEOBNRv4TsurrogateHS}{0.01}{SEOBNRv4TsurrogateLS}{0.008}{SEOBNRv4TsurrogatehighspinRIFT}{0.0}{SEOBNRv4TsurrogatelowspinRIFT}{0.0}{TEOBResumS-HS}{0.0}{TEOBResumS-LS}{0.0}{TaylorF2-HS}{0.007}{TaylorF2-LS}{0.008}{PrecessingSpinIMRTidalHS}{0.009}{PrecessingSpinIMRTidalLS}{0.01}{PublicationSamples}{0.009}}}
\newcommand{\geocenttimefourtwofivemed}[1]{\IfEqCase{#1}{{AlignedSpinInspiralTidalHS}{1240215503.0}{AlignedSpinInspiralTidalLS}{1240215503.0}{AlignedSpinTidalHS}{1240215503.0}{AlignedSpinTidalLS}{1240215503.0}{IMRPhenomDNRTidal-HS}{1240215503.0}{IMRPhenomDNRTidal-LS}{1240215503.0}{IMRPhenomPv2NRTidal-HS}{1240215503.0}{IMRPhenomPv2NRTidal-LS}{1240215503.0}{SEOBNRv4TsurrogateHS}{1240215503.0}{SEOBNRv4TsurrogateLS}{1240215503.0}{SEOBNRv4TsurrogatehighspinRIFT}{1240215503.0}{SEOBNRv4TsurrogatelowspinRIFT}{1240215503.0}{TEOBResumS-HS}{1240215503.0}{TEOBResumS-LS}{1240215503.0}{TaylorF2-HS}{1240215503.0}{TaylorF2-LS}{1240215503.0}{PrecessingSpinIMRTidalHS}{1240215503.0}{PrecessingSpinIMRTidalLS}{1240215503.0}{PublicationSamples}{1240215503.0}}}
\newcommand{\geocenttimefourtwofiveplus}[1]{\IfEqCase{#1}{{AlignedSpinInspiralTidalHS}{0.04}{AlignedSpinInspiralTidalLS}{0.03}{AlignedSpinTidalHS}{0.02}{AlignedSpinTidalLS}{0.02}{IMRPhenomDNRTidal-HS}{0.04}{IMRPhenomDNRTidal-LS}{0.03}{IMRPhenomPv2NRTidal-HS}{0.03}{IMRPhenomPv2NRTidal-LS}{0.03}{SEOBNRv4TsurrogateHS}{0.04}{SEOBNRv4TsurrogateLS}{0.04}{SEOBNRv4TsurrogatehighspinRIFT}{0.0}{SEOBNRv4TsurrogatelowspinRIFT}{0.0}{TEOBResumS-HS}{0.0}{TEOBResumS-LS}{0.0}{TaylorF2-HS}{0.04}{TaylorF2-LS}{0.03}{PrecessingSpinIMRTidalHS}{0.03}{PrecessingSpinIMRTidalLS}{0.03}{PublicationSamples}{0.03}}}
\newcommand{\costilttwofourtwofiveminus}[1]{\IfEqCase{#1}{{AlignedSpinInspiralTidalHS}{2.00}{AlignedSpinInspiralTidalLS}{2.00}{AlignedSpinTidalHS}{2.00}{AlignedSpinTidalLS}{2.00}{IMRPhenomDNRTidal-HS}{2.00}{IMRPhenomDNRTidal-LS}{2.00}{IMRPhenomPv2NRTidal-HS}{0.87}{IMRPhenomPv2NRTidal-LS}{1.13}{SEOBNRv4TsurrogateHS}{2.00}{SEOBNRv4TsurrogateLS}{2.00}{SEOBNRv4TsurrogatehighspinRIFT}{2.00}{SEOBNRv4TsurrogatelowspinRIFT}{2.00}{TEOBResumS-HS}{2.00}{TEOBResumS-LS}{2.00}{TaylorF2-HS}{2.00}{TaylorF2-LS}{2.00}{PrecessingSpinIMRTidalHS}{0.87}{PrecessingSpinIMRTidalLS}{1.12}{PublicationSamples}{0.86}}}
\newcommand{\costilttwofourtwofivemed}[1]{\IfEqCase{#1}{{AlignedSpinInspiralTidalHS}{1.00}{AlignedSpinInspiralTidalLS}{1.00}{AlignedSpinTidalHS}{1.00}{AlignedSpinTidalLS}{1.00}{IMRPhenomDNRTidal-HS}{1.00}{IMRPhenomDNRTidal-LS}{1.00}{IMRPhenomPv2NRTidal-HS}{0.16}{IMRPhenomPv2NRTidal-LS}{0.46}{SEOBNRv4TsurrogateHS}{1.00}{SEOBNRv4TsurrogateLS}{1.00}{SEOBNRv4TsurrogatehighspinRIFT}{1.00}{SEOBNRv4TsurrogatelowspinRIFT}{1.00}{TEOBResumS-HS}{1.00}{TEOBResumS-LS}{1.00}{TaylorF2-HS}{1.00}{TaylorF2-LS}{1.00}{PrecessingSpinIMRTidalHS}{0.16}{PrecessingSpinIMRTidalLS}{0.46}{PublicationSamples}{0.16}}}
\newcommand{\costilttwofourtwofiveplus}[1]{\IfEqCase{#1}{{AlignedSpinInspiralTidalHS}{0.00}{AlignedSpinInspiralTidalLS}{0.00}{AlignedSpinTidalHS}{0.00}{AlignedSpinTidalLS}{0.00}{IMRPhenomDNRTidal-HS}{0.00}{IMRPhenomDNRTidal-LS}{0.00}{IMRPhenomPv2NRTidal-HS}{0.70}{IMRPhenomPv2NRTidal-LS}{0.49}{SEOBNRv4TsurrogateHS}{0.00}{SEOBNRv4TsurrogateLS}{0.00}{SEOBNRv4TsurrogatehighspinRIFT}{0.00}{SEOBNRv4TsurrogatelowspinRIFT}{0.00}{TEOBResumS-HS}{0.00}{TEOBResumS-LS}{0.00}{TaylorF2-HS}{0.00}{TaylorF2-LS}{0.00}{PrecessingSpinIMRTidalHS}{0.70}{PrecessingSpinIMRTidalLS}{0.49}{PublicationSamples}{0.70}}}
\newcommand{\luminositydistancefourtwofiveminus}[1]{\IfEqCase{#1}{{AlignedSpinInspiralTidalHS}{0.08}{AlignedSpinInspiralTidalLS}{0.07}{AlignedSpinTidalHS}{0.07}{AlignedSpinTidalLS}{0.07}{IMRPhenomDNRTidal-HS}{0.07}{IMRPhenomDNRTidal-LS}{0.07}{IMRPhenomPv2NRTidal-HS}{0.07}{IMRPhenomPv2NRTidal-LS}{0.07}{SEOBNRv4TsurrogateHS}{0.07}{SEOBNRv4TsurrogateLS}{0.07}{SEOBNRv4TsurrogatehighspinRIFT}{0.07}{SEOBNRv4TsurrogatelowspinRIFT}{0.07}{TEOBResumS-HS}{0.07}{TEOBResumS-LS}{0.07}{TaylorF2-HS}{0.08}{TaylorF2-LS}{0.08}{PrecessingSpinIMRTidalHS}{0.07}{PrecessingSpinIMRTidalLS}{0.07}{PublicationSamples}{0.07}}}
\newcommand{\luminositydistancefourtwofivemed}[1]{\IfEqCase{#1}{{AlignedSpinInspiralTidalHS}{0.16}{AlignedSpinInspiralTidalLS}{0.16}{AlignedSpinTidalHS}{0.16}{AlignedSpinTidalLS}{0.16}{IMRPhenomDNRTidal-HS}{0.16}{IMRPhenomDNRTidal-LS}{0.16}{IMRPhenomPv2NRTidal-HS}{0.16}{IMRPhenomPv2NRTidal-LS}{0.16}{SEOBNRv4TsurrogateHS}{0.16}{SEOBNRv4TsurrogateLS}{0.16}{SEOBNRv4TsurrogatehighspinRIFT}{0.16}{SEOBNRv4TsurrogatelowspinRIFT}{0.16}{TEOBResumS-HS}{0.16}{TEOBResumS-LS}{0.16}{TaylorF2-HS}{0.16}{TaylorF2-LS}{0.16}{PrecessingSpinIMRTidalHS}{0.16}{PrecessingSpinIMRTidalLS}{0.16}{PublicationSamples}{0.16}}}
\newcommand{\luminositydistancefourtwofiveplus}[1]{\IfEqCase{#1}{{AlignedSpinInspiralTidalHS}{0.07}{AlignedSpinInspiralTidalLS}{0.07}{AlignedSpinTidalHS}{0.07}{AlignedSpinTidalLS}{0.07}{IMRPhenomDNRTidal-HS}{0.07}{IMRPhenomDNRTidal-LS}{0.07}{IMRPhenomPv2NRTidal-HS}{0.07}{IMRPhenomPv2NRTidal-LS}{0.07}{SEOBNRv4TsurrogateHS}{0.07}{SEOBNRv4TsurrogateLS}{0.07}{SEOBNRv4TsurrogatehighspinRIFT}{0.08}{SEOBNRv4TsurrogatelowspinRIFT}{0.07}{TEOBResumS-HS}{0.07}{TEOBResumS-LS}{0.07}{TaylorF2-HS}{0.07}{TaylorF2-LS}{0.07}{PrecessingSpinIMRTidalHS}{0.07}{PrecessingSpinIMRTidalLS}{0.07}{PublicationSamples}{0.07}}}
\newcommand{\spinonezfourtwofiveminus}[1]{\IfEqCase{#1}{{AlignedSpinInspiralTidalHS}{0.14}{AlignedSpinInspiralTidalLS}{0.02}{AlignedSpinTidalHS}{0.15}{AlignedSpinTidalLS}{0.02}{IMRPhenomDNRTidal-HS}{0.22}{IMRPhenomDNRTidal-LS}{0.02}{IMRPhenomPv2NRTidal-HS}{0.12}{IMRPhenomPv2NRTidal-LS}{0.02}{SEOBNRv4TsurrogateHS}{0.11}{SEOBNRv4TsurrogateLS}{0.02}{SEOBNRv4TsurrogatehighspinRIFT}{0.14}{SEOBNRv4TsurrogatelowspinRIFT}{0.02}{TEOBResumS-HS}{0.13}{TEOBResumS-LS}{0.02}{TaylorF2-HS}{0.14}{TaylorF2-LS}{0.02}{PrecessingSpinIMRTidalHS}{0.12}{PrecessingSpinIMRTidalLS}{0.02}{PublicationSamples}{0.12}}}
\newcommand{\spinonezfourtwofivemed}[1]{\IfEqCase{#1}{{AlignedSpinInspiralTidalHS}{0.04}{AlignedSpinInspiralTidalLS}{0.01}{AlignedSpinTidalHS}{0.04}{AlignedSpinTidalLS}{0.01}{IMRPhenomDNRTidal-HS}{0.06}{IMRPhenomDNRTidal-LS}{0.01}{IMRPhenomPv2NRTidal-HS}{0.06}{IMRPhenomPv2NRTidal-LS}{0.01}{SEOBNRv4TsurrogateHS}{0.04}{SEOBNRv4TsurrogateLS}{0.01}{SEOBNRv4TsurrogatehighspinRIFT}{0.04}{SEOBNRv4TsurrogatelowspinRIFT}{0.01}{TEOBResumS-HS}{0.04}{TEOBResumS-LS}{0.01}{TaylorF2-HS}{0.04}{TaylorF2-LS}{0.01}{PrecessingSpinIMRTidalHS}{0.06}{PrecessingSpinIMRTidalLS}{0.01}{PublicationSamples}{0.06}}}
\newcommand{\spinonezfourtwofiveplus}[1]{\IfEqCase{#1}{{AlignedSpinInspiralTidalHS}{0.19}{AlignedSpinInspiralTidalLS}{0.03}{AlignedSpinTidalHS}{0.20}{AlignedSpinTidalLS}{0.03}{IMRPhenomDNRTidal-HS}{0.26}{IMRPhenomDNRTidal-LS}{0.03}{IMRPhenomPv2NRTidal-HS}{0.18}{IMRPhenomPv2NRTidal-LS}{0.03}{SEOBNRv4TsurrogateHS}{0.16}{SEOBNRv4TsurrogateLS}{0.03}{SEOBNRv4TsurrogatehighspinRIFT}{0.16}{SEOBNRv4TsurrogatelowspinRIFT}{0.03}{TEOBResumS-HS}{0.16}{TEOBResumS-LS}{0.03}{TaylorF2-HS}{0.19}{TaylorF2-LS}{0.03}{PrecessingSpinIMRTidalHS}{0.18}{PrecessingSpinIMRTidalLS}{0.03}{PublicationSamples}{0.18}}}
\newcommand{\networkmatchedfiltersnrfourtwofiveminus}[1]{\IfEqCase{#1}{{AlignedSpinInspiralTidalHS}{0.4}{AlignedSpinInspiralTidalLS}{0.4}{IMRPhenomDNRTidal-HS}{0.4}{IMRPhenomDNRTidal-LS}{0.4}{IMRPhenomPv2NRTidal-HS}{0.4}{IMRPhenomPv2NRTidal-LS}{0.4}{SEOBNRv4TsurrogateHS}{0.4}{SEOBNRv4TsurrogateLS}{0.4}{TaylorF2-HS}{0.4}{TaylorF2-LS}{0.4}{PrecessingSpinIMRTidalHS}{0.4}{PrecessingSpinIMRTidalLS}{0.4}{PublicationSamples}{0.4}}}
\newcommand{\networkmatchedfiltersnrfourtwofivemed}[1]{\IfEqCase{#1}{{AlignedSpinInspiralTidalHS}{12.4}{AlignedSpinInspiralTidalLS}{12.5}{IMRPhenomDNRTidal-HS}{12.3}{IMRPhenomDNRTidal-LS}{12.4}{IMRPhenomPv2NRTidal-HS}{12.4}{IMRPhenomPv2NRTidal-LS}{12.5}{SEOBNRv4TsurrogateHS}{12.4}{SEOBNRv4TsurrogateLS}{12.4}{TaylorF2-HS}{12.4}{TaylorF2-LS}{12.5}{PrecessingSpinIMRTidalHS}{12.4}{PrecessingSpinIMRTidalLS}{12.5}{PublicationSamples}{12.4}}}
\newcommand{\networkmatchedfiltersnrfourtwofiveplus}[1]{\IfEqCase{#1}{{AlignedSpinInspiralTidalHS}{0.3}{AlignedSpinInspiralTidalLS}{0.2}{IMRPhenomDNRTidal-HS}{0.3}{IMRPhenomDNRTidal-LS}{0.3}{IMRPhenomPv2NRTidal-HS}{0.3}{IMRPhenomPv2NRTidal-LS}{0.3}{SEOBNRv4TsurrogateHS}{0.3}{SEOBNRv4TsurrogateLS}{0.3}{TaylorF2-HS}{0.3}{TaylorF2-LS}{0.2}{PrecessingSpinIMRTidalHS}{0.3}{PrecessingSpinIMRTidalLS}{0.3}{PublicationSamples}{0.3}}}
\newcommand{\chirpmasssourcefourtwofiveminus}[1]{\IfEqCase{#1}{{AlignedSpinInspiralTidalHS}{0.02}{AlignedSpinInspiralTidalLS}{0.02}{AlignedSpinTidalHS}{0.02}{AlignedSpinTidalLS}{0.02}{IMRPhenomDNRTidal-HS}{0.02}{IMRPhenomDNRTidal-LS}{0.02}{IMRPhenomPv2NRTidal-HS}{0.02}{IMRPhenomPv2NRTidal-LS}{0.02}{SEOBNRv4TsurrogateHS}{0.02}{SEOBNRv4TsurrogateLS}{0.02}{SEOBNRv4TsurrogatehighspinRIFT}{0.02}{SEOBNRv4TsurrogatelowspinRIFT}{0.02}{TEOBResumS-HS}{0.02}{TEOBResumS-LS}{0.02}{TaylorF2-HS}{0.02}{TaylorF2-LS}{0.02}{PrecessingSpinIMRTidalHS}{0.02}{PrecessingSpinIMRTidalLS}{0.02}{PublicationSamples}{0.02}}}
\newcommand{\chirpmasssourcefourtwofivemed}[1]{\IfEqCase{#1}{{AlignedSpinInspiralTidalHS}{1.44}{AlignedSpinInspiralTidalLS}{1.44}{AlignedSpinTidalHS}{1.44}{AlignedSpinTidalLS}{1.44}{IMRPhenomDNRTidal-HS}{1.44}{IMRPhenomDNRTidal-LS}{1.44}{IMRPhenomPv2NRTidal-HS}{1.44}{IMRPhenomPv2NRTidal-LS}{1.44}{SEOBNRv4TsurrogateHS}{1.44}{SEOBNRv4TsurrogateLS}{1.44}{SEOBNRv4TsurrogatehighspinRIFT}{1.44}{SEOBNRv4TsurrogatelowspinRIFT}{1.44}{TEOBResumS-HS}{1.44}{TEOBResumS-LS}{1.44}{TaylorF2-HS}{1.44}{TaylorF2-LS}{1.44}{PrecessingSpinIMRTidalHS}{1.44}{PrecessingSpinIMRTidalLS}{1.44}{PublicationSamples}{1.44}}}
\newcommand{\chirpmasssourcefourtwofiveplus}[1]{\IfEqCase{#1}{{AlignedSpinInspiralTidalHS}{0.02}{AlignedSpinInspiralTidalLS}{0.02}{AlignedSpinTidalHS}{0.02}{AlignedSpinTidalLS}{0.02}{IMRPhenomDNRTidal-HS}{0.02}{IMRPhenomDNRTidal-LS}{0.02}{IMRPhenomPv2NRTidal-HS}{0.02}{IMRPhenomPv2NRTidal-LS}{0.02}{SEOBNRv4TsurrogateHS}{0.02}{SEOBNRv4TsurrogateLS}{0.02}{SEOBNRv4TsurrogatehighspinRIFT}{0.02}{SEOBNRv4TsurrogatelowspinRIFT}{0.02}{TEOBResumS-HS}{0.02}{TEOBResumS-LS}{0.02}{TaylorF2-HS}{0.02}{TaylorF2-LS}{0.02}{PrecessingSpinIMRTidalHS}{0.02}{PrecessingSpinIMRTidalLS}{0.02}{PublicationSamples}{0.02}}}
\newcommand{\phionefourtwofiveminus}[1]{\IfEqCase{#1}{{AlignedSpinInspiralTidalHS}{0.00}{AlignedSpinInspiralTidalLS}{0.00}{AlignedSpinTidalHS}{0.00}{AlignedSpinTidalLS}{0.00}{IMRPhenomDNRTidal-HS}{0.00}{IMRPhenomDNRTidal-LS}{0.00}{IMRPhenomPv2NRTidal-HS}{2.73}{IMRPhenomPv2NRTidal-LS}{2.85}{SEOBNRv4TsurrogateHS}{0.00}{SEOBNRv4TsurrogateLS}{0.00}{SEOBNRv4TsurrogatehighspinRIFT}{0.00}{SEOBNRv4TsurrogatelowspinRIFT}{0.00}{TEOBResumS-HS}{0.00}{TEOBResumS-LS}{0.00}{TaylorF2-HS}{0.00}{TaylorF2-LS}{0.00}{PrecessingSpinIMRTidalHS}{2.73}{PrecessingSpinIMRTidalLS}{2.85}{PublicationSamples}{2.73}}}
\newcommand{\phionefourtwofivemed}[1]{\IfEqCase{#1}{{AlignedSpinInspiralTidalHS}{0.00}{AlignedSpinInspiralTidalLS}{0.00}{AlignedSpinTidalHS}{0.00}{AlignedSpinTidalLS}{0.00}{IMRPhenomDNRTidal-HS}{0.00}{IMRPhenomDNRTidal-LS}{0.00}{IMRPhenomPv2NRTidal-HS}{3.05}{IMRPhenomPv2NRTidal-LS}{3.15}{SEOBNRv4TsurrogateHS}{0.00}{SEOBNRv4TsurrogateLS}{0.00}{SEOBNRv4TsurrogatehighspinRIFT}{0.00}{SEOBNRv4TsurrogatelowspinRIFT}{0.00}{TEOBResumS-HS}{0.00}{TEOBResumS-LS}{0.00}{TaylorF2-HS}{0.00}{TaylorF2-LS}{0.00}{PrecessingSpinIMRTidalHS}{3.05}{PrecessingSpinIMRTidalLS}{3.15}{PublicationSamples}{3.06}}}
\newcommand{\phionefourtwofiveplus}[1]{\IfEqCase{#1}{{AlignedSpinInspiralTidalHS}{0.00}{AlignedSpinInspiralTidalLS}{0.00}{AlignedSpinTidalHS}{0.00}{AlignedSpinTidalLS}{0.00}{IMRPhenomDNRTidal-HS}{0.00}{IMRPhenomDNRTidal-LS}{0.00}{IMRPhenomPv2NRTidal-HS}{2.90}{IMRPhenomPv2NRTidal-LS}{2.83}{SEOBNRv4TsurrogateHS}{0.00}{SEOBNRv4TsurrogateLS}{0.00}{SEOBNRv4TsurrogatehighspinRIFT}{0.00}{SEOBNRv4TsurrogatelowspinRIFT}{0.00}{TEOBResumS-HS}{0.00}{TEOBResumS-LS}{0.00}{TaylorF2-HS}{0.00}{TaylorF2-LS}{0.00}{PrecessingSpinIMRTidalHS}{2.90}{PrecessingSpinIMRTidalLS}{2.83}{PublicationSamples}{2.90}}}
\newcommand{\symmetricmassratiofourtwofiveminus}[1]{\IfEqCase{#1}{{AlignedSpinInspiralTidalHS}{0.03}{AlignedSpinInspiralTidalLS}{0.005}{AlignedSpinTidalHS}{0.03}{AlignedSpinTidalLS}{0.005}{IMRPhenomDNRTidal-HS}{0.04}{IMRPhenomDNRTidal-LS}{0.005}{IMRPhenomPv2NRTidal-HS}{0.03}{IMRPhenomPv2NRTidal-LS}{0.005}{SEOBNRv4TsurrogateHS}{0.03}{SEOBNRv4TsurrogateLS}{0.004}{SEOBNRv4TsurrogatehighspinRIFT}{0.02}{SEOBNRv4TsurrogatelowspinRIFT}{0.005}{TEOBResumS-HS}{0.03}{TEOBResumS-LS}{0.005}{TaylorF2-HS}{0.03}{TaylorF2-LS}{0.005}{PrecessingSpinIMRTidalHS}{0.03}{PrecessingSpinIMRTidalLS}{0.005}{PublicationSamples}{0.03}}}
\newcommand{\symmetricmassratiofourtwofivemed}[1]{\IfEqCase{#1}{{AlignedSpinInspiralTidalHS}{0.242}{AlignedSpinInspiralTidalLS}{0.249}{AlignedSpinTidalHS}{0.245}{AlignedSpinTidalLS}{0.249}{IMRPhenomDNRTidal-HS}{0.243}{IMRPhenomDNRTidal-LS}{0.249}{IMRPhenomPv2NRTidal-HS}{0.240}{IMRPhenomPv2NRTidal-LS}{0.249}{SEOBNRv4TsurrogateHS}{0.246}{SEOBNRv4TsurrogateLS}{0.249}{SEOBNRv4TsurrogatehighspinRIFT}{0.246}{SEOBNRv4TsurrogatelowspinRIFT}{0.249}{TEOBResumS-HS}{0.245}{TEOBResumS-LS}{0.249}{TaylorF2-HS}{0.242}{TaylorF2-LS}{0.249}{PrecessingSpinIMRTidalHS}{0.240}{PrecessingSpinIMRTidalLS}{0.249}{PublicationSamples}{0.240}}}
\newcommand{\symmetricmassratiofourtwofiveplus}[1]{\IfEqCase{#1}{{AlignedSpinInspiralTidalHS}{0.007}{AlignedSpinInspiralTidalLS}{0.0008}{AlignedSpinTidalHS}{0.005}{AlignedSpinTidalLS}{0.0008}{IMRPhenomDNRTidal-HS}{0.007}{IMRPhenomDNRTidal-LS}{0.0008}{IMRPhenomPv2NRTidal-HS}{0.010}{IMRPhenomPv2NRTidal-LS}{0.0007}{SEOBNRv4TsurrogateHS}{0.004}{SEOBNRv4TsurrogateLS}{0.0007}{SEOBNRv4TsurrogatehighspinRIFT}{0.004}{SEOBNRv4TsurrogatelowspinRIFT}{0.0008}{TEOBResumS-HS}{0.005}{TEOBResumS-LS}{0.0009}{TaylorF2-HS}{0.007}{TaylorF2-LS}{0.0008}{PrecessingSpinIMRTidalHS}{0.010}{PrecessingSpinIMRTidalLS}{0.0007}{PublicationSamples}{0.010}}}

\newcommand{\dMf}{\ensuremath{\Delta M_{\rm f} / \bar{M}_{\rm f}}}
\newcommand{\dchif}{\ensuremath{\Delta \chi_{\rm f} / \bar{\chi}_{\rm f}}}
\newcommand{\QGR}{\ensuremath{\mathcal{Q}_{\rm GR}^{\rm{2D}}}}

\acrodef{MDC}[MDC]{mock data challenge}
\acrodef{MECO}[MECO]{minimum energy circular orbit}
\acrodef{QNM}[QNM]{quasi-normal mode}

\begin{document}

\newcommand{\TGRINUMTESTS}{\reviewed{four}\xspace}

\newcommand{\TGRImrctEVENTSTATS}[1]{\IfEqCase{#1}{{S200225qDMFGWTC4PHENOM}{\reviewed{\ensuremath{0.09^{+0.43}_{-0.59}}}}{S200225qDCHIFGWTC4PHENOM}{\reviewed{\ensuremath{0.05^{+0.61}_{-0.38}}}}{S200225qGRQUANTGWTC4}{\reviewed{\ensuremath{1.3}}}{S230811nDMFGWTC4PHENOM}{\reviewed{\ensuremath{0.01^{+0.39}_{-0.31}}}}{S230811nDCHIFGWTC4PHENOM}{\reviewed{\ensuremath{-0.13^{+0.55}_{-0.35}}}}{S230811nGRQUANTGWTC4}{\reviewed{\ensuremath{14.7}}}{GW170814DMFGWTC4PHENOM}{\reviewed{\ensuremath{0.50^{+0.68}_{-0.70}}}}{GW170814DCHIFGWTC4PHENOM}{\reviewed{\ensuremath{-0.01^{+0.55}_{-0.24}}}}{GW170814GRQUANTGWTC4}{\reviewed{\ensuremath{9.9}}}{S231206ccDMFGWTC4PHENOM}{\reviewed{\ensuremath{-0.02^{+0.23}_{-0.19}}}}{S231206ccDCHIFGWTC4PHENOM}{\reviewed{\ensuremath{0.01^{+0.31}_{-0.27}}}}{S231206ccGRQUANTGWTC4}{\reviewed{\ensuremath{24.5}}}{S231226avDMFGWTC4PHENOM}{\reviewed{\ensuremath{0.03^{+0.16}_{-0.15}}}}{S231226avDCHIFGWTC4PHENOM}{\reviewed{\ensuremath{0.00^{+0.24}_{-0.24}}}}{S231226avGRQUANTGWTC4}{\reviewed{\ensuremath{14.8}}}{S231206caDMFGWTC4PHENOM}{\reviewed{\ensuremath{0.51^{+0.46}_{-0.52}}}}{S231206caDCHIFGWTC4PHENOM}{\reviewed{\ensuremath{-0.14^{+0.47}_{-0.72}}}}{S231206caGRQUANTGWTC4}{\reviewed{\ensuremath{88.6}}}{S230628axDMFGWTC4PHENOM}{\reviewed{\ensuremath{0.82^{+0.19}_{-0.73}}}}{S230628axDCHIFGWTC4PHENOM}{\reviewed{\ensuremath{0.16^{+0.33}_{-0.37}}}}{S230628axGRQUANTGWTC4}{\reviewed{\ensuremath{90.1}}}{S190814bvDMFGWTC4PHENOM}{\reviewed{\ensuremath{-0.34^{+0.48}_{-1.11}}}}{S190814bvDCHIFGWTC4PHENOM}{\reviewed{\ensuremath{-0.91^{+0.27}_{-0.19}}}}{S190814bvGRQUANTGWTC4}{\reviewed{\ensuremath{99.9}}}{S190828jDMFGWTC4PHENOM}{\reviewed{\ensuremath{0.03^{+0.36}_{-0.25}}}}{S190828jDCHIFGWTC4PHENOM}{\reviewed{\ensuremath{-0.07^{+0.44}_{-0.34}}}}{S190828jGRQUANTGWTC4}{\reviewed{\ensuremath{21.0}}}{S190408anDMFGWTC4PHENOM}{\reviewed{\ensuremath{0.02^{+0.40}_{-0.27}}}}{S190408anDCHIFGWTC4PHENOM}{\reviewed{\ensuremath{-0.01^{+0.48}_{-0.46}}}}{S190408anGRQUANTGWTC4}{\reviewed{\ensuremath{11.5}}}{GW170809DMFGWTC4PHENOM}{\reviewed{\ensuremath{-0.05^{+0.38}_{-0.29}}}}{GW170809DCHIFGWTC4PHENOM}{\reviewed{\ensuremath{-0.19^{+0.51}_{-0.46}}}}{GW170809GRQUANTGWTC4}{\reviewed{\ensuremath{24.7}}}{S190513bmDMFGWTC4PHENOM}{\reviewed{\ensuremath{-0.03^{+0.33}_{-0.27}}}}{S190513bmDCHIFGWTC4PHENOM}{\reviewed{\ensuremath{-0.21^{+0.56}_{-0.42}}}}{S190513bmGRQUANTGWTC4}{\reviewed{\ensuremath{34.6}}}{S190503bfDMFGWTC4PHENOM}{\reviewed{\ensuremath{0.90^{+0.29}_{-0.59}}}}{S190503bfDCHIFGWTC4PHENOM}{\reviewed{\ensuremath{0.40^{+0.89}_{-0.30}}}}{S190503bfGRQUANTGWTC4}{\reviewed{\ensuremath{94.3}}}{S200311bgDMFGWTC4PHENOM}{\reviewed{\ensuremath{0.06^{+0.38}_{-0.25}}}}{S200311bgDCHIFGWTC4PHENOM}{\reviewed{\ensuremath{-0.24^{+0.49}_{-0.63}}}}{S200311bgGRQUANTGWTC4}{\reviewed{\ensuremath{15.2}}}{S230920alDMFGWTC4PHENOM}{\reviewed{\ensuremath{-0.01^{+1.54}_{-1.14}}}}{S230920alDCHIFGWTC4PHENOM}{\reviewed{\ensuremath{0.39^{+1.31}_{-0.79}}}}{S230920alGRQUANTGWTC4}{\reviewed{\ensuremath{59.9}}}{S230919bjDMFGWTC4PHENOM}{\reviewed{\ensuremath{0.11^{+0.36}_{-0.28}}}}{S230919bjDCHIFGWTC4PHENOM}{\reviewed{\ensuremath{0.10^{+0.67}_{-0.28}}}}{S230919bjGRQUANTGWTC4}{\reviewed{\ensuremath{7.6}}}{S200224caDMFGWTC4PHENOM}{\reviewed{\ensuremath{0.30^{+0.62}_{-0.37}}}}{S200224caDCHIFGWTC4PHENOM}{\reviewed{\ensuremath{0.18^{+0.30}_{-0.33}}}}{S200224caGRQUANTGWTC4}{\reviewed{\ensuremath{20.7}}}{S230606dDMFGWTC4PHENOM}{\reviewed{\ensuremath{0.40^{+0.54}_{-0.81}}}}{S230606dDCHIFGWTC4PHENOM}{\reviewed{\ensuremath{0.34^{+0.85}_{-0.68}}}}{S230606dGRQUANTGWTC4}{\reviewed{\ensuremath{38.1}}}{GW170104DMFGWTC4PHENOM}{\reviewed{\ensuremath{-0.08^{+0.34}_{-0.23}}}}{GW170104DCHIFGWTC4PHENOM}{\reviewed{\ensuremath{-0.08^{+0.41}_{-0.34}}}}{GW170104GRQUANTGWTC4}{\reviewed{\ensuremath{28.9}}}{GW170823DMFGWTC4PHENOM}{\reviewed{\ensuremath{0.82^{+0.21}_{-0.38}}}}{GW170823DCHIFGWTC4PHENOM}{\reviewed{\ensuremath{0.20^{+0.63}_{-0.43}}}}{GW170823GRQUANTGWTC4}{\reviewed{\ensuremath{95.1}}}{S230922gDMFGWTC4PHENOM}{\reviewed{\ensuremath{0.21^{+0.38}_{-0.33}}}}{S230922gDCHIFGWTC4PHENOM}{\reviewed{\ensuremath{0.16^{+0.95}_{-0.44}}}}{S230922gGRQUANTGWTC4}{\reviewed{\ensuremath{38.3}}}{S190630agDMFGWTC4PHENOM}{\reviewed{\ensuremath{-0.13^{+0.27}_{-0.20}}}}{S190630agDCHIFGWTC4PHENOM}{\reviewed{\ensuremath{-0.15^{+0.45}_{-0.28}}}}{S190630agGRQUANTGWTC4}{\reviewed{\ensuremath{58.5}}}{S231108uDMFGWTC4PHENOM}{\reviewed{\ensuremath{0.20^{+0.42}_{-0.34}}}}{S231108uDCHIFGWTC4PHENOM}{\reviewed{\ensuremath{0.02^{+0.67}_{-0.30}}}}{S231108uGRQUANTGWTC4}{\reviewed{\ensuremath{32.0}}}{GW150914DMFGWTC4PHENOM}{\reviewed{\ensuremath{0.22^{+0.32}_{-0.22}}}}{GW150914DCHIFGWTC4PHENOM}{\reviewed{\ensuremath{0.16^{+0.61}_{-0.41}}}}{GW150914GRQUANTGWTC4}{\reviewed{\ensuremath{54.3}}}{S230924anDMFGWTC4PHENOM}{\reviewed{\ensuremath{0.06^{+0.33}_{-0.24}}}}{S230924anDCHIFGWTC4PHENOM}{\reviewed{\ensuremath{-0.02^{+0.41}_{-0.31}}}}{S230924anGRQUANTGWTC4}{\reviewed{\ensuremath{9.9}}}{S230927beDMFGWTC4PHENOM}{\reviewed{\ensuremath{-0.04^{+0.19}_{-0.16}}}}{S230927beDCHIFGWTC4PHENOM}{\reviewed{\ensuremath{-0.13^{+0.27}_{-0.17}}}}{S230927beGRQUANTGWTC4}{\reviewed{\ensuremath{37.0}}}{S200208qDMFGWTC4PHENOM}{\reviewed{\ensuremath{0.17^{+1.46}_{-0.50}}}}{S200208qDCHIFGWTC4PHENOM}{\reviewed{\ensuremath{0.03^{+1.66}_{-0.66}}}}{S200208qGRQUANTGWTC4}{\reviewed{\ensuremath{10.5}}}{GW170818DMFGWTC4PHENOM}{\reviewed{\ensuremath{0.18^{+0.49}_{-0.32}}}}{GW170818DCHIFGWTC4PHENOM}{\reviewed{\ensuremath{-0.03^{+0.60}_{-0.47}}}}{GW170818GRQUANTGWTC4}{\reviewed{\ensuremath{24.5}}}{S200129mDMFGWTC4PHENOM}{\reviewed{\ensuremath{0.04^{+0.26}_{-0.19}}}}{S200129mDCHIFGWTC4PHENOM}{\reviewed{\ensuremath{-0.03^{+0.36}_{-0.30}}}}{S200129mGRQUANTGWTC4}{\reviewed{\ensuremath{1.5}}}{S190521rDMFGWTC4PHENOM}{\reviewed{\ensuremath{0.10^{+0.36}_{-0.22}}}}{S190521rDCHIFGWTC4PHENOM}{\reviewed{\ensuremath{0.07^{+0.30}_{-0.28}}}}{S190521rGRQUANTGWTC4}{\reviewed{\ensuremath{0.4}}}}}

\newcommand{\EVENTSELECTION}[1]{\IfEqCase{#1}{{S230606dFCIMR}{\reviewed{\ensuremath{96}}}{S230606dOPTSNRPREIMR}{\reviewed{\ensuremath{7.4}}}{S230606dOPTSNRPOSTIMR}{\reviewed{\ensuremath{8.0}}}{S230606dOPTSNR}{\reviewed{\ensuremath{10.9}}}{S230609uFCIMR}{\reviewed{\ensuremath{94}}}{S230609uOPTSNRPREIMR}{\reviewed{\ensuremath{6.0}}}{S230609uOPTSNRPOSTIMR}{\reviewed{\ensuremath{7.4}}}{S230609uOPTSNR}{\reviewed{\ensuremath{9.5}}}{S230628axFCIMR}{\reviewed{\ensuremath{118}}}{S230628axOPTSNRPREIMR}{\reviewed{\ensuremath{10.9}}}{S230628axOPTSNRPOSTIMR}{\reviewed{\ensuremath{9.5}}}{S230628axOPTSNR}{\reviewed{\ensuremath{14.4}}}{S230811nFCIMR}{\reviewed{\ensuremath{121}}}{S230811nOPTSNRPREIMR}{\reviewed{\ensuremath{12.4}}}{S230811nOPTSNRPOSTIMR}{\reviewed{\ensuremath{7.7}}}{S230811nOPTSNR}{\reviewed{\ensuremath{14.6}}}{S230919bjFCIMR}{\reviewed{\ensuremath{174}}}{S230919bjOPTSNRPREIMR}{\reviewed{\ensuremath{14.2}}}{S230919bjOPTSNRPOSTIMR}{\reviewed{\ensuremath{7.6}}}{S230919bjOPTSNR}{\reviewed{\ensuremath{16.1}}}{S230920alFCIMR}{\reviewed{\ensuremath{115}}}{S230920alOPTSNRPREIMR}{\reviewed{\ensuremath{8.5}}}{S230920alOPTSNRPOSTIMR}{\reviewed{\ensuremath{7.2}}}{S230920alOPTSNR}{\reviewed{\ensuremath{11.1}}}{S230922gFCIMR}{\reviewed{\ensuremath{111}}}{S230922gOPTSNRPREIMR}{\reviewed{\ensuremath{8.6}}}{S230922gOPTSNRPOSTIMR}{\reviewed{\ensuremath{8.0}}}{S230922gOPTSNR}{\reviewed{\ensuremath{11.7}}}{S230924anFCIMR}{\reviewed{\ensuremath{133}}}{S230924anOPTSNRPREIMR}{\reviewed{\ensuremath{9.9}}}{S230924anOPTSNRPOSTIMR}{\reviewed{\ensuremath{7.1}}}{S230924anOPTSNR}{\reviewed{\ensuremath{12.1}}}{S230927beFCIMR}{\reviewed{\ensuremath{207}}}{S230927beOPTSNRPREIMR}{\reviewed{\ensuremath{17.7}}}{S230927beOPTSNRPOSTIMR}{\reviewed{\ensuremath{9.4}}}{S230927beOPTSNR}{\reviewed{\ensuremath{20.0}}}{S231108uFCIMR}{\reviewed{\ensuremath{166}}}{S231108uOPTSNRPREIMR}{\reviewed{\ensuremath{11.3}}}{S231108uOPTSNRPOSTIMR}{\reviewed{\ensuremath{6.9}}}{S231108uOPTSNR}{\reviewed{\ensuremath{13.2}}}{S231206caFCIMR}{\reviewed{\ensuremath{96}}}{S231206caOPTSNRPREIMR}{\reviewed{\ensuremath{8.4}}}{S231206caOPTSNRPOSTIMR}{\reviewed{\ensuremath{8.2}}}{S231206caOPTSNR}{\reviewed{\ensuremath{11.7}}}{S231206ccFCIMR}{\reviewed{\ensuremath{113}}}{S231206ccOPTSNRPREIMR}{\reviewed{\ensuremath{16.5}}}{S231206ccOPTSNRPOSTIMR}{\reviewed{\ensuremath{14.4}}}{S231206ccOPTSNR}{\reviewed{\ensuremath{21.9}}}{S231226avFCIMR}{\reviewed{\ensuremath{102}}}{S231226avOPTSNRPREIMR}{\reviewed{\ensuremath{26.7}}}{S231226avOPTSNRPOSTIMR}{\reviewed{\ensuremath{21.8}}}{S231226avOPTSNR}{\reviewed{\ensuremath{34.5}}}}}

\newcommand{\gwOhEightFourteenSMAMedian}{\reviewed{-0.21}}
\newcommand{\gwOhEightFourteenSMAUpperError}{\reviewed{1.82}}
\newcommand{\gwOhEightFourteenSMALowerError}{\reviewed{3.39}}

\newcommand{\gwOhFourTwelveSMAMedian}{\reviewed{0.53}}
\newcommand{\gwOhFourTwelveSMAUpperError}{\reviewed{3.43}}
\newcommand{\gwOhFourTwelveSMALowerError}{\reviewed{5.85}}

\title{\paperscommonname I. Overview and General Tests}

\iftoggle{endauthorlist}{
 \let\mymaketitle\maketitle
 \let\myauthor\author
 \let\myaffiliation\affiliation
 \author{\LVKCollabAuthors}
 {\def\thefootnote{}\footnotetext{\LVKCorrespondence}}
 \email{~~~~~~lvc.publications@ligo.org}
}{
 \iftoggle{fullauthorlist}{
 %% LVK authorlist in AAS format

%\collaboration{3000}{\LVKCollabAuthors}
%\affiliation{LIGO Hanford Observatory, Richland, WA 99352, USA}
%\affiliation{LIGO Livingston Observatory, Livingston, LA 70754, USA}
%\affiliation{Institute for Cosmic Ray Research, KAGRA Observatory, The University of Tokyo, 238 Higashi-Mozumi, Kamioka-cho, Hida City, Gifu 506-1205, Japan}

\author[0000-0003-4786-2698]{A.~G.~Abac}
\affiliation{Max Planck Institute for Gravitational Physics (Albert Einstein Institute), D-14476 Potsdam, Germany}
\author{I.~Abouelfettouh}
\affiliation{LIGO Hanford Observatory, Richland, WA 99352, USA}
\author{F.~Acernese}
\affiliation{Dipartimento di Farmacia, Universit\`a di Salerno, I-84084 Fisciano, Salerno, Italy}
\affiliation{INFN, Sezione di Napoli, I-80126 Napoli, Italy}
\author[0000-0002-8648-0767]{K.~Ackley}
\affiliation{University of Warwick, Coventry CV4 7AL, United Kingdom}
\author[0000-0001-5525-6255]{C.~Adamcewicz}
\affiliation{OzGrav, School of Physics \& Astronomy, Monash University, Clayton 3800, Victoria, Australia}
\author[0009-0004-2101-5428]{S.~Adhicary}
\affiliation{The Pennsylvania State University, University Park, PA 16802, USA}
\author{D.~Adhikari}
\affiliation{Max Planck Institute for Gravitational Physics (Albert Einstein Institute), D-30167 Hannover, Germany}
\affiliation{Leibniz Universit\"{a}t Hannover, D-30167 Hannover, Germany}
\author[0000-0002-4559-8427]{N.~Adhikari}
\affiliation{University of Wisconsin-Milwaukee, Milwaukee, WI 53201, USA}
\author[0000-0002-5731-5076]{R.~X.~Adhikari}
\affiliation{LIGO Laboratory, California Institute of Technology, Pasadena, CA 91125, USA}
\author{V.~K.~Adkins}
\affiliation{Louisiana State University, Baton Rouge, LA 70803, USA}
% ...
 }{
 \author{\LVKCollabAuthors}
}
}

\date[\relax]{Compiled: \today}

\begin{abstract}

The worldwide LIGO--Virgo--KAGRA network of gravitational-wave (GW) detectors
    continues to increase in sensitivity, thus increasing the quantity and quality of the detected GW signals from compact binary coalescences.
These signals allow us to perform ever-more sensitive tests of general relativity (GR) in the dynamical and strong-field regime of gravity.
This paper is the first of three,
    where we present the results of a suite of tests of GR using the binary signals included in
    the fourth GW Transient Catalog (GWTC-4.0), i.e., up to and including the first part of the fourth observing run of the detectors (O4a).
We restrict our analysis to the \TGRNUMEVENTSPREVPLUSOFOURA confident signals,
    henceforth called \emph{events},
    that were measured by at least two detectors,
    and have false alarm rates \TGRFARTHRESH.
These include \TGRNUMEVENTS events from O4a.
This first paper presents an overview of the methods, selection of events and GR tests, and serves as a guidemap for all three papers.
Here we focus on the \TGRINUMTESTS general tests of consistency, where we find no evidence for deviations from our models.
Specifically, for all the events considered, we find consistency of the residuals with noise.
    The final mass and final spin as inferred from the low- and high-frequency parts of the waveform are consistent with each other.
    We also find no evidence for deviations from the GR predictions for the amplitudes of subdominant GW multipole moments,
    or for non-GR modes of polarization.
We thus find that GR, without new physics beyond it, is still consistent with these GW events.
The results of the two additional papers in this trio also find overall consistency with vacuum GR,
    with more than 90\% of the events being consistent with GR at the 90\% credible level.
While one of the ringdown analyses finds the GR value in the tails for its combined results, this may be due in part to catalog variance.

\end{abstract}

\subpapersection{Introduction}\label{sec:intro}

\acused{BBH} \acused{BNS} \acused{NSBH}

The past decade has seen a surge of precision measurements of \acp{BH} and \acp{NS}.
For \acp{GW}, this started with the direct observation of \acp{GW} from the merging of \aclp{BBH} \citep[BBHs;][]{GW150914_paper},
    \aclp{BNS} \citep[BNSs;][]{GW170817_DetectionPaper}, and \aclp{NSBH} \citep[NSBHs;][]{LIGOScientific:2021qlt}. These observations have enabled
    a wide variety of new tests of Einstein's theory of \ac{GR}, starting with \citet{LIGOScientific:2016lio} for \acp{BBH}, \citet{LIGOScientific:2018dkp}
    for \acp{BNS}, and \citet{LIGOScientific:2021sio} for \acp{NSBH}.

Precision electromagnetic
    observations of \acp{BH} and \acp{NS} include measurements of short-period stars around Sgr~A${}^*$, which provide constraints on relativistic precession, redshift, and a putative fifth force \citep{Do:2019txf, GRAVITY:2020gka, GRAVITY:2025ahf}.
    Additionally, observations of \ac{BH} shadows \citep{Akiyama:2019cqa, EventHorizonTelescope:2022wkp} have enabled tests of the Kerr metric \citep{EventHorizonTelescope:2019ggy, EventHorizonTelescope:2022xqj}, while
    measurements of isolated \ac{NS} masses and radii \citep{Bogdanov:2019ixe,Miller:2019cac,Riley:2019yda} provide constraints on parity violation in gravity \citep{Silva:2020acr}.
These tests complement standard laboratory and astrophysical tests of \ac{GR}~\citep{Will:2014kxa,Berti:2015itd}, including tests using the double pulsar \citep{Kramer:2021jcw}
    and cosmological observations \citep[reviewed in, e.g.,][]{Ferreira:2019xrr,Ishak:2018his}.

Among these precision measurements, the first three observing runs of the \ac{GW} observatories
    Advanced LIGO~\citep{TheLIGOScientific:2014jea} and Advanced Virgo~\citep{TheVirgo:2014hva} have provided new tests of \ac{GR}
    across regimes previously untestable, by analyzing \ac{GW} signals from \acp{CBC} \citep{LIGOScientific:2016lio,Abbott:2017oio,Abbott:2017vtc,LIGOScientific:2018dkp,LIGOScientific:2019fpa,LIGOScientific:2020stg,LIGOScientific:2020tif, LIGOScientific:2021sio}.
The tests reported to date have found consistency with \ac{GR} in all but a few cases where data quality was suspected to be problematic \citep{LIGOScientific:2021sio}.
We now update these results by including the significant compact binary signals from the \ac{O4a} of the advanced-detector network, as reported by the \ac{LVK} in \citet{GWTC:Results}, and adding additional tests.
Specifically, this paper (Paper~I), along with its two subsequent parts, \citet{GWTC:TGR-II} and \citet{GWTC:TGR-III}, henceforth Paper~II and Paper~III,
    report the results of 19 tests of GR as well as a test for an acceleration along the line of sight. 
    We combine together the results of the tests on all significant signals to date whenever possible.
\clearpage

\begin{table*}
\caption{\label{tab:summary}
Summary of TGR pipelines and new results across Papers I--III
} 
\begin{center}
\scriptsize
\begin{tabular}{c c c c c c c}
\toprule
{Test} & {\!\!Paper\!\!} & {\!\!Section\!\!} & {Quantity }
   &  {Parameter} & {Main Result} & {Improvement} \\

\midrule

RT & I &\ref{sec:residual}&  \textit{p}-value for the presence of a residual signal
    & \textit{p}-value & Consistent with uniform dist. & $\cdots$
\\

IMRCT & I & \ref{sec:imrct} & Fractional deviation in remnant mass and spin
   & $\left\{\dfrac{\Delta M_{\rm f}}{\bar{M}_{\rm f}}, \dfrac{\Delta \chi_{\rm f}}{\bar{\chi}_{\rm f}}\right\}$ & $\left\{\TGRImrctHierPopGWTCFOURResults{DMFGWTC4HIERPOP}, \TGRImrctHierPopGWTCFOURResults{DCHIFGWTC4HIERPOP}\right\}$ & $\left\{\ImrMfHierImprovement, \ImrChifHierImprovement\right\}$
 \\

SMA & I & \ref{sec:sma} & Frac. deviation in amplitude of higher multipole moments  & $\delta A_{33}$  & ${\gwOhEightFourteenSMAMedian}^{+\gwOhEightFourteenSMAUpperError}_{-\gwOhEightFourteenSMALowerError}$ & New
\\

    POL & I & \ref{sec:polarization}& Bayes factors between different polarization hypotheses
    & $\log_{10}{\cal B}^{X}_{\rm T}$ &  $\TGRPOLlogBResults{min}_{-\TGRPOLlogBResults{minerr}}^{+\TGRPOLlogBResults{minerr}}$ -- $\TGRPOLlogBResults{max}_{-\TGRPOLlogBResults{maxerr}}^{+\TGRPOLlogBResults{maxerr}}$  &  $\TGRPOLlogBImprovement{min}$ -- $\TGRPOLlogBImprovement{max}$
\\
\noalign{\vspace{0.5mm}} \hline

PAR & II & 2.1 & FTI: PN deformation params
   & $\left|\delta{\hat{\varphi}_k}\right|$ &  $\leq \TGRFTIBound{min}$ -- $\TGRFTIBound{max}$ &  $\TGRFTIBoundImprovement{min}$ -- $\TGRFTIBoundImprovement{max}$
\\ 
& & 2.1 &   TIGER: PN deformation params
    & $\left|\delta{\hat{\varphi}_k}\right|$ &  $\leq \TGRTIGERBound{min}$ -- $\TGRTIGERBound{max}$ &  $\TGRTIGERBoundImprovement{min}$ -- $\TGRTIGERBoundImprovement{max}$
\\ 
& & 2.1 &   TIGER: Post-inspiral deformation params
    & $\left|\delta{\hat{b}_k}\right|, \left|\delta{\hat{c}_k}\right|$ &  $\leq \TGRTIGERBoundPI{min}$ -- $\TGRTIGERBoundPI{max}$  &  Maj. Update
\\ 
& & 2.2 & PCA: Best-constrained combination of PN deformation params
    & $\delta{\hat{\varphi}^{(1)}}_\mathrm{PCA, FTI}$, $\delta{\hat{\varphi}^{(1)}}_\mathrm{PCA, TIGER}$ &  \FirstPCAFTIjointbound, \FirstPCATIGERjointbound & New
 \\

SIM & II & 2.3 & Phenom: Deformation in spin-induced multipole parameter
	& $\delta \kappa_s$ & \TGRSIMPhenomCombinedCI{HIER_POP} & \TGRSIMPhenomImprovement \\
& & 2.3 & EOB: Deformation in spin-induced multipole param
    & $\delta \kappa_s$ & \TGRSIMEOBCombinedCI{HIER_POP} &  New  \\

    LOSA & II & 2.4 & Line-of-sight acceleration   & $a/c$ [${\rm s}^{-1}$] &  $\TGRLOSAResults{GW170817}$  & New \\

MDR & II & 3.1 & Magnitude of dispersion
    & $\left|A_{\alpha}\right|$ [$\mathrm{peV}^{2-\alpha}$] &  $\leq (\TGRMDRAmplitudeBoundPeV{amplitude_min}$ -- $\TGRMDRAmplitudeBoundPeV{amplitude_max})\times 10^{-22}$ &  $\TGRMDRBoundImprovement{amplitude_min}$ -- $\TGRMDRBoundImprovement{amplitude_max}$
 \\

  &  & 3.1 & Graviton mass bound& $m_{g}$ [$\mathrm{eV}/c^2$]
    &  $\leq \TGRMDRGravitonBoundNoUnits{gwtc4}$ &  \TGRMDRBoundImprovement{mg}
 \\

 SSB 
 & II & 3.2 &  Constraints on anisotropic birefringent propagation
     &  $\left| k^{(5)}_{00} \right|$ [$\mathrm{m}$]  &  $\leq \TGRSSBKFiveVZeroZeroNoUnits{gwtc4}$ & New \\
\noalign{\vspace{0.5mm}} \hline

RD & III & 2.1 &
    \soft{pyRing} (\texttt{KerrPostmerger}): Frac. dev. in freq. \& damp. time
      &  $\left\{\delta \hat{f}_{220}, \delta \hat{\tau}_{220}\right\}$ &  $\bigl\{ \pyRingHierarchicalDeltaFMedian_{-\pyRingHierarchicalDeltaFMinus}^{+\pyRingHierarchicalDeltaFPlus}, \pyRingHierarchicalDeltaTauMedian_{-\pyRingHierarchicalDeltaTauMinus}^{+\pyRingHierarchicalDeltaTauPlus} \bigr\}$ &  New \\
 & & 2.2 &
    pSEOBNR: Frac. deviations in frequency \& damping time
       & $\left\{\delta \hat{f}_{220}, \delta \hat{\tau}_{220}\right\}$ &  $\bigl\{ \pseobHierarchicalDeltaFMedian_{-\pseobHierarchicalDeltaFMinus}^{+\pseobHierarchicalDeltaFPlus}, \pseobHierarchicalDeltaTauMedian_{-\pseobHierarchicalDeltaTauMinus}^{+\pseobHierarchicalDeltaTauPlus} \bigr\}$ &  $\bigl\{\pseobImprovementDeltaF, \pseobImprovementDeltaTau\bigr\}$ \\
 & & 2.3 &
    QNMRF: Detection statistic for subdominant ringdown modes
       & $\mathcal{D} - \mathcal{D}_{1\%}$ &  221: $\leq \highestDTwoTwoOneoverThreshold$ &  New \\

E-WFM & III & 3.1 &
    ADA:  Bayes factor for IMR plus echoes to IMR only
       & $\log_{10} \mathcal{B}^{\mathrm{IMRE}}_{\mathrm{IMR}} $ &  $\leq \highestlogTenBFIMREvsIMR$ & $\cdots$ \\

 & & 3.1 &
        BHP:  Bayes factor for IMR plus echoes to IMR only  &
        $\log_{10} \mathcal{B}^{\mathrm{IMRE}}_{\mathrm{IMR}} $ &  $\leq \BHPhighestlogTenBFIMREvsIMR$ & New   \\

    E-MM & III & 3.2 & \BAYESWAVE: Signal-to-noise Bayes factor for echoes
    &  $\log_{10} \mathcal{B}^{\mathrm{signal}}_{\mathrm{noise}} $  & $\leq \BWEchoeshighestlogTenBFSvsN$ & $\cdots$ \\

 & & 3.3 & \CWB: \textit{p}-value for the presence of echoes
    &  \textit{p}-value  & Consistent with uniform dist. & New   \\

\bottomrule
\end{tabular}
\end{center}
  \tablecomments{Analysis abbreviations are defined in the text.
    The main results and improvements are calculated using the combined results unless otherwise specified, and use the hierarchically combined results if available.
    For analyses where we quote bounds on two parameters inferred simultaneously, we enclose the parameters in curly brackets.
    The improvements are computed relative to the previous analyses reported in the GWTC-3.0 test of \ac{GR} paper \citep{LIGOScientific:2020tif},
            when applicable, or for updated analyses applied to GWTC-3.0; otherwise, ``New'' is given.
            For most tests, the improvement is defined as $X_{\mathrm{GWTC}\text{-}\mathrm{3.0}}/X_{\mathrm{GWTC}\text{-}\mathrm{4.0}}$,
                where $X$ denotes the width of the 90\% credible interval for the parameters for each test.
            For the tests where we give ranges for the main result and improvements,
                we quote the minimum and maximum over all different cases of that test independently. Thus, the endpoints of the improvement range do not necessarily correspond to the endpoints of the main result range.
    For two new tests, the main results were obtained with individual pre-O4 events: The SMA result is from GW190814, while the LOSA result is from GW170817.
    For IMRCT and pSEOBNR, the improvement is with respect to the two-dimensional hierarchical analysis \citep{Zhong:2024pwb}, not the one-dimensional hierarchical analysis performed in the GWTC-3.0 testing GR paper.
    For POL, the improvement is the difference of the $\log_{10}$ Bayes factors (GWTC-3.0 minus GWTC-4.0), to illustrate the increased support for \GR.
    The upper bounds for FTI are only for O4a, and the improvement factors are obtained by comparing O4a with results reported in the GWTC-3.0 testing GR paper.
        As no TIGER results were reported in the latter, its improvement factors are obtained by comparing GWTC-4.0 with a recent analysis of GWTC-3.0 events~\citep{Roy:2025gzv}.
    The \soft{pyRing} results are also only for O4a, as are the upper bounds on Bayes factors for E-WFM and E-MM-\BAYESWAVE.
    The QNMRF parameter given is the detection statistic for the 221 mode minus the threshold corresponding to a 1\% false alarm probability,
            computed over a period when including the 221 mode improves consistency with the IMR results, here for \FULLNAME{GW231028_153006}.
    The $\cdots$ marker for improvements is used for RT, where the result of overall consistency is maintained;
        for ADA, a test absent from the GWTC-3.0 analysis, but previously run (in an older implementation) in the GWTC-2.0 testing GR paper;
        and for \BAYESWAVE, which quoted $p$-values rather than Bayes factors for GWTC-3.0.
}

\end{table*}

The signals analyzed constitute the fourth Gravitational-Wave Transient Catalog \citep[GWTC-4.0;][]{GWTC:Results,GWTC:Introduction}, and include the new \ac{O4a} observations (from \OfourAStartDate{}
                                            to \OfourAEndDate)
    together with those from the previous runs,
    O1 \citep[the first observing run, from \OoneStartDate{}
                                        to \OoneEndDate%
                                        ;][]{LIGOScientific:2016dsl},
    O2 \citep[the second observing run, from \OtwoStartDate{}
                                        to \OtwoEndDate%
                                        ;][]{GWTC1},
    O3a \citep[the first part of the third observing run, from \OthreeAStartDate{}
                                                            to \OthreeAEndDate%
                                                            ;][]{GWTC2,GWTC2p1},
    and O3b \citep[the second part of the third observing run, from \OthreeBStartDate{}
                                                                to \OthreeBEndDate%
                                                                ;][]{GWTC3}.
All the candidates in the catalog are consistent with being \ac{CBC} signals,
    generated by either \acp{BBH}, \acp{BNS}, or \acp{NSBH}.

In these papers, we restrict ourselves to candidates that have passed the selection criteria of
    having a confident \ac{FAR} \TGRFARTHRESH \citep[the same as in][]{LIGOScientific:2020tif,LIGOScientific:2021sio},
    and having been seen by at least two detectors.
We henceforth refer to these in the context of the test of GR papers as \emph{events}.
Thus, of the \ac{O4a} candidates, this paper covers \TGRNUMEVENTS new events,
    while specifically excluding the single-detector events \FULLNAME{GW230529_181500} \citep[shortened to \COMMONNAME{GW230529};][]{GW230529}
    and \FULLNAME{GW230814_230901} \citep[shortened to \COMMONNAME{GW230814single} to avoid confusion with \FULLNAME{GW230814_061920};][]{GW230814}.
    The tests of GR on those two events are reported elsewhere \citep{Sanger:2024axs, GW230814}.
Many of the tests covered have additional, narrower selection criteria for choosing which events are relevant for them,
    based on the required and supported physical parameters, available data, etc.; these are described in the respective sections.

\Ac{GW} observations enable testing many different aspects of \ac{GR}
    or its alternatives and extensions \citep{Will:2014kxa, Colleoni:2024lpj, Yunes:2024lzm, Gupta:2025utd},
    among them the linearized theory of the \acp{GW} themselves,
    and the dynamical and highly nonlinear theory of the two-body system generating them.
    
In this series of papers, we perform \TGRNUMTESTS distinct tests of \ac{GR}.
We give each test an abbreviated (or acronym) uppercase name,
    as used for them in Tables~\ref{tab:summary}--\ref{tab:selectionO1+O2}.
When a single test uses or compares multiple models, we write the model names in the typewriter font (as in \texttt{KerrPostmerger}),
    and software packages are indicated with their own font (as in \BILBY).
We subdivide our tests
    into three papers:
\begin{enumerate}
    \item This paper (Paper I) includes tests of consistency,
        either consistency of each signal's residual with the noise,
        or internal consistency of the signal with itself, according to \ac{GR} expectations.
        We present the results of the
            residuals test \citep[RT;][]{LIGOScientific:2019fpa};
            the inspiral--merger--ringdown (IMR) consistency test \citep[IMRCT;][]{Hughes:2004vw, Ghosh:2016qgn, Ghosh:2017gfp};
            the subdominant multipole amplitudes (SMA) test \citep{Puecher:2022sfm};
            and the test of the polarization content \citep[POL;][]{Wong:2021cmp}.
    \item Paper II features parameterized tests, using quantifiable parameters for various imaginable deviations from \ac{GR} signals in possible alternatives of extensions of \ac{GR}.
        These parameters can refer to any physics involved with either the generating system or the propagation of the waves.
        The parametrized tests of \ac{GW} generation, grouped in the tables as PAR,
            constrain deviations in \ac{PN} coefficients \citep{Blanchet:1994ex,Blanchet:1994ez,Arun:2006hn, Arun:2006yw,Maggiore:2007ulw,Will:2014kxa,Blanchet:2013haa}
                in the inspiral and phenomenological coefficients in the post-inspiral.
        These include both the FTI \citep[inspiral only;][]{Mehta:2022pcn} and TIGER \citep{Li:2011cg, Agathos:2013upa, Meidam:2017dgf, Roy:2025gzv} single-parameter tests,
            as well as the principal component analysis (PCA) applied to their multi-parameter inspiral versions \citep{Mahapatra:2025cwk}.
        The parameterized tests of \ac{GW} generation also include
            the spin-induced moments (SIM) tests \citep{Mehta:2022pcn, Divyajyoti:2023izl}, using either Phenom or EOB waveforms.
        Additionally, we test for the presence of a line-of-sight acceleration \citep[LOSA;][]{Vijaykumar:2023tjg, Tiwari:2025aec}, which is not a beyond-GR effect, but could be confused for a GR deviation.
        The tests for parameterized deviations in the propagation of the \acp{GW}, grouped under the abbreviation PRP, are
                the MDR test for a modified dispersion relation \citep{Baka:2025drk},
                and the SSB test for spacetime symmetry breaking, specifically anisotropic birefringent propagation \citep{Haegel:2022ymk}.

    \item Paper III presents tests of the final remnant object.
        These include tests comparing the object's immediate ringdown to that expected
            from the quasi-normal modes (QNMs) of a Kerr \ac{BH} in vacuum (for the appropriate events),
        and searches for any post-ringdown (echo) content.
        The ringdown analyses, grouped as RD, are \soft{pyRing} \citep{Carullo:2019flw},
            pSEOBNR \citep{Brito:2018rfr, Ghosh:2021mrv, Pompili:2025cdc},
            and QNM rational filter \citep[QNMRF;][]{Ma:2022wpv,Ma:2023vvr,Ma:2023cwe,Lu:2025mwp}.
        The echo searches are subdivided into searches for echoes using proposed waveform models, grouped as E-WFM,
                namely the ADA templates \citep{Abedi:2016hgu, Lo:2018sep} and BHP templates \citep{Nakano:2017fvh,Uchikata:2019frs,Uchikata:2023zcu},
            and minimally modeled searches for echoes, grouped as E-MM,
                namely the \BAYESWAVE~\citep{Tsang:2018uie} and coherent WaveBurst \citep[\CWB;][]{Miani:2023} analyses.
\end{enumerate}
Of these, almost half of the tests are new, namely SMA, PCA, SIM-EOB, LOSA, SSB, \soft{pyRing} (\texttt{KerrPostmerger}), QNMRF, E-WFM-BHP, and E-MM-\CWB;
    additionally TIGER is significantly updated \citep{Roy:2025gzv} compared to the version last used in an LVK testing \ac{GR} paper \citep[][]{LIGOScientific:2020tif}.
The companion paper about constraints on cosmology \citep{GWTC:Cosmology} also presents constraints on dissipative propagation effects (\ac{GW} friction) obtained using that paper's methods.

All the tests and their specific paper and section,
    as well as highlights of the improvements with respect to GWTC-3.0 (see \citealt{GWTC:Introduction} for catalog designations) and main results, are summarized in Table~\ref{tab:summary}.
We find improvements in the constraints due to including the additional events, as expected, notably improvements in the \ac{PN} coefficient constraints of up to $\TGRFTIBoundImprovement{max}$ (FTI)
   and $\TGRTIGERBoundImprovement{max}$ (TIGER), though some of this improvement is due to changes to the portion of the signal where the \ac{PN} coefficient deviations are implemented.
The IMRCT analysis has improvements of up to $\ImrChifHierImprovement$, while the MDR analysis has improvements of
   up to almost a factor of three.
Additionally, some O4a events give notable constraints on their own.
For instance, the likely \ac{NSBH} \FULLNAME{GW230518_125908} provides a constraint on the $-1$PN coefficient that
    is comparable to the combined GWTC-4.0 constraint, and provides constraints on many higher \ac{PN} coefficients that are
    significantly better than those from all of GWTC-3.0 (see Section~2.1 of Paper~II).
For the pSEOBNR ringdown analysis, the loud event \FULLNAME{GW231226_101520}
    provides a constraint on deviations in the damping time of the dominant QNM that is slightly
    better than the GWTC-3.0 hierarchical constraint (see Section~2.2 of Paper~III).
However, the loud event GW250114~\citep{GW250114} from the \ac{O4b} 
\clearpage
\clearpage

\startlongtable
\begin{deluxetable*}{l ccc cccc c cccc c ccc}
\centerwidetable

\tablecaption{Event selection table for the analyses in this paper, from the first part of the fourth observing run \citep[O4a;][]{GWTC:Results}
}

\tablehead{
\colhead{O4a} & \colhead{} & \colhead{} & \colhead{} &
\multicolumn{4}{c}{Paper I} & \colhead{} &
\multicolumn{4}{c}{Paper II} & \colhead{} &
\multicolumn{3}{c}{Paper III} \\
\cline{5-8}\cline{10-13}\cline{15-17}
\colhead{Event Name} & 
 \colhead{SNR} & 
 \colhead{IFOs} & 
 \colhead{$(1\!\!+\!\!z)\mathcal{M}/\Msun$} & 
 \colhead{RT} & 
 \colhead{\!\!IMRCT\!\!} & 
 \colhead{\!\!SMA\!\!} & 
 \colhead{\!\!POL\!\!} & 
 \colhead{} & 
 \colhead{\!\!PAR\!\!} & 
 \colhead{\!\!SIM\!\!} & 
 \colhead{\!\!LOSA\!\!} & 
 \colhead{\!\!PRP\!\!} & 
 \colhead{} & 
 \colhead{\!\!RD\!\!} & 
 \colhead{\!\!E-WFM\!\!} & 
 \colhead{\!\!E-MM\!\!}
}
\startdata
 \cellcolor{lightgray}{GW230518\_125908} & 
 \cellcolor{lightgray}{$\networkmatchedfiltersnruncert{GW230518_125908}$} & 
 \cellcolor{lightgray}{\INSTRUMENTS{GW230518_125908}} & 
 \cellcolor{lightgray}{$\chirpmassdetuncert{GW230518_125908}$} & 
 \cellcolor{lightgray}{\checkmark} & 
 \cellcolor{lightgray}{\checknodata} & 
 \cellcolor{lightgray}{\checknodata} & 
 \cellcolor{lightgray}{\checkmark} & 
 \cellcolor{lightgray}{ } & 
 \cellcolor{lightgray}{\checkmark} & 
 \cellcolor{lightgray}{\checknodata} & 
 \cellcolor{lightgray}{\checknodata} & 
 \cellcolor{lightgray}{\checknodata} & 
 \cellcolor{lightgray}{ } & 
 \cellcolor{lightgray}{\checknodata} & 
 \cellcolor{lightgray}{\checknodata} & 
 \cellcolor{lightgray}{\checkmark} \\
 GW230601\_224134 & 
 $\networkmatchedfiltersnruncert{GW230601_224134}$ & 
 \INSTRUMENTS{GW230601_224134} & 
 $\chirpmassdetuncert{GW230601_224134}$ & 
 \checkmark & 
 \checknodata & 
 \checknodata & 
 \checkmark & 
   & 
 $\checkmark$ & 
 \checknodata & 
 \checknodata & 
 \checkmark & 
   & 
 \checkmark & 
 \checkmark & 
 \checkmark \\
 \cellcolor{lightgray}{GW230605\_065343} & 
 \cellcolor{lightgray}{$\networkmatchedfiltersnruncert{GW230605_065343}$} & 
 \cellcolor{lightgray}{\INSTRUMENTS{GW230605_065343}} & 
 \cellcolor{lightgray}{$\chirpmassdetuncert{GW230605_065343}$} & 
 \cellcolor{lightgray}{\checkmark} & 
 \cellcolor{lightgray}{\checknodata} & 
 \cellcolor{lightgray}{\checknodata} & 
 \cellcolor{lightgray}{\checkmark} & 
 \cellcolor{lightgray}{ } & 
 \cellcolor{lightgray}{\checkmark} & 
 \cellcolor{lightgray}{\checknodata} & 
 \cellcolor{lightgray}{\checknodata} & 
 \cellcolor{lightgray}{\checkmark} & 
 \cellcolor{lightgray}{ } & 
 \cellcolor{lightgray}{\checknodata} & 
 \cellcolor{lightgray}{\checkmark} & 
 \cellcolor{lightgray}{\checkmark} \\
\enddata

\tablecomments{
For each event we give the matched-filter network \ac{SNR},
  the redshifted chirp mass, $(1\!+\!z)\mathcal{M}$ \citep[e.g.,][]{GWTC:Introduction},
  and the interferometers (IFOs) used for the analysis of the events,
  which are Hanford (H) and Livingston (L) for all O4a events,
  and may include Virgo (V) for previous runs, in Tables~\ref{tab:selectionO3b}--\ref{tab:selectionO1+O2} below.
More details about the events appear in the relevant GWTC papers \citep[for O4a events,][]{GWTC:Results}.
  A \checkmark{} indicates an event meeting our selection criteria for an analysis, and thus included in our results.
     A \checkminus{} indicates an event that originally met the selection criteria, but technical difficulties in its analysis prevented obtaining meaningful results.
     A \checknodata indicates that an event does not meet an analysis's selection criteria.
}

\label{tab:selection}

\end{deluxetable*}

\startlongtable
\begin{deluxetable*}{l c c c c c c c c c c c c c c c c}
\centerwidetable
\tablewidth{97.5pt}

\tablecaption{Updated table of events from the second half of the third observing run \citep[O3b;][]{GWTC3}
}

\tablehead{
\colhead{O3b} & \colhead{} & \colhead{} & \colhead{} &
\multicolumn{4}{c}{Paper I} & \colhead{} &
\multicolumn{4}{c}{Paper II} & \colhead{} &
\multicolumn{3}{c}{Paper III} \\
\cline{5-8}\cline{10-13}\cline{15-17}
\colhead{Event Name} & 
 \colhead{SNR} & 
 \colhead{IFOs} & 
 \colhead{$(1\!\!+\!\!z)\mathcal{M}/\Msun$} & 
 \colhead{RT} & 
 \colhead{\!\!IMRCT\!\!} & 
 \colhead{\!\!SMA\!\!} & 
 \colhead{\!\!POL\!\!} & 
 \colhead{} & 
 \colhead{\!\!PAR\!\!} & 
 \colhead{\!\!SIM\!\!} & 
 \colhead{\!\!LOSA\!\!} & 
 \colhead{\!\!PRP\!\!} & 
 \colhead{} & 
 \colhead{\!\!RD\!\!} & 
 \colhead{\!\!E-WFM\!\!} & 
 \colhead{\!\!E-MM\!\!}
}%%
\startdata
 \cellcolor{lightgray}{GW191109\_010717} & 
 \cellcolor{lightgray}{$\networkmatchedfiltersnrgwtcthreeuncert{GW191109_010717}$} & 
 \cellcolor{lightgray}{\INSTRUMENTSOTHREEB{GW191109A}} & 
 \cellcolor{lightgray}{$\chirpmassdetgwtcthreeuncert{GW191109_010717}$} & 
 \cellcolor{lightgray}{\checkmark} & 
 \cellcolor{lightgray}{\checknodata} & 
 \cellcolor{lightgray}{\checknodata} & 
 \cellcolor{lightgray}{\checkmark} & 
 \cellcolor{lightgray}{ } & 
 \cellcolor{lightgray}{\checkminus} & 
 \cellcolor{lightgray}{\checknodata} & 
 \cellcolor{lightgray}{\checknodata} & 
 \cellcolor{lightgray}{\checkminus} & 
 \cellcolor{lightgray}{ } & 
 \cellcolor{lightgray}{\checkmark} & 
 \cellcolor{lightgray}{\checkprevext} & 
 \cellcolor{lightgray}{\checkprevlvk} \\
 GW191129\_134029 & 
 $\networkmatchedfiltersnrgwtcthreeuncert{GW191129_134029}$ & 
 \INSTRUMENTSOTHREEB{GW191129G} & 
 $\chirpmassdetgwtcthreeuncert{GW191129_134029}$ & 
 \checkmark & 
 \checknodata & 
 \checknodata & 
 \checkmark & 
   & 
 \checkmark & 
 \checknodata & 
 \checknodata & 
 \checkmark & 
   & 
 \checknodata & 
 \checkprevext & 
 \checkprevlvk \\
 \cellcolor{lightgray}{GW191204\_171526} & 
 \cellcolor{lightgray}{$\networkmatchedfiltersnrgwtcthreeuncert{GW191204_171526}$} & 
 \cellcolor{lightgray}{\INSTRUMENTSOTHREEB{GW191204G}} & 
 \cellcolor{lightgray}{$\chirpmassdetgwtcthreeuncert{GW191204_171526}$} & 
 \cellcolor{lightgray}{\checkmark} & 
 \cellcolor{lightgray}{\checknodata} & 
 \cellcolor{lightgray}{\checknodata} & 
 \cellcolor{lightgray}{\checkmark} & 
 \cellcolor{lightgray}{ } & 
 \cellcolor{lightgray}{\checkmark} & 
 \cellcolor{lightgray}{\checkmark} & 
 \cellcolor{lightgray}{\checknodata} & 
 \cellcolor{lightgray}{\checkmark} & 
 \cellcolor{lightgray}{ } & 
 \cellcolor{lightgray}{\checknodata} & 
 \cellcolor{lightgray}{\checkprevext} & 
 \cellcolor{lightgray}{\checkprevlvk} \\
\enddata

\tablecomments{In addition to the notation used in Table~\ref{tab:selection}, \checkprevext{} indicates existing results from previous runs by external works,
    while \checkprevlvk{} indicates previous results by the \ac{LVK} exist, but are not used for the combined bounds here.}

\label{tab:selectionO3b}

\end{deluxetable*}

\startlongtable
\begin{deluxetable*}{l c c c c c c c c c c c c c c c c}
\centerwidetable
\tablewidth{97.5pt}

\tablecaption{Updated table of events from the first half of the third observing run \citep[O3a;][]{GWTC2, GWTC2p1}
}

\tablehead{
\colhead{O3a} & \colhead{} & \colhead{} & \colhead{} &
\multicolumn{4}{c}{Paper I} & \colhead{\!\!\!\!} &
\multicolumn{4}{c}{Paper II} & \colhead{\!\!\!\!} &
\multicolumn{3}{c}{Paper III} \\
\cline{5-8}\cline{10-13}\cline{15-17}
\colhead{Event Name} & 
 \colhead{SNR} & 
 \colhead{IFOs} & 
 \colhead{$(1\!\!+\!\!z)\mathcal{M}/\Msun$} & 
 \colhead{RT} & 
 \colhead{\!\!IMRCT\!\!} & 
 \colhead{\!\!SMA\!\!} & 
 \colhead{\!\!POL\!\!} & 
 \colhead{\!\!\!\!} & 
 \colhead{\!\!PAR\!\!} & 
 \colhead{\!\!SIM\!\!} & 
 \colhead{\!\!LOSA\!\!} & 
 \colhead{\!\!PRP\!\!} & 
 \colhead{\!\!\!\!} & 
 \colhead{\!\!RD\!\!} & 
 \colhead{\!\!E-WFM\!\!} & 
 \colhead{\!\!E-MM\!\!}
}
\startdata
 \cellcolor{lightgray}{GW190408\_181802} & 
 \cellcolor{lightgray}{$\networkmatchedfiltersnrgwtctwopointoneuncert{GW190408_181802}$} & 
 \cellcolor{lightgray}{\INSTRUMENTSOTHREEA{GW190408A}} & 
 \cellcolor{lightgray}{$\chirpmassdetgwtctwopointoneuncert{GW190408_181802}$} & 
 \cellcolor{lightgray}{\checkmark} & 
 \cellcolor{lightgray}{\checkmark} & 
 \cellcolor{lightgray}{\checknodata} & 
 \cellcolor{lightgray}{\checkmark} & 
 \cellcolor{lightgray}{\!\!\!\!} & 
 \cellcolor{lightgray}{\checkmark} & 
 \cellcolor{lightgray}{\checknodata} & 
 \cellcolor{lightgray}{\checknodata} & 
 \cellcolor{lightgray}{\checkmark} & 
 \cellcolor{lightgray}{\!\!\!\!} & 
 \cellcolor{lightgray}{\checkprevlvk} & 
 \cellcolor{lightgray}{\checkprevlvk} & 
 \cellcolor{lightgray}{\checkprevext} \\
 GW190412 & 
 $\networkmatchedfiltersnrgwtctwopointoneuncert{GW190412_053044}$ & 
 \INSTRUMENTSOTHREEA{GW190412A} & 
 $\chirpmassdetgwtctwopointoneuncert{GW190412_053044}$ & 
 \checkmark & 
 \checkprevlvk & 
 \checkmark & 
 \checkmark & 
   \!\!\!\! & 
 \checkmark & 
 \checkmark & 
 \checknodata & 
 \checkmark & 
  \!\!\!\! & 
 \checknodata & 
 \checkprevlvk & 
\checkprevext \\
 \cellcolor{lightgray}{GW190421\_213856} & 
 \cellcolor{lightgray}{$\networkmatchedfiltersnrgwtctwopointoneuncert{GW190421_213856}$} & 
 \cellcolor{lightgray}{\INSTRUMENTSOTHREEA{GW190421A}} & 
 \cellcolor{lightgray}{$\chirpmassdetgwtctwopointoneuncert{GW190421_213856}$} & 
 \cellcolor{lightgray}{\checkmark} & 
 \cellcolor{lightgray}{\checkprevlvk} & 
 \cellcolor{lightgray}{\checknodata} & 
 \cellcolor{lightgray}{\checkmark} & 
 \cellcolor{lightgray}{\!\!\!\!} & 
 \cellcolor{lightgray}{\checkmark} & 
 \cellcolor{lightgray}{\checknodata} & 
 \cellcolor{lightgray}{\checknodata} & 
 \cellcolor{lightgray}{\checkmark} & 
 \cellcolor{lightgray}{\!\!\!\!} & 
 \cellcolor{lightgray}{\checknodata} & 
 \cellcolor{lightgray}{\checkprevlvk} & 
 \cellcolor{lightgray}{\checknodata} \\
\enddata

\tablecomments{Notation is as in Table~\ref{tab:selectionO3b}.
              The IMR results for GW190412 appeared in \citet{LIGOScientific:2020stg}.}

\label{tab:selectionO3a}

\end{deluxetable*}

\startlongtable
\begin{deluxetable*}{l c c c c c c c c c c c c c c c c}
\centerwidetable
\tablewidth{97.5pt}

\tablecaption{Updated table of events from the first \citep[O1;][]{LIGOScientific:2016dsl} and second \citep[O2;][]{GWTC1} observing runs}

\tablehead{
\colhead{O1+O2} & \colhead{} & \colhead{} & \colhead{} &
\multicolumn{4}{c}{Paper I} & \colhead{} &
\multicolumn{4}{c}{Paper II} & \colhead{} &
\multicolumn{3}{c}{Paper III} \\
\cline{5-8}\cline{10-13}\cline{15-17}
\colhead{Event Name} & 
 \colhead{SNR} & 
 \colhead{IFOs} & 
 \colhead{$(1\!\!+\!\!z)\mathcal{M}/\Msun$} & 
 \colhead{RT} & 
 \colhead{\!\!IMRCT\!\!} & 
 \colhead{\!\!SMA\!\!} & 
 \colhead{\!\!POL\!\!} & 
 \colhead{} & 
 \colhead{\!\!PAR\!\!} & 
 \colhead{\!\!SIM\!\!} & 
 \colhead{\!\!LOSA\!\!} & 
 \colhead{\!\!PRP\!\!} & 
 \colhead{} & 
 \colhead{\!\!RD\!\!} & 
 \colhead{\!\!E-WFM\!\!} & 
 \colhead{\!\!E-MM\!\!}
}
\startdata
 \cellcolor{lightgray}{GW150914} & 
 \cellcolor{lightgray}{$\networkmatchedfiltersnrgwtctwopointoneuncert{GW150914_095045}$} & 
 \cellcolor{lightgray}{\INSTRUMENTSOONEOTWO{GW150914}} & 
 \cellcolor{lightgray}{$\chirpmassdetgwtctwopointoneuncert{GW150914_095045}$} & 
 \cellcolor{lightgray}{\checkmark} & 
 \cellcolor{lightgray}{\checkmark} & 
 \cellcolor{lightgray}{\checknodata} & 
 \cellcolor{lightgray}{\checkmark} & 
 \cellcolor{lightgray}{ } & 
 \cellcolor{lightgray}{\checkmark} & 
 \cellcolor{lightgray}{\checknodata} & 
 \cellcolor{lightgray}{\checknodata} & 
 \cellcolor{lightgray}{\checkmark} & 
 \cellcolor{lightgray}{ } & 
 \cellcolor{lightgray}{\checkmark} & 
 \cellcolor{lightgray}{\checkprevlvk} & 
 \cellcolor{lightgray}{\checkprevext} \\
 GW151012* & 
 $\networkmatchedfiltersnrgwtctwopointoneuncert{GW151012_095443}$ & 
 \INSTRUMENTSOONEOTWO{GW151012} & 
 $\chirpmassdetgwtctwopointoneuncert{GW151012_095443}$ & 
 \checkprevlvk & 
 \checknodata & 
 \checknodata & 
 \checknodata & 
   & 
 \checkprevlvk & 
 \checknodata & 
 \checknodata & 
 \checkprevlvk &
   & 
 \checknodata & 
 \checkprevext & 
 \checkprevext \\
 \cellcolor{lightgray}{GW151226} & 
 \cellcolor{lightgray}{$\networkmatchedfiltersnrgwtctwopointoneuncert{GW151226_033853}$} & 
 \cellcolor{lightgray}{\INSTRUMENTSOONEOTWO{GW151226}} & 
 \cellcolor{lightgray}{$\chirpmassdetgwtctwopointoneuncert{GW151226_033853}$} & 
 \cellcolor{lightgray}{\checkmark} & 
 \cellcolor{lightgray}{\checknodata} & 
 \cellcolor{lightgray}{\checknodata} & 
 \cellcolor{lightgray}{\checkmark} & 
 \cellcolor{lightgray}{ } & 
 \cellcolor{lightgray}{\checkmark} & 
 \cellcolor{lightgray}{\checkmark} & 
 \cellcolor{lightgray}{\checknodata} & 
 \cellcolor{lightgray}{\checkmark} & 
 \cellcolor{lightgray}{ } & 
 \cellcolor{lightgray}{\checknodata} & 
 \cellcolor{lightgray}{\checkprevlvk} & 
 \cellcolor{lightgray}{\checkprevext} \\
\enddata

\tablecomments{Notation is as in Table~\ref{tab:selectionO3b}.
            The starred events GW151012 and GW170729 had previously been used for the indicated tests of GR in \citet{LIGOScientific:2016dsl,LIGOScientific:2019fpa},
	            but do not meet the current selection criteria.
            GW170817's PAR and PRP results appeared in \citet{LIGOScientific:2018dkp}.
              }

\label{tab:selectionO1+O2}

\end{deluxetable*}

\vspace{-20pt}

\noindent has now provided even better constraints on both PN coefficients and QNM deviations \citep{GW250114_TGR}.
Similarly, the loud \ac{O4b} event GW241011~\citep{GW241011/GW241110},
    whose source was an unequal-mass binary,
    provides the best constraints for the SMA and SIM analyses.

Most of the tests are null tests, placing bounds on deviations from \ac{GR} (or more generally, deviations from the waveform models for isolated, quasi-circular binaries used in the analysis)
    instead of constraining a specific alternative theory
    as modeling \acp{CBC} (or even single objects) in modified theories is much more complicated than in \ac{GR} \citep{Will:1994fb,Berti:2015itd,Abac:2025saz,Yunes:2024lzm}.
    Illustrative translations of the bounds on deviations in \ac{PN} coefficients to constraints on specific alternative theories (with many caveats) are given in Section~2.1 of Paper~II.

Overall, the results of tests in all three of these papers show consistency with \ac{GR}.
There are a few events which indicate inconsistency for one or more of the analyses,
    and these cases are discussed for in the papers describing each respective analysis.
The standard consistency measure we use is finding \ac{GR} within the 90\% credible interval.
Thus, given enough events and analyses, some are expected to be inconsistent, statistically.
We therefore find that our results for individual events are still fully explainable assuming \ac{GR} and statistical noise.
We additionally find that certain apparent deviations from \ac{GR} are due to prior effects or systematic uncertainties in waveform modeling, as discussed in Paper~II.
There are also certain analyses for which we find that there are more significant deviations from \ac{GR} when combining together events than for the individual events,
specifically SIM, where the apparent deviation is driven by correlations with the effective inspiral spin, and both \soft{pyRing} and pSEOBNR, where there are apparently quite significant deviations in the combined results:
For \soft{pyRing}, \ac{GR} is only found at $\pyRingHierarchicalQuantileFourDValue\%$ credibility when including all the O4a events analyzed, though if one restricts to only the events with a significant
ringdown signal, \ac{GR} is found at $\pyRingHierarchicalQuantileOneDMuDeltaTauValuelnBcut\%$ credibility, and a bootstrapping analysis finds that even this deviation may not be as significant, due to the finite number of events considered \citep[cf.][]{Pacilio:2023uef}.
For pSEOBNR, \ac{GR} is found at $\QuantileJointGWTCFour$ or $\QuantileHierGWTCFourTauOneD$ credibility depending on how the events are combined, with similar reductions in significance from the bootstrapping analysis.
Indeed, GR is found at a credibility of $\QuantileJointGWTCFourPlusTwoFiveZeroOneOneFour$ or $\QuantileHierGWTCFourTauOneDPlusTwoFiveZeroOneOneFour$
when including the loud \ac{O4b} event GW250114 \citep{GW250114, GW250114_TGR}.
Thus, these combined results do not provide significant evidence of a deviation from \ac{GR}.

The rest of this paper is organized as follows:
Section~\ref{sec:catalogref} discusses the events included, and details which tests were performed for each.
Section~\ref{sec:pemethods} describes the general methods common between many of the tests.
Section~\ref{sec:consistency} describes our consistency tests, namely the residuals test (Section~\ref{sec:residual}),
    the inspiral--merger--ringdown consistency (Section~\ref{sec:imrct}),
    and the subdominant multipole amplitude test (Section~\ref{sec:sma}).
Section~\ref{sec:polarization} describes the test for possible non-GR polarizations.
Section~\ref{sec:conclusion} provides overall concluding remarks.   \\

\subpapersection{The Gravitational-wave Events}\label{sec:catalogref}

All the events passing the selection criteria, and an enumeration of which events contributed to which sort of test,
    and appear in which paper (I/II/III),
    are given in Tables~\ref{tab:selection}--\ref{tab:selectionO1+O2},
    subdivided by observing run.
We also give some basic information about the events, including their matched-filter network \ac{SNR}
    and redshifted chirp mass, $(1+z)\mathcal{M}$,
    as a well-measured combination of the masses \citep[e.g.,][]{GWTC:Introduction}. These are obtained from the parameter estimation analyses in the cited GWTC papers
    \citep[for O4a events, the GWTC-4.0 results paper;][]{GWTC:Results}.
We further list the interferometers (IFOs) used for the analysis of the events (H = Hanford, L = Livingston, V = Virgo).
For the O4a run, all events use data from only the Hanford and Livingston detectors.

The events analyzed in this work are drawn from \gwtc[\thisgwtcversion] \citep{GWTC:Results}.
This catalog includes all events detected by the \ac{LVK} before \GWTCfourENDDate{}.
Compared to previous versions of the catalog, this version additionally contains events from the first part of \ac{O4a}.
During \ac{O4a}, the LIGO Hanford and Livingston observatories accumulated {\sisetup{round-mode=places,round-precision=0}\OfouraDurationHL{}}
of coincident data across a period of {\sisetup{round-mode=places,round-precision=0}\OfouraDuration{}} in 2023 and early 2024;
    Virgo and KAGRA were mostly offline for upgrades.
For inclusion in the general relativity tests described in this work,
    an event must have a false alarm rate \TGRFARTHRESH{} in at least one search and must have at least two observatories used for their analysis.
Thus, for \ac{O4a}, both LIGO observatories must be used in the analysis of the events.
For GWTC\nobreakdash-1.0,
    the tests of \ac{GR}~\citep{LIGOScientific:2019fpa} were applied to all events in that initial catalog,
    due to the smaller number of total events.
The events GW151012 and GW170729 were analyzed in that paper and appear in the event tables here,
    but they do not satisfy the significance criterion. Thus, they are not included in combined results.%

The selection criteria used for applying tests of \GR to subsequent editions of the catalog,
    GWTC-2.0 and GWTC-3.0 \citep[][]{LIGOScientific:2020tif,LIGOScientific:2021sio},
    are the same as the ones for the present paper, though the multiple-detector criterion was not applied explicitly in the GWTC-2.0 testing \ac{GR} paper, because it was already satisfied for all events considered.
    Additionally, while we include the likely \acp{BNS} GW170817 and GW190425 in this paper, they were previously excluded from the testing GR catalog papers, though GW170817 had a paper dedicated to testing GR results \citep{LIGOScientific:2018dkp}.
    As in the GWTC-3.0 testing \ac{GR} paper \citep{LIGOScientific:2021sio}, we also keep the three events (GW190421\_213856, GW190521, and GW190910\_112807) included in the GWTC-2.0 testing \ac{GR} paper \citep{LIGOScientific:2020tif} whose significance
    dropped slightly below the threshold in the reanalysis in GWTC-2.1 \citep{GWTC2p1}.

All events covered appear in the respective Tables~\ref{tab:selection}--\ref{tab:selectionO1+O2}.
Further details about \FULLNAME{GW231123_135430} (shortened to \COMMONNAME{GW231123})
    appear in its discovery paper \citep{GW231123};
    there are also separate papers describing tests of \ac{GR} on the exceptional single-detector events
        \COMMONNAME{GW230529} \citep{Sanger:2024axs} and \COMMONNAME{GW230814single} \citep{GW230814}
        that do not appear in this paper.

One might worry that the requirement that the events to which we apply the tests of \GR are detected with a high significance means that we are a priori excluding events that have significant deviations from the waveform models used in the searches.
However, the minimally modeled \CWB search \citep[][]{GWTC:Methods} is able to detect signals that have the generic chirp structure expected for \acp{CBC} but differ from the exact predictions for \acp{BBH} in \GR.
This sensitivity of \CWB is illustrated in \citet{Mishra:2022ott} for the effects of precession or eccentricity that are not included in the template banks used in the modeled searches \citep[][]{GWTC:Methods}, or the training set used for \CWB.
While \CWB is mostly sensitive to signals from high-mass binaries, the inclusion of the \CWB significance in the selection criteria means that we are at least not missing a significant population of non-\GR signals in the set of events to which we apply tests of \GR.

Previous catalog papers suffered from the incorrect implementation of the detector calibration uncertainty.
As described in \citet{GWTC:Methods}, this affected parameter estimation, but produced only a negligible effect on the posteriors, with changes typically within the error from statistical sampling
\citep{Baka:2025bbb}.
For \ac{O4a} events, we performed all the tests of \ac{GR} with the correct calibration uncertainty, but how we handle pre-O4a results differs between analyses.
SMA, PCA, SIM-EOB, LOSA, and E-WFM all have used the correct calibration uncertainty in analyzing pre-O4a events,
    while \soft{pyRing} and QNMRF do not use the calibration uncertainties.
Post-inspiral TIGER, SIM-Phenom, and pSEOBNR have rerun the analysis with the correct calibration priors, while
the MDR analysis reweights the old results to the correct calibration.
The FTI and SSB analyses currently only present results for O4a events.
POL has used the correct calibration uncertainty except for using older maximum-likelihood parameter-estimation results obtained with the incorrect calibration uncertainty to determine the time--frequency cluster used in the analysis.
However, the POL analysis has a fairly broad frequency resolution, so using the corrected parameter estimation results is not expected to make a significant difference in the POL results.
E-MM-\CWB uses simulated signals generated using posterior samples from parameter estimation to compute a background,
    but this procedure is insensitive to the fine details of the waveform and is thus unaffected by the error.
RT, IMRCT, and inspiral TIGER currently use the previous results with the incorrect calibration, though these will be updated later.
E-MM-\BAYESWAVE uses the incorrect calibration as well, but the calibration enters only through the parameter-estimation results used to determine the stretch of data for the analysis. As such, this analysis should be only minimally affected.
Analyses corrected to the proper calibration showed only a negligible difference in the results, even for tests combining multiple observations.

Previous catalog papers also employed a likelihood function with a small error due to an incorrectly applied window correction, as discussed in more detail in Section~\ref{sec:pemethods}.

Finally, all events were checked for data quality issues, as discussed in Section~4 of \citet{GWTC:Methods} and glitch mitigation has been performed for events when appropriate,
as discussed in Appendix~B of \citet{GWTC:Results}. We have no evidence for data quality affecting the tests presented here.\\

\subpapersection{Methods for Testing GR}\label{sec:pemethods}

\subpapersubsection{Bayesian Inference}
\label{sec:bayes}

The detection of the \ac{GW} events is followed by Bayesian inference
of parameters assuming one or several waveform models within \GR.
This is usually carried out using the \BILBY package \citep{Ashton:2018jfp,Romero-Shaw:2020owr}, which many of our tests also employ, often with the \BILBYTGR plugin \citep{ashton_2025_15676285}.
The detailed description of the parameter estimation within \GR is given in the companion paper \cite{GWTC:Methods}.
We consider a short data segment $d(t)$ around each detected signal and model it as a sum of Gaussian colored noise and the \ac{GW} signal. This model defines the likelihood $p(d| \boldsymbol{\theta}, \mathbb{M})$, such that the residuals should have a Gaussian distribution after subtracting the correctly modeled \ac{GW} signal
(see Section \ref{sec:residual} below).  Here we assume that the observed data are described by
a model $\mathbb{M}$ parameterized by $\boldsymbol{\theta}$. For example, in several tests
we consider a model extended beyond \GR parameterized by $\boldsymbol{\theta}_{\rm{GR}}, \boldsymbol{\theta}_{\rm{nGR}}$.
The likelihood for a network of detectors is given as a product of
likelihoods, assuming that the measurements are independent.

We use agnostic priors $p(\boldsymbol{\theta}|\mathbb{M})$.  In general, we choose uniform priors over a range that is wide enough to cover the region of parameter space where the posterior has support, while not compromising the efficiency of the sampling. Specifically, we choose the prior ranges to be wide enough to avoid \emph{railing}, where there is significant posterior probability density right up to at least one prior boundary.
The priors for GR parameters are described in Section~5.5 of \citet{GWTC:Methods}, while the priors for non-GR parameters are specified in the description of each analysis.
The detector calibration uncertainties translate into a possible frequency-dependent shift in the \ac{GW} amplitude and phase which is modeled as a spline with Gaussian prior over its coefficients and marginalized over, as discussed in Section~5.4 of \citet{GWTC:Methods}.
The posterior distribution 
\begin{equation}
	p(\boldsymbol{\theta}| d, \mathbb{M})  = \frac{p(d |\boldsymbol{\theta}, \mathbb{M}) p(\boldsymbol{\theta} | \mathbb{M})}{\mathcal{Z}(d|\mathbb{M})} 
\end{equation}
is sampled using numerical techniques, where $\mathcal{Z}(d|\mathbb{M})$ is the evidence of the considered model.
A central part of several analyses presented here and in the companion papers is to evaluate which model fits better the observed data on the basis of the Bayes factor.  The Bayes factor is equal to the posterior odds with equal prior given to all models, which we always assume:

\begin{equation}
	\mathcal{B}^{\mathbb{M}_1}_{\mathbb{M}_2} = 
	\frac{\mathcal{Z}(d|\mathbb{M}_1)}{\mathcal{Z}(d|\mathbb{M}_2)} =
	\frac
	{	\int p(d |\boldsymbol{\theta}_{\mathbb{M}_1} ,\mathbb{M}_1) p(\boldsymbol{\theta}_{\mathbb{M}_1}  | \mathbb{M}_1)\;
		 \mathrm{d}\boldsymbol{\theta}_{\mathbb{M}_1} }
	{	\int p(d |\boldsymbol{\theta}_{\mathbb{M}_2} , \mathbb{M}_2) p(\boldsymbol{\theta}_{\mathbb{M}_2}  | \mathbb{M}_2 )\; \mathrm{d}\boldsymbol{\theta}_{\mathbb{M}_2} }.
\end{equation}
In particular, the Bayes factor between GR and extended beyond GR (nGR) models is given as
\begin{equation}
	\label{eq:GR_nGR_bayes_factor}
\mathcal{B}^{\rm{nGR}}_{\rm{GR}} = \frac
	{	
		\int p(d |\boldsymbol{\theta}_{\rm{GR}}, \boldsymbol{\theta}_{\rm{nGR}}) p(\boldsymbol{\theta}_{\rm{GR}},\boldsymbol{\theta}_{\rm{nGR}})\;
		\!\mathrm{d}\boldsymbol{\theta}_{\rm{GR}}\; \!\mathrm{d}\boldsymbol{\theta}_{\rm{nGR}} 
		}
	{
		\int p(d |\boldsymbol{\theta}_{\rm{GR}}) p(\boldsymbol{\theta}_{\rm{GR}})\;
		\!\mathrm{d}\boldsymbol{\theta}_{\rm{GR}}
	}
\end{equation}

We typically use nested sampling \citep{Skilling:2006} to evaluate the evidence for the considered models.
We use two implementations of the nested-sampling algorithm.
Most of the tests used \DYNESTY \citep{Speagle:2020spe} embedded in the \BILBY{} package \citep{Ashton:2018jfp, Romero-Shaw:2020owr}.
 Another  implementation  of nested sampling, \soft{cpnest} \citep{cpnest}, was used only by \soft{pyRing}.
 During sampling, SMA, TIGER, and SIM analytically marginalized out the distance and time of coalescence \citep{Romero-Shaw:2020owr}; LOSA marginalized out the distance and coalescence phase \citep{Veitch:2014wba};
	while IMRCT, FTI, PCA, and pSEOBNR marginalized out only the distance \citep{GWTC:Methods}.
Among all the tests working within the Bayesian framework, only the two analyses using \BAYESWAVE (RT and E-MM-\BAYESWAVE) performed the trans-dimensional sampling using a reversible-jump Markov Chain Monte Carlo algorithm \citep{Cornish:2014kda}.

The results of the parameter estimation are presented as a set of samples for each model
  $\{\theta_i(\mathbb{M})\}$
  as well as a single number that quantifies the agreement with \GR.
For some analyses this number is a Bayes factor, while for others where \GR is a nested model of a more general model (like in the example above),
        it is the quantile at which GR is found for the marginalized distributions,
			with smaller values indicating better consistency with \GR.
	For some tests examining a two-dimensional deviation space, we also quote the two-dimensional \GR quantile $\QGR$,
		denoting the fraction of the posterior enclosed by the isoprobability contour that passes through the \GR value.
        Thus, smaller values of $\QGR$ indicate better consistency with \GR. The same definition holds for higher-dimensional \GR quantiles.

\subpapersubsection{Waveform Models and Automation}
\label{sec:waveform models}

The deviations from GR are implemented on top of the GR waveform models.
 RT, IMRCT, SMA, TIGER, PCA, SIM, MDR, SSB, and E-WFM have used the \IMRPhenomXPHM model~\citep{Pratten:2020ceb} as the default, usually in its \IMRPhenomXPHMST version~\citep{Colleoni:2024knd}.
 Thus, we will use \IMRPhenomXPHM as a shorthand for \IMRPhenomXPHMST in these papers, and use \IMRPhenomXPHMMSA to specify the original model from \citet{Pratten:2020ceb} which is used by SIM.
 FTI, PCA, and SIM have used the \SEOBNRFIVEHMROM model~\citep{Pompili:2023tna} and pSEOBNR uses \SEOBNRFIVEPHM~\citep{Ramos-Buades:2023ehm}.
 MDR has also used \SURSEVENDQFOUR \citep{Varma:2019csw} for tests of waveform systematic errors.
 In addition, the FTI analysis also used  \SEOBNRFOURNRtidalTWONSBH~\citep{Matas:2020wab}  for GW230518\_125908.
 The analysis performed by LOSA is based on \IMRPhenomXP~\citep[in its original version, which we refer to as \IMRPhenomXPMSA, for clarity]{Pratten:2020ceb}, \IMRPhenomXPNRTidalTWO~\citep{Colleoni:2023ple}, and \IMRPhenomNSBH~\citep{Thompson:2020nei}.
 A detailed description of these models and references is given in \citet{GWTC:Methods}.
 The pre-O4 analysis used earlier versions of these waveform models, as described in \citet{GWTC3}.

Some tests of GR require multiple runs probing a discrete set of models.
In order to avoid human error in preparing and conducting these runs, we have used automation through the
  \soft{Asimov} software library \citep{Williams:2022pgn} together with \soft{CBCFlow}~\citep{cbcflow} for fetching the metadata for some analyses.

We also use \emph{injections}, i.e., simulated signals (either with or without noise), to check the performance of these analyses, and in particular to assess potential deviations from \ac{GR}.

\subpapersubsection{Windowing and the Likelihood}
\label{sec:windowing and likelihood}

Late in the paper preparation process, we discovered an error in the likelihood functions used in our
analysis \citep[see Section~5.10 in][]{GWTC:Methods,Talbot:2025vth}.
We incorrectly applied a correction factor intended to compensate for the power loss when applying the Tukey window to our data.
Due to the late discovery of the likelihood issue, we were not able to correct all analyses. POL, \soft{pyRing}, QNMRF, and E-MM-\CWB are unaffected.
We were able to correct SMA, FTI, TIGER, PCA, SIM, MDR, SSB, and pSEOBNR by reweighting the posteriors. IMRCT, LOSA, and E-WFM analyses were rerun
with the correct likelihood, although pre-O4 IMRCT results are not yet corrected.
The RES and E-MM-\BAYESWAVE analyses are left with the incorrect likelihood and will be corrected later.
However, using the correct likelihood leads only to a small difference in
the posteriors, although they are systematically wider and the Bayes factor with respect to the noise systematically lower.
As an example, for the MDR analysis, the Bayes factors are $21\%$ lower, the posteriors have standard deviations larger by $7\%$,
and their means move by $2\%$ of standard deviation on average for O4a results.
The error also applies to
previous analyses from O1 to O3, but with lesser effects due to different window settings used.
For MDR, the Bayes factors, standard deviations, and means for the GWTC-3.0 results move by $1\%$, $4\%$, and $1\%$ respectively.
These changes do not modify the overall conclusion of our tests, i.e., that we have no evidence that GR is violated.

\subpapersubsection{Hierarchical Inference}
\label{sec:hier}

With the ever-increasing number of \ac{GW} observations, it is important to accurately combine information
across the detections made to place the most stringent bounds on deviations from \ac{GR}.
With this in mind, we employ hierarchical inference techniques to construct a summary of the underlying distribution of possible 
deviations present in the data~\citep{Isi:2019asy,Mandel:2018mve,Zimmerman:2019wzo}. For cases where the parameterized effect is
identical across observations, such as in propagation tests,
the combined constraint is summarized by single measure of the parameterized effect from all observations. However, in the case where
individual observations may not share a common parameter, such as in null tests like those for deviations in \ac{PN} coefficients, we instead infer the
structure of a Gaussian distribution fit to the collection of measured deviation parameters from all events, 
parameterized with a mean $\mu$ and standard deviation $\sigma$
\citep{Isi:2019asy}. If multiple deviation parameters are measured and hierarchically combined, we infer the full multi-dimensional
Gaussian structure and thus have correlation coefficients $\rho$ along with the means and standard deviations~\citep{Zhong:2024pwb}.
This is the first application of the multi-dimensional hierarchical inference technique in an LVK testing GR analysis.

\vfill \break

\subpapersection{Tests of Consistency}\label{sec:consistency}

\subpapersubsection{Residual test}\label{sec:residual}

The residual test involves checking for excess coherent power remaining in the detector network after the best-fit GR template is subtracted from the data.
If GR is the correct theory of gravitation, a GR-based template should capture all the features of the astrophysical signal and thus the residuals should be consistent with instrumental noise.
We use the same method as in previous analyses \citep{LIGOScientific:2019fpa,LIGOScientific:2020tif, LIGOScientific:2021sio}.

The noise in ground based GW detectors comes from various sources \citep{TheLIGOScientific:2014jea,TheLIGOScientific:2016zmo}. Detector noise is assumed to be stationary and Gaussian,
so the data time series $\boldsymbol{d}(t)$ is composed of the Gaussian noise $\boldsymbol{n}(t)$ and the GW model waveform $\boldsymbol{h}(t)$, and can be modeled as:
\begin{equation}
  \boldsymbol{d}(t) = \boldsymbol{h}(t) + \boldsymbol{n}(t),
\end{equation}
where the boldface notation indicates a quantity with a component corresponding to each detector.
Here we choose for the best-fit model of the signal the set of parameters that maximize the likelihood of observing the recorded data under the assumption that the signal is present in the data. These maximum-likelihood parameters minimize the difference between the data $\boldsymbol{d}(t)$ and the template $\boldsymbol{h}(t)$.%

Once the best-fit (maximum-likelihood) waveform $\boldsymbol{h}_{\mathrm{maxL}}(t)$ is inferred using a GR template, where we use \IMRPhenomXPHM for the O4a events, the residual $\boldsymbol{r}(t) = \boldsymbol{d}(t) - \boldsymbol{h}_{\mathrm{maxL}}(t)$ can be obtained. If $\boldsymbol{h}_{\mathrm{maxL}}$ is an accurate estimate of the signal based on GR, the residual $\boldsymbol{r}(t)$ should be consistent with noise. This is tested by analyzing the residual using \BAYESWAVE \citep{Cornish:2014kda,Cornish:2020dwh,bayeswave}. \BAYESWAVE uses a template-independent model based on Morlet--Gabor wavelets to characterize any coherent feature in the detector network. \BAYESWAVE produces a discretized probability distribution in the parameter space of the wavelets, with each point corresponding to a residual that has a well defined SNR.  Any loud multi-detector coherent features in the data $\boldsymbol{d}(t)$ that were not captured by the GR based model $\boldsymbol{h}(t)$ are reconstructed. If the true underlying signal was reliably reconstructed by the maximum-likelihood waveform $\boldsymbol{h}_{\mathrm{maxL}}$, the \BAYESWAVE reconstruction will have a median that is consistent with noise \citep{Johnson-McDaniel:2021yge}.

\begin{figure}
        \begin{center}
        \includegraphics[width=\TGRFigureWidth]{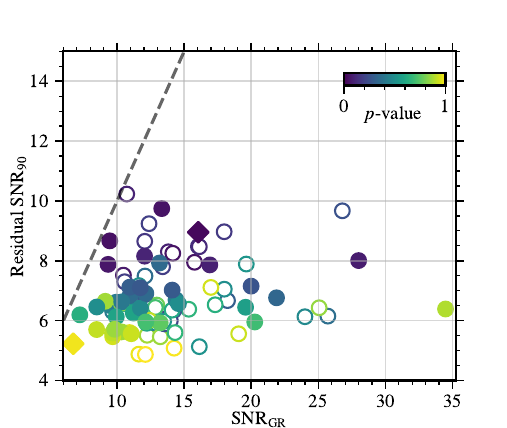}
        \end{center}
        \caption{
        Results of the residuals analysis. Scatter plot of the \ac{SNR} of the maximum-likelihood template ($\mathrm{SNR_{GR}}$) and the upper limit on the residual network \ac{SNR} ($\mathrm{SNR_{90}}$) for each event. The color bar denotes the \textit{p}-values of individual 
        events. Solid (empty) circles represent the O4a (pre-O4a) events, where the pre-O4a events are from \citet{LIGOScientific:2019fpa,LIGOScientific:2020tif, LIGOScientific:2021sio}. The events with the highest and lowest \textit{p}-values are shown by diamonds. The dashed 
        line represents $\mathrm{SNR_{GR}}=\mathrm{SNR_{90}}$. We see that $\mathrm{SNR_{GR}} \geq \mathrm{SNR_{90}}$ for all events considered.
  }
\label{fig:res_scatter}
\end{figure}

To quantify the results, we describe the loudness of the residual by calculating the $90$th percentile of the network \ac{SNR} distribution produced by \BAYESWAVE, $\mathrm{SNR}_{\mathrm{90}}$. That is, there is a $90\%$ probability that after subtracting $\boldsymbol{h}_{\mathrm{maxL}}$, the residual signal has an optimal network SNR $\leq$ $\mathrm{SNR}_{\mathrm{90}}$. For the case of Gaussian noise in the LIGO Hanford, LIGO Livingston, and Virgo network, this tends to be $\lesssim 5$ \citep{Johnson-McDaniel:2021yge}.

We perform a more systematic assessment by defining a \textit{p}-value under the null hypothesis that the residual is consistent with noise. This is done by running identical \BAYESWAVE analysis on $200$ randomly selected data segments around the event that cover a time window of $4096$~s symmetric about the event time and do not contain a known GW signal or any transient non-Gaussian features, and then calculating the probability that instrumental noise alone could produce an $\mathrm{SNR}_{\mathrm{90}}^{n}$, where $\mathrm{SNR}_{\mathrm{90}}^{n} \geq \mathrm{SNR}_{\mathrm{90}}$, giving the \textit{p}-value $= P(\mathrm{SNR}_{\mathrm{90}}^{n} \geq \mathrm{SNR}_\mathrm{90})$. A higher chance for the residual power to originate from instrumental noise will be reflected in a higher \textit{p}-value, whereas if the residual power is less likely to come from noise alone, the \textit{p}-value would be smaller.

Finally, we can quantify how well the GR based template fits the signal in the data. Since our GR templates are not perfect, and we have a probabilistic measurement of the signal in the data, we can quantify this fit using the $90\%$ lower bound on the fitting factor between the model and the signal:
\begin{equation}
    \mathrm{FF}_{\mathrm{90}} = \dfrac{\mathrm{SNR}_{\mathrm{GR}}}{\sqrt{\mathrm{SNR}^2_{\mathrm{GR}} + \mathrm{SNR}^2_{\mathrm{90}}}},
\end{equation}
where $\mathrm{SNR}_{\mathrm{GR}}$ is the optimal network SNR for $\boldsymbol{h}_{\mathrm{maxL}}$.
A value of $\mathrm{FF_{\mathrm{90}}} = 1$ indicates perfect agreement between the GR template and the signal in the data.

\begin{figure}
	\begin{center}
	\includegraphics[width=\TGRFigureWidth]{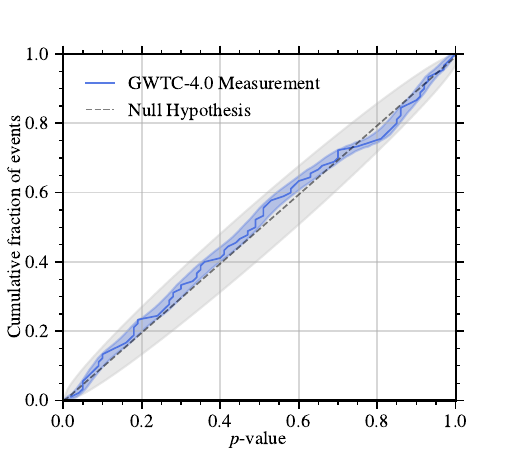}
	\end{center}
	\caption{
	Results of the residuals analysis. The blue line
  shows the fraction of events with \textit{p}-values of the residual SNR less
  than or equal to the value of the abscissa (PP plot). The 90\% credible interval is represented by the shaded blue region due to the finite number of noise-only instantiations. The null hypothesis that the \textit{p}-value is uniformly distributed over $[0,1]$ is shown by the diagonal dashed line, with the shaded light grey area denoting the 90\% credible interval around the null hypothesis due to the finite number of events.
  }
\label{fig:res_pp}
\end{figure}

All events that pass this paper's selection criteria and are analyzed with \BBH waveforms are considered. We summarize the results for $\mathrm{SNR_{GR}}$, the residual $\mathrm{SNR_{90}}$, $\mathrm{FF_{90}}$, and \textit{p}-values for each event in Table~\ref{residuals_table}.
Figure~\ref{fig:res_scatter} shows $\mathrm{SNR_{90}}$ versus $\mathrm{SNR_{GR}}$ for both O4a events and pre-O4 events. We confirm that the residual $\mathrm{SNR_{90}}$ is always smaller than the original $\mathrm{SNR_{GR}}$. For O4a events, GW230919\_215712 has the lowest \textit{p}-value of $0.02$ with a residual $\mathrm{SNR_{90}} = 8.95$, while GW231223\_032836 has the highest \textit{p}-value of $0.98$ with a residual $\mathrm{SNR_{90}} = 5.23$. GW231001\_140220 has the lowest $\mathrm{FF_{90}}$ of 0.75, and a low \textit{p}-value of $0.04$. The mean $\mathrm{FF_{90}}$ for O4a events is $0.87$, which indicates agreement between the GR templates and the signals in the data.

If our best-fit template has successfully recovered the GW signals, then we expect that the contribution of the residual $\mathrm{SNR_{90}}$ comes from the detector noise.
This means that the \textit{p}-values should be distributed uniformly within $[0,1]$.
To confirm the distribution of \textit{p}-values, we show the probability--probability (PP) plot in Figure~\ref{fig:res_pp}. The measurement of \textit{p}-values is subject to uncertainty due to finite number of background runs. If the \textit{p}-value $\hat{p} = n/N $ is estimated by running on $N$ background segments, of which $n$ produce $\mathrm{SNR_{90}}$ greater than that of the event, the likelihood of the estimated \textit{p}-values is given by a binomial function,
\begin{equation}
  \mathcal{L}(\hat{p}) = \left( \begin{array}{c}
    N \\
    n
    \end{array} \right) p^n (1-p)^{N-n} .
\end{equation}

Assuming a uniform prior distribution for the \textit{p}-value, the posterior is given by a beta distribution,
\begin{equation}
  P(p|N,n) = \mathrm{Beta}(n+1, N-n+1).
\end{equation}
In Figure~\ref{fig:res_pp}, the shaded region around the PP curve represents the $90\%$ credible interval of the \textit{p}-value posterior.

\begin{table}[H]
    \caption{\label{residuals_table}%
    Results from the residuals analysis for O4a events.
    }
    \begin{ruledtabular}
        \begin{tabular}{lcccc}
        \textrm{Events}&
        \textrm{$\mathrm{SNR_{GR}}$}&
        \textrm{Res. $\mathrm{SNR_{90}}$}&
        \textrm{$\mathrm{FF_{90}}$}&
        \textrm{\textit{p}-value}\\
        \colrule
        GW230518\_125908 & 12.01 & 6.86 & 0.87 & 0.28 \\
        GW230601\_224134 & 13.33 & 9.74 & 0.81 & 0.05 \\
        GW230605\_065343 & 10.65 & 6.56 & 0.85 & 0.53 \\
        GW230606\_004305 & 10.92 & 5.60 & 0.89 & 0.95 \\
        GW230609\_064958 & 9.51 & 6.50 & 0.83 & 0.44 \\
        GW230624\_113103 & 9.93 & 6.61 & 0.83 & 0.41 \\
        GW230627\_015337 & 28.00 & 8.01 & 0.96 & 0.08 \\
        GW230628\_231200 & 14.39 & 6.70 & 0.91 & 0.42 \\
        GW230630\_234532 & 10.33 & 5.62 & 0.88 & 0.86 \\
        GW230702\_185453 & 8.47 & 5.70 & 0.83 & 0.91 \\
        GW230731\_215307 & 10.97 & 6.84 & 0.85 & 0.35 \\
        GW230811\_032116 & 14.57 & 6.57 & 0.91 & 0.49 \\
        GW230814\_061920 & 9.12 & 6.65 & 0.81 & 0.86 \\
        GW230824\_033047 & 10.32 & 6.57 & 0.84 & 0.82 \\
        GW230904\_051013 & 10.35 & 6.63 & 0.84 & 0.40 \\
        GW230914\_111401 & 16.91 & 7.86 & 0.91 & 0.09 \\
        GW230919\_215712 & 16.05 & 8.95 & 0.87 & 0.02 \\
        GW230920\_071124 & 11.21 & 6.25 & 0.87 & 0.58 \\
        GW230922\_020344 & 11.74 & 6.42 & 0.88 & 0.47 \\
        GW230922\_040658 & 9.94 & 6.13 & 0.85 & 0.58 \\
        GW230924\_124453 & 12.14 & 6.91 & 0.87 & 0.34 \\
        GW230927\_043729 & 9.33 & 7.88 & 0.77 & 0.06 \\
        GW230927\_153832 & 19.99 & 7.15 & 0.94 & 0.26 \\
        GW230928\_215827 & 7.22 & 6.19 & 0.77 & 0.63 \\
        GW231001\_140220 & 9.45 & 8.67 & 0.75 & 0.04 \\
        GW231020\_142947 & 10.95 & 7.12 & 0.84 & 0.30 \\
        GW231028\_153006 & 20.26 & 5.95 & 0.96 & 0.69 \\
        GW231102\_071736 & 13.16 & 7.93 & 0.86 & 0.34 \\
        GW231104\_133418 & 9.67 & 5.46 & 0.87 & 0.93 \\
        GW231108\_125142 & 13.21 & 5.92 & 0.91 & 0.75 \\
        GW231110\_040320 & 9.66 & 5.59 & 0.87 & 0.90 \\
        GW231113\_200417 & 11.06 & 5.56 & 0.89 & 0.92 \\
        GW231114\_043211 & 8.49 & 6.46 & 0.80 & 0.49 \\
        GW231118\_005626 & 11.73 & 7.10 & 0.86 & 0.27 \\
        GW231118\_090602 & 12.05 & 8.15 & 0.83 & 0.10 \\
        GW231123\_135430 & 19.58 & 6.45 & 0.95 & 0.56 \\
        GW231206\_233134 & 12.14 & 5.96 & 0.90 & 0.70 \\
        GW231206\_233901 & 21.88 & 6.76 & 0.96 & 0.33 \\
        GW231213\_111417 & 9.82 & 5.71 & 0.86 & 0.85 \\
        GW231223\_032836 & 6.74 & 5.23 & 0.80 & 0.98 \\
        GW231224\_024321 & 14.11 & 7.02 & 0.90 & 0.28 \\
        GW231226\_101520 & 34.48 & 6.39 & 0.98 & 0.88 \\
        \end{tabular}
    \end{ruledtabular}

    \tablecomments{For each event we list the SNR of the best-fit waveform ($\mathrm{SNR_{GR}}$),
                    the 90\% credible upper limit on the residual coherent network SNR ($\mathrm{SNR_{90}}$),
                    the 90\% credible lower limit on the fitting factor $\mathrm{FF_{90}}$,
                    and the \textit{p}-value calculated from the background analysis.}
\end{table}

While the residuals analysis is in principle sensitive to any of the deviations from \GR to which the other tests considered are sensitive, in practice it is less sensitive to a given \GR deviation than a more specific test, particularly when the deviation is not well localized in time, as is the case for low-mass compact binary signals \citep{Johnson-McDaniel:2021yge}.

\subpapersubsection{Inspiral--merger--ringdown consistency test}
\label{sec:imrct}
The IMR consistency test checks for the consistency of the estimates of the final mass and final spin of the remnant BH inferred independently from the low- and high-frequency parts of the GW signal emitted by a BBH merger \citep{Hughes:2004vw, Ghosh:2016qgn, Ghosh:2017gfp}.
For the current implementation of the IMR consistency test, we first estimate the component masses, dimensionless spins, and spin angles from both the low- and high-frequency portions of the signal. Next, we apply NR-calibrated fits on the masses and spins inferred for each frequency regime to compute the final mass and final spin of the remnant BH \citep{Hofmann:2016yih, Healy:2016lce, PhysRevD.95.064024}, where the specifics of the calculation are described in \citet{GWTC:Methods}. If the observed signal is correctly described by a quasi-circular \BBH coalescence in GR, then the estimates of the mass and spin of the remnant BH from the two frequency regimes will be in agreement. However, any deviations from GR could lead to discrepancies between the two estimates. Some studies have shown that the IMR consistency test leads to false violations of GR for simulated GW signals with unequal masses along with either high spin or near-face-on inclinations \citep{Narayan:2023vhm, Narayan:2024rat}. The fraction of events affected by this bias is expected to be negligible in GWTC-4.0, and could be quantified by a performing a comprehensive set of simulations for future catalogs.

The separation between the low- and high-frequency parts of the signal is rather arbitrary as long as both parts have sufficient SNRs. In this analysis, we divide the signal in the frequency domain at the cutoff frequency $f_\text{c}^{\rm IMR}$, the dominant multipole GW frequency of the innermost stable circular orbit of the remnant Kerr BH \citep{Bardeen:1972fi}. To determine this, we compute the posterior distribution on the cutoff frequency from the posteriors on the masses and spins inferred using the full IMR signal. The median of the cutoff frequency is then used to split the GW signal. This is different than previous analyses where the cutoff frequency was calculated from the medians of the component masses and spins inferred from the full IMR signal \citep{LIGOScientific:2019fpa, LIGOScientific:2020tif, LIGOScientific:2021sio}, though it only leads to a $\lesssim 1$~Hz difference in the results.

The low- and high-frequency regimes roughly correspond to the inspiral and post-inspiral, respectively, of the dominant mode of the waveform. To make sure that the two regimes of the signal have enough information, we calculate the SNR of the inspiral and the post-inspiral parts of the waveform for each event using their maximum a posteriori parameter values obtained from the full IMR signal.

We only perform the test on those events that have SNRs greater than $6$ in both the inspiral and the post-inspiral parts \citep{LIGOScientific:2019fpa, LIGOScientific:2020tif, LIGOScientific:2021sio}. We also impose an extra constraint on the median value of the total redshifted mass $(1+z) M < 100\, M_{\odot}$ to ensure enough inspiral signal for heavier BBHs \citep{LIGOScientific:2020tif, LIGOScientific:2021sio}.
We find that 13 events satisfy the selection criteria in O4a. The SNRs for the full IMR,  inspiral, and post-inspiral regimes (for the maximum a posteriori waveform) along with the cutoff frequency $f_\text{c}^{\rm IMR}$ of the events analyzed are reported in Table~\ref{tab:imrct_params}.

We start by estimating the posterior distributions on the mass $M_{\rm f}$ and the dimensionless spin $\chi_{\rm f}$ of the remnant BH from both the inspiral and the post-inspiral parts of the signal.
To quantify the consistency between the two estimates, or a possible deviation from GR,
	we define two fractional deviation parameters $\Delta M_{\rm f}/ \bar{M}_{\rm f}$ and $\Delta \chi_{\rm f}/\bar{\chi}_{\rm f}$:
\begin{align}
\dfrac{\Delta M_{\rm f}}{\bar{M}_{\rm f}} &= 2 \dfrac{M_{\rm f}^{\rm I}-M_{\rm f}^{\rm PI}}{M_{\rm f}^{\rm I}+M_{\rm f}^{\rm PI}} \,, \quad \dfrac{\Delta \chi_{\rm f}}{\bar{\chi}_{\rm f}} = 2 \dfrac{\chi_{\rm f}^{\rm I} - \chi_{\rm f}^{\rm PI}}{\chi_{\rm f}^{\rm I} + \chi_{\rm f}^{\rm PI}},
\end{align}
where $\bar{M}_{\rm f}$ and $\bar{\chi}_{\rm f}$ denote the averages of the final mass and final spin obtained from analyzing the inspiral and post-inspiral parts of the signal, respectively.
The superscripts here refer to the inspiral (I) and the post-inspiral (PI) portions of the signal. 
The two-dimensional posterior distribution of these fractional deviation parameters should peak around $(0,0)$ if the observed signal corresponds to a quasi-circular BBH merger described by GR.

\begin{table}
\caption{
    Results from the IMR consistency test for O4a events.
}
\label{tab:imrct_params}
\begin{center}
\begin{tabular}{l@{~} c@{\quad}c@{\quad}r@{\quad}r@{\quad} r@{\quad} r}
\toprule
Event           & $(1\!+\!z)M$ & $f_\text{c}^{\rm IMR}$ & $\rho_\mathrm{IMR}$ & $\rho_\mathrm{I}$ & $\rho_\mathrm{PI}$ & $\QGR$ \\
 & [$M_\odot$] & [Hz] & & & & [\%] \\
\midrule

\!\!GW230606\_004305 & \totalmassdetuncert{GW230606_004305} & \EVENTSELECTION{S230606dFCIMR}   &   \EVENTSELECTION{S230606dOPTSNR}  & \EVENTSELECTION{S230606dOPTSNRPREIMR} &    \EVENTSELECTION{S230606dOPTSNRPOSTIMR}   &  \TGRImrctEVENTSTATS{S230606dGRQUANTGWTC4}\phantom{${}^*$} \\
\!\!GW230609\_064958* & \totalmassdetuncert{GW230609_064958} &  \EVENTSELECTION{S230609uFCIMR}   &   \EVENTSELECTION{S230609uOPTSNR}  & \EVENTSELECTION{S230609uOPTSNRPREIMR} &    \EVENTSELECTION{S230609uOPTSNRPOSTIMR}  &    \nodata   \\
\!\!GW230628\_231200 & \totalmassdetuncert{GW230628_231200} &  \EVENTSELECTION{S230628axFCIMR}  &   \EVENTSELECTION{S230628axOPTSNR}  & \EVENTSELECTION{S230628axOPTSNRPREIMR} &    \EVENTSELECTION{S230628axOPTSNRPOSTIMR}    &  \TGRImrctEVENTSTATS{S230628axGRQUANTGWTC4}\phantom{${}^*$} \\
\!\!GW230811\_032116 & \totalmassdetuncert{GW230811_032116} &  \EVENTSELECTION{S230811nFCIMR}   &   \EVENTSELECTION{S230811nOPTSNR}  & \EVENTSELECTION{S230811nOPTSNRPREIMR} &    \EVENTSELECTION{S230811nOPTSNRPOSTIMR}    &  \TGRImrctEVENTSTATS{S230811nGRQUANTGWTC4}\phantom{${}^*$} \\
\!\!GW230919\_215712 & \totalmassdetuncert{GW230919_215712} &  \EVENTSELECTION{S230919bjFCIMR}  &   \EVENTSELECTION{S230919bjOPTSNR}  & \EVENTSELECTION{S230919bjOPTSNRPREIMR} &    \EVENTSELECTION{S230919bjOPTSNRPOSTIMR}    &  \TGRImrctEVENTSTATS{S230919bjGRQUANTGWTC4}\phantom{${}^*$} \\
\!\!GW230920\_071124 & \totalmassdetuncert{GW230920_071124} &  \EVENTSELECTION{S230920alFCIMR}  &   \EVENTSELECTION{S230920alOPTSNR}  & \EVENTSELECTION{S230920alOPTSNRPREIMR} &    \EVENTSELECTION{S230920alOPTSNRPOSTIMR}  &  \TGRImrctEVENTSTATS{S230920alGRQUANTGWTC4}\phantom{${}^*$} \\
\!\!GW230922\_020344 & \totalmassdetuncert{GW230922_020344} &  \EVENTSELECTION{S230922gFCIMR}   &   \EVENTSELECTION{S230922gOPTSNR}  & \EVENTSELECTION{S230922gOPTSNRPREIMR} &    \EVENTSELECTION{S230922gOPTSNRPOSTIMR}   &  \TGRImrctEVENTSTATS{S230922gGRQUANTGWTC4}\phantom{${}^*$} \\
\!\!GW230924\_124453 & \totalmassdetuncert{GW230924_124453} &  \EVENTSELECTION{S230924anFCIMR}  &   \EVENTSELECTION{S230924anOPTSNR}  & \EVENTSELECTION{S230924anOPTSNRPREIMR} &    \EVENTSELECTION{S230924anOPTSNRPOSTIMR}  & \TGRImrctEVENTSTATS{S230924anGRQUANTGWTC4}\phantom{${}^*$} \\
\!\!GW230927\_153832 & \totalmassdetuncert{GW230927_153832} &  \EVENTSELECTION{S230927beFCIMR}  &   \EVENTSELECTION{S230927beOPTSNR}  & \EVENTSELECTION{S230927beOPTSNRPREIMR} &    \EVENTSELECTION{S230927beOPTSNRPOSTIMR}  & \TGRImrctEVENTSTATS{S230927beGRQUANTGWTC4}\phantom{${}^*$} \\
\!\!GW231108\_125142 & \totalmassdetuncert{GW231108_125142} &  \EVENTSELECTION{S231108uFCIMR}   &   \EVENTSELECTION{S231108uOPTSNR}  & \EVENTSELECTION{S231108uOPTSNRPREIMR} &    \EVENTSELECTION{S231108uOPTSNRPOSTIMR}  &  \TGRImrctEVENTSTATS{S231108uGRQUANTGWTC4}\phantom{${}^*$} \\
\!\!GW231206\_233134 & \totalmassdetuncert{GW231206_233134} &  \EVENTSELECTION{S231206caFCIMR}  &   \EVENTSELECTION{S231206caOPTSNR}  & \EVENTSELECTION{S231206caOPTSNRPREIMR} &    \EVENTSELECTION{S231206caOPTSNRPOSTIMR}   &  \TGRImrctEVENTSTATS{S231206caGRQUANTGWTC4}\phantom{${}^*$} \\
\!\!GW231206\_233901 & \totalmassdetuncert{GW231206_233901} &  \EVENTSELECTION{S231206ccFCIMR}  &   \EVENTSELECTION{S231206ccOPTSNR}  & \EVENTSELECTION{S231206ccOPTSNRPREIMR} &    \EVENTSELECTION{S231206ccOPTSNRPOSTIMR} & \TGRImrctEVENTSTATS{S231206ccGRQUANTGWTC4}\phantom{${}^*$}  \\
\!\!GW231226\_101520 & \totalmassdetuncert{GW231226_101520} &  \EVENTSELECTION{S231226avFCIMR}  &   \EVENTSELECTION{S231226avOPTSNR}  & \EVENTSELECTION{S231226avOPTSNRPREIMR} &    \EVENTSELECTION{S231226avOPTSNRPOSTIMR}  & \TGRImrctEVENTSTATS{S231226avGRQUANTGWTC4}\phantom{${}^*$} \\
\bottomrule
\end{tabular}
\end{center}
\tablecomments{The median and $90\%$ credible interval of the redshifted total mass come from the analysis in \citet{GWTC:Results}; $f_\text{c}^{\rm IMR}$ denotes the cutoff frequency between the inspiral and post-inspiral regimes;
					$\rho_\mathrm{IMR}$, $\rho_\mathrm{I}$, and $\rho_\mathrm{PI}$ are the SNR in the full signal,
					the inspiral part, and the post-inspiral part respectively;
					and the GR quantile $\QGR$ is defined in Section \ref{sec:bayes} (and here is obtained from the reweighted posterior).
                                        As discussed in the text, we were unable to obtain reliable results for GW230609\_064958 due to its low $\rho_\mathrm{I}$.
}

\end{table}

\begin{figure}
	\begin{center}
	\includegraphics[width=\TGRFigureWidth]{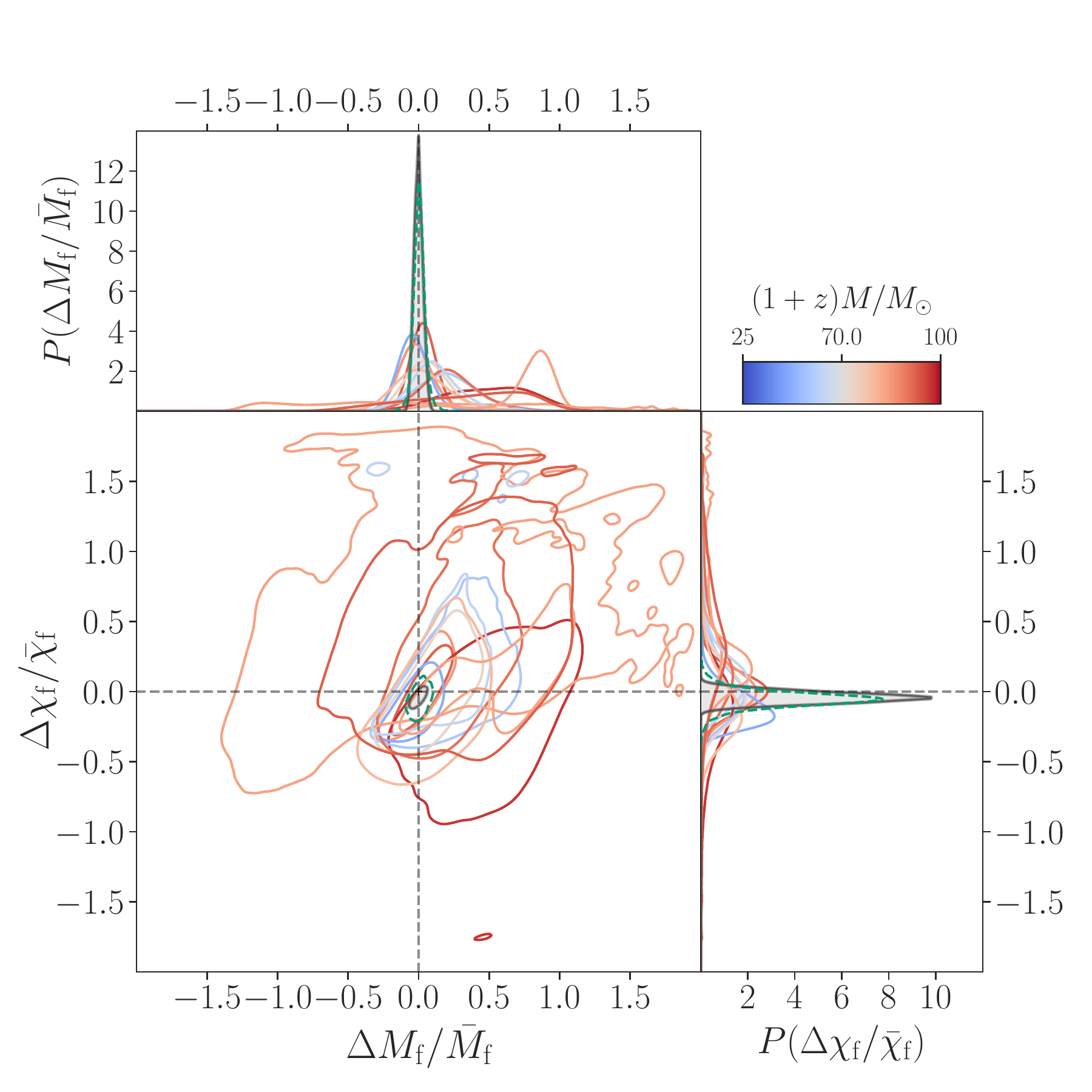}
	\end{center}
	\caption{
		The main panel shows the 90\% credible regions of the 2D posteriors on ($\Delta M_{\rm f}/ \bar{M}_{\rm f}, \Delta \chi_{\rm f}/\bar{\chi}_{\rm f})$ assuming a uniform prior, with $(0, 0)$ being the expected value for GR.
		The side panels show the marginalized posterior on $\Delta M_{\rm f}/ \bar{M}_{\rm f}$ and $\Delta \chi_{\rm f}/\bar{\chi}_{\rm f}$. The unfilled solid contours correspond to the individual O4a events, colored by their median redshifted total mass, while the filled contours are the results from the joint posterior.
		The dashed green contours correspond to hierarchically inferred bounds on the fractional deviation parameters.
  }
\label{fig:imrct_posteriors}
\end{figure}

\begin{figure}[tb] 
	\begin{center}
	\includegraphics[width=\columnwidth]{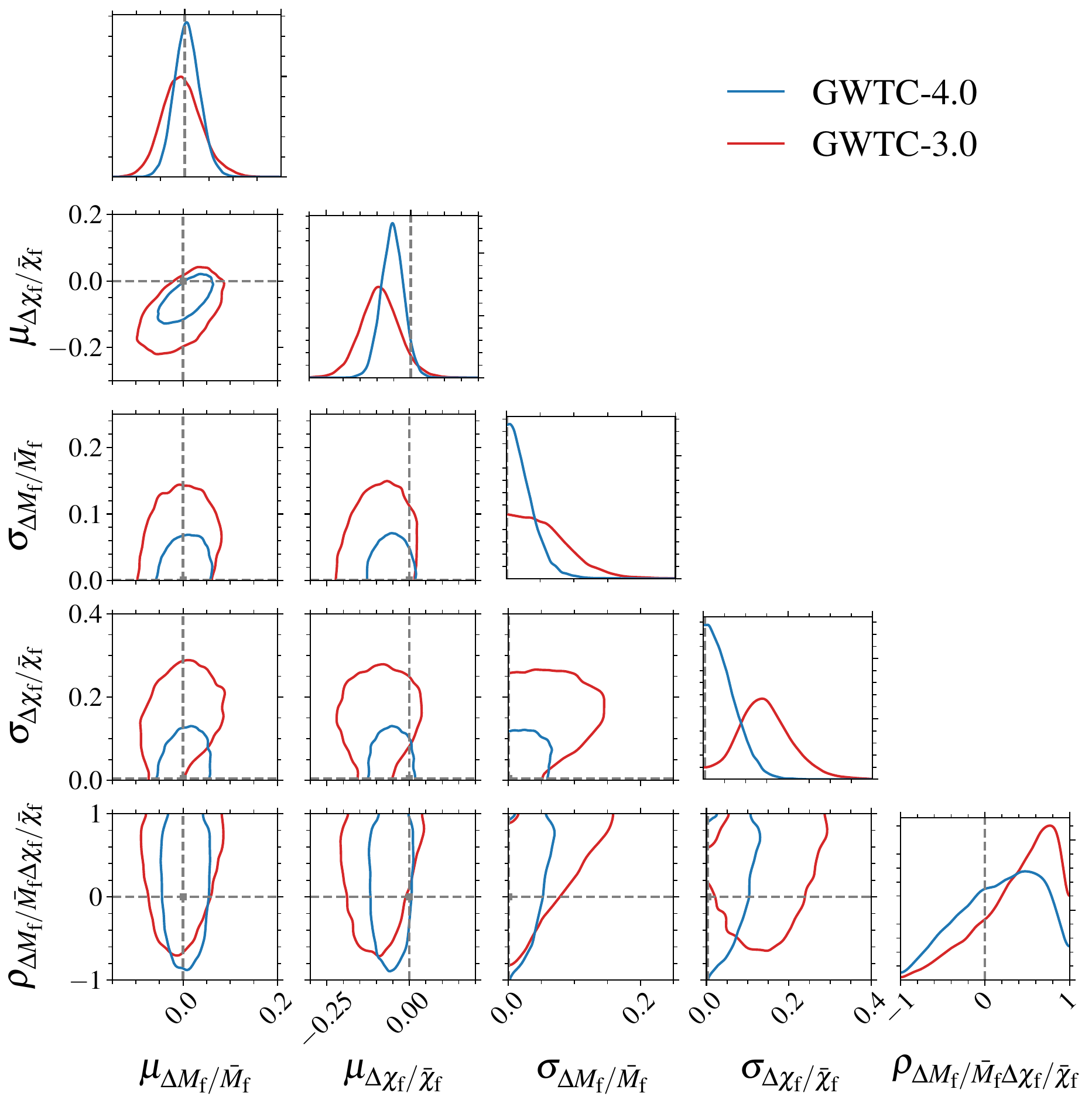}
	\end{center}
	\caption{%
		Result of the two-dimensional hierarchical analyses on $\dMf$ and $\dchif$ joint measurements from GWTC-4.0 events including GW190814.
		The five hyperparameters characterizing the population model are defined in the main text.
		The contours enclose 90\% probability mass.
		The blue and red contours correspond to the estimates using GWTC-4.0 and GWTC-3.0 \citep{Zhong:2024pwb, LIGOScientific:2021sio}.
		GR corresponds to $\mu_{\dMf} = \sigma_{\dMf} = \mu_{\dchif} = \sigma_{\dchif} = 0$ (gray dashed line).
  }
	\label{fig:imrct_hier}
\end{figure}

We use a phenomenological quasi-circular, precessing frequency domain waveform model, \IMRPhenomXPHM, to perform parameter estimation on the inspiral and post-inspiral parts of the signals. We assume uniform priors on the detector-frame component masses and spins. These priors translate into nontrivial priors on $\dMf$ and $\dchif$. Thus, similar to previous analyses \citep{LIGOScientific:2020tif, LIGOScientific:2021sio}, we reweight the posteriors to obtain uniform priors on the deviation parameters. The results are plotted in Figure~\ref{fig:imrct_posteriors}, where the contours correspond to $90\%$ credible regions of the two-dimensional posteriors on the fractional deviation parameters for the O4a events which satisfy our selection criteria. We do not find any deviation from GR for any of the events. The two-dimensional GR quantile values $\QGR$  for the events are given in Table~\ref{tab:imrct_params}. $\QGR$ is defined as the fraction of the posterior enclosed by the isoprobability contour that passes through $(0,0)$, the GR value.
Smaller values indicate better consistency with GR. All events in O4a have $\QGR \leq \TGRImrctEVENTSTATS{S230628axGRQUANTGWTC4}\%$.

We were unable to obtain reliable results for GW230609\_064958 due to its low inspiral SNR, which led to significant railing in the chirp mass and luminosity distance posteriors even when extending the prior range to values well beyond those expected physically. We also found significant railing in the mass-ratio posterior even when extending the prior range to the minimum value of $0.02$ for which \IMRPhenomXPHM is deemed to be able to be extrapolated without committing large errors \citep{GWTC3}. Railing against the lower bound on the mass ratio also affects GW230920\_071124 post-inspiral significantly, where there is also a long tail in the luminosity distance that
has some slight railing against a prior boundary that is well above what is expected physically. However, the railing in the distance is slight enough that we quote results for this event even though the results are affected by our choice of prior boundaries. The railing in these cases with SNRs close to the cutoff suggests that we should increase the minimum SNR in the IMRCT selection criteria.

The deviations are only constrained modestly for most of the events with moderate SNRs. However, one can combine the likelihoods on the deviation parameters for multiple events to improve these constraints. We do this by first interpolating the reweighted posteriors on the fractional deviation parameters $\dMf$ and $\dchif$ of individual events on a grid with bounds $[-2,2]$ for both the parameters, and then multiplying the interpolated posteriors to obtain the joint posterior. Here, we assume the deviation does not lie outside the above range for all events. As shown in gray in Figure~\ref{fig:imrct_posteriors}, the joint posterior on the fractional deviation parameters of GWTC-4.0 events is consistent with the GR prediction with $(\dMf)_\mathrm{joint} = \TGRImrctGWTCFOURResults{DMFGWTC4PHENOM}$ and $(\dchif)_\mathrm{joint} = \TGRImrctGWTCFOURResults{DCHIFGWTC4PHENOM}$.
The GR quantile of the joint posterior is $\TGRImrctGWTCFOURResults{GRQUANTGWTC4}\%$, an increase from the result of $\TGRImrctGWTCFOURResults{GRQUANTGWTC3}\%$ obtained for GWTC-3.0 \citep{LIGOScientific:2021sio}.

Figure~\ref{fig:imrct_hier} shows the posterior distribution for the hyperparameters from the multi-dimensional hierarchical analysis, with contours indicating the 90\% credible regions. The five hyperparameters are the mean and standard deviation of $\dMf$ ($\mu_{\dMf}$, $\sigma_{\dMf}$), the mean and standard deviation of $\dchif$ ($\mu_{\dchif}$, $\sigma_{\dchif}$), and the correlation between the two ($\rho_{\dMf\dchif}$). The GR prediction corresponds to $\mu_{\dMf} = \sigma_{\dMf} = \mu_{\dchif} = \sigma_{\dchif} = 0$, while $\rho_{\dMf\dchif}$ indicates the population-level correlation between $\dMf$ and $\dchif$. The addition of new events from the latest observation run yields a more stringent constraints on the hyperparameters. In the previous analysis using GWTC-3.0 events, the inclusion of the event GW190814 produced a non-zero peak in the $\sigma_{\dchif}$ posterior \citep{Zhong:2024pwb}, as illustrated in Figure~\ref{fig:imrct_hier}. With the new events incorporated, however, the peak shifts back toward zero. We also show the population-marginalized constraint from the hierarchical analysis in Figure~\ref{fig:imrct_posteriors}. The hierarchical inference finds that $(\dMf)_\mathrm{hier} = \TGRImrctHierPopGWTCFOURResults{DMFGWTC4HIERPOP}$ and $(\dchif)_\mathrm{hier} = \TGRImrctHierPopGWTCFOURResults{DCHIFGWTC4HIERPOP}$, and the GR quantile for the hierarchically combined distribution is $\TGRImrctHierPopGWTCFOURResults{GRQUANTHyperParamsFourDGWTC4HIERPOP}\%$, considering both means and standard deviations. However, if one just considers the two means, which show a shift away from GR in Figure~\ref{fig:imrct_hier} similar to that seen in the joint posterior in Figure~\ref{fig:imrct_posteriors}, then one obtains a GR quantile of $\TGRImrctHierPopGWTCFOURResults{GRQUANTMeansTwoDGWTC4HIERPOP}\%$, larger than the one obtained from the joint posterior. We still find that this apparent deviation from GR is driven by GW190814, since if we exclude that event from the combined results, the GR quantile from the two means reduces to $\TGRImrctHierPopGWTCFOURResultsWOGWOneNineZeroEightOneFour{GRQUANTMeansTwoDGWTC4HIERPOP}\%$, while the one from the combined posteriors reduces to $\TGRImrctGWTCFOURResultsWOGWOneNineZeroEightOneFour{GRQUANTGWTC4}\%$. The apparent GR deviation from GW190814 itself (GR quantile of $99.9\%$) is due to a prior effect, from its low post-inspiral SNR, as discussed in \citet{LIGOScientific:2020tif}. Specifically, the final spin inferred from the post-inspiral is around $0.7$, while the final spin inferred from the many inspiral cycles is well constrained around $0.28$. Thus, since this is purely due to the prior, there is no evidence for a GR deviation here.

\subpapersubsection{Subdominant multipole amplitudes}
\label{sec:sma}

\Ac{GW} radiation from compact binaries can be expressed using $s=-2$ spin-weighted spherical harmonics $Y^{\ell m}_{-2} (\theta,\phi)$ \citep{Gelfand1958,Newman:1966ub,Creighton:2011zz} as
\begin{equation}
    h(t, \theta_{JN}, \boldsymbol{\lambda}) = \sum_{\ell=2}^{\infty} \sum_{m=-\ell}^{\ell} Y_{-2}^{\ell m} (\theta_{JN},0)\,h_{\ell m}(t,\boldsymbol{\lambda}),
\end{equation}
where $h(t)$ is the time-domain strain, $(\theta,\phi)$ denotes the direction of radiation in the source-frame, and $\boldsymbol{\lambda}$ represents the source parameters, including component masses and spins. 
Here we have used the convention in the \IMRPhenomXPHM waveform model, so $(\theta,\phi) = (\theta_{JN},0)$~\citep{Pratten:2020ceb}, where $\theta_{JN}$ is the angle between the total angular momentum of the binary and the line of sight.
The quadrupolar $(\ell, m)=(2,\pm 2)$ multipole moments dominate the signal, but subdominant higher-order multipole moments (HOMs) become significant in asymmetric-mass systems, particularly for orientations that are not close to face-on or face-off ($\theta_{JN} = 0$ or $\pi$).

The SMA test evaluates the consistency of the HOM amplitudes with general relativity predictions~\citep{Islam:2019dmk, Puecher:2022sfm, Gupta:2025paz}. Using the \IMRPhenomXPHM waveform model, this test constrains amplitude deviations $\delta A_{\ell m}$ in the $(2,\pm 1)$ and $(3,\pm 3)$ multipole moments (currently just considered separately):
\begin{align}
h(t, \theta_{JN}, \boldsymbol{\lambda}) 
&= \sum_{m=\pm 2} 
Y^{2m}_{-2}(\theta_{JN}, 0)\,
h_{2m}(t,\,\boldsymbol{\lambda}) \notag \\
&\quad + \sum_{m=\pm 1} 
\bigl(1 + \delta A_{2 1}\bigr) Y^{2m}_{-2}(\theta_{JN}, 0)\,
h_{2m}(t,\,\boldsymbol{\lambda}) \notag \\
&\quad + \sum_{m=\pm 3} 
\bigl(1 + \delta A_{3 3}\bigr) Y^{3m}_{-2}(\theta_{JN}, 0)\,
h_{3m}(t,\,\boldsymbol{\lambda}) \notag \\
&\quad+\sum_{\text{other HOM}}
Y^{\ell m}_{-2}(\theta_{JN}, 0)\,
h_{\ell m}(t,\,\boldsymbol{\lambda}).
\end{align}

For a GW event, meaningful constraints on $\delta A_{\ell m}$ require sufficient \ac{SNR} in the relevant $(\ell, m)$ multipole moment. 
The multipole moment-wise \ac{SNR}, $\rho_{\ell m}$, is computed by using the component of $h_{\ell m}$ orthogonal to the $(2,2)$ multipole moment, and calculating the corresponding optimal \ac{SNR} \citep{Mills:2020thr}. 
In the absence of a $(\ell,m)$ multipole moment in data, $\rho_{\ell m}$, in Gaussian noise, follows a $\chi$ distribution (i.e., the square root of a $\chi^2$-distributed variable) with two degrees of freedom~\citep{LIGOScientific:2020stg,Fairhurst:2019vut,Mills:2020thr}.
Deviations from this distribution indicate the presence of the multipole moment in the data.
Events are selected for this test when the lower bound of the $68\%$ credible interval of the $\rho_{\ell m}$ distribution exceeds $2.145$ (the $90$-th percentile of a $\chi$ distribution).

\COMMONNAME{GW231123}~\citep{GW231123} is the only event from the \ac{O4a} observing run that satisfies our selection criteria, exhibiting significant \ac{SNR} in both the $(2,\pm 1)$ and $(3,\pm 3)$ multipoles.
However, substantial waveform systematics were previously identified for this signal~\citep{GW231123}.
These waveform systematics are expected to impact tests of \GR, and indeed, as discussed in Appendix~A of Paper~II, the
MDR test finds significant apparent \GR deviations when analyzing this event using the same \IMRPhenomXPHM model used for the SMA test.
When the SMA test is applied to this event, the posterior distributions for both $\delta A_{21}$ and $\delta A_{33}$ accumulate near the edges of the prior range (we use uniform priors in the range $[-10, 10]$), indicating pronounced (spurious) deviations from \GR~\citep{Gupta:2025paz}.

\begin{figure}
    \centering
	\includegraphics[width=\TGRFigureWidth]{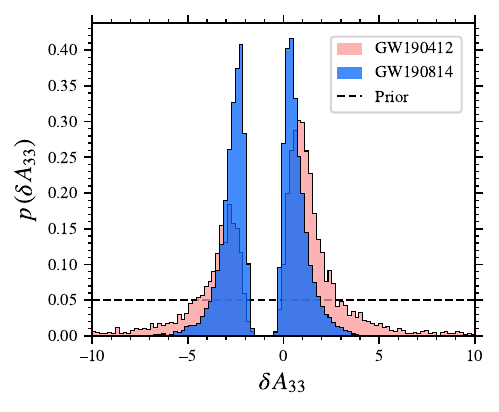}
    \caption{Constraints on amplitude deviation parameters $\delta A_{\ell m}$ for GW190412 and GW190814.}
    \label{fig:sma}
\end{figure}

Additionally, events GW190412~\citep{LIGOScientific:2020stg} and GW190814~\citep{LIGOScientific:2020zkf} from O3a exhibit sufficient \ac{SNR} in the $(3,\pm 3)$ multipole moments to enable the application of this analysis.
Figure~\ref{fig:sma} presents the resulting constraints on $\delta A_{33}$ for these events. 
We find $\delta A_{33} = {\gwOhFourTwelveSMAMedian}^{+\gwOhFourTwelveSMAUpperError}_{-\gwOhFourTwelveSMALowerError}$ for GW190412, and $\delta A_{33} = {\gwOhEightFourteenSMAMedian}^{+\gwOhEightFourteenSMAUpperError}_{-\gwOhEightFourteenSMALowerError}$ for GW190814.
GW190814 provides stronger constraints due to the higher SNR in the $(3, \pm3)$ multipole moment, resulting from the significant asymmetry in the masses of the binary components.
In both events the \GR prediction ($\delta A_{33}=0$) lies well within the $90\%$ credible interval.
However, in both cases, the posterior distributions for $\delta A_{33}$ are characteristically bimodal, arising from degeneracies between $\delta A_{33}$, the binary inclination, and the reference orbital phase~\citep{Puecher:2022sfm}.  
This bimodality reflects the limited ability of current data to disentangle multipole amplitude deviations from source geometry, rather than indicating a true departure from \GR. Indeed, even for the loud \ac{O4b} event GW241011 \citep{GW241011/GW241110}, which allows this test to place the best constraints so far, due to the unequal masses of its source, the bimodality is reduced significantly, but not eliminated.

\bigskip

\subpapersection{Tests of Polarization}\label{sec:polarization}

Polarization tests~\citep{Wong:2021cmp} involve constructing a linear combination of strain data,
called the null stream,
to eliminate excess power and compare the residual data to noise, allowing one to constrain possible non-tensorial polarizations of \acp{GW}.
The null stream is determined geometrically,
relying solely on the beam pattern function \citep{Forward:1978zm,GuerselTinto1989,Nishizawa:2009bf,Blaut:2012zz,Isi:2015cva,Isi:2017fbj},
which describes detector responses to \acp{GW},
independent of the waveform morphology.

In the frequency domain,
the observation model is expressed as
\begin{equation}
    \tilde{\boldsymbol{d}}(f;\Delta \boldsymbol{t}) = 
    \boldsymbol{F}(\alpha, \delta, \psi, t_{\mathrm{event}})\tilde{\boldsymbol{h}}(f)+
    \tilde{\boldsymbol{n}}(f;\Delta \boldsymbol{t})\,,
\end{equation}
where $\tilde{\boldsymbol{d}}(f;\Delta \boldsymbol{t})$ is the time-shifted strain data at the geocenter,
$\boldsymbol{F}(\alpha, \delta, \psi, t_{\mathrm{event}})$ is the beam pattern matrix,
$\tilde{\boldsymbol{h}}(f)$ represents polarization modes at the geocentric time,
and $\tilde{\boldsymbol{n}}(f;\Delta \boldsymbol{t})$ is the detector noise,
also time-shifted to the geocenter.
$\alpha$ and $\delta$ denote the source's right ascension and declination,
$\psi$ is the polarization angle,
and $\Delta \boldsymbol{t} = \Delta \boldsymbol{t}(\alpha, \delta, t_{\mathrm{event}})$ represents the time delays from the geocenter to the detectors.
Since the beam pattern function varies negligibly over the duration of the \ac{GW} transients considered,
it is approximated as constant throughout.

The null stream construction can be formulated as a projection using the projector
constructed from the beam pattern matrix $\boldsymbol{F}$~\citep{Wong:2021cmp}.
The projector removes data within the hyperplane spanned by the column vectors of $\boldsymbol{F}$,
irrespective of the waveform morphology of $\boldsymbol{h}(f)$.
Here $\boldsymbol{F}$ defines the polarization model;
for instance,
in a scalar--tensor model:
\begin{equation}
\boldsymbol{F}=
\begin{bmatrix}
\boldsymbol{f}_{+} &
\boldsymbol{f}_{\times} &
\boldsymbol{f}_\mathrm{b}
\end{bmatrix},
\end{equation}
where the column vectors $\boldsymbol{f}_{+}$, $\boldsymbol{f}_{\times}$,
and $\boldsymbol{f}_\mathrm{b}$ represent the plus, cross, and scalar breathing polarization modes,
respectively.
The $i$th entry of the column vectors, $\boldsymbol{f}_{+/\times/\mathrm{b}}$, corresponds to the beam pattern function of the $i$th detector for the $+$/$\times$/scalar breathing polarization mode.
The scalar longitudinal mode, being degenerate with the scalar breathing mode,
is excluded from the matrix.

\begin{table}
    \caption{\label{tab:pol_configurations}
    Polarization modes and their corresponding basis choices for the polarization analysis.
    }
    \begin{ruledtabular}
        \begin{tabular}{lcc}
        \textrm{Hypothesis} &
        \textrm{Polarization modes} &
        \textrm{Pol. basis} \\
        & & \textrm{modes} \\
        \colrule
        Scalar (S) & b & b \\
        Vector (V) & x, y & x \\
        Tensor (T) & $+$, $\times$ &  $+$ \\
        Tensor--scalar (TS) & $+$, $\times$, b & $+$ \\
        Tensor--vector (TV) & $+$, $\times$, x, y & $+$ \\
        Vector--scalar (VS) & x, y, b & x \\
        Tensor--vector--scalar (TVS) & $+$, $\times$, x, y, b & $+$ \\
        \end{tabular}
    \end{ruledtabular}

\tablecomments{The tensor plus and cross modes are denoted as ``$+$'' and ``$\times$,'' respectively,
    while the vector x and y modes are labeled ``x'' and ``y.''
    The scalar breathing mode is represented by ``b.''
    Only the breathing mode is considered for scalar polarizations,
    as the longitudinal mode is degenerate with it.
}

\end{table}

\begin{figure}
	\begin{center}
	\includegraphics[width=\TGRFigureWidth]{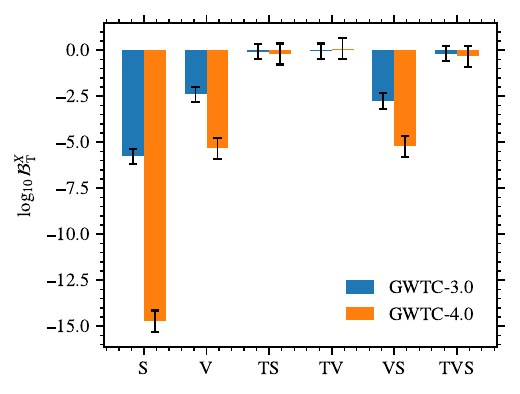}
	\end{center}
	\caption{
        Comparison of $\log_{10}$ Bayes factors $\mathcal{B}^{X}_\mathrm{T}$ for various polarization hypotheses (S, V, TS, TV, VS, TVS) against the tensor hypothesis, for GWTC-3.0 and GWTC-4.0. Polarization states are projected onto one basis mode as detailed in Table~\ref{tab:pol_configurations}. Positive (negative) values indicate that the hypothesis in the superscript is favored (disfavored) relative to the tensor hypothesis. Error bars represent 90\% credible intervals.
  }
\label{fig:pol_result}
\end{figure}

The number of detectors must exceed the number of columns in the beam pattern matrix.
However,
for O1 and O4a,
only two detectors were operational,
constraining the analysis to a single effective polarization mode.
We assume the polarization modes are similar to chosen basis modes indicated in Table~\ref{tab:pol_configurations},
differing at most by a complex scaling factor~\citep{Wong:2021cmp},
in a manner similar to the elliptical parameterization in~\cite{Lee:2025qpu}.
This assumption defines the effective beam pattern matrix used in the analysis and imposes a simplifying structure on the polarization
subspace (e.g., in the single basis mode case for tensorial signals, it is equivalent to testing consistency with $h_{\times} = \mathcal{C}h_{+}$
for a complex constant $\mathcal{C}$). However, the null-stream approach remains robust to such assumptions because it is a projection-based
test that searches for excess power in the direction orthogonal to the assumed subspace, making it more tolerant to inaccuracies in the
polarization model than fully modeled (template-based) searches, which require precise agreement in both polarization content and waveform
evolution. A concrete illustration for non-tensorial signals is provided by scalar--tensor models, where one can choose the plus mode as the
basis and assume similarity with the cross and scalar breathing modes, leading to the effective beam pattern matrix
\begin{equation}
    \boldsymbol{F}_{\mathrm{scalar-tensor,eff}}
    =
    \begin{bmatrix}
        \boldsymbol{f}_{+} +
        \mathcal{C}_{\times}\boldsymbol{f}_{\times} +
        \mathcal{C}_\mathrm{b}\boldsymbol{f}_\mathrm{b}
    \end{bmatrix} \,,
\end{equation}
where $\mathcal{C}_{\times}$ and $\mathcal{C}_\mathrm{b}$ are complex scaling factors.
This method,
independent of the waveform of the basis mode $\tilde{h}_{+}$,
is sensitive only to the geometric projection of the polarization modes in the detector responses.
Assuming similarity between polarization modes addresses the limitation posed by the number of detectors.
Any violation of this assumption would indicate a signal component orthogonal to the signal space of the \ac{GR} waveform model,
which should appear in the residual test
described in Section~\ref{sec:residual}.
However, as discussed there, dedicated analyses are more sensitive to given \GR deviations than the residuals test is.

\begin{table*}[ht]
    \caption{\label{tab:pol_result}Results from the polarizations analysis.}
    \centering
    \begin{tabular}{c c c c c c c}
    \toprule
    Events & $\log_{10}\mathcal{B}_{\mathrm{T}}^{\mathrm{S}}$ & $\log_{10}\mathcal{B}_{\mathrm{T}}^{\mathrm{V}}$ & $\log_{10}\mathcal{B}_{\mathrm{T}}^{\mathrm{TS}}$ & $\log_{10}\mathcal{B}_{\mathrm{T}}^{\mathrm{TV}}$ & $\log_{10}\mathcal{B}_{\mathrm{T}}^{\mathrm{VS}}$ & $\log_{10}\mathcal{B}_{\mathrm{T}}^{\mathrm{TVS}}$\\
    \midrule
    GWTC-3.0 & $-5.76_{-0.42}^{+0.42}$ & $-2.39_{-0.41}^{+0.41}$ & $-0.07_{-0.41}^{+0.41}$ & $-0.04_{-0.41}^{+0.41}$ & $-2.75_{-0.41}^{+0.41}$ & $-0.19_{-0.41}^{+0.41}$ \\
    O4a & $-8.97_{-0.41}^{+0.41}$ & $-2.94_{-0.40}^{+0.40}$ & $-0.13_{-0.40}^{+0.40}$ & $0.14_{-0.40}^{+0.40}$ & $-2.46_{-0.40}^{+0.40}$ & $-0.13_{-0.40}^{+0.40}$ \\
    \midrule
    GWTC-4.0 & $-14.72_{-0.59}^{+0.59}$ & $-5.33_{-0.58}^{+0.58}$ & $-0.20_{-0.57}^{+0.57}$ & $0.10_{-0.57}^{+0.57}$ & $-5.21_{-0.58}^{+0.58}$ & $-0.31_{-0.57}^{+0.57}$ \\
    \bottomrule
    \end{tabular}

\tablecomments{Combined $\log_{10}$ Bayes factors $\mathcal{B}$ for various polarization hypotheses against the tensor hypothesis, using both 2-detector and 3-detector events. Polarization states have been projected onto one basis mode as detailed in Table~\ref{tab:pol_configurations}. The combined values are obtained by summing the individual $\log_{10}\mathcal{B}_{\mathrm{T}}^{X}$ under the assumption that the events are independent. Positive (negative) values indicate that the hypothesis indicated in the superscript is favored (disfavored) with respect to the tensor hypothesis. Error bars refer to $90\%$ credible intervals.}

\end{table*}

We use a Bayesian framework to compute the evidence for each polarization hypothesis
by sampling parameters $\boldsymbol{\theta} = \{\alpha, \delta, \psi, t_{\mathrm{event}},\boldsymbol{\mathcal{C}}\}$ using the \DYNESTY sampler.
Here,
$\boldsymbol{\mathcal{C}}$ are the complex scaling factors of the polarization modes relative to the basis mode(s).
The number of scaling factors depends on the assumed number of polarization and basis modes.
For each polarization hypothesis,
we choose a particular polarization mode as the basis,
but its waveform morphology is not assumed or involved in the analysis.
Instead, only the relative amplitudes $\boldsymbol{\mathcal{C}}$ are used
to construct the effective beam pattern matrix for null stream projection.
After constructing the null stream time-shifted to the geocenter,
we then perform the time--frequency transform \citep{Necula:2012zz} to obtain
the time--frequency representation of the null stream.

The Bayes factor,
defined in Equation~\eqref{eq:GR_nGR_bayes_factor},
quantifies the relative plausibility of the hypotheses.
We adopt the same uniform priors for sky position, polarization angle, and event time as in previous analyses.
For GW170817, instead of adopting a uniform sky prior as for other events, we fix the sky position to the coordinates measured from electromagnetic counterparts \citep{GBM:2017lvd}.
For the complex scaling factors $\boldsymbol{\mathcal{C}}$,
the amplitude prior is uniform from 0 to 1,
and the phase prior is uniform from 0 to $2\pi$.
We report the base-10 logarithm of the Bayes factor $\log_{10}\mathcal{B}_\mathrm{T}^{X}$ comparing the non-tensor hypothesis $X$ to the tensor hypothesis T.

The likelihood is calculated by summing the contributions of a predefined set of time--frequency pixels $\mathcal{I}$,
known as the time--frequency cluster.
The time--frequency cluster is constructed by performing the time--frequency transform on the maximum-likelihood \ac{GR} waveform.
The advantages of this approach include both improving the sensitivity of the tests and mitigating the impacts of instrumental noise.
However, it reduces sensitivity to some non-tensorial polarizations,
such as dominant dipolar scalar radiation at the orbital frequency during the inspiral in scalar--tensor theories \citep{Bernard:2022noq},
when the non-tensorial radiation frequency significantly deviates from that of the tensorial radiation
by more than the analysis's frequency resolution of $\Delta f = 16$~Hz in the
time--frequency representation. This frequency resolution was selected as a practical
choice that still sufficiently resolves the merger phase containing the dominant
excess power in \ac{CBC} signals, providing a corresponding time resolution of
$\Delta t \simeq 1/(2 \Delta f) \simeq 31$~ms, which is commensurate with the
characteristic duration of the merger phase in stellar-mass \ac{BBH}
coalescences, thereby enabling the analysis to capture excess power during coalescence
with minimal temporal smearing.

We analyze all the O4a events satisfying the selection criteria of this paper
and reanalyze all events from GWTC-3.0 satisfying those criteria using the updated implementation.
Figure~\ref{fig:pol_result} compares the results for events from GWTC-3.0 and GWTC-4.0.
Table~\ref{tab:pol_result} lists the combined log Bayes factors for GWTC-3.0 events, O4a events, and GWTC-4.0 events.
The scalar, vector, and vector--scalar hypotheses are strongly disfavored compared to the tensor hypothesis,
indicating a clear preference for the tensorial polarization modes.
After including O4a events,
these hypotheses are even more disfavored,
with improved constraints on their viability due to the additional data.
In contrast,
the tensor--scalar,
tensor--vector,
and tensor--vector--scalar hypotheses show only marginal improvements in their constraints with the inclusion of O4a events,
but their results remain uninformative.
Since the tensor hypothesis is a subset of these combined hypotheses,
they are not expected to be strongly disfavored when the tensor hypothesis represents the ground truth.

\subpapersection{Conclusions}\label{sec:conclusion}

The first part of the fourth observing run of the LVK has provided \TGRNUMEVENTS confident events to which we have applied a set of tests of \GR.
We subjected each of these events, as well as the other confident events from previous observing runs in GWTC-4.0, as appropriate,
to some or all of our \TGRNUMTESTS test pipelines,
    probing different regimes and potential deviations from \GR.

Overall, we find that the data are consistent with \GR for all of our gauntlet of tests.
More specifically, regarding the \TGRINUMTESTS tests of consistency described in this paper,
    we find consistency of the residuals with noise for all the events.
The final mass and final spin inferred using the low- and high-frequency parts of each signal
    are consistent with each other.
We find no evidence for deviations from the GR predictions for the amplitudes of subdominant GW multipole moments,
    nor for non-GR modes of polarization.
The quantitative bounds on deviations are summarized in Table \ref{tab:summary}, as well as their improvements over previous bounds,
if appropriate.

The table also records the bounds and improvements for the tests described in Papers~II and~III,
    which are further discussed in those two papers.
Example constraints on specific alternative theories from applying some of the \PN coefficient bounds obtained in Paper~II
    are given in that paper's Table~3, and that paper also discusses the caveats involved in such translations.

A subset of parameterized tests (from Paper~II) and ringdown tests (from Paper~III) have found
    that there are a few events which place the GR parameters outside the $90\%$ posterior probability distribution.
This is to be expected statistically with such a high number of tests and independent events,
    and these are analyzed further in the respective papers II \& III.
Additionally, there is an apparent GR deviation found in the combined results for the pSEOBNR ringdown test,
    whose significance is discussed in Paper III, along with studies that indicate a reduced significance.
All this indicates the importance of controlling false GR deviations \citep{Gupta:2024gun} going forward.
    The effects of non-Gaussian noise \citep[e.g.,][]{Kwok:2021zny} and waveform modeling uncertainties \citep[e.g.,][]{Dhani:2024jja} are expected to be particularly important,
    with the expected growth in both event numbers and individual signal amplitudes relative to the noise arising from
    the increase in the detectors' sensitivities \citep{Aasi:2013wya},\footnote{LVK observing run plans \url{https://observing.docs.ligo.org/plan}}
    and thus their expanded reach out into the Universe.
This is especially pertinent regarding the exceptional single detector events \COMMONNAME{GW230529} \citep{GW230529}
    and \COMMONNAME{GW230814single} \citep{GW230814},
    both excluded from these three papers for failing the multiple-detector criterion,
        but discussed in the separate dedicated papers (\COMMONNAME{GW230529} is analyzed with tests of \GR in \citealt{Sanger:2024axs}),
    as well as for the two highest mass BBH events that do meet the selection criteria,
    \COMMONNAME{GW231123} \citep{GW231123} and \FULLNAME{GW231028_153006} \citep{GWTC:Results}.
These two high-mass events simultaneously offer the data most appropriate for ringdown analyses (see Section~2 in Paper~III),
      and challenge the systematic accuracy of our waveform models and modeling of detector noise. The results obtained for \COMMONNAME{GW231123} may be
      due to inaccuracies in waveform modeling, wave-optics lensing, or other features, as described in our dedicated paper on that event \citep{GW231123}.
Future detector upgrades will provide even higher SNR events and a larger catalog of high-significance signals. These, combined with the more refined
       analyses required to exploit the data, will allow us to perform ever tighter tests of \GR.

All strain data analyzed in this paper are available from the Gravitational Wave Open Science Center \citep{OpenData}. The data and scripts used to prepare the figures and tables are available at \citet{PaperI_DCC_release}.

\section*{Acknowledgements}

This material is based upon work supported by NSF's LIGO Laboratory, which is a
major facility fully funded by the National Science Foundation.
The authors also gratefully acknowledge the support of
the Science and Technology Facilities Council (STFC) of the
United Kingdom, the Max-Planck-Society (MPS), and the State of
Niedersachsen/Germany for support of the construction of Advanced LIGO 
and construction and operation of the GEO\,600 detector. 
Additional support for Advanced LIGO was provided by the Australian Research Council.
The authors gratefully acknowledge the Italian Istituto Nazionale di Fisica Nucleare (INFN),  
the French Centre National de la Recherche Scientifique (CNRS) and
the Netherlands Organization for Scientific Research (NWO)
for the construction and operation of the Virgo detector
and the creation and support  of the EGO consortium. 
The authors also gratefully acknowledge research support from these agencies as well as by 
the Council of Scientific and Industrial Research of India, 
the Department of Science and Technology, India,
the Science \& Engineering Research Board (SERB), India,
the Ministry of Human Resource Development, India,
the Spanish Agencia Estatal de Investigaci\'on (AEI),
the Spanish Ministerio de Ciencia, Innovaci\'on y Universidades,
the European Union NextGenerationEU/PRTR (PRTR-C17.I1),
the ICSC - CentroNazionale di Ricerca in High Performance Computing, Big Data
and Quantum Computing, funded by the European Union NextGenerationEU,
the Comunitat Auton\`oma de les Illes Balears through the Conselleria d'Educaci\'o i Universitats,
the Conselleria d'Innovaci\'o, Universitats, Ci\`encia i Societat Digital de la Generalitat Valenciana and
the CERCA Programme Generalitat de Catalunya, Spain,
the Polish National Agency for Academic Exchange,
the National Science Centre of Poland and the European Union - European Regional
Development Fund;
the Foundation for Polish Science (FNP),
the Polish Ministry of Science and Higher Education,
the Swiss National Science Foundation (SNSF),
the Russian Science Foundation,
the European Commission,
the European Social Funds (ESF),
the European Regional Development Funds (ERDF),
the Royal Society, 
the Scottish Funding Council, 
the Scottish Universities Physics Alliance, 
the Hungarian Scientific Research Fund (OTKA),
the French Lyon Institute of Origins (LIO),
the Belgian Fonds de la Recherche Scientifique (FRS-FNRS), 
Actions de Recherche Concert\'ees (ARC) and
Fonds Wetenschappelijk Onderzoek - Vlaanderen (FWO), Belgium,
the Paris \^{I}le-de-France Region, 
the National Research, Development and Innovation Office of Hungary (NKFIH), 
the National Research Foundation of Korea,
the Natural Sciences and Engineering Research Council of Canada (NSERC),
the Canadian Foundation for Innovation (CFI),
the Brazilian Ministry of Science, Technology, and Innovations,
the International Center for Theoretical Physics South American Institute for Fundamental Research (ICTP-SAIFR), 
the Research Grants Council of Hong Kong,
the National Natural Science Foundation of China (NSFC),
the Israel Science Foundation (ISF),
the US-Israel Binational Science Fund (BSF),
the Leverhulme Trust, 
the Research Corporation,
the National Science and Technology Council (NSTC), Taiwan,
the United States Department of Energy,
and
the Kavli Foundation.
The authors gratefully acknowledge the support of the NSF, STFC, INFN and CNRS for provision of computational resources.

This work was supported by MEXT,
the JSPS Leading-edge Research Infrastructure Program,
JSPS Grant-in-Aid for Specially Promoted Research 26000005,
JSPS Grant-in-Aid for Scientific Research on Innovative Areas 2402: 24103006,
24103005, and 2905: JP17H06358, JP17H06361 and JP17H06364,
JSPS Core-to-Core Program A.\ Advanced Research Networks,
JSPS Grants-in-Aid for Scientific Research (S) 17H06133 and 20H05639,
JSPS Grant-in-Aid for Transformative Research Areas (A) 20A203: JP20H05854,
the joint research program of the Institute for Cosmic Ray Research,
University of Tokyo,
the National Research Foundation (NRF),
the Computing Infrastructure Project of the Global Science experimental Data hub
Center (GSDC) at KISTI,
the Korea Astronomy and Space Science Institute (KASI),
the Ministry of Science and ICT (MSIT) in Korea,
Academia Sinica (AS),
the AS Grid Center (ASGC) and the National Science and Technology Council (NSTC)
in Taiwan under grants including the Science Vanguard Research Program,
the Advanced Technology Center (ATC) of NAOJ,
and the Mechanical Engineering Center of KEK.

Additional acknowledgements for support of individual authors may be found in the following document: \\
\texttt{https://dcc.ligo.org/LIGO-M2300033/public}.

For the purpose of open access, the authors have applied a Creative Commons Attribution (CC BY)
license to any Author Accepted Manuscript version arising.
We request that citations to this article use `A. G. Abac {\it et al.} (LIGO-Virgo-KAGRA Collaboration), ...' or similar phrasing, depending on journal convention.

\emph{The following open-source software has been used:}

Calibration of the \ac{LIGO} strain data was performed with a \GSTLAL{}-based
    calibration software pipeline~\citep{Viets:2017yvy}.
    Calibration of the Virgo strain data is performed with C-based software~\citep{VIRGO:2021kfv}.
Data-quality products and event-validation results were computed using the
    \soft{DMT}{}~\citep{DMTdocumentation}, \soft{DQR}{}~\citep{DQRdocumentation},
    \soft{DQSEGDB}{}~\citep{Fisher:2020pnr}, \soft{gwdetchar}{}~\citep{gwdetchar-software},
    \soft{hveto}{}~\citep{Smith:2011an}, \soft{iDQ}{}~\citep{Essick:2020qpo},
    \soft{Omicron}{}~\citep{Robinet:2020lbf} and
    \soft{PythonVirgoTools}{}~\citep{pythonvirgotools} software packages and contributing
    software tools.
Analyses in this catalog relied upon the \LALSUITE{} software library~\citep{lalsuite-software,Wette:2020air}.
The detection of the signals and subsequent significance evaluations in this catalog were performed with the
    \GSTLAL{}-based inspiral software pipeline~\citep{Messick:2016aqy,Sachdev:2019vvd,Hanna:2019ezx,Cannon:2020qnf},
    with the \MBTA{} pipeline~\citep{Adams:2015ulm,Aubin:2020goo}, and with the
    \PYCBC{}~\citep{Usman:2015kfa,Nitz:2017svb,Davies:2020tsx} and the
    \CWB{}~\citep{Klimenko:2004qh,Klimenko:2011hz,Klimenko:2015ypf} packages.
Estimates of the noise spectra and glitch models were obtained using
    \BAYESWAVE{}~\citep{Cornish:2014kda,Littenberg:2015kpb,Cornish:2020dwh,Gupta:2023jrn}.
Noise subtraction for one candidate was also performed with \soft{gwsubtract}{}~\citep{Davis:2022ird}.
Source-parameter estimation was performed with the \BILBY{}
    and \PBILBY{} libraries~\citep{Ashton:2018jfp,Romero-Shaw:2020owr,Smith:2019ucc} using the
    \DYNESTY{} nested sampling package~\citep{Speagle:2020spe}.

\SEOBNRFIVEPHM waveforms used in parameter estimation were generated using \soft{pySEOBNR}~\citep{Mihaylov:2023bkc}.
SMA, FTI, TIGER, SIM, LOSA, MDR, SSB, and pSEOBNR waveforms used for testing \GR were generated using \BILBYTGR~\citep{ashton_2025_15676285}.
E-M waveforms used for constraining echoes were generated using \soft{echoes\_waveform\_models}~\citep{echowfm}. 
Other analyses used \soft{cpnest}~\citep{cpnest}, \soft{IGWN Wave Compare}~\citep{iwc}, \soft{nullpol}~\citep{nullpol}, and \soft{pyRing}~\citep{pyRing}.
Quasinormal mode frequencies were computed using \soft{QNM}~\citep{Stein:2019mop}.
The QNMRF analysis used \citet{qnmrf-soft}. The multi-dimensional hierarchical analysis results were produced using \soft{hierfit}~\citep{hierfit}.
\PESUMMARY{} was used to postprocess and collate parameter-estimation
results~\citep{Hoy:2020vys}.  The various stages of the parameter-estimation
analysis were managed with the \ASIMOV{} library~\citep{Williams:2022pgn} together with \soft{CBCFlow}~\citep{cbcflow}.
Plots were prepared with \soft{Matplotlib}~\citep{Hunter:2007ouj},
\SEABORN{}~\citep{Waskom:2021psk}, and \GWPY{}~\citep{gwpy-software}.
\NUMPY~\citep{Harris:2020xlr}, \soft{scikit-learn}~\citep{scikit-learn}, and \SCIPY~\citep{Virtanen:2019joe} were used for analyses in the manuscript.

\bibliography{}

\begin{thebibliography}{}
\expandafter\ifx\csname natexlab\endcsname\relax\def\natexlab#1{#1}\fi
\providecommand{\url}[1]{\href{#1}{#1}}
\providecommand{\dodoi}[1]{doi:~\href{http://doi.org/#1}{\nolinkurl{#1}}}
\providecommand{\doeprint}[1]{\href{http://ascl.net/#1}{\nolinkurl{http://ascl.net/#1}}}
\providecommand{\doarXiv}[1]{\href{https://arxiv.org/abs/#1}{\nolinkurl{https://arxiv.org/abs/#1}}}

\bibitem[{Aasi {et~al.}(2015)}]{TheLIGOScientific:2014jea}
Aasi, J., {et~al.} 2015, Class. Quantum Grav., 32, 074001,
  \dodoi{10.1088/0264-9381/32/7/074001}

\bibitem[{Abac {et~al.}(2025{\natexlab{a}})}]{Abac:2025saz}
Abac, A., {et~al.} 2025{\natexlab{a}}.
\newblock \doarXiv{2503.12263}

\bibitem[{Abac {et~al.}(2024)}]{GW230529}
Abac, A.~G., {et~al.} 2024, Astrophys. J. Lett., 970, L34,
  \dodoi{10.3847/2041-8213/ad5beb}

\bibitem[{Abac {et~al.}(2025{\natexlab{b}})}]{GWTC:Results}
---. 2025{\natexlab{b}}.
\newblock \doarXiv{2508.18082}

\bibitem[{Abac {et~al.}(2025{\natexlab{c}})}]{GWTC:TGR-II}
---. 2025{\natexlab{c}}.
\newblock \url{https://dcc.ligo.org/LIGO-P2500066/public}

\bibitem[{Abac {et~al.}(2025{\natexlab{d}})}]{GWTC:TGR-III}
---. 2025{\natexlab{d}}.
\newblock \url{https://dcc.ligo.org/LIGO-P2500067/public}

\bibitem[{Abac {et~al.}(2025{\natexlab{e}})}]{GWTC:Introduction}
---. 2025{\natexlab{e}}, Astrophys. J. Lett., 995, L18,
  \dodoi{10.3847/2041-8213/ae0c06}

\bibitem[{Abac {et~al.}(2025{\natexlab{f}})}]{GW230814}
---. 2025{\natexlab{f}}.
\newblock \doarXiv{2509.07348}

\bibitem[{Abac {et~al.}(2025{\natexlab{g}})}]{GWTC:Cosmology}
---. 2025{\natexlab{g}}.
\newblock \doarXiv{2509.04348}

\bibitem[{Abac {et~al.}(2025{\natexlab{h}})}]{GW250114}
---. 2025{\natexlab{h}}, Phys. Rev. Lett., 135, 111403,
  \dodoi{10.1103/kw5g-d732}

\bibitem[{Abac {et~al.}(2025{\natexlab{i}})}]{GW241011/GW241110}
---. 2025{\natexlab{i}}, Astrophys. J. Lett., 993, L21,
  \dodoi{10.3847/2041-8213/ae0d54}

\bibitem[{Abac {et~al.}(2025{\natexlab{j}})}]{GW231123}
---. 2025{\natexlab{j}}, Astrophys. J. Lett., 993, L25,
  \dodoi{10.3847/2041-8213/ae0c9c}

\bibitem[{Abac {et~al.}(2025{\natexlab{k}})}]{GWTC:Methods}
---. 2025{\natexlab{k}}.
\newblock \doarXiv{2508.18081}

\bibitem[{Abac {et~al.}(2025{\natexlab{l}})}]{OpenData}
---. 2025{\natexlab{l}}.
\newblock \doarXiv{2508.18079}

\bibitem[{Abac {et~al.}(2026)}]{GW250114_TGR}
---. 2026, Phys. Rev. Lett., 136, 041403, \dodoi{10.1103/6c61-fm1n}

\bibitem[{Abbott {et~al.}(2016{\natexlab{a}})}]{GW150914_paper}
Abbott, B.~P., {et~al.} 2016{\natexlab{a}}, Phys. Rev. Lett., 116, 061102,
  \dodoi{10.1103/PhysRevLett.116.061102}

\bibitem[{Abbott {et~al.}(2016{\natexlab{b}})}]{LIGOScientific:2016lio}
---. 2016{\natexlab{b}}, Phys. Rev. Lett., 116, 221101,
  \dodoi{10.1103/PhysRevLett.116.221101}

\bibitem[{Abbott {et~al.}(2016{\natexlab{c}})}]{LIGOScientific:2016dsl}
---. 2016{\natexlab{c}}, Phys. Rev. X, 6, 041015,
  \dodoi{10.1103/PhysRevX.6.041015}

\bibitem[{Abbott {et~al.}(2016{\natexlab{d}})}]{TheLIGOScientific:2016zmo}
---. 2016{\natexlab{d}}, Class. Quantum Grav., 33, 134001,
  \dodoi{10.1088/0264-9381/33/13/134001}

\bibitem[{Abbott {et~al.}(2017{\natexlab{a}})}]{GW170817_DetectionPaper}
---. 2017{\natexlab{a}}, Phys. Rev. Lett., 119, 161101,
  \dodoi{10.1103/PhysRevLett.119.161101}

\bibitem[{Abbott {et~al.}(2017{\natexlab{b}})}]{Abbott:2017oio}
---. 2017{\natexlab{b}}, Phys. Rev. Lett., 119, 141101,
  \dodoi{10.1103/PhysRevLett.119.141101}

\bibitem[{Abbott {et~al.}(2017{\natexlab{c}})}]{Abbott:2017vtc}
---. 2017{\natexlab{c}}, Phys. Rev. Lett., 118, 221101,
  \dodoi{10.1103/PhysRevLett.118.221101}

\bibitem[{Abbott {et~al.}(2017{\natexlab{d}})}]{GBM:2017lvd}
---. 2017{\natexlab{d}}, Astrophys. J. Lett., 848, L12,
  \dodoi{10.3847/2041-8213/aa91c9}

\bibitem[{Abbott {et~al.}(2019{\natexlab{a}})}]{LIGOScientific:2018dkp}
---. 2019{\natexlab{a}}, Phys. Rev. Lett., 123, 011102,
  \dodoi{10.1103/PhysRevLett.123.011102}

\bibitem[{Abbott {et~al.}(2019{\natexlab{b}})}]{LIGOScientific:2019fpa}
---. 2019{\natexlab{b}}, Phys. Rev. D, 100, 104036,
  \dodoi{10.1103/PhysRevD.100.104036}

\bibitem[{Abbott {et~al.}(2019{\natexlab{c}})}]{GWTC1}
---. 2019{\natexlab{c}}, Phys. Rev. X, 9, 031040,
  \dodoi{10.1103/PhysRevX.9.031040}

\bibitem[{Abbott {et~al.}(2020{\natexlab{a}})}]{Aasi:2013wya}
---. 2020{\natexlab{a}}, Living Rev. Relativity, 23, 3,
  \dodoi{10.1007/s41114-020-00026-9}

\bibitem[{Abbott {et~al.}(2020{\natexlab{b}})}]{LIGOScientific:2020stg}
Abbott, R., {et~al.} 2020{\natexlab{b}}, Phys. Rev. D, 102, 043015,
  \dodoi{10.1103/PhysRevD.102.043015}

\bibitem[{Abbott {et~al.}(2020{\natexlab{c}})}]{LIGOScientific:2020zkf}
---. 2020{\natexlab{c}}, Astrophys. J. Lett., 896, L44,
  \dodoi{10.3847/2041-8213/ab960f}

\bibitem[{Abbott {et~al.}(2021{\natexlab{a}})}]{LIGOScientific:2021qlt}
---. 2021{\natexlab{a}}, Astrophys. J. Lett., 915, L5,
  \dodoi{10.3847/2041-8213/ac082e}

\bibitem[{Abbott {et~al.}(2021{\natexlab{b}})}]{LIGOScientific:2020tif}
---. 2021{\natexlab{b}}, Phys. Rev. D, 103, 122002,
  \dodoi{10.1103/PhysRevD.103.122002}

\bibitem[{Abbott {et~al.}(2021{\natexlab{c}})}]{GWTC2}
---. 2021{\natexlab{c}}, Phys. Rev. X, 11, 021053,
  \dodoi{10.1103/PhysRevX.11.021053}

\bibitem[{Abbott {et~al.}(2023)}]{GWTC3}
---. 2023, Phys. Rev. X, 13, 041039, \dodoi{10.1103/PhysRevX.13.041039}

\bibitem[{Abbott {et~al.}(2024)}]{GWTC2p1}
---. 2024, Phys. Rev. D, 109, 022001, \dodoi{10.1103/PhysRevD.109.022001}

\bibitem[{Abbott {et~al.}(2025)}]{LIGOScientific:2021sio}
---. 2025, Phys. Rev. D, 112, 084080, \dodoi{10.1103/PhysRevD.112.084080}

\bibitem[{Abd El~Dayem {et~al.}(2025)}]{GRAVITY:2025ahf}
Abd El~Dayem, K., {et~al.} 2025, Astron. Astrophys., 698, L15,
  \dodoi{10.1051/0004-6361/202554676}

\bibitem[{Abedi {et~al.}(2017)Abedi, Dykaar, \& Afshordi}]{Abedi:2016hgu}
Abedi, J., Dykaar, H., \& Afshordi, N. 2017, Phys. Rev. D, 96, 082004,
  \dodoi{10.1103/PhysRevD.96.082004}

\bibitem[{Abuter {et~al.}(2020)}]{GRAVITY:2020gka}
Abuter, R., {et~al.} 2020, Astron. Astrophys., 636, L5,
  \dodoi{10.1051/0004-6361/202037813}

\bibitem[{Acernese {et~al.}(2015)}]{TheVirgo:2014hva}
Acernese, F., {et~al.} 2015, Class. Quantum Grav., 32, 024001,
  \dodoi{10.1088/0264-9381/32/2/024001}

\bibitem[{Acernese {et~al.}(2022)}]{VIRGO:2021kfv}
---. 2022, Class. Quantum Grav., 39, 045006, \dodoi{10.1088/1361-6382/ac3c8e}

\bibitem[{Adams {et~al.}(2016)Adams, Buskulic, Germain, Guidi, Marion, Montani,
  Mours, Piergiovanni, \& Wang}]{Adams:2015ulm}
Adams, T., Buskulic, D., Germain, V., {et~al.} 2016, Class. Quantum Grav., 33,
  175012, \dodoi{10.1088/0264-9381/33/17/175012}

\bibitem[{Agathos {et~al.}(2014)Agathos, Del~Pozzo, Li, Van Den~Broeck, Veitch,
  \& Vitale}]{Agathos:2013upa}
Agathos, M., Del~Pozzo, W., Li, T. G.~F., {et~al.} 2014, Phys. Rev. D, 89,
  082001, \dodoi{10.1103/PhysRevD.89.082001}

\bibitem[{Akiyama {et~al.}(2019{\natexlab{a}})}]{Akiyama:2019cqa}
Akiyama, K., {et~al.} 2019{\natexlab{a}}, Astrophys. J. Lett., 875, L1,
  \dodoi{10.3847/2041-8213/ab0ec7}

\bibitem[{Akiyama {et~al.}(2019{\natexlab{b}})}]{EventHorizonTelescope:2019ggy}
---. 2019{\natexlab{b}}, Astrophys. J. Lett., 875, L6,
  \dodoi{10.3847/2041-8213/ab1141}

\bibitem[{Akiyama {et~al.}(2022{\natexlab{a}})}]{EventHorizonTelescope:2022wkp}
---. 2022{\natexlab{a}}, Astrophys. J. Lett., 930, L12,
  \dodoi{10.3847/2041-8213/ac6674}

\bibitem[{Akiyama {et~al.}(2022{\natexlab{b}})}]{EventHorizonTelescope:2022xqj}
---. 2022{\natexlab{b}}, Astrophys. J. Lett., 930, L17,
  \dodoi{10.3847/2041-8213/ac6756}

\bibitem[{Arun {et~al.}(2006{\natexlab{a}})Arun, Iyer, Qusailah, \&
  Sathyaprakash}]{Arun:2006hn}
Arun, K.~G., Iyer, B.~R., Qusailah, M. S.~S., \& Sathyaprakash, B.~S.
  2006{\natexlab{a}}, Phys. Rev. D, 74, 024006,
  \dodoi{10.1103/PhysRevD.74.024006}

\bibitem[{Arun {et~al.}(2006{\natexlab{b}})Arun, Iyer, Qusailah, \&
  Sathyaprakash}]{Arun:2006yw}
---. 2006{\natexlab{b}}, Class. Quantum Grav., 23, L37,
  \dodoi{10.1088/0264-9381/23/9/L01}

\bibitem[{Ashton {et~al.}(2025{\natexlab{a}})Ashton, Udall, \&
  Yarbrough}]{cbcflow}
Ashton, G., Udall, R., \& Yarbrough, Z. 2025{\natexlab{a}}, CBC Workflow.
\newblock \url{https://github.com/Rhiannon-Udall/cbcflow}

\bibitem[{Ashton {et~al.}(2019)}]{Ashton:2018jfp}
Ashton, G., {et~al.} 2019, Astrophys. J. Suppl. Ser., 241, 27,
  \dodoi{10.3847/1538-4365/ab06fc}

\bibitem[{Ashton {et~al.}(2025{\natexlab{b}})Ashton, Talbot, Roy, Pratten,
  Pang, Agathos, Baka, Sänger, Mehta, Steinhoff, Maggio, Ghosh, Vijaykumar,
  Enficiaud, \& Pompili}]{ashton_2025_15676285}
Ashton, G., Talbot, C., Roy, S., {et~al.} 2025{\natexlab{b}}, Bilby TGR, v0.2,
  Zenodo, \dodoi{10.5281/zenodo.15676285}

\bibitem[{Aubin {et~al.}(2021)}]{Aubin:2020goo}
Aubin, F., {et~al.} 2021, Class. Quantum Grav., 38, 095004,
  \dodoi{10.1088/1361-6382/abe913}

\bibitem[{Baka {et~al.}(2025{\natexlab{a}})Baka, Cirok, Haris, Noller, \&
  Krishnendu}]{Baka:2025drk}
Baka, T., Cirok, B., Haris, K., Noller, J., \& Krishnendu, N.~V.
  2025{\natexlab{a}}.
\newblock \doarXiv{2511.00497}

\bibitem[{Baka {et~al.}(2025{\natexlab{b}})Baka, Wright, Romero-Shaw, Berry,
  Haster, Hoy, Talbot, Pang, Raymond, Williams, Ashton, Bossilkov, Dartez,
  Manning, Veitch, Zimmerman, Sun, \& Farr}]{Baka:2025bbb}
Baka, T., Wright, M., Romero-Shaw, I., {et~al.} 2025{\natexlab{b}}, Correcting
  misspecification of calibration uncertainties in gravitational-wave data
  analysis with efficient reweighting, Tech. Rep. {LIGO}-T2500295, {LIGO}
  Project.
\newblock \url{https://dcc.ligo.org/LIGO-T2500295/public}

\bibitem[{Bardeen {et~al.}(1972)Bardeen, Press, \& Teukolsky}]{Bardeen:1972fi}
Bardeen, J.~M., Press, W.~H., \& Teukolsky, S.~A. 1972, Astrophys. J., 178,
  347, \dodoi{10.1086/151796}

\bibitem[{Bernard {et~al.}(2022)Bernard, Blanchet, \&
  Trestini}]{Bernard:2022noq}
Bernard, L., Blanchet, L., \& Trestini, D. 2022, JCAP, 08, 008,
  \dodoi{10.1088/1475-7516/2022/08/008}

\bibitem[{Berti {et~al.}(2015)}]{Berti:2015itd}
Berti, E., {et~al.} 2015, Class. Quantum Grav., 32, 243001,
  \dodoi{10.1088/0264-9381/32/24/243001}

\bibitem[{Blanchet(2024)}]{Blanchet:2013haa}
Blanchet, L. 2024, Living Rev. Relativity, 27, 4,
  \dodoi{10.1007/s41114-024-00050-z}

\bibitem[{Blanchet \& Sathyaprakash(1994)}]{Blanchet:1994ex}
Blanchet, L., \& Sathyaprakash, B.~S. 1994, Class. Quantum Grav., 11, 2807,
  \dodoi{10.1088/0264-9381/11/11/020}

\bibitem[{Blanchet \& Sathyaprakash(1995)}]{Blanchet:1994ez}
---. 1995, Phys.\ Rev.\ Lett., 74, 1067, \dodoi{10.1103/PhysRevLett.74.1067}

\bibitem[{B{\l}aut(2012)}]{Blaut:2012zz}
B{\l}aut, A. 2012, Phys. Rev. D, 85, 043005, \dodoi{10.1103/PhysRevD.85.043005}

\bibitem[{Bogdanov {et~al.}(2019)}]{Bogdanov:2019ixe}
Bogdanov, S., {et~al.} 2019, Astrophys. J. Lett., 887, L25,
  \dodoi{10.3847/2041-8213/ab53eb}

\bibitem[{Brito {et~al.}(2018)Brito, Buonanno, \& Raymond}]{Brito:2018rfr}
Brito, R., Buonanno, A., \& Raymond, V. 2018, Phys. Rev. D, 98, 084038,
  \dodoi{10.1103/PhysRevD.98.084038}

\bibitem[{{Cannon} {et~al.}(2021){Cannon}, {Caudill}, {Chan}, {Cousins},
  {Creighton}, {Ewing}, {Fong}, {Godwin}, {Hanna}, {Hooper}, {Huxford},
  {Magee}, {Meacher}, {Messick}, {Morisaki}, {Mukherjee}, {Ohta}, {Pace},
  {Privitera}, {de Ruiter}, {Sachdev}, {Singer}, {Singh}, {Tapia}, {Tsukada},
  {Tsuna}, {Tsutsui}, {Ueno}, {Viets}, {Wade}, \& {Wade}}]{Cannon:2020qnf}
{Cannon}, K., {Caudill}, S., {Chan}, C., {et~al.} 2021, SoftwareX, 14, 100680,
  \dodoi{10.1016/j.softx.2021.100680}

\bibitem[{Carullo {et~al.}(2019)Carullo, Del~Pozzo, \&
  Veitch}]{Carullo:2019flw}
Carullo, G., Del~Pozzo, W., \& Veitch, J. 2019, Phys. Rev. D, 99, 123029,
  \dodoi{10.1103/PhysRevD.99.123029}

\bibitem[{Carullo {et~al.}(2025)Carullo, Del~Pozzo, \& Veitch}]{pyRing}
---. 2025, pyRing, 2.7.0,  Zenodo, \dodoi{10.5281/zenodo.8165507}

\bibitem[{Colleoni {et~al.}(2024)Colleoni, Krishnendu, Mourier, Bera, \&
  Jim\'enez-Forteza}]{Colleoni:2024lpj}
Colleoni, M., Krishnendu, N.~V., Mourier, P., Bera, S., \& Jim\'enez-Forteza,
  X. 2024, {Testing Gravity with~Binary Black Hole Gravitational Waves}, ed.
  C.~Bambi \& A.~C{\'a}rdenas-Avenda{\~{n}}o (Singapore: Springer Nature
  Singapore), 239--274, \dodoi{10.1007/978-981-97-2871-8_7}

\bibitem[{Colleoni {et~al.}(2025{\natexlab{a}})Colleoni, Ramis~Vidal,
  Johnson-McDaniel, Dietrich, Haney, \& Pratten}]{Colleoni:2023ple}
Colleoni, M., Ramis~Vidal, F.~A., Johnson-McDaniel, N.~K., {et~al.}
  2025{\natexlab{a}}, Phys. Rev. D, 111, 064025,
  \dodoi{10.1103/PhysRevD.111.064025}

\bibitem[{Colleoni {et~al.}(2025{\natexlab{b}})Colleoni, Vidal,
  Garc\'\i{}a-Quir\'os, Ak\c{c}ay, \& Bera}]{Colleoni:2024knd}
Colleoni, M., Vidal, F. A.~R., Garc\'\i{}a-Quir\'os, C., Ak\c{c}ay, S., \&
  Bera, S. 2025{\natexlab{b}}, Phys. Rev. D, 111, 104019,
  \dodoi{10.1103/PhysRevD.111.104019}

\bibitem[{Cornish \& Littenberg(2015)}]{Cornish:2014kda}
Cornish, N.~J., \& Littenberg, T.~B. 2015, Class. Quantum Grav., 32, 135012,
  \dodoi{10.1088/0264-9381/32/13/135012}

\bibitem[{Cornish {et~al.}(2021)Cornish, Littenberg, B\'ecsy, Chatziioannou,
  Clark, Ghonge, \& Millhouse}]{Cornish:2020dwh}
Cornish, N.~J., Littenberg, T.~B., B\'ecsy, B., {et~al.} 2021, Phys. Rev. D,
  103, 044006, \dodoi{10.1103/PhysRevD.103.044006}

\bibitem[{Cornish {et~al.}(2024)}]{bayeswave}
Cornish, N.~J., {et~al.} 2024, {BayesWave software}.
\newblock \url{https://git.ligo.org/lscsoft/bayeswave}

\bibitem[{Creighton \& Anderson(2011)}]{Creighton:2011zz}
Creighton, J. D.~E., \& Anderson, W.~G. 2011, {Gravitational-wave physics and
  astronomy: An introduction to theory, experiment and data analysis}
  (Weinheim: John Wiley \& Sons, Ltd), \dodoi{10.1002/9783527636037}

\bibitem[{Davies {et~al.}(2020)Davies, Dent, T\'apai, Harry, McIsaac, \&
  Nitz}]{Davies:2020tsx}
Davies, G.~S., Dent, T., T\'apai, M., {et~al.} 2020, Phys. Rev. D, 102, 022004,
  \dodoi{10.1103/PhysRevD.102.022004}

\bibitem[{Davis {et~al.}(2022)Davis, Littenberg, Romero-Shaw, Millhouse,
  McIver, Di~Renzo, \& Ashton}]{Davis:2022ird}
Davis, D., Littenberg, T.~B., Romero-Shaw, I.~M., {et~al.} 2022, Class. Quantum
  Grav., 39, 245013, \dodoi{10.1088/1361-6382/aca238}

\bibitem[{Dhani {et~al.}(2025)Dhani, V{\"o}lkel, Buonanno, Estelles, Gair,
  Pfeiffer, Pompili, \& Toubiana}]{Dhani:2024jja}
Dhani, A., V{\"o}lkel, S.~H., Buonanno, A., {et~al.} 2025, Phys. Rev. X, 15,
  031036, \dodoi{10.1103/5pks-qz6b}

\bibitem[{Divyajyoti {et~al.}(2024)Divyajyoti, Krishnendu, Saleem, Colleoni,
  Vijaykumar, Arun, \& Mishra}]{Divyajyoti:2023izl}
Divyajyoti, Krishnendu, N.~V., Saleem, M., {et~al.} 2024, Phys. Rev. D, 109,
  023016, \dodoi{10.1103/PhysRevD.109.023016}

\bibitem[{Do {et~al.}(2019)}]{Do:2019txf}
Do, T., {et~al.} 2019, Science, 365, 664, \dodoi{10.1126/science.aav8137}

\bibitem[{Essick {et~al.}(2020)Essick, Godwin, Hanna, Blackburn, \&
  Katsavounidis}]{Essick:2020qpo}
Essick, R., Godwin, P., Hanna, C., Blackburn, L., \& Katsavounidis, E. 2020,
  Mach. Learn. Sci. Technol., 2, 015004, \dodoi{10.1088/2632-2153/abab5f}

\bibitem[{Fairhurst {et~al.}(2020)Fairhurst, Green, Hoy, Hannam, \&
  Muir}]{Fairhurst:2019vut}
Fairhurst, S., Green, R., Hoy, C., Hannam, M., \& Muir, A. 2020, Phys. Rev. D,
  102, 024055, \dodoi{10.1103/PhysRevD.102.024055}

\bibitem[{Ferreira(2019)}]{Ferreira:2019xrr}
Ferreira, P.~G. 2019, Ann. Rev. Astron. Astrophys., 57, 335,
  \dodoi{10.1146/annurev-astro-091918-104423}

\bibitem[{Fisher {et~al.}(2021)Fisher, Hemming, Bizouard, Brown, Couvares,
  Robinet, \& Verkindt}]{Fisher:2020pnr}
Fisher, R.~P., Hemming, G., Bizouard, M.-A., {et~al.} 2021, SoftwareX, 14,
  100677, \dodoi{10.1016/j.softx.2021.100677}

\bibitem[{Forward(1978)}]{Forward:1978zm}
Forward, R.~L. 1978, Phys. Rev. D, 17, 379, \dodoi{10.1103/PhysRevD.17.379}

\bibitem[{Gelfand {et~al.}(1958)Gelfand, Minlos, \& Shapiro}]{Gelfand1958}
Gelfand, I.~M., Minlos, R.~A., \& Shapiro, Z.~J. 1958, Predstavleniya gruppy
  vrashcheni i gruppy Lorentsa, ikh primeneniya [Representations of the
  rotation and Lorentz groups and their applications] (Gosudarstv. Izdat.
  Fiz.-Mat. Lit., Moscow, MR 0114876)

\bibitem[{Ghosh {et~al.}(2021)Ghosh, Brito, \& Buonanno}]{Ghosh:2021mrv}
Ghosh, A., Brito, R., \& Buonanno, A. 2021, Phys. Rev. D, 103, 124041,
  \dodoi{10.1103/PhysRevD.103.124041}

\bibitem[{Ghosh {et~al.}(2016)Ghosh, Ghosh, Johnson-McDaniel,
  {et~al.}}]{Ghosh:2016qgn}
Ghosh, A., Ghosh, A., Johnson-McDaniel, N.~K., {et~al.} 2016, Phys. Rev. D, 94,
  021101(R), \dodoi{10.1103/PhysRevD.94.021101}

\bibitem[{Ghosh {et~al.}(2018)Ghosh, Johnson-McDaniel, Ghosh, Mishra, Ajith,
  Del~Pozzo, Berry, Nielsen, \& London}]{Ghosh:2017gfp}
Ghosh, A., Johnson-McDaniel, N.~K., Ghosh, A., {et~al.} 2018, Class. Quantum
  Grav., 35, 014002, \dodoi{10.1088/1361-6382/aa972e}

\bibitem[{Gupta(2026)}]{Gupta:2025utd}
Gupta, A. 2026, Class. Quantum Grav., 43, 053001,
  \dodoi{10.1088/1361-6382/ae470b}

\bibitem[{Gupta {et~al.}(2024)}]{Gupta:2024gun}
Gupta, A., {et~al.} 2024, SciPost Phys. Comm. Rep., 5,
  \dodoi{10.21468/SciPostPhysCommRep.5}

\bibitem[{Gupta {et~al.}(2025)Gupta, Narayan, London, Tiwari, \&
  Sathyaprakash}]{Gupta:2025paz}
Gupta, I., Narayan, P., London, L., Tiwari, S., \& Sathyaprakash, B. 2025.
\newblock \doarXiv{2511.11886}

\bibitem[{Gupta \& Cornish(2024)}]{Gupta:2023jrn}
Gupta, T., \& Cornish, N.~J. 2024, Phys. Rev. D, 109, 064040,
  \dodoi{10.1103/PhysRevD.109.064040}

\bibitem[{G{\"u}rsel \& Tinto(1989)}]{GuerselTinto1989}
G{\"u}rsel, Y., \& Tinto, M. 1989, Phys. Rev. D, 40, 3884,
  \dodoi{10.1103/PhysRevD.40.3884}

\bibitem[{Haegel {et~al.}(2023)Haegel, O'Neal-Ault, Bailey, Tasson, Bloom, \&
  Shao}]{Haegel:2022ymk}
Haegel, L., O'Neal-Ault, K., Bailey, Q.~G., {et~al.} 2023, Phys. Rev. D, 107,
  064031, \dodoi{10.1103/PhysRevD.107.064031}

\bibitem[{Hanna {et~al.}(2020)}]{Hanna:2019ezx}
Hanna, C., {et~al.} 2020, Phys. Rev. D, 101, 022003,
  \dodoi{10.1103/PhysRevD.101.022003}

\bibitem[{Harris {et~al.}(2020)}]{Harris:2020xlr}
Harris, C.~R., {et~al.} 2020, Nature, 585, 357,
  \dodoi{10.1038/s41586-020-2649-2}

\bibitem[{Healy \& Lousto(2017)}]{Healy:2016lce}
Healy, J., \& Lousto, C.~O. 2017, Phys. Rev. D, 95, 024037,
  \dodoi{10.1103/PhysRevD.95.024037}

\bibitem[{Hofmann {et~al.}(2016)Hofmann, Barausse, \&
  Rezzolla}]{Hofmann:2016yih}
Hofmann, F., Barausse, E., \& Rezzolla, L. 2016, Astrophys. J. Lett., 825, L19,
  \dodoi{10.3847/2041-8205/825/2/L19}

\bibitem[{Hoy \& Raymond(2021)}]{Hoy:2020vys}
Hoy, C., \& Raymond, V. 2021, SoftwareX, 15, 100765,
  \dodoi{10.1016/j.softx.2021.100765}

\bibitem[{Hughes \& Menou(2005)}]{Hughes:2004vw}
Hughes, S.~A., \& Menou, K. 2005, Astrophys. J., 623, 689,
  \dodoi{10.1086/428826}

\bibitem[{Hunter(2007)}]{Hunter:2007ouj}
Hunter, J.~D. 2007, Comput. Sci. Eng., 9, 90, \dodoi{10.1109/MCSE.2007.55}

\bibitem[{Ishak(2019)}]{Ishak:2018his}
Ishak, M. 2019, Living Rev. Relativity, 22, 1,
  \dodoi{10.1007/s41114-018-0017-4}

\bibitem[{Isi {et~al.}(2019)Isi, Chatziioannou, \& Farr}]{Isi:2019asy}
Isi, M., Chatziioannou, K., \& Farr, W.~M. 2019, Phys. Rev. Lett., 123, 121101,
  \dodoi{10.1103/PhysRevLett.123.121101}

\bibitem[{Isi \& Weinstein(2017)}]{Isi:2017fbj}
Isi, M., \& Weinstein, A.~J. 2017, {Tech. Note, LIGO-P1700276}.
\newblock \doarXiv{1710.03794}

\bibitem[{Isi {et~al.}(2015)Isi, Weinstein, Mead, \& Pitkin}]{Isi:2015cva}
Isi, M., Weinstein, A.~J., Mead, C., \& Pitkin, M. 2015, Phys. Rev. D, 91,
  082002, \dodoi{10.1103/PhysRevD.91.082002}

\bibitem[{Islam {et~al.}(2020)Islam, Mehta, Ghosh, Varma, Ajith, \&
  Sathyaprakash}]{Islam:2019dmk}
Islam, T., Mehta, A.~K., Ghosh, A., {et~al.} 2020, Phys. Rev. D, 101, 024032,
  \dodoi{10.1103/PhysRevD.101.024032}

\bibitem[{Jim\'enez-Forteza {et~al.}(2017)Jim\'enez-Forteza, Keitel, Husa,
  Hannam, Khan, \& P\"urrer}]{PhysRevD.95.064024}
Jim\'enez-Forteza, X., Keitel, D., Husa, S., {et~al.} 2017, Phys. Rev. D, 95,
  064024, \dodoi{10.1103/PhysRevD.95.064024}

\bibitem[{Johnson-McDaniel {et~al.}(2022)Johnson-McDaniel, Ghosh, Ghonge,
  Saleem, Krishnendu, \& Clark}]{Johnson-McDaniel:2021yge}
Johnson-McDaniel, N.~K., Ghosh, A., Ghonge, S., {et~al.} 2022, Phys. Rev. D,
  105, 044020, \dodoi{10.1103/PhysRevD.105.044020}

\bibitem[{Klimenko {et~al.}(2004)Klimenko, Yakushin, Rakhmanov, \&
  Mitselmakher}]{Klimenko:2004qh}
Klimenko, S., Yakushin, I., Rakhmanov, M., \& Mitselmakher, G. 2004, Class.
  Quantum Grav., 21, S1685, \dodoi{10.1088/0264-9381/21/20/011}

\bibitem[{Klimenko {et~al.}(2011)Klimenko, Vedovato, Drago, Mazzolo,
  Mitselmakher, Pankow, Prodi, Re, Salemi, \& Yakushin}]{Klimenko:2011hz}
Klimenko, S., Vedovato, G., Drago, M., {et~al.} 2011, Phys. Rev. D, 83, 102001,
  \dodoi{10.1103/PhysRevD.83.102001}

\bibitem[{Klimenko {et~al.}(2016)}]{Klimenko:2015ypf}
Klimenko, S., {et~al.} 2016, Phys. Rev. D, 93, 042004,
  \dodoi{10.1103/PhysRevD.93.042004}

\bibitem[{Kramer {et~al.}(2021)}]{Kramer:2021jcw}
Kramer, M., {et~al.} 2021, Phys. Rev. X, 11, 041050,
  \dodoi{10.1103/PhysRevX.11.041050}

\bibitem[{Kwok {et~al.}(2022)Kwok, Lo, Weinstein, \& Li}]{Kwok:2021zny}
Kwok, J. Y.~L., Lo, R. K.~L., Weinstein, A.~J., \& Li, T. G.~F. 2022, Phys.
  Rev. D, 105, 024066, \dodoi{10.1103/PhysRevD.105.024066}

\bibitem[{Lee {et~al.}(2025)Lee, Doshi, Millhouse, \& Melatos}]{Lee:2025qpu}
Lee, Y. S.~C., Doshi, S., Millhouse, M., \& Melatos, A. 2025, Phys. Rev. D,
  111, 082002, \dodoi{10.1103/PhysRevD.111.082002}

\bibitem[{Li {et~al.}(2012)Li, Del~Pozzo, Vitale, Van Den~Broeck, Agathos,
  {et~al.}}]{Li:2011cg}
Li, T. G.~F., Del~Pozzo, W., Vitale, S., {et~al.} 2012, Phys. Rev. D, 85,
  082003, \dodoi{10.1103/PhysRevD.85.082003}

\bibitem[{{LIGO Scientific Collaboration and Virgo
  Collaboration}(2018)}]{DQRdocumentation}
{LIGO Scientific Collaboration and Virgo Collaboration}. 2018, {Data quality
  report user documentation},
  \href{https://docs.ligo.org/detchar/data-quality-report/}{docs.ligo.org/detchar/data-quality-report/}

\bibitem[{{LIGO Scientific, Virgo, and KAGRA
  Collaboration}(2025)}]{lalsuite-software}
{LIGO Scientific, Virgo, and KAGRA Collaboration}. 2025, {LVK} {A}lgorithm
  {L}ibrary - {LALS}uite, \dodoi{10.7935/GT1W-FZ16}

\bibitem[{{LIGO Scientific, Virgo, and KAGRA
  Collaboration}(2026)}]{PaperI_DCC_release}
---. 2026, Data release for GWTC-4.0: Tests of General Relativity. I. Overview
  and General Tests.
\newblock \url{https://dcc.ligo.org/P2600128/public}

\bibitem[{Littenberg {et~al.}(2016)Littenberg, Kanner, Cornish, \&
  Millhouse}]{Littenberg:2015kpb}
Littenberg, T.~B., Kanner, J.~B., Cornish, N.~J., \& Millhouse, M. 2016, Phys.
  Rev. D, 94, 044050, \dodoi{10.1103/PhysRevD.94.044050}

\bibitem[{Lo {et~al.}(2019)Lo, Li, \& Weinstein}]{Lo:2018sep}
Lo, R. K.~L., Li, T. G.~F., \& Weinstein, A.~J. 2019, Phys. Rev. D, 99, 084052,
  \dodoi{10.1103/PhysRevD.99.084052}

\bibitem[{Lo {et~al.}(2025)Lo, Uchikata, Narikawa, Wang, \&
  Birnholtz}]{echowfm}
Lo, R. K.~L., Uchikata, N., Narikawa, T., Wang, Y., \& Birnholtz, O. 2025,
  \soft{echoes\_waveform\_models}, v0.0.5.
\newblock
  \url{https://git.ligo.org/echoes_template_search/echoes_waveform_models}

\bibitem[{Lu {et~al.}(2025)Lu, Ma, Piccinni, Sun, \& Finch}]{Lu:2025mwp}
Lu, N., Ma, S., Piccinni, O.~J., Sun, L., \& Finch, E. 2025, Phys. Rev. D, 112,
  064047, \dodoi{10.1103/h6jy-sd94}

\bibitem[{Ma {et~al.}(2023{\natexlab{a}})Ma, Sun, \& Chen}]{Ma:2023vvr}
Ma, S., Sun, L., \& Chen, Y. 2023{\natexlab{a}}, Phys. Rev. D, 107, 084010,
  \dodoi{10.1103/PhysRevD.107.084010}

\bibitem[{Ma {et~al.}(2023{\natexlab{b}})Ma, Sun, \& Chen}]{Ma:2023cwe}
---. 2023{\natexlab{b}}, Phys. Rev. Lett., 130, 141401,
  \dodoi{10.1103/PhysRevLett.130.141401}

\bibitem[{Ma {et~al.}(2022)Ma, Mitman, Sun, Deppe, H\'ebert, Kidder, Moxon,
  Throwe, Vu, \& Chen}]{Ma:2022wpv}
Ma, S., Mitman, K., Sun, L., {et~al.} 2022, Phys. Rev. D, 106, 084036,
  \dodoi{10.1103/PhysRevD.106.084036}

\bibitem[{Ma {et~al.}(2025)}]{qnmrf-soft}
Ma, S., {et~al.} 2025, {QNMRF software},
  \url{https://github.com/Sizheng-Ma/qnm_filter},  GitHub

\bibitem[{Macleod {et~al.}(2021)}]{gwpy-software}
Macleod, D., {et~al.} 2021, {gwpy/gwpy},  Zenodo, \dodoi{10.5281/zenodo.597016}

\bibitem[{Maggiore(2007)}]{Maggiore:2007ulw}
Maggiore, M. 2007, {Gravitational Waves. Vol. 1: Theory and Experiments}
  (Oxford University Press), \dodoi{10.1093/acprof:oso/9780198570745.001.0001}

\bibitem[{Mahapatra {et~al.}(2025)}]{Mahapatra:2025cwk}
Mahapatra, P., {et~al.} 2025, Phys. Rev. D, 112, 104007,
  \dodoi{10.1103/c1sj-jc4v}

\bibitem[{Mandel {et~al.}(2019)Mandel, Farr, \& Gair}]{Mandel:2018mve}
Mandel, I., Farr, W.~M., \& Gair, J.~R. 2019, Mon. Not. R. Astron. Soc., 486,
  1086, \dodoi{10.1093/mnras/stz896}

\bibitem[{Matas {et~al.}(2020)}]{Matas:2020wab}
Matas, A., {et~al.} 2020, Phys. Rev. D, 102, 043023,
  \dodoi{10.1103/PhysRevD.102.043023}

\bibitem[{Mehta {et~al.}(2023)Mehta, Buonanno, Cotesta, Ghosh, Sennett, \&
  Steinhoff}]{Mehta:2022pcn}
Mehta, A.~K., Buonanno, A., Cotesta, R., {et~al.} 2023, Phys. Rev. D, 107,
  044020, \dodoi{10.1103/PhysRevD.107.044020}

\bibitem[{Meidam {et~al.}(2018)}]{Meidam:2017dgf}
Meidam, J., {et~al.} 2018, Phys. Rev. D, 97, 044033,
  \dodoi{10.1103/PhysRevD.97.044033}

\bibitem[{Messick {et~al.}(2017)}]{Messick:2016aqy}
Messick, C., {et~al.} 2017, Phys. Rev. D, 95, 042001,
  \dodoi{10.1103/PhysRevD.95.042001}

\bibitem[{Miani {et~al.}(2023)Miani, Lazzaro, Prodi, Tiwari, Drago, Milotti, \&
  Vedovato}]{Miani:2023}
Miani, A., Lazzaro, C., Prodi, G.~A., {et~al.} 2023, Phys. Rev. D, 108, 064018,
  \dodoi{10.1103/PhysRevD.108.064018}

\bibitem[{Mihaylov {et~al.}(2025)Mihaylov, Ossokine, Buonanno, Estelles,
  Pompili, P\"urrer, \& Ramos-Buades}]{Mihaylov:2023bkc}
Mihaylov, D.~P., Ossokine, S., Buonanno, A., {et~al.} 2025, SoftwareX, 30,
  102080, \dodoi{10.1016/j.softx.2025.102080}

\bibitem[{Miller {et~al.}(2019)}]{Miller:2019cac}
Miller, M.~C., {et~al.} 2019, Astrophys. J. Lett., 887, L24,
  \dodoi{10.3847/2041-8213/ab50c5}

\bibitem[{Mills \& Fairhurst(2021)}]{Mills:2020thr}
Mills, C., \& Fairhurst, S. 2021, Phys. Rev. D, 103, 024042,
  \dodoi{10.1103/PhysRevD.103.024042}

\bibitem[{Mishra {et~al.}(2022)}]{Mishra:2022ott}
Mishra, T., {et~al.} 2022, Phys. Rev. D, 105, 083018,
  \dodoi{10.1103/PhysRevD.105.083018}

\bibitem[{Nakano {et~al.}(2017)Nakano, Sago, Tagoshi, \&
  Tanaka}]{Nakano:2017fvh}
Nakano, H., Sago, N., Tagoshi, H., \& Tanaka, T. 2017, PTEP, 2017, 071E01,
  \dodoi{10.1093/ptep/ptx093}

\bibitem[{Narayan {et~al.}(2023)Narayan, Johnson-McDaniel, \&
  Gupta}]{Narayan:2023vhm}
Narayan, P., Johnson-McDaniel, N.~K., \& Gupta, A. 2023, Phys. Rev. D, 108,
  064003, \dodoi{10.1103/PhysRevD.108.064003}

\bibitem[{Narayan {et~al.}(2026)Narayan, Johnson-McDaniel, \&
  Gupta}]{Narayan:2024rat}
---. 2026, Phys. Rev. D, 113, 044007, \dodoi{10.1103/5hbl-hr2x}

\bibitem[{Necula {et~al.}(2012)Necula, Klimenko, \&
  Mitselmakher}]{Necula:2012zz}
Necula, V., Klimenko, S., \& Mitselmakher, G. 2012, J. Phys. Conf. Ser., 363,
  012032, \dodoi{10.1088/1742-6596/363/1/012032}

\bibitem[{Newman \& Penrose(1966)}]{Newman:1966ub}
Newman, E.~T., \& Penrose, R. 1966, J. Math. Phys., 7, 863,
  \dodoi{10.1063/1.1931221}

\bibitem[{Nishizawa {et~al.}(2009)Nishizawa, Taruya, Hayama, Kawamura, \&
  Sakagami}]{Nishizawa:2009bf}
Nishizawa, A., Taruya, A., Hayama, K., Kawamura, S., \& Sakagami, M.-a. 2009,
  Phys. Rev. D, 79, 082002, \dodoi{10.1103/PhysRevD.79.082002}

\bibitem[{Nitz {et~al.}(2017)Nitz, Dent, Dal~Canton, Fairhurst, \&
  Brown}]{Nitz:2017svb}
Nitz, A.~H., Dent, T., Dal~Canton, T., Fairhurst, S., \& Brown, D.~A. 2017,
  Astrophys. J., 849, 118, \dodoi{10.3847/1538-4357/aa8f50}

\bibitem[{Pacilio {et~al.}(2024)Pacilio, Gerosa, \& Bhagwat}]{Pacilio:2023uef}
Pacilio, C., Gerosa, D., \& Bhagwat, S. 2024, Phys. Rev. D, 109, L081302,
  \dodoi{10.1103/PhysRevD.109.L081302}

\bibitem[{Pedregosa {et~al.}(2011)}]{scikit-learn}
Pedregosa, F., {et~al.} 2011, J. Machine Learning Res., 12, 2825

\bibitem[{Pompili {et~al.}(2025)Pompili, Maggio, Silva, \&
  Buonanno}]{Pompili:2025cdc}
Pompili, L., Maggio, E., Silva, H.~O., \& Buonanno, A. 2025, Phys. Rev. D, 111,
  124040, \dodoi{10.1103/ng8w-98sz}

\bibitem[{Pompili {et~al.}(2023)}]{Pompili:2023tna}
Pompili, L., {et~al.} 2023, Phys. Rev. D, 108, 124035,
  \dodoi{10.1103/PhysRevD.108.124035}

\bibitem[{Pratten {et~al.}(2021)}]{Pratten:2020ceb}
Pratten, G., {et~al.} 2021, Phys. Rev. D, 103, 104056,
  \dodoi{10.1103/PhysRevD.103.104056}

\bibitem[{Puecher {et~al.}(2022)Puecher, Kalaghatgi, Roy, Setyawati, Gupta,
  Sathyaprakash, \& Van Den~Broeck}]{Puecher:2022sfm}
Puecher, A., Kalaghatgi, C., Roy, S., {et~al.} 2022, Phys. Rev. D, 106, 082003,
  \dodoi{10.1103/PhysRevD.106.082003}

\bibitem[{Ramos-Buades {et~al.}(2023)Ramos-Buades, Buonanno, Estell\'es,
  Khalil, Mihaylov, Ossokine, Pompili, \& Shiferaw}]{Ramos-Buades:2023ehm}
Ramos-Buades, A., Buonanno, A., Estell\'es, H., {et~al.} 2023, Phys. Rev. D,
  108, 124037, \dodoi{10.1103/PhysRevD.108.124037}

\bibitem[{Riley {et~al.}(2019)}]{Riley:2019yda}
Riley, T.~E., {et~al.} 2019, Astrophys. J. Lett., 887, L21,
  \dodoi{10.3847/2041-8213/ab481c}

\bibitem[{Robinet {et~al.}(2020)Robinet, Arnaud, Leroy, Lundgren, Macleod, \&
  McIver}]{Robinet:2020lbf}
Robinet, F., Arnaud, N., Leroy, N., {et~al.} 2020, SoftwareX, 12, 100620,
  \dodoi{10.1016/j.softx.2020.100620}

\bibitem[{Romero-Shaw {et~al.}(2020)}]{Romero-Shaw:2020owr}
Romero-Shaw, I.~M., {et~al.} 2020, Mon. Not. R. Astron. Soc., 499, 3295,
  \dodoi{10.1093/mnras/staa2850}

\bibitem[{Roy {et~al.}(2026)Roy, Haney, Pratten, Pang, \& Van
  Den~Broeck}]{Roy:2025gzv}
Roy, S., Haney, M., Pratten, G., Pang, P. T.~H., \& Van Den~Broeck, C. 2026,
  Phys. Rev. D, 113, 024016, \dodoi{10.1103/855k-sys5}

\bibitem[{Sachdev {et~al.}(2019)}]{Sachdev:2019vvd}
Sachdev, S., {et~al.} 2019.
\newblock \doarXiv{1901.08580}

\bibitem[{S\"anger {et~al.}(2024)}]{Sanger:2024axs}
S\"anger, E.~M., {et~al.} 2024.
\newblock \doarXiv{2406.03568}

\bibitem[{Silva {et~al.}(2021)Silva, Holgado, C\'ardenas-Avenda\~no, \&
  Yunes}]{Silva:2020acr}
Silva, H.~O., Holgado, A.~M., C\'ardenas-Avenda\~no, A., \& Yunes, N. 2021,
  Phys. Rev. Lett., 126, 181101, \dodoi{10.1103/PhysRevLett.126.181101}

\bibitem[{Skilling(2006)}]{Skilling:2006}
Skilling, J. 2006, Bayesian Analysis, 1, 833, \dodoi{10.1214/06-BA127}

\bibitem[{Smith {et~al.}(2011)Smith, Abbott, Hirose, Leroy, Macleod, McIver,
  Saulson, \& Shawhan}]{Smith:2011an}
Smith, J.~R., Abbott, T., Hirose, E., {et~al.} 2011, Class. Quantum Grav., 28,
  235005, \dodoi{10.1088/0264-9381/28/23/235005}

\bibitem[{Smith {et~al.}(2020)Smith, Ashton, Vajpeyi, \&
  Talbot}]{Smith:2019ucc}
Smith, R., Ashton, G., Vajpeyi, A., \& Talbot, C. 2020, Mon. Not. R. Astron.
  Soc., 498, 4492, \dodoi{10.1093/mnras/staa2483}

\bibitem[{Speagle(2020)}]{Speagle:2020spe}
Speagle, J.~S. 2020, Mon. Not. R. Astron. Soc., 493, 3132,
  \dodoi{10.1093/mnras/staa278}

\bibitem[{Stein(2019)}]{Stein:2019mop}
Stein, L.~C. 2019, J. Open Source Softw., 4, 1683, \dodoi{10.21105/joss.01683}

\bibitem[{Sullivan {et~al.}(2025)Sullivan, Ranjan, \& Millhouse}]{iwc}
Sullivan, J., Ranjan, S., \& Millhouse, M. 2025, IGWN Wave Compare.
\newblock \url{https://git.ligo.org/bayeswave/igwn-wave-compare}

\bibitem[{Talbot {et~al.}(2025)}]{Talbot:2025vth}
Talbot, C., {et~al.} 2025, Class. Quantum Grav., 42, 235023,
  \dodoi{10.1088/1361-6382/ae1ac7}

\bibitem[{Thompson {et~al.}(2020)Thompson, Fauchon-Jones, Khan, Nitoglia,
  Pannarale, Dietrich, \& Hannam}]{Thompson:2020nei}
Thompson, J.~E., Fauchon-Jones, E., Khan, S., {et~al.} 2020, Phys. Rev. D, 101,
  124059, \dodoi{10.1103/PhysRevD.101.124059}

\bibitem[{Tiwari {et~al.}(2025)Tiwari, Vijaykumar, Kapadia, Ghosh, \&
  Nielsen}]{Tiwari:2025aec}
Tiwari, A., Vijaykumar, A., Kapadia, S.~J., Ghosh, S., \& Nielsen, A.~B. 2025.
\newblock \doarXiv{2506.22272}

\bibitem[{Tsang {et~al.}(2018)Tsang, Rollier, Ghosh, Samajdar, Agathos,
  Chatziioannou, Cardoso, Khanna, \& Van Den~Broeck}]{Tsang:2018uie}
Tsang, K.~W., Rollier, M., Ghosh, A., {et~al.} 2018, Phys. Rev. D, 98, 024023,
  \dodoi{10.1103/PhysRevD.98.024023}

\bibitem[{Uchikata {et~al.}(2019)Uchikata, Nakano, Narikawa, Sago, Tagoshi, \&
  Tanaka}]{Uchikata:2019frs}
Uchikata, N., Nakano, H., Narikawa, T., {et~al.} 2019, Phys. Rev. D, 100,
  062006, \dodoi{10.1103/PhysRevD.100.062006}

\bibitem[{Uchikata {et~al.}(2023)Uchikata, Narikawa, Nakano, Sago, Tagoshi, \&
  Tanaka}]{Uchikata:2023zcu}
Uchikata, N., Narikawa, T., Nakano, H., {et~al.} 2023, Phys. Rev. D, 108,
  104040, \dodoi{10.1103/PhysRevD.108.104040}

\bibitem[{Urban {et~al.}(2021)}]{gwdetchar-software}
Urban, A.~L., {et~al.} 2021, {gwdetchar/gwdetchar},  Zenodo,
  \dodoi{10.5281/zenodo.597016}

\bibitem[{Usman {et~al.}(2016)}]{Usman:2015kfa}
Usman, S.~A., {et~al.} 2016, Class. Quantum Grav., 33, 215004,
  \dodoi{10.1088/0264-9381/33/21/215004}

\bibitem[{Varma {et~al.}(2019)Varma, Field, Scheel, Blackman, Gerosa, Stein,
  Kidder, \& Pfeiffer}]{Varma:2019csw}
Varma, V., Field, S.~E., Scheel, M.~A., {et~al.} 2019, Phys. Rev. Research., 1,
  033015, \dodoi{10.1103/PhysRevResearch.1.033015}

\bibitem[{Veitch {et~al.}(2015)}]{Veitch:2014wba}
Veitch, J., {et~al.} 2015, Phys. Rev. D, 91, 042003,
  \dodoi{10.1103/PhysRevD.91.042003}

\bibitem[{Veitch {et~al.}(2025)Veitch, Pozzo, Lyttle, Williams, Talbot, Pitkin,
  Ashton, Cody, Hübner, Macleod, Nitz, Mihaylov, Carullo, Davies, Maenaut, \&
  Wang}]{cpnest}
Veitch, J., Pozzo, W.~D., Lyttle, A., {et~al.} 2025, johnveitch/cpnest:
  v0.11.8, v0.11.8,  Zenodo, \dodoi{10.5281/zenodo.592884}

\bibitem[{Viets {et~al.}(2018)}]{Viets:2017yvy}
Viets, A., {et~al.} 2018, Class. Quantum Grav., 35, 095015,
  \dodoi{10.1088/1361-6382/aab658}

\bibitem[{Vijaykumar {et~al.}(2023)Vijaykumar, Tiwari, Kapadia, Arun, \&
  Ajith}]{Vijaykumar:2023tjg}
Vijaykumar, A., Tiwari, A., Kapadia, S.~J., Arun, K.~G., \& Ajith, P. 2023,
  Astrophys. J., 954, 105, \dodoi{10.3847/1538-4357/acd77d}

\bibitem[{{Virgo Collaboration}(2021)}]{pythonvirgotools}
{Virgo Collaboration}. 2021, {PythonVirgoTools}, v5.1.1,
  \href{https://git.ligo.org/virgo/virgoapp/PythonVirgoTools}{git.ligo.org/virgo/virgoapp/PythonVirgoTools}

\bibitem[{Virtanen {et~al.}(2020)}]{Virtanen:2019joe}
Virtanen, P., {et~al.} 2020, Nat. Meth., 17, 261,
  \dodoi{10.1038/s41592-019-0686-2}

\bibitem[{Waskom(2021)}]{Waskom:2021psk}
Waskom, M.~L. 2021, J. Open Source Softw., 6, 3021, \dodoi{10.21105/joss.03021}

\bibitem[{Wette(2020)}]{Wette:2020air}
Wette, K. 2020, SoftwareX, 12, 100634, \dodoi{10.1016/j.softx.2020.100634}

\bibitem[{Will(1994)}]{Will:1994fb}
Will, C.~M. 1994, Phys. Rev. D, 50, 6058, \dodoi{10.1103/PhysRevD.50.6058}

\bibitem[{Will(2014)}]{Will:2014kxa}
---. 2014, Living Rev. Relativity, 17, 4, \dodoi{10.12942/lrr-2014-4}

\bibitem[{Williams {et~al.}(2023)Williams, Veitch, Chiofalo, Schmidt, Udall,
  Vajpeji, \& Hoy}]{Williams:2022pgn}
Williams, D., Veitch, J., Chiofalo, M.~L., {et~al.} 2023, J. Open Source
  Softw., 8, 4170, \dodoi{10.21105/joss.04170}

\bibitem[{Wong {et~al.}(2025)Wong, Ng, \& Cirok}]{nullpol}
Wong, I. C.~F., Ng, T. C.~K., \& Cirok, B. 2025, nullpol software,  GitHub.
\newblock \url{https://github.com/isaac-cf-wong/nullpol}

\bibitem[{Wong {et~al.}(2021)Wong, Pang, Lo, Li, \& Van
  Den~Broeck}]{Wong:2021cmp}
Wong, I. C.~F., Pang, P. T.~H., Lo, R. K.~L., Li, T. G.~F., \& Van Den~Broeck,
  C. 2021.
\newblock \doarXiv{2105.09485}

\bibitem[{Yunes {et~al.}(2025)Yunes, Siemens, \& Yagi}]{Yunes:2024lzm}
Yunes, N., Siemens, X., \& Yagi, K. 2025, Living Rev. Relativity, 28, 3,
  \dodoi{10.1007/s41114-024-00054-9}

\bibitem[{Zhong {et~al.}(2024)Zhong, Isi, Chatziioannou, \&
  Farr}]{Zhong:2024pwb}
Zhong, H., Isi, M., Chatziioannou, K., \& Farr, W.~M. 2024, Phys. Rev. D, 110,
  044053, \dodoi{10.1103/PhysRevD.110.044053}

\bibitem[{Zhong {et~al.}(2026)Zhong, Isi, Farr, \& Chatziioannou}]{hierfit}
Zhong, H., Isi, M., Farr, W.~M., \& Chatziioannou, K. 2026, \soft{hierfit}, ND
  Hierarchical Analysis.
\newblock \url{https://git.ligo.org/haowen.zhong/2d-nd-hierarchical-analysis}

\bibitem[{Zimmerman {et~al.}(2019)Zimmerman, Haster, \&
  Chatziioannou}]{Zimmerman:2019wzo}
Zimmerman, A., Haster, C.-J., \& Chatziioannou, K. 2019, Phys. Rev. D, 99,
  124044, \dodoi{10.1103/PhysRevD.99.124044}

\bibitem[{{Zweizig, J.}(2006)}]{DMTdocumentation}
{Zweizig, J.} 2006, {The Data Monitor Tool Project},
  \href{https://labcit.ligo.caltech.edu/~jzweizig/DMT-Project.html}{labcit.ligo.caltech.edu/\~{}jzweizig/DMT-Project.html}

\end{thebibliography}

\clearpage

\iftoggle{endauthorlist}{
 \let\author\myauthor
 \let\affiliation\myaffiliation
 \let\maketitle\mymaketitle
 \title{Authors}
 \pacs{}
 
 \newpage
 \maketitle
}

\end{document}